\documentclass[letterpaper]{nature}

\usepackage{booktabs}
\usepackage{lineno}
\usepackage{lscape}
\usepackage{amsfonts}
\usepackage{wrapfig}
\usepackage{afterpage}
\usepackage{listings}
\usepackage{soul}
\usepackage[utf8]{inputenc}
\usepackage[left=1in,right=1in,bottom=1.1in,top=1in]{geometry}
\usepackage{algpseudocode}
\usepackage{amssymb}
\usepackage{amsmath} 
\usepackage{amsthm}
\usepackage{amsmath}
\usepackage{mathtools}
\usepackage{amssymb}
\usepackage{color}
\usepackage{longtable}
\usepackage{enumitem}
\usepackage{lscape}
\usepackage[table,xcdraw,dvipsnames]{xcolor}
\usepackage[normalem]{ulem}
\usepackage[colorlinks,citecolor=blue,urlcolor=magenta,linkcolor=blue]{hyperref}
\usepackage{multirow}
\usepackage{xspace}
\usepackage{graphicx}
\usepackage{pbox}
\usepackage{float}
\usepackage{chngpage}
\usepackage{url}
\usepackage{booktabs} 
\usepackage{times} 
\usepackage[innercaption]{sidecap}
\usepackage[font={stretch=1, footnotesize},labelfont=bf]{caption}
\usepackage[CaptionAfterwards]{fltpage}
\usepackage{lineno}
\usepackage{subcaption}
\usepackage{placeins}
\usepackage{multibib}

\newcites{Methods}{Methods References}

\newcommand{\metric}{exposure segregation\xspace}
\newcommand{\crosscount}{1.6 billion }
\newcommand{\usercount}{9.6 million }
\newcommand{\crosscountx}{1,570,782,460\xspace}
\newcommand{\usercountx}{9,567,559\xspace}
\newcommand{\countycount}{2829 }
\newcommand{\msacount}{382 }
\newcommand{\edgename}{exposure\xspace}
\newcommand{\edgenamecaps}{Exposure\xspace}

 \newcommand{\new}[1]{{\color{black}#1}} %

\makeatletter
\let\LN@align\align
\let\LN@endalign\endalign
\renewcommand{\align}{\linenomath\LN@align}
\renewcommand{\endalign}{\LN@endalign\endlinenomath}
\makeatother

\newcommand\refmsec[1]{Methods~\ref{sec:#1}}

\ifthenelse{\isundefined{\definition}}{\newtheorem{definition}{Definition}}{}
\ifthenelse{\isundefined{\assumption}}{\newtheorem{assumption}{Assumption}}{}
\ifthenelse{\isundefined{\hypothesis}}{\newtheorem{hypothesis}{Hypothesis}}{}
\ifthenelse{\isundefined{\proposition}}{\newtheorem{proposition}{Proposition}}{}
\ifthenelse{\isundefined{\theorem}}{\newtheorem{theorem}{Theorem}}{}
\ifthenelse{\isundefined{\lemma}}{\newtheorem{lemma}{Lemma}}{}
\ifthenelse{\isundefined{\corollary}}{\newtheorem{corollary}{Corollary}}{}
\ifthenelse{\isundefined{\alg}}{\newtheorem{alg}{Algorithm}}{}
\ifthenelse{\isundefined{\example}}{\newtheorem{example}{Example}}{}

\title{\Large 
    \begin{center}
    Human mobility networks reveal \\increased segregation in large cities 

    \end{center}}

\author
{\begin{center}
\vspace{-5mm}
Hamed Nilforoshan$^{*,1}$, Wenli Looi$^{*,1}$, Emma Pierson$^{*, 2}$, Blanca Villanueva$^{3}$, Nic Fishman$^{1}$,\\[1mm] Yiling Chen$^{1}$, John Sholar$^{1}$, Beth Redbird$^{4,5}$, David Grusky$^{6}$, Jure Leskovec$^{\dagger,1}$\\[5mm]
\footnotesize{$^{1}\;$Department of Computer Science, Stanford University, Stanford, CA 94305, USA} \\[1mm]
\footnotesize{$^{2}\;$Department of Computer Science, Cornell Tech, New York, NY 10044, USA} \\[1mm]
\footnotesize{$^{3}\;$Department of Biomedical Data Science, Stanford University, Stanford, CA 94305, USA} \\[1mm]
\footnotesize{$^{4}\;$Institute for Policy Research, Northwestern University, Evanston, IL 60208, USA} \\[1mm]
\footnotesize{$^{5}\;$Department of Sociology, Northwestern University, Evanston, IL 60208, USA} \\[1mm]
\footnotesize{$^{6}\;$Department of Sociology, Stanford University, Stanford, CA 94305, USA} \\[1mm]
\footnotesize{$^{*}\;$These authors contributed equally to this work} \\[1mm]
\footnotesize{$^{\dagger}\;$Corresponding author. Email: jure@cs.stanford.edu} \\[1mm]
\end{center}
}

\begin{document}
\maketitle
\begin{abstract}
\linespread{1.0}\selectfont
\vspace{1em}

A long-standing expectation is that large, dense, and cosmopolitan areas support socioeconomic mixing \new{and exposure between diverse individuals}\cite{jacobs1961, wirth1938urbanism, milgram1970,derex2013experimental,gomez2016explaining, stier2021evidence}.
It has been difficult to assess this hypothesis because past approaches to measuring socioeconomic mixing have relied on static residential housing data rather than real-life \new{exposures between people} at work, in places of leisure, and in home neighborhoods\cite{jargowsky1996take,massey1993spatial}. Here we develop a new measure of \metric (ES) that captures the \new{socio}economic diversity of everyday \new{encounters}.
Leveraging cell phone mobility data to represent \crosscount \new{exposures} among \usercount people in the United States, we measure \new{exposure segregation} across \msacount Metropolitan Statistical Areas (MSAs) and \countycount counties. 
\new{We discover that exposure segregation} is 67\% higher in the 10 largest Metropolitan Statistical Areas (MSAs) than in small MSAs with fewer than 100,000 residents. 
\new{This means that, contrary to expectation, residents of large cosmopolitan areas have significantly less exposure to diverse individuals.}
\new{Second,} we find evidence that large cities offer a greater choice of differentiated spaces targeted to specific socioeconomic groups, \new{a dynamic that accounts for this increase} in everyday socioeconomic segregation. 
\new{Third,} we discover that this segregation-increasing effect is countered when a city's hubs (e.g. shopping malls) are positioned to bridge diverse neighborhoods and thus
attract people of all socioeconomic statuses.
Overall, our findings challenge a long-standing conjecture in human geography and urban design, and highlight how built environment can both prevent and facilitate  \new{exposure between diverse individuals}.

\end{abstract}

\clearpage

\pagenumbering{arabic}

{
\linespread{1.6}\selectfont
\section*{Introduction}\label{sec:introduction}

In the U.S., economic segregation is very high, with income affecting where one lives\cite{reardon2018has}, who one marries\cite{schwartz2010earnings}, and who one meets and befriends\cite{chetty2022soccap2}. %
This extreme segregation is costly: it reduces economic mobility\cite{massey1993american, chetty2016effects,chetty2018impacts, chetty2022soccap1}, fosters a wide range of health problems\cite{bor2017population, do2019triple}, and increases political polarization\cite{brown2021childhood}. Although there are all manner of reforms designed to reduce economic segregation (e.g., subsidized housing), it has long been argued that one of the most powerful segregation-reducing dynamics is rising urbanization\cite{united2018world} and the  happenstance mixing that it induces\cite{jacobs1961, milgram1970,derex2013experimental,gomez2016explaining, wirth1938urbanism, stier2021evidence}. \new{This ``cosmopolitan mixing hypothesis'' anticipates that in large cities the combination of increased population diversity, constrained space, and accessible public transportation will bring diverse individuals into close physical proximity with one another\cite{wirth1938urbanism}---reducing everyday socioeconomic segregation. The NYC subway has been lauded, for example, as a mixing bowl of diversity\cite{ocejo2014subway}.}

As plausible as the cosmopolitan mixing hypothesis %
might seem, big cities also provide new opportunities for self-segregation, given that they're large enough to allow people to seek out and find others like themselves\cite{fischer1982dwell}. %
These contrasting hypotheses about the \new{relationship between} urbanization \new{and socioeconomic mixing} remain untested because it has been difficult to measure real-world exposure among individuals\cite{jargowsky1996take, jargowsky2005measure, massey1988dimensions, brown2021measurement}. 
Cell phone geolocation data provide an exciting opportunity to do so, as they have been used for many research purposes\cite{matthews2013spatial, zenk2011activity,moro2021mobility,athey2020experienced,candipan2021residence,phillips2021social,wang2018urban,levy2020triple,levy2022neighborhood,xu2019quantifying, abbiasov2020urban}, but a nationwide study of socioeconomic mixing and urbanization has not been undertaken because of difficulties in  ascertaining individual-level socioeconomic status, determining when dyadic exposures occur, and amassing the data needed to compare across cities or counties\cite{matthews2013spatial, zenk2011activity,moro2021mobility,candipan2021residence,phillips2021social,wang2018urban,levy2020triple,levy2022neighborhood}.

\new{Here, we undertake the first credible test of the ``cosmopolitan mixing hypothesis’’ and the dynamics  underlying it.} To assess this hypothesis and understand the relationship between urbanization and segregation, we use cell phone data to \new{link geolocated individual-level exposures between people with individual-level socioeconomic status.} This allows us to develop a measure of segregation \new{for an area (e.g., MSA or county)} that captures where people go, when they go there, and whom they encounter on the way. We first construct a dynamic exposure network that captures each individual's exposures to other individuals in the real world. Our network contains \crosscount exposures among \usercount people across \msacount Metropolitan Statistical Areas (MSA) and \countycount counties in the United States. 
\new{Then, we determine the socioeconomic status of a person (i.e., the cell phone) by identifying their home location and its monthly rent value. We link these data} to define a measure of \emph{\metric} (ES) that extends a traditional static segregation measure by capturing the \new{diversity of person-to-person exposures localized in space and time.}

\section*{Results}\label{sec:results} 

\begin{figure}

\centering
\vspace{-2.5cm}
\hspace*{-0.75cm}  
\includegraphics[width=1.1\textwidth]{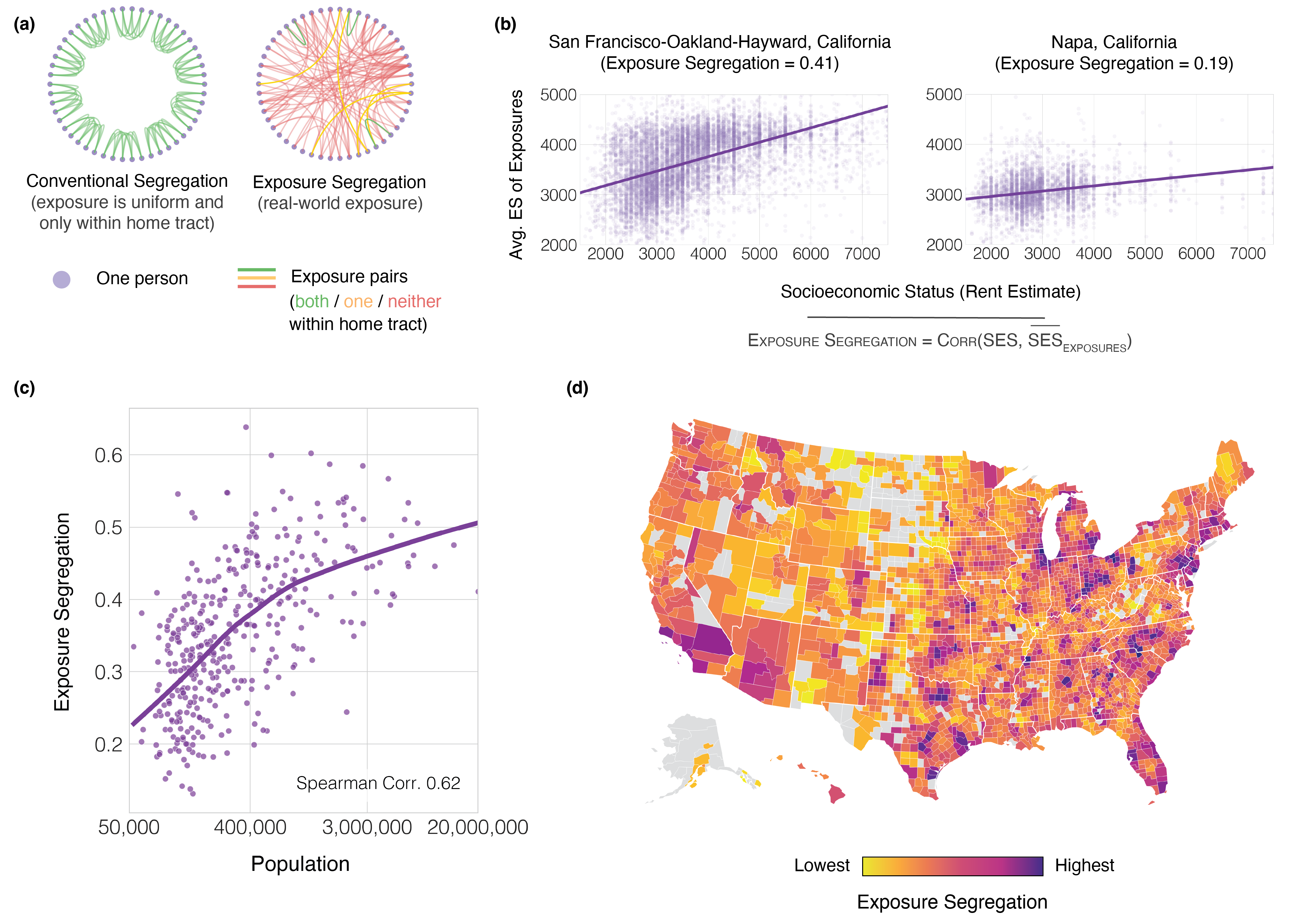}
\caption{\textbf{Exposure segregation (ES) captures the likelihood of exposure between people of different socioeconomic backgrounds---and reveals increased segregation in highly-populated metropolitan areas}
\textbf{(a)} The nationwide network of \crosscount exposures spans \countycount counties and \msacount Metropolitan Statistical Areas (MSAs). Our exposure network contrasts with a conventional measure of economic segregation, the Neighborhood Sorting Index, which assumes that people cross paths uniformly and only with other residents of their home Census tract. Graphs based on a sample community of 50 individuals from San Francisco, CA residing in 10 different census tracts. Nodes are individuals; edges are exposures. As this sample illustrates, most exposures happen when both persons are away from their home tract. These cross-tract exposures are undetected by conventional segregation measures.
\textbf{(b)} For each geographic region (e.g. MSA, county), we 
estimate \metric, defined as the correlation between an individual's socioeconomic status (SES) and the mean SES of those with whom they are exposed to. 1 signifies perfect segregation; 0 signifies no segregation. This definition is equivalent to the conventional Neighborhood Sorting Index (NSI), but with the key difference that it leverages real-life exposure from mobility data instead of synthetic exposures from individuals grouped by Census tracts. For two MSAs, we show the raw data; each point represents one individual. San Francisco-Oakland-Hayward, CA is $2.2\times$ (${p}{<}{10}^{-4}$, 95\% CI ${1.6}{-}{2.8}{\times}$) more segregated than Napa, CA.  
\textbf{(c-d)} Contrary to the hypothesis that highly-populated metropolitan areas support diverse exposures and socioeconomic mixing, we find that \textbf{(c)} Larger MSAs are more segregated. Exposure segregation presented as a function of population size; each dot represents one MSA; purple line indicates LOWESS fit. Upward trend reveals that urbanization is associated with higher exposure segregation (Spearman Correlation 0.62, N=382, p$<10^{-4}$). The top 10 largest MSAs, by population size, are 67\% more segregated than small MSAs with less than 100,000 residents. Associations are robust to controlling for potential confounders and are similar for population density and ES (Extended Data Table \ref{tab:segregation_population}, Supplementary Table \ref{tab:segregation_density}).
\textbf{(d)} Exposure segregation across the \countycount US counties. Analysis limited to counties with at least 50 individuals \new{present in the dataset}. Exposure segregation varies significantly across counties in the United States. Moreover, as with MSA-level segregation, county-level \metric is also positively associated with both population size and population density (Extended Data Figure \ref{fig:county_correlations}).}
\label{fig:main_1}
\end{figure}

We estimate \emph{\metric} (ES), defined as the extent to which individuals of different economic statuses are exposed to one another, for each geographic area in the U.S. (e.g.,  MSA, county).
This entails building a dynamic \emph{exposure network} with \usercountx nodes (representing individuals) and \crosscountx edges (representing exposures in physical space).

\paragraph{Developing a more realistic measure of socioeconomic segregation.} To estimate each person's socioeconomic status (SES), we first infer their night-time home location from cell phone mobility data, and we then recover the estimated monthly rent value of the home at this location (Methods \ref{sec:methods_dataprocessing}). This method is more accurate than the convention of proxying individual SES with neighborhood-level Census averages\cite{moro2021mobility,athey2020experienced}. The economic segregation of each geographic region is measured by the correlation between a person's SES and the mean SES of everyone they cross paths with \new{(unweighted by edge strength)}. This correlation is estimated by fitting a linear mixed effects model that eliminates attenuation bias \new{and secures unbiased estimates of exposure segregation even when observed exposures are sparse} (Methods \ref{sec:methods_analysis}). The resulting measure of \metric (Figure \ref{fig:main_1}a-b), which ranges from 0 (perfect integration) to 1 (complete segregation), is a generalization of a widely used measure of socioeconomic segregation, the Neighborhood Sorting Index (NSI)\cite{jargowsky1996take}. The NSI is equivalent to the correlation between each person's SES and the mean SES of all people in their Census tract, whereas the ES is equivalent to the correlation between each person's SES and the mean SES of \textit{all} people they have crossed paths with, either inside or outside their census tract. Like the NSI, the ES measures cross-class exposure of any type, rather than \new{focusing on} ties that are persistent \new{or} strong (such as friendship ties)\cite{chetty2022soccap2}. \new{We opt for this approach because the cosmopolitan mixing hypothesis dictates this expansive test, and we find that our results are robust to tie strength (Supplementary Figure \ref{fig:robust_6})}. %

\paragraph{\new{Exposure segregation is extremely high in large cities}.} 

We discover that, contrary to the ``cosmopolitan mixing hypothesis’’, \metric is higher in large MSAs (Figure \ref{fig:main_1}c). The Spearman correlation between MSA population and MSA segregation is 0.62 ($p$${<}$$10^{-4}$), and the 10 largest MSAs by population size are 67\% ($p$${<}$$10^{-4}$, 95\% CI 49-87\%) more segregated than small MSAs with less than 100,000 residents. This result is robust: we validate it by recalculating the correlation with a measure of density rather than population size (Spearman Correlation 0.45, $p$${<}$$10^{-4}$, Supplementary Table \ref{tab:segregation_density}), by controlling for relevant covariates (Extended Data Table \ref{tab:segregation_population} and Supplementary Table \ref{tab:segregation_density}), by varying the granularity of the analysis (Figure \ref{fig:main_1}d, Extended Data Figure \ref{fig:county_correlations}), and by testing a variety of specifications of \metric (Supplementary Table \ref{tab:correlations_between_segregation_measures}, Supplementary Figures \ref{fig:robust_1}-\ref{fig:robust_8}). The consistent result that larger, denser cities are more segregated runs counter to the hypothesis that such cities \new{promote socioeconomic mixing} by attracting diverse individuals and  constraining space in ways that oblige them to cross paths with each other\cite{jacobs1961, milgram1970,derex2013experimental,gomez2016explaining, wirth1938urbanism, stier2021evidence}. Our results support the opposite hypothesis: big cities allow their inhabitants to seek out people who are more like themselves. \new{The key advancement that enables this discovery is our realistic fine-grained measure of proximity with respect to both time and space (Extended Data Figure \ref{fig:atheyreproduce}).}

\begin{figure}
\centering
\includegraphics[width=1.0\textwidth]{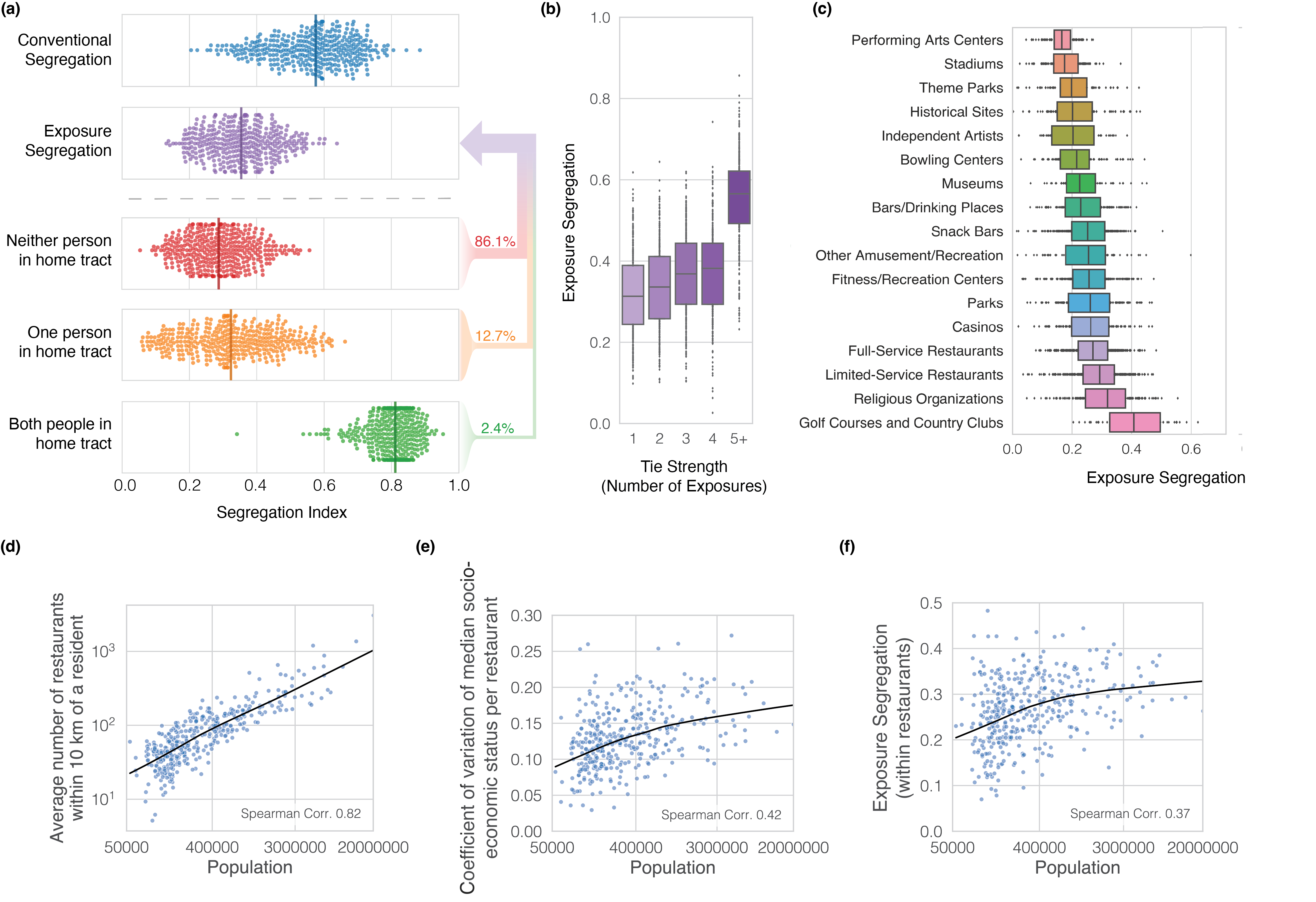}
\caption{\new{\textbf{Exploring the dynamics of exposure segregation} reveals that socioeconomic differentiation of spaces accounts for increased segregation in large cities.}
\textbf{(a)} Top: exposure segregation is 38\% (${p}{<}{10}^{-4}$, 95\% CI ${37}{\%}{-}{41}{\%}$) lower than the conventional segregation measure NSI. Each point represents the \metric estimate in one MSA; vertical colored lines represent median across MSAs. Bottom: breaking down exposure segregation into component parts. Exposures where both people are within their home Census tract (green) are most segregated, reflecting the \emph{homophily effect} in which people preferentially cross paths with those of similar SES in their home tracts. Out-of-tract exposures (orange and red) are less segregated, reflecting the \emph{visitor effect} in which visiting other tracts exposes individuals to economically diverse individuals. Because a small minority (2.4\%, 95\% CI ${2.4}{\%}{-}{2.4}{\%}$) of exposures happen within home tract, the visitor effect dominates the homophily effect and thus \metric is lower than conventional NSI. 
\textbf{(\new{b-c})} Exposure segregation varies by tie strength and location type. Each point represents segregation in one MSA using only exposure pairs occurring \new{(d) with a specific tie strength} or (e) in a given location type. Boxes indicate the interquartile range across MSAs.  \new{Segregation increases with tie strength and is especially high for the strongest ties (5+ exposures; median ES 0.57)}. Segregation is highest at golf courses and country clubs (median ES 0.42), and lowest at performing arts centers (median ES 0.16) and stadiums (median ES 0.17). 
\textbf{(d-f)} A case study of full-service restaurants \new{illustrates the relationship between urbanization and \metric}. Highly-populated metropolitan areas are more segregated not only because they offer a wider choice of venues but also because these venues are more socioeconomically differentiated. %
\textbf{(d)} Larger MSAs have more restaurants within 10 kilometers of the average resident, giving residents more options to self-segregate. \textbf{(e)} Moreover, restaurants in larger MSAs vary more in the median SES of their visitors, offering a greater choice of socioeconomically differentiated restaurants. The coefficient of variation across restaurant SES in 10 largest MSAs is 63\% ($p$${<}$$10^{-4}$, 95\% CI 37-100\%) more  than the coefficient of variation in small MSAs (with fewer than 100,000 residents). 
\textbf{(f)} Consequently, exposure segregation within restaurants is higher in larger MSAs. These relationships are also detectable at the scale of city hubs (i.e. higher-level clusters of POIs such as plazas and shopping malls) as well as at the  neighborhood level (Extended Data Figures \ref{fig:hub_segregation}-\ref{fig:home_segregation}).
}
 \label{fig:main_2}
\end{figure}

\paragraph{Exploring exposure segregation.} 
\new{Our methodology further allows for comparisons between a conventional static measure of segregation (Neighborhood Sorting Index, NSI) and our dynamic measure.} The median level of exposure segregation across all MSAs is 38\% lower (${p}{<}{10}^{-4}$, 95\% confidence interval ${37}{\%}{-}{41}{\%}$) than the corresponding value for a conventional static estimate (NSI; Figure \ref{fig:main_2}a top)\cite{athey2020experienced}. Exposure segregation is lower because when people venture outside their home tract, they experience more diversity (Figure \ref{fig:main_2}a bottom). For instance, exposures are 50\% less segregated when both people are outside the home tract (${p}{<}{10}^{-4}$, 95\% CI ${48}{\%}{-}{53}{\%}$). An important caveat is that exposures that occur when both people are within their home Census tract are 41\% more segregated (${p}{<}{10}^{-4}$, 95\% CI ${38}{\%}{-}{44}{\%}$) than under the hypothetical that residents are exposed uniformly to all people in the same tract (as the NSI assumes). This implies that even within their own neighborhood, people interact with neighbors who are socioeconomically most similar to them. \new{In prior work pertaining to racial segregation (rather than economic segregation), it was likewise found that a static measure overstates segregation, while segregation within home tracts exceeds static estimates\cite{athey2020experienced}.} \new{Unique to our study, we find that each component of exposure segregation is elevated in large cities (Extended Data Figure \ref{fig:home_segregation}, Supplementary Figure \ref{fig:robust_3}), showing the robustness of our rejection of the cosmopolitan mixing hypothesis. }

\new{We also quantify variability in exposure segregation both by tie strength (Figure \ref{fig:main_2}b-c) and across different points of interest (POIs)}. \new{Stronger ties are more segregated (Figure \ref{fig:main_2}b)\cite{mcpherson2001birds, granovetter1973strength}.} \new{We also find much  variability in POI-level segregation\cite{moro2021mobility, chetty2022soccap2}} (Figure  \ref{fig:main_2}c). We explain variability in POI-level segregation (Figure  \ref{fig:main_2}c) by the degree to which POIs service small and thereby socioeconomically homogeneous communities, operationalized as average travel distance to nearest POI\cite{moro2021mobility} and \# of POIs  (Spearman Corr. -0.75, ${p}{<}{0.001}$ for travel distance,  Spearman Corr. 0.69, ${p}{<}{0.01}$ for \# of POIs, Extended Data Figure \ref{fig:poi_disparity_explanation}a,b).  For instance, in the median MSA, religious organizations require 92\% (${p}{<}{10}^{-4}$, 95\% CI ${92}{\%}{-}{93}{\%}$) less travel distance and are 16$\times$ (${p}{<}{10}^{-4}$, 95\% CI ${8}{\times}{-}{18}{\times}$) more numerous than stadiums, and are thus 75\% (${p}{<}{10}^{-4}$, 95\% CI ${58}{\%}{-}{87}{\%}$) more segregated. In rare cases, a POI with few venues may still be highly segregated (e.g., golf courses) because cross-venue economic differentiation is generated through other mechanisms, such as a public-private distinction (Extended Data Figure \ref{fig:poi_disparity_explanation}c). 

\paragraph{\new{Differentiation of space in large cities}.} 
To understand why large metropolitan areas support homophily, we present an example of segregation within leisure POIs. Full-service restaurants provide an illustrative example (Figure \ref{fig:main_2}d,e,f) of a segregation-inducing dynamic that holds widely across other leisure sites (Supplementary Figure \ref{fig:robustness_check_leisure_pois}) and other scales of analysis (Extended Data Figure \ref{fig:hub_segregation}-\ref{fig:home_segregation}). We find that larger MSAs offer their residents a greater number of leisure choices: the average resident of one of the 10 largest MSAs has 22$\times$ ($p$${<}$$10^{-4}$, 95\% CI 11-39$\times$) more restaurants within 10 kilometers of their home than an average resident of a small MSA (where a ``small MSA'' is defined as one with less than 100,000 residents; Figure \ref{fig:main_2}d). These choices are also more socioeconomically differentiated. When a restaurant's SES is defined as the median SES of all people who cross paths within it, the coefficient of variation of ``restaurant SES’’ in the 10 largest MSAs is 63\% ($p$${<}$$10^{-4}$, 95\% CI 37-100\%) larger than that in small ones (Figure \ref{fig:main_2}e). Thus, large MSAs not only offer their residents a larger choice of restaurants, but these restaurants are also more socioeconomically differentiated. These processes \new{are associated with} a 29\% ($p$${<}$$10^{-3}$, 95\% CI 8-49\%) increase in \metric at restaurants in the 10 largest MSAs relative to those in small MSAs (Figure \ref{fig:main_2}f). 
We also find analogous results at higher levels of scale: hubs (e.g. plazas, malls, shopping centers, boardwalks) (Extended Data Figure \ref{fig:hub_segregation}) as well as neighborhoods (Extended Data Figure \ref{fig:home_segregation}) and across different many POI types (Supplementary Figure \ref{fig:robustness_check_leisure_pois}). 

\paragraph{Mitigating segregation via urban design.}

Our results so far suggest that segregation could be mitigated via urban design by placing \new{frequently-visited} POIs to act as bridges between diverse neighborhoods, which would allow residents of nearby high-SES and low-SES neighborhoods to easily visit and \new{cross paths} (Figure \ref{fig:main_3}c)\cite{zipf1946p, simini2012universal, schlapfer2021universal}. 
We develop the \emph{Bridging Index} (BI; Methods \ref{sec:ccbi}) to measure whether hubs (i.e., highly-visited POIs) are located in such bridging positions. This index, which measures the economic diversity of the groups that would \new{be exposed to each other} if everybody visited only their nearest hub, is computed by clustering individuals by the nearest hub to their home and then measuring the economic diversity within these clusters (Extended Data Figure \ref{fig:explain_metrics}).
The resulting index ranges from 0 to 1, where 0 means that individuals near each hub have uniform SES, and 1 means that individuals near each hub are as diverse as the overall area (Extended Data Figure \ref{fig:ccbi_intuition}). We compute BI for commercial centers (e.g. plazas, malls, shopping centers, boardwalks) because we find that they are common hubs of exposure: the majority (56.9\%, 95\% CI 56.9\%-56.9\%) of \new{exposures} across all 382 MSAs occur in close proximity (within 1km) of a commercial center, even though only 2.5\% of land area is within 1km of a commercial center. (see Figure \ref{fig:main_3}c).
The results show that BI is strongly associated with \metric (Spearman Correlation -0.78, p$<10^{-4}$; Figure \ref{fig:main_3}d). The top 10 MSAs with the highest BI are 53.1\% ($p$${<}$$10^{-4}$, 95\% CI 44-60\%) less segregated than the 10 MSAs with the lowest BI. This finding is again robust: the effect of BI is strong and significant ($p$${<}$$10^{-4}$) even after including controls for race, population size, economic inequality, and many other variables (Extended Data Tables \ref{tab:segregation_ccbi} and  \ref{tab:segregation_corrs}; see also Supplementary Table \ref{tab:correlations_between_segregation_measures}; Supplementary Figures \ref{fig:robust_1} and \ref{fig:robust_7}; Supplementary Figure \ref{fig:ccbi_ablation}). It follows that policies (e.g., zoning laws) that encourage developers to locate hubs, such as shopping malls, between diverse residential neighborhoods may reduce \metric. 
We have identified several large cities that increase integration in this way (Supplementary Table S8) and present an illustrative example (Figure \ref{fig:main_3}c-d) in which well-placed hubs bridge diverse individuals in Fayetteville, North Carolina.

\begin{figure}

\centering
\vspace{-2.0cm}
\hspace*{-0.5cm}  
\includegraphics[width=1.0\textwidth]{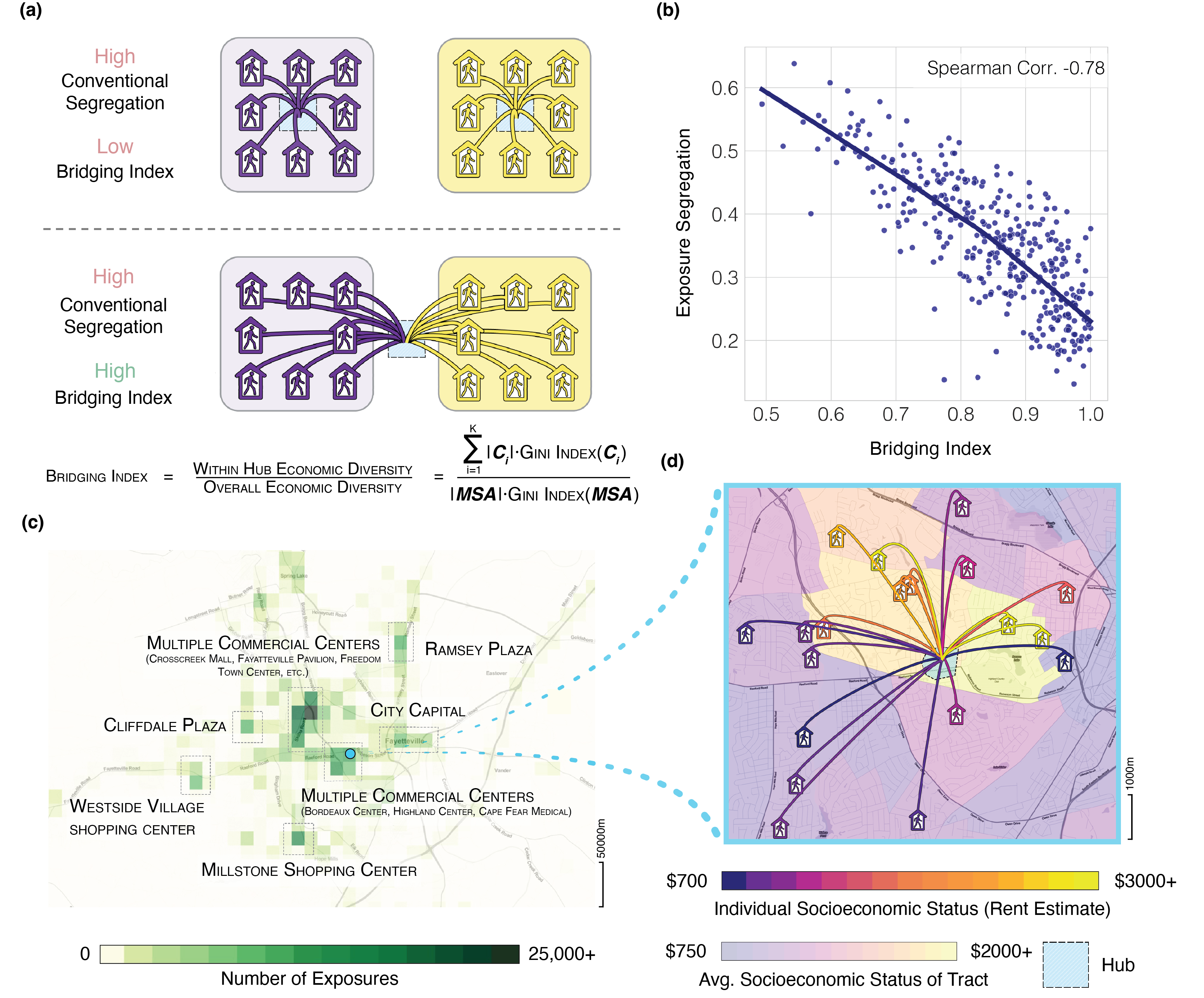}
\caption{\textbf{Exposure segregation is lower when frequently-visited hubs bridge socioeconomically-diverse neighborhoods.}
\textbf{(a)} We develop a Bridging Index (BI) that quantifies the extent to which highly-visited hubs (i.e., exposure hubs) bridge socioeconomically diverse neighborhoods. The metric is constructed by clustering homes by nearest exposure hub, then measuring the within-cluster diversity of ES (Methods \ref{sec:ccbi}).  Two plots illustrate that BI is distinct from a conventional residential measure of segregation (i.e. Neighborhood Sorting Index). 
BI ranges from 0 (no bridging; top) to 1 (perfect bridging; bottom) while residential segregation is constant (high and low-ES individuals are highly segregated by census tract, denoted by purple and yellow bounding boxes). We compute BI with hubs of exposure defined as commercial centers (e.g. shopping malls, plazas) because the majority (56.9\%, 95\% CI 56.9\%-56.9\%) of exposures across all 382 MSAs occur in close proximity (within 1km) of a commercial center, even though only 2.5\% of land area is within 1km of a commercial center. \textbf{(b)} BI strongly predicts \metric (Spearman Correlation $-0.78$, N=382, p$<10^{-4}$). The top 10 MSAs with the highest BI are 53.1\% ($p$${<}$$10^{-4}$, 95\% CI 44-60\%) less segregated than the 10 MSAs with lowest BI. Bridging Index predicts segregation more accurately ($p$${<}$$10^{-4}$) than population size, ES inequality, NSI, and racial demographics, and is significantly ($p$${<}$$10^{-4}$) associated with exposure segregation after controlling for these variables and other potential confounders (Extended Data Tables \ref{tab:segregation_ccbi}-\ref{tab:segregation_corrs}). \textbf{(c-d)} A case study of Fayetteville, NC, an MSA with low \metric (21st percentile) despite having above-median population size (64th percentile) and income inequality (60th percentile). \textbf{(c)} Exposure heat map of Fayetteville; all visually discernible hubs are associated with one or more commercial centers. \textbf{(d)} Exposure hubs are located in accessible  proximity to both high and low ES census tracts (Bridging Index = 0.90, 62nd percentile), leading to diverse exposures. An illustrative example of one hub (Highland Center) in Fayetteville and a random sample of 10 exposures occurring inside of it. Home icons demarcate individual home location (up to 100m of random noise added for anonymity); colors denote individual and mean tract SES.}
\label{fig:main_3}
\end{figure}

\section*{Discussion}\label{sec:discussion} 

As big cities continue to grow and spread, it is important to ask whether they are increasing socioeconomic mixing. Although it is often argued that big cities promote mixing by increasing density, in fact we find that exposure diversity and city size are \textit{negatively} related. This result means that scale matters. Because large cities can sustain venues that are targeted to thin socioeconomic slices of the population, they have become homophily-generating machines that are more segregated than small cities.
We also find that some cities are able to mitigate this segregative effect because their hubs are located in bridging zones that can draw in people from diverse neighborhoods. We were able to detect these pockets of homophily (and the counteracting effects of bridging hubs) because we have developed a dynamic measure of economic segregation that captures everyday \new{socioeconomic mixing} at home, work, and leisure. 

This new methodology for measuring \new{exposure} segregation, while an improvement over static approaches, of course has limitations. 
Because we use physical proximity as a proxy for exposure, \new{it is difficult to ascertain exactly how weak or strong the ties are\cite{dietze2016social}.}
It is reassuring, however, that our core results persist under stricter time, distance, and tie-strength thresholds (Supplementary Table \ref{tab:correlations_between_segregation_measures}, Supplementary Figures \ref{fig:robust_4}-\ref{fig:robust_7})\new{, and are associated with downstream outcomes (Extended Data Figure \ref{fig:external_validation}, Supplementary Figure \ref{fig:external_validation_2})}. It is likewise important to locate and analyze supplementary datasets that cover subpopulations (e.g., homeless subpopulations) that aren't as well represented in our dataset\cite{coston2021leveraging}. The available evidence indicates that our sample is \new{well-balanced} on many key racial, economic, and demographic variables\cite{squire2019}, but cellphone market penetration is still not complete, \new{and GPS ping data is unevenly distributed by time}. Lastly, our measure of SES relies on housing consumption, an indicator that does not exhaust the concept of SES. It is again reassuring that our analytic approach, which improves on conventional neighborhood-level imputations, is robust under a range of alternative measures of SES (Supplementary Figure \ref{fig:robust_2}).

This is all to suggest that dynamic segregation data are rich enough to overcome many seeming limitations. The dynamic approach that we have taken here could further be extended to examine cross-population differences in the sources of segregation and to develop a more complete toolkit of approaches to reducing segregation and improving urban design. %

\clearpage
}

\nolinenumbers %
\section*{References}
\linespread{1.0}\selectfont
\bibliographystyle{naturemag}
\bibliography{main}
\clearpage

\linespread{1.6}\selectfont
\renewcommand*{\thesubsection}{M\arabic{subsection}}

\section*{Methods}
\label{sec:methods}
\setcounter{section}{1}
\setcounter{subsection}{0}

\captionsetup[table]{name=Methods Table}
\captionsetup[figure]{name=Methods Figure}

\renewcommand{\thefigure}{\arabic{figure}}
\setcounter{figure}{1}
\renewcommand{\thetable}{\arabic{table}} 
\setcounter{table}{1}
\renewcommand{\theequation}{Supplementary Equation \arabic{equation}} 
\setcounter{equation}{1}

In \refmsec{methods_datasets}, we explain the datasets used in our analysis; in \refmsec{methods_dataprocessing}, we explain data processing procedures we leverage to infer socioeconomic status and \edgename{s}; and in \refmsec{methods_analysis}, we explain the analysis underlying our main results. 

\subsection{Datasets} 
\label{sec:methods_datasets}

\subsubsection*{SafeGraph}
Our primary mobility and location data comprise GPS locations from a sample of adult smartphone users in the United States, provided by the company SafeGraph. The data are de-identified GPS location pings from smartphone applications which are collected and transmitted to SafeGraph by participating users\citeMethods{safegraph}. While the sample is not random sample, prior work has demonstrated that SafeGraph data is geographically \new{well-balanced} (i.e. an approximately unbiased sample of different census tracts within each State), and well-balanced along the dimensions of race, income, and education\citeMethods{squire2019, chang2021mobility}. Furthermore, SafeGraph data is a widely used standard in large-scale studies of human mobility across many different areas including COVID-19 modeling\citeMethods{chang2021mobility}, political polarization\citeMethods{chen2018effect}, and tracking consumer preferences\citeMethods{athey2018estimating}. All data provided by SafeGraph was \new{de-identified} and stored on a secure server behind a firewall. Data handling and analysis was conducted in accordance with SafeGraph policies and in accordance with the guidelines of the Stanford University Institutional Review Board. 

The raw data consists of 91,755,502 users and 61,730,645,084 pings (one latitude and longitude for one user at one timestamp) from three evenly spaced months in 2017: March, July, and November. The mean number of raw pings associated with a user is 667 and the median number of pings is 12. We apply several filters to improve the reliability of the SafeGraph data, and subsequently link each user to an estimated rent (i.e. Zillow Zestimate) using their inferred home location (i.e. CoreLogic address), as described in \refmsec{methods_dataprocessing}.

We apply several filters to improve the reliability of the SafeGraph data. To ensure locations are reliable, we remove pings whose location is estimated with accuracy worse than 100 meters as recommended by SafeGraph\citeMethods{safegraph_visit}. We filter out users with fewer than 500 pings, as these are largely noise. Since we incorporate a user's home value and rent in measuring their socioeconomic status, we filter out users for whom we are unable to infer a home. Finally, to avoid duplicate users, we remove users if more than 80\% of their pings have identical latitudes, longitudes, and timestamps to those of another user; this could potentially occur if, for example, a single person in the real world carries multiple mobile devices. After the initial filters on ping counts and reliability, we are able to infer home locations for 12,805,490 users in the United States (50 states and Washington D.C.), leveraging the CoreLogic database. Of users for whom we can infer a home location, we are able to successfully link 9,576,650 to an estimated rent value via the Zillow API. Section \refmsec{methods_dataprocessing} provides full details on the use of CoreLogic database to infer home locations and the use of the Zillow API to link these home locations to estimated rent values. Finally, after removing users where $>80$\% of their pings are duplicates with another user, we reduce the number of users from 9,576,650 to 9,567,559 (i.e., we remove about 0.1\% of users through de-duplication).

\subsubsection*{CoreLogic}
We use the CoreLogic real estate database to link users to home locations\citeMethods{corelogic}. The database provides information covering over 99\% of US residential properties (145 million properties), over 99\% of commercial real estate properties (26 million properties), and 100\% of US county, municipal, and special tax districts (3141 counties). The CoreLogic real estate database includes the latitude and longitude of each home, in addition to its full address: street name, number, county, state, and zip code. 

\subsubsection*{Zillow}
We use the Zillow property database to query for rent estimates\citeMethods{zillow} (our primary measure of socioeconomic status). The Zillow database contains rent data (``rent Zestimate'') for 119 million US residential properties. 
We were able to determine a rent Zestimate, the primary measure of socioeconomic status (SES) used in our analysis, for 9,576,650 out of 12,183,523 inferred SafeGraph user homes (a 79\% hit rate).

\subsubsection*{SafeGraph Places}
Our database of US business establishment boundaries and annotations comes from the SafeGraph Places database\citeMethods{safegraph}, which  indexes the names, addresses, categories, latitudes, longitudes, and geographical boundary polygons of 5.5 million US points of interest (POIs) in the United States. SafeGraph includes the NAICS (North American Industry Classification System) category of each POI, which is standard taxonomy used by the Federal government to classify business establishments\citeMethods{kelton2008using}. For instance, the NAICS code 722511 indicates full service restaurants. We identify relevant leisure sites using the prefixes 7, which includes arts, entertainment, recreation, accommodation, and food services, and supplement these POIs with the prefix 8131 to include religious organizations such as churches. We restrict our analysis of leisure sites to the top most frequently visited POI categories within these NAICS code prefixes (Figure \ref{fig:main_1}d): full service restaurants, snack bars, limited-service restaurants, stadiums, etc. SafeGraph Places also includes higher-level ``parent'' POI polygons  which encapsulate smaller POIs. Specifically, we identified exposure hubs  with the NAICS code 531120 (lessors of non-residential real estate) which we find in practice corresponds to commercial centers such as shopping malls, plazas, boardwalks, and other clusters of businesses. We provide illustrative examples of such exposure hubs in Supplementary Figures \ref{fig:ccbi_example1}-\ref{fig:ccbi_example4}.

\subsubsection*{US Census}
We extract demographic and geographic features from the 5-year 2013-2017 American Community Survey (ACS)\citeMethods{censusreporter}. This allows us, as described below, to link cell phone locations to geographic areas including census block group, census tract, and Metropolitan Statistical Area (MSA), as well as to infer demographic features corresponding to those demographic areas including median household income.

A census block group (CBG) is a statistical division of a census tract. CBGs are generally defined to contain between 600 and 3,000 people. A CBG can be identified on the national level by the unique combination of state, county, tract, and block group codes.

A census tract is a statistical subdivision of a county containing an average of roughly 4,000 inhabitants. Census tracts range in population from 1,200 to 8,000 inhabitants. Each tract is identified by a unique numeric code within a county. A tract can be identified on the national level by the unique combination of state, county, and tract codes.

Census tracts and block groups typically cover a contiguous geographic area, though this is not a constraint on the shape of the tract or block group. Census tract and block group boundaries generally persist over time so that temporal and geographical analysis is possible across multiple censuses.

Most census tracts and CBGs are delineated by inhabitants who participate in the Census Bureau's Participant Statistical Areas Program. The Census Bureau determines the boundaries of the remaining tracts and block groups when delineation by inhabitants, local governments, or regional organizations is not possible \citeMethods{census_glossary}.

A Metropolitan Statistical Area (MSA) is a US geographic area defined by the Office of Management and Budget (OMB) and is one of two types of Core Based Statistical Area (CBSA). A CBSA comprises a county or counties associated with a core urbanized area with a population of at least 10,000 inhabitants and adjacent counties with a high degree of social and economic integration with the core area. Social and economic integration is measured through commuting ties between the adjacent counties and the core. A Micropolitan Statistical Area is a CBSA whose core has a population of between 10,000 and 50,000; a Metropolitan Statistical Area is a CBSA whose core has a population of over 50,000. In our primary analysis, we follow Athey et al\citeMethods{athey} and focus on Metropolitan Statistical Areas, excluding Micropolitan Statistical Areas due to data sparsity concerns.

\subsubsection*{TIGER}
Road and transportation feature annotations come from the Census-curated Topologically Integrated Geographic Encoding and Referencing system (TIGER) database\citeMethods{tiger}. The TIGER databases are an extract of selected geographic and cartographic information from the U.S. Census Bureau's Master Address File / Topologically Integrated Geographic Encoding and Referencing (MAF/TIGER) Database (MTDB). We use the MAF/TIGER Feature Class Code (MTFCC) from the TIGER Roads and TIGER Rails databases to identify road and railways.  TIGER data is in the format of Shapefiles, which provide the exact boundaries of roads and railways as latitude/longitude coordinates.

\subsection{Data processing} 
\label{sec:methods_dataprocessing}

For each individual, we first infer their home location and subsequently estimate socioeconomic status based on their home rent value (see \emph{Inferring home location} and subsequently \emph{Inferring socioeconomic status}). We then calculate all \edgename{s} between individuals (see \emph{Constructing exposure network}), which we then annotate based on the location, i.e. if the \edgename occurred in both, one, or neither individual's home tract, and whether it occurred inside of a fine-grained POI such as a specific restaurant or a  ``parent'' POI such as an exposure hub (see \emph{Annotating \edgename{s}}). Details on all inferences and \edgename calculations are provided below.  

\subsubsection*{Inferring home location}

We first infer a user's home latitude and longitude using the latitude and longitude coordinates of their pings during local nighttime hours, based on best practices established by SafeGraph\cite{safegraph_pattern}.
We first remove users with fewer than 500 pings to ensure that we have enough data to reliably infer home locations.
We then interpolate each person's location \new{for each one-hour window (eg, 6-7 PM, 7-8 PM, and 8-9 PM)} using linear interpolation of latitudes and longitudes, to ensure we have timeseries at constant time resolution. \new{We perform interpolation using the $interpolate$ package of the $scipy$ library.} We filter for hours  between 6 PM and 9 AM where the person moves less than 50 meters until the next hour; these stationary nighttime observations represent cases when the person is more likely to be at home. We filter for users who have stationary nighttime observations on at least 3 nights and with at least 60\% of observations within a 50 meter radius. Finally, we infer home latitude and longitude as the median latitude and longitude of these nighttime home locations (after removing outliers outside the 50 meter radius). We choose the thresholds above because they yield a good compromise between inferring the home location of most users and inferring home locations with high confidence. Overall, we are able to infer home locations for 70\% of users with more than 500 pings, and these locations are inferred with high confidence; 89\% of stationary nighttime observations are within 50 meters of the inferred home latitude and longitude.

\subsubsection*{Inferring socioeconomic status from home latitude and longitude}

Having inferred home location from nighttime GPS pings, we link their latitude and longitude to a large-scale housing database (Zillow) to infer the estimated rent of each user's home, which we use as a measure of socioeconomic status. We do this in two steps. First, we link the inferred user's home latitude and longitude to the CoreLogic property database (\refmsec{methods_datasets}), a comprehensive database of properties in the United States, by taking the closest CoreLogic residential property (single family residence, condominium, duplex, or apartment) to the user's inferred home latitude and longitude. Second, we use the CoreLogic address to query the Zillow database, which provides estimated home rent and price for each user. (The Zillow database does not allow for queries using raw latitude and longitude, which it is necessary to leverage to CoreLogic to obtain an  address for each user.) We use Zillow's estimated rent for the user's home as our main measure of socioeconomic status. We apply several quality control filters to ensure that the final set of users we use in our main analyses have reliably inferred home locations and socioeconomic statuss: 1) we remove a small number of users whose inferred nighttime home latitude and longitude are identical to another user's, since we empirically observe that these people have unusual ping patterns; 2) we remove users for whom we are lacking an Zillow rent estimate, since this constitutes our primary socioeconomic status measure; 3) we winsorize Zillow rent estimates which are greater than \$20,000 to avoid spurious results from a small number of outliers; 4) we remove a small number of users who are missing Census demographic information for their inferred home location; 5) we remove users whose Zillow home location is further than 100 meters from their CoreLogic home location, or whose CoreLogic home location is further than 100 meters from their nighttime latitude and longitude; 6) we remove a small number of users in single family residences who are mapped to the exact same single family residence as more than 10 other people, since this may indicate a data error in the Zillow database.

The set of users who pass these filters constitute our final analysis set of \usercountx users. We confirm that the Census demographic statistics of these users' inferred home locations are similar to those of the US population in terms of income, age, sex, and race.%

Any individual quantitative measure provides only a partial picture of a person's socioeconomic status. Recognizing this, we conduct robustness checks in which rather than using the Zillow estimated rent of the user's home as a proxy for socioeconomic status, we use 1) the median Census Block Group household income in that area; and 2) the percentile-scored rent of the home, to account for long-tailed rent distributions. Our main results are robust to using these alternate measures of socioeconomic status (Supplementary Figure \ref{fig:robust_2}).

\subsubsection*{Constructing exposure network} %
We construct a fine-grained, dynamic exposure network $\mathcal{G}$ between all \usercountx individuals across \msacount MSAs and \countycount counties, which is represented as an undirected graph $\mathcal{G} = (\mathcal{V}, \mathcal{E})$ with time-varying edges. Each node $v_i \in \mathcal{V} $ in the graph represents one of the $N=\usercountx$ individuals in our study, such that the set of nodes is $V = \{v_1,v_2,...,v_N\}$. Each node $v_i$ has a single attribute $x_i$, representing the inferred socioeconomic status (estimated rent) of the individual. 

Individuals $v_i$ and $v_j$ are connected by one edge $e_{i,j,k} \in \mathcal{E}$ per \edgename, with $k$ indicating the $k$th \edgename between individuals $v_i$ and $v_j$. Each edge $e_{i,j,k}$ has three attributes $t_{i,j,k}$, $lat_{i,j,k}$, $lon_{i,j,k}$ indicating the timestamp, latitude, and longitude of the \edgename respectively. We now focus our discussion on explaining how each of the \edgename{s} edges of the network is calculated. 

We define an \emph{\edgename} to occur when two users have GPS pings which are close (according to a fixed threshold) in both physical proximity and time. Specifically if user  $v_i$ has a GPS ping with $t_i, lat_i, lon_i$ (indicating the timestamp, latitude, and longitude of the ping respectively), and user $v_j$ has a GPS ping with $t_j, lat_j, lon_j$, then we users are said to have shared an \edgename if $|t_i - t_j| < T$ \and{and} $distance((lat_i,lat_i),(lat_j,lat_j)) < D$, where $T$ represents the time threshold (i.e. maximum time distance the two pings can be apart to count as an \edgename) and $D$ represents the distance threshold (i.e. maximum physical distance the two pings can be apart to count as an \edgename). We filter for both distance and time simultaneously to ensure that our exposure network only includes pairs of users who are likely to have come into contact with each other. This contrasts to other methods which consider all individuals to visit the same location, irrespective of time\citeMethods{moro2021mobility}, to have an equal likelihood of exposure, an assumption which may prove unrealistic in many cities (e.g. demographics of individuals visiting public parks varies starkly by time of day\citeMethods{madge1997public}). We use a threshold $T$ of 5 minutes, which is a stringent threshold on time as the mean number of pings per person per hour during day time is approximately one ping; we use a distance threshold $D$ of 50 meters, following prior work which shows that even exposure to individuals from afar is linked to long term outcomes\citeMethods{brown2021childhood}. Our network is validated by correlation to external, gold-standard datasets (Extended Data Figure \ref{fig:external_validation}). Furthermore, we show through a series of robustness checks that our key results in Figure 1, Figure 2, and Figure 3 are highly robust to varying thresholds (i.e. 1 minute or 2 minutes time threshold, as well as 10 meters or 25 meters distance threshold), as well as additional criteria to increase tie strength (i.e. requiring multiple consecutive exposures, or multiple exposures on unique days)---and under all observed circumstances the main findings remain consistent (see Supplementary Table \ref{tab:correlations_between_segregation_measures}, Supplementary Figures \ref{fig:robust_1}-\ref{fig:robust_7}). 

To efficiently calculate the \edgename{s} between all users, we implement our \edgename threshold as a k-d tree\citeMethods{bentley1975multidimensional}, a data structure which allows one to efficiently identify all pairs of points within a given distance of each other. In total, we identify \crosscountx \edgename{s}. The timestamp $t_{i,j,k}$ of the \edgename is the minimum ping timestamp in the pair of individuals' ping timestamps ($t_i$,$t_j$), and the location $lat_{i,j,k}$, $lon_{i,j,k}$ of the \edgename is the average latitude and longitude of pair of pings belonging to the two individuals ($lat_i$,$lat_j$) and ($lon_i$,$lon_j$). \new{We implement our \edgename detection system to parallelize across multiple cores, allowing us to efficiently construct the network using a single supercomputer (with 12TB RAM, and 288 cores) in under a week. By contrast, a naive implementation (without k-d trees or parallelization) would necessitate on the order of \~10 years of compute time. }

\subsubsection*{Annotating \edgename{s}} \label{sec:annotating} %
\edgenamecaps{s} are annotated to indicate whether they occurred at or near features of interest: e.g., near a user's home. Annotations are not mutually exclusive in that an \edgename may be simultaneously tagged as having occurred near multiple features from multiple data sources. We describe the specific annotations below. 

We annotate a user's \edgename as having occurred in their home if it occurs within 50 meters of the user's home location. 
An \edgename is annotated with a TIGER road/railway if it occurs within 20 meters from that feature. An \edgename is annotated as having occurred within a SafeGraph Places point-of-interest (POI) if the \edgename occurs within the polygon defined for the POI. Polygons are provided by the SafeGraph Places database for both fine-grained POIs (e.g. individual restaurants) as well as ``parent'' POIs (e.g. exposure hubs). We focus our analysis of fine-grained POIs (Figure \ref{fig:main_1}e, Extended Data Figure \ref{fig:poi_disparity_explanation}) on the most visited fine-grained POIs: full-service restaurants, snack bars, limited-service restaurants (e.g. fast food), stadiums, etc (see Figure \ref{fig:main_1}e for full list). These categories roughly align with those used by prior work \citeMethods{athey}.

\subsection{Analysis} 
\label{sec:methods_analysis}

\subsubsection*{Exposure Segregation}

We define the \metric (ES) of a specified geographical area (i.e. Metropolitan Statistical Area, County) as the Pearson correlation between the socioeconomic status (SES)  of individuals residing in that geographical area, and the mean SES those that they cross paths with. 

\begin{align*}
Exposure\:Segregation &= Corr(SES, \overline{SES}_{exposures}) = \new{ \frac{\mathrm{cov} \left( SES, \overline{SES}_{exposures} \right) }{\sigma_{SES}\sigma_{\overline{SES}_{exposures}}} }\\
\end{align*}

Our metric captures the extent to which an individual's SES predicts the SES of their immediate exposure network. Thus, in a perfectly integrated area in which individuals cross paths randomly with others regardless of SES, \metric would equal 0.0. In a perfectly segregated area in which individuals cross paths with only those of the exact same SES, \metric would equal 1.0.

Exposure segregation nests a classic definition residential segregation, the Neighborhood Sorting Index\citeMethods{jargowsky1996take, jargowsky2005measure} (NSI), which is equivalent to the Pearson correlation across between each person's SES and the mean SES in their Census tract. The NSI is widely used because it can be calculated directly from Census data on the SES of people living in each tract. However, a fundamental limitation of NSI as a measure segregation is that the Census tract in which people live is a weak proxy for who they cross paths with. Census tracts are  static and artificial boundaries which fail to capture socioeconomic mixing as individuals move throughout the cityscape during work, leisure time, and schooling. 

We design our \metric (ES) metric such that it accommodates any exposure network, and thus NSI is a special case of our metric. Specifically, if \metric is computed for a synthetic exposure network under the unrealistic assumptions that a) people only cross paths with those in their home Census tract and b) they do so uniformly at random---then it is equivalent to NSI (Supplementary Figure \ref{fig:nsi_vs_mixed_model}). However, constructing such a synthetic exposure network from Census tracts has limited applicability to measuring segregation in the real world, because people may also be exposed to more heterogeneous populations as they visit other Census tracts for work, leisure, or other activities, a phenomenon we refer to as the \emph{visitor effect}. Furthermore, even within home tract, individuals may cross paths non-uniformly as they seek out people of similar socioeconomic status; we refer to this as the \emph{homophily effect}. Thus, we instead leverage dynamic mobility data from cell phones to captures the extent of contact between diverse individuals throughout the day, and apply our metric, \metric (ES), to this real-world exposure network. An advantage of \metric is that it allows for direct comparability to NSI, because both measures are of the same underlying statistical quantity, but differ in their definition of the exposure network. Our results indicate that this choice of exposure network matters; ES is a stronger predictor of upward economic mobility (Extended Data Figure \ref{fig:external_validation}) as the two metrics are shown to be distinct (Supplementary Figure \ref{fig:low_correlation_figure_1}). 

To calculate the \metric of a specified geographical area (i.e. Metropolitan Statistical Area, County), we first select the set of all individuals who reside in area: $\mathcal{V}_{A} \subset \mathcal{V}$. For instance, to calculate exposure segregation for Napa, California (Figure \ref{fig:main_1}c Top), $\mathcal{V}_{A}$ is the 3707 individuals with home locations inside the geographical boundary of the Napa, CA MSA. Subsequently, for each individual resident of the area $v_i \in \mathcal{V}_{A}$ we query the population exposure network ($\mathcal{G} = (\mathcal{V}, \mathcal{E})$) for the SES of the set of individuals they cross paths with, $\mathcal{Y}_i$: $\{x_j \in \mathcal{V} | e_{i,j,k} \in \mathcal{E} \}$. We then aim to estimate the Pearson correlation between the SES of each individual $x_i$ and the mean SES of those they are exposed to, $y_i = mean(\mathcal{Y}_i)$.

\subsubsection*{\new{Estimating exposure segregation}}

\newcommand{\SES}{ES}

\new{Here, we first motivate why a ``naive'' approach to estimating exposure segregation via a standard Pearson correlation on the observed exposure network is problematic (resulting in downwardly biased estimates of exposure segregation). We then elaborate on how we leverage a linear mixed effects model to compute a \emph{corrected} Pearson correlation, allowing us to obtain unbiased estimates of exposure segregation even in areas where data is sparse.  } 

\new{A ``naive'' approach to estimate exposure segregation would be to compute the observed sample mean SES of who each person is exposed to. Then, exposure segregation could be estimated using a sample Pearson correlation\footnote{ \begin{align*} r_{xy} = \frac{{}\sum_{i=1}^{n} (x_i - \overline{x})(y_i - \overline{y})}
{\sqrt{\sum_{i=1}^{n} (x_i - \overline{x})^2(y_i - \overline{y})^2}} \end{align*}} between individual SES ($x_i$) and the sample mean SES of those they are exposed to ($y_i$). This approach is problematic because naively computing such a correlation based on limited data (in counties or MSAs with low population sizes) will result in estimates that are downward biased.} To illustrate why naive estimates of \metric are downward biased, imagine that we compute the correlation between a person's SES and the ``true'' mean SES of the people they are exposed to. Now, we add noise to the mean SES values, which represents the noisy mean estimates given limited data. As the noise is increased, the correlation is decreased. Thus, because estimates of each person's mean SES will be more noisy in geographical areas with less data, there will be a downward bias to naive estimates of the Pearson correlation in these areas.

\begin{figure}[htbp]\ContinuedFloat
  \centering
  \includegraphics[width=0.8\textwidth]{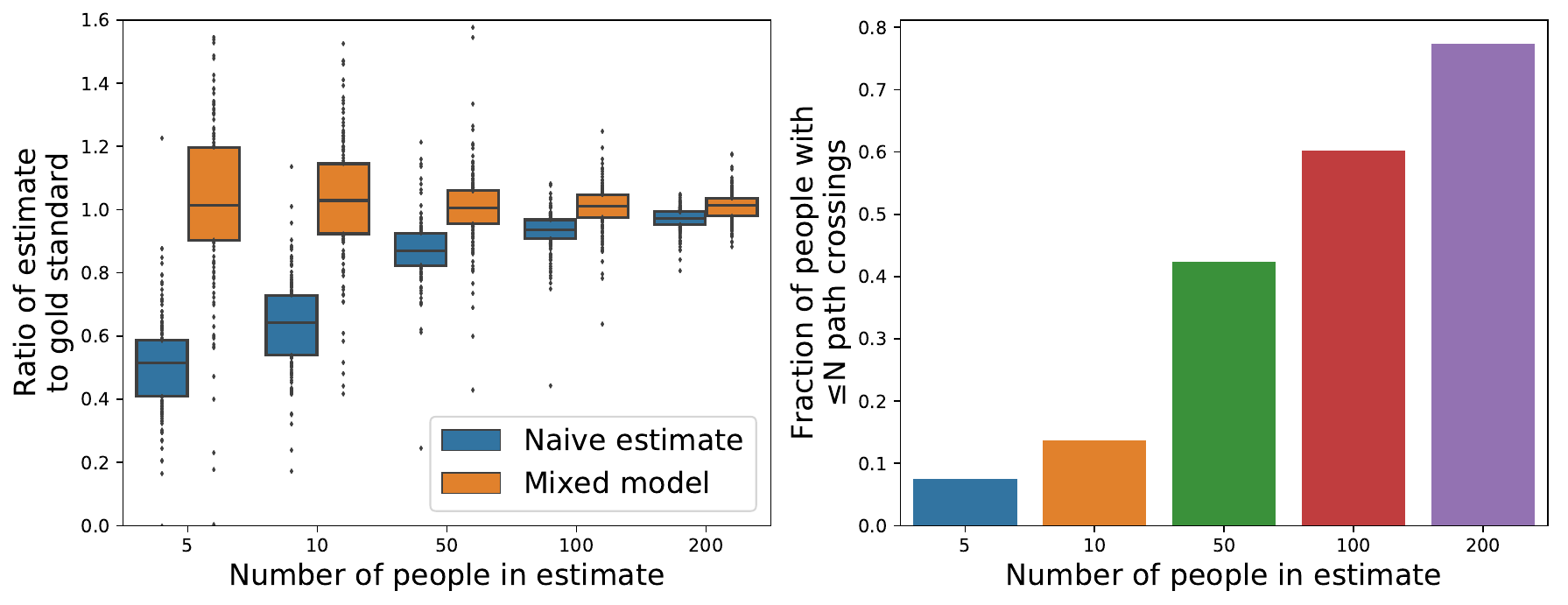}
  \caption{Our estimates compared with naive estimates of the Pearson correlation. We took people who crossed paths with at least 500 other people and computed the Pearson correlation coefficient (the ``gold standard estimate''). Then, for each person we randomly sampled 5, 10, 50, 100, and 200 people from the 500+ people and computed segregation estimates based on the reduced sets of people. The left plot shows the ratio of the estimates to the gold standard, for each MSA. The right plot shows the overall number of people in the dataset with $\leq N$ \edgename{s}.}
  \label{fig:mixed_model_validation}
\end{figure}

\new{We instead compute a corrected Pearson correlation, using a linear mixed effects model to accurately estimate \metric: the correlation  between a person's SES and the mean SES of the people they are exposed to. Our linear mixed effects model is an unbiased estimator of the Pearson correlation.} We compare the unbiased estimates from our linear mixed effects model to naive estimates of \metric in Methods Figure~\ref{fig:mixed_model_validation}.

Our mixed model models the distribution of datapoints ($x_i$, $y_{ij})$ through the following equation:

\begin{align*}
y_{ij} &= ax_i + b + \epsilon_i^{(1)} + \epsilon_{ij}^{(2)} \\
\text{where } x_i&=\text{SES of person $i$} \\
y_{ij} &= \text{SES of person $j$ who was exposed to person $i$} \\
a, b &= \text{model parameters} \\
\epsilon_i^{(1)} &= \text{person-specific noise term} \\
\epsilon_{ij}^{(2)} &= \text{noise for each data point}
\end{align*}

\new{Above, the true mean SES of the exposure set for each person is modeled as $ax_i + b + \epsilon_i^{(1)}$ . Individual exposures $y_{ij}$, are then modeled as noisy draws from a distribution centered at this true mean.} The Pearson correlation coefficient between person $i$'s SES and the mean SES of the people they were exposed to is then computed as follows. We assume that $x_i$ has a variance of $1$ through data preprocessing and that $x_i$ is uncorrelated with $\epsilon_i^{(1)}$.

\begin{align*}
\mathrm{corr} \left( x_i, ax_i + b + \epsilon_i^{(1)} \right) &= \mathrm{corr} \left( x_i, ax_i + \epsilon_i^{(1)} \right) \\
&=\frac{\mathrm{cov} \left( x_i, ax_i + \epsilon_i^{(1)} \right) }{\sqrt{\mathrm{Var} \left( x_i \right) Var \left( ax_i + \epsilon_i^{(1)} \right)}} \\
&=\frac{\mathrm{cov} \left( x_i, ax_i \right) }{\sqrt{\mathrm{Var} \left( ax_i + \epsilon_i^{(1)} \right)}} \\
&=\frac{a}{\sqrt{ a^2 + \mathrm{Var} \left( \epsilon_i^{(1)} \right)}}
\end{align*}

We estimate $a$ and $\mathrm{Var} \left( \epsilon_i^{(1)} \right)$ by fitting the mixed model using the R lme4 package, optimizing the restricted maximum likelihood (REML) objective.

\subsubsection*{Decomposing segregation by time}
\label{sec:segregation_by_time}

Each \edgename edge ($e_{i,j,k}$) in our exposure network is timestamped with a time of exposure $t_{i,j,k}$. This allows us to decompose our overall \metric into fine-grained estimates of segregation during different hours of the day, by filtering for exposures that occurred within a specific hour. In Supplementary Figures \ref{fig:segregation_by_hour}, we partition estimates of segregation by 3 hour windows to illustrate how segregation varies throughout the day (see Supplementary Information).

\subsubsection*{Decomposing segregation by activity}
\label{sec:segregation_by_poi}

Each \edgename edge ($e_{i,j,k}$) in our exposure network occurs at a specific location  $lat_{i,j,k}$, $lon_{i,j,k}$. Thus, it is possible to annotate exposures by the fine-grained POI (e.g. specific restaurant) they occurred in, as well as the by the higher-level ``parent'' POI (e.g. commercial center) in which the POI was located (Methods \ref{sec:methods_dataprocessing}). This allows us to decompose our overall \metric into fine-grained estimates of segregation by specific leisure activity. We do so by filtering the network for all exposures that occurred in a specific POI category, and re-calculating \metric for the MSA or county, using only those exposures. In Figure \ref{fig:main_1}e, we show the variation in \metric by leisure site, and further explain these variations in Extended Data Figure \ref{fig:poi_disparity_explanation}.

\subsubsection*{Bridging Index}
\label{sec:ccbi}

We seek to identify a modifiable, extrinsic aspect of a city's built environment which may reduce \metric. One promising candidate is the location of a city's hubs of exposure. We define a new measure, the \emph{Bridging Index} (BI), which measures the extent to which a particular set of exposure hubs (i.e. high-exposure POIs, $\mathcal{P}$) \new{may} facilitate the integration of individuals of diverse socioeconomic status within a geographic area (i.e. MSA or county). Specifically, BI measures the economic diversity of the groups that would cross paths if everybody visited only their nearest hub from $\mathcal{P}$---based on the observation that physical proximity significantly influences which hubs individuals visit\cite{zipf1946p, simini2012universal, schlapfer2021universal}. 

We compute the Bridging Index (BI) via two steps (Extended Data Figure \ref{fig:explain_metrics}). 
\begin{enumerate}
\item Cluster all individuals who live in an area (i.e. MSA or county residents, $\mathcal{V}_{A}$) into $K$ clusters ($\mathcal{C}_1,\mathcal{C}_2,...,\mathcal{C}_K$) according to the exposure hub from $\mathcal{P}$ closest to their home location. $K$ is the number of hubs in $\mathcal{P}$. 
\item Bridging Index is computed as the weighted average of the economic diversity (i.e. Gini Index) of these clusters of people, relative to the area's overall economic diversity.
\end{enumerate}

\begin{align*}
Bridging\:Index\:(BI) &= \frac{Within\:Hub\:Economic\:Diversity}{Overall\:Economic\:Diversity} &= \frac{\sum_{i=1}^{K}|\mathcal{C}_i| \cdot Gini\:Index(\mathcal{C}_i)}{|\mathcal{V}_{A}| \cdot Gini\:Index(\mathcal{V}_{A})} \\
\end{align*}

 We illustrate the intuition for BI and how it captures the relationship between home and hub locations in Extended Data Figure \ref{fig:ccbi_intuition}. A BI of 1.0 indicates that if everybody visits their nearest exposure hub, each person will be exposed to a set of people \emph{as economically diverse as the overall city they reside in}. Thus, a BI of 1.0 signifies perfect bridging, i.e. even if individuals live in segregated neighborhoods, hubs are located such that individuals must leave their neighborhoods and cross paths with diverse others. On the other hand, a BI of 0.0 signifies the opposite extreme; a city with a BI of 0.0 is one in which, if everybody visits the nearest exposure hub, each person will be exposed to only people of the exact same socioeconomic status. 
 
 The economic diversity of each cluster $\mathcal{C}_i$ is quantified using the Gini Index: $Gini\:Index(\mathcal{C}_i)$, a well-established measure of economic statistical dispersion (Extended Data Figure \ref{fig:explain_metrics}c)\citeMethods{dorfman1979formula}, although results are robust to choice of economic diversity measure such as using variance instead of Gini Index (Supplementary Figure \ref{fig:ccbi_variance}). The denominator of BI normalizes for the baseline economic diversity observed in the city, allowing for direct comparisons between cities.

In our primary analysis, we identify hubs of exposure via commercial centers (e.g. shopping malls, plazas, etc. which are higher-level clusters of individual POIs) because they are associated with a high density of exposures. Specifically, the majority (56.9\%) of exposures happen inside of or within 1km of a commercial center (e.g. shopping mall, plaza, etc.) even though only 2.5\% of the land area of MSAs is within 1km of a commercial center. We thus compute BI using the set $\mathcal{P}$ of all commercial centers within each MSA. We discover that BI strongly predicts exposure segregation (Spearman Correlation $-0.78$, Figure \ref{fig:main_3}d). The top 10 MSAs with the highest BI are 53.1\% less segregated than the 10 MSAs with the lowest BI. BI predicts segregation more accurately than population size, racial demographics SES inequality, NSI, and racial demographics, and is significantly associated with segregation $({p}{<}{10^{-8}})$ after controlling for all aforementioned variables (Extended Data Tables \ref{tab:segregation_ccbi}-\ref{tab:segregation_corrs}).

\subsubsection*{Hypothesis Testing and Confidence Intervals } 
Unless otherwise noted, confidence intervals and hypothesis tests were conducted using a bootstrap with 10,000 replications\cite{bootstrap}. Steiger's Z-test was used to compare different predictors of segregation indices, and hypothesis tests for Spearman correlation coefficients were computed using two-sided Student's t-tests\cite{myers2004spearman,steiger,corder2014nonparametric}.
\clearpage

\nolinenumbers
 \linespread{1.0}\selectfont
 \bibliographystyleMethods{naturemag}
 \bibliographyMethods{main}
\clearpage

\linespread{1.6}\selectfont
\renewcommand*{\thesubsection}{M\arabic{subsection}}

\section*{Extended Data}
\label{sec:extended_data}
\vspace{7cm}
\setcounter{section}{1}
\setcounter{subsection}{0}

\captionsetup[table]{name=Extended Data Table}
\captionsetup[figure]{name=Extended Data Figure}

\renewcommand{\thefigure}{\arabic{figure}}
\setcounter{figure}{0}
\renewcommand{\thetable}{\arabic{table}} 
\setcounter{table}{0}
\renewcommand{\theequation}{Supplementary Equation \arabic{equation}} 
\setcounter{equation}{0}

\begin{figure}[htbp]
\vspace*{-20mm}
   \centering
   \hspace*{-15mm}
\begin{subfigure}[t]{.5\textwidth}
\includegraphics[width=\textwidth,keepaspectratio]{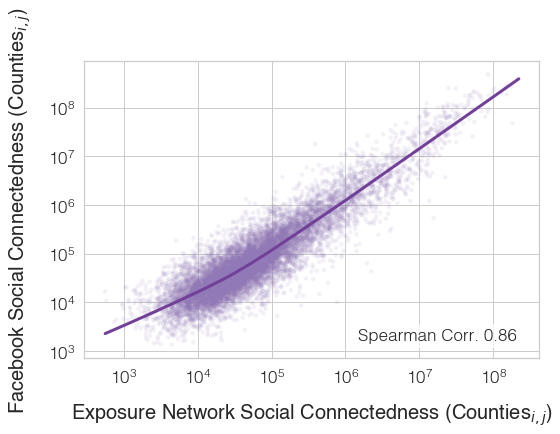}
   \caption{}
\end{subfigure}
\begin{subfigure}[t]{.5\textwidth}
\includegraphics[width=\textwidth,keepaspectratio]{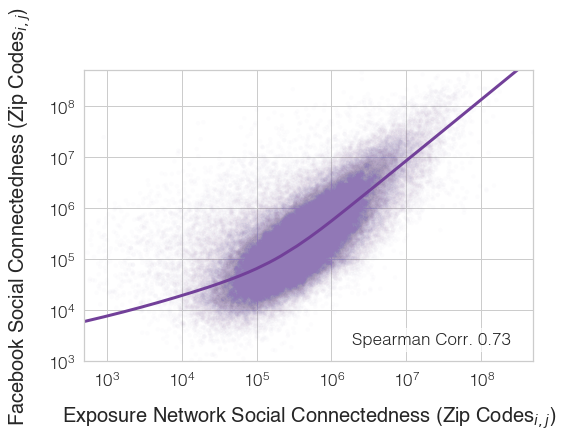}
   \caption{}
\end{subfigure}
\hspace*{-15mm}
\begin{subfigure}[t]{.5\textwidth}
  \includegraphics[width=\textwidth,keepaspectratio]{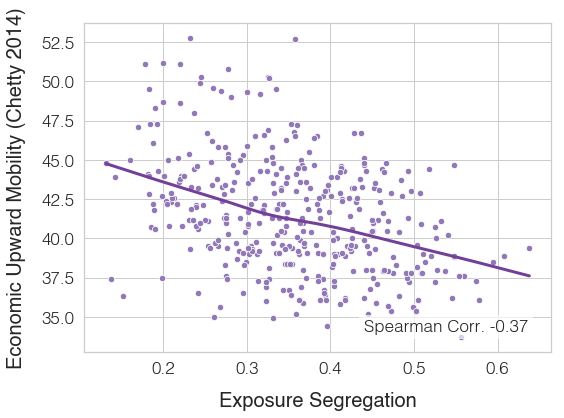}
  \caption{}
\end{subfigure}
\begin{subfigure}[t]{.5\textwidth}
  \includegraphics[width=\textwidth,keepaspectratio]{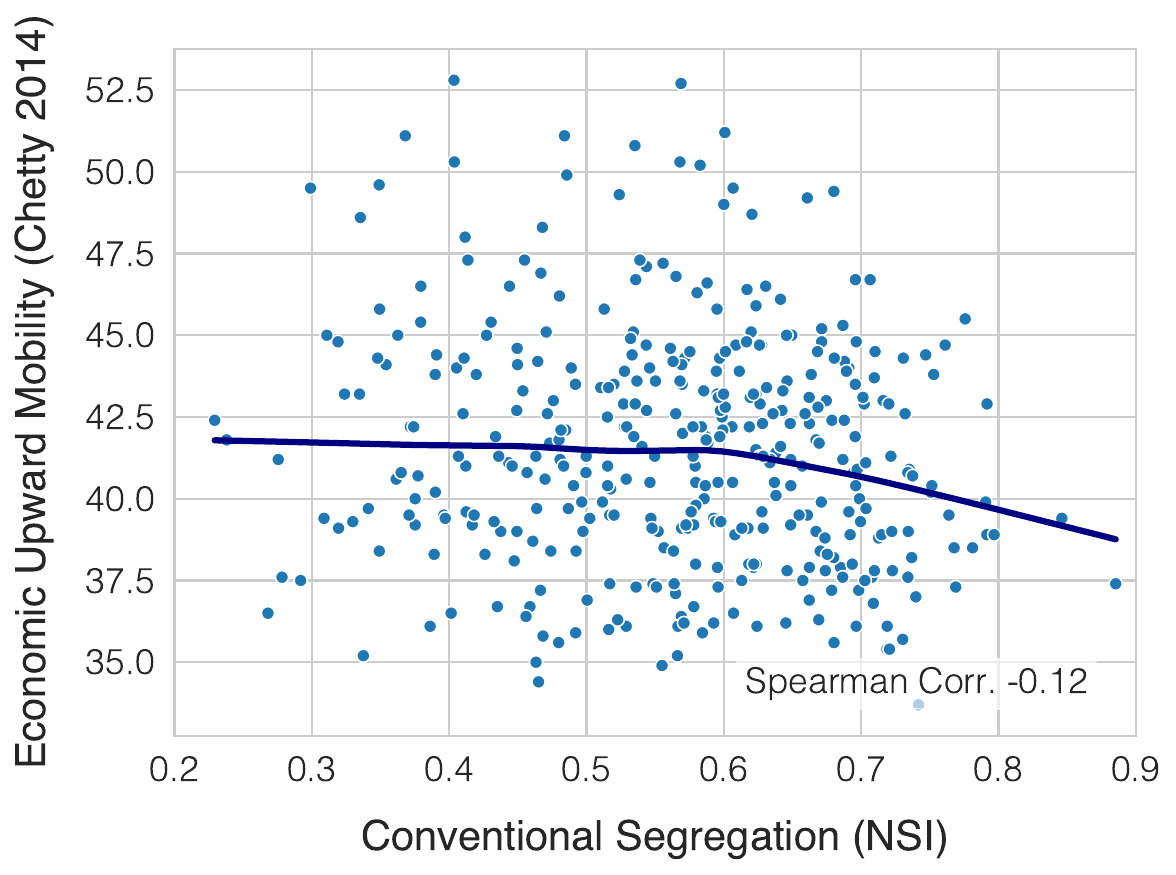}
  \caption{}
\end{subfigure}
   
   \caption{\textbf{This studies' exposure network predicts population-scale friendship formation and upward economic mobility outcomes}. We measure the external validity of our definition of exposure, by linking our exposure network to  outcomes across two gold-standard, large-scale, datasets. We find at the zip code, county, and MSA-level, our exposure network mirrors population-scale outcomes resulting from dynamic human processes: \textbf{(a-b)} the Facebook Social Connectedness Index\cite{bailey2018social} measures the relative probability of a Facebook friendship link between a given Facebook user in location $i$ and a given user in location $j$. FB Social Connectedness Index has been used social segregation\cite{bailey2020social}, and has also been linked to economic\cite{bailey2021international,kuchler2020social} and public health outcomes\cite{kuchler2021jue}. We reproduce the Social Connectness Index using our exposure network ($\frac{\# Exposure Pairs_{i,j}}{\# Individuals_{i} \cdot \# Individuals_{j}}$) at the county \textbf{(a)} and zip code \textbf{(b)} level, and find strong correlations across county pairs (Spearman Correlation 0.85, $N=121,595$, $p  <10^{-4}$) and zip code pairs (Spearman Correlation 0.73, $N=1,053,539$, $p  <10^{-4}$). \textbf{(c-d)} The Chetty et al. Intergenerational Mobility dataset quantifies upward economic mobility from federal income tax records for each MSA as the mean income rank of children with parents in the bottom half of the income distribution~\cite{chetty2014land}. We find that \metric at the MSA-level \textbf{(c)} correlates to (absolute) upward economic mobility (Spearman Correlation -0.37, $N=379$, $p  <10^{-4}$), and does so significantly more strongly ($p  <10^{-4}$) than \textbf{(d)} the conventional segregation measure NSI (Spearman Correlation -0.12, $N=379$, $p  < 0.05$)}.
   \label{fig:external_validation}
\end{figure}

\begin{figure}[htbp]

\vspace{-20mm}
   \centering
   \hspace*{-17mm}
   \includegraphics[width=1.0\textwidth,keepaspectratio]{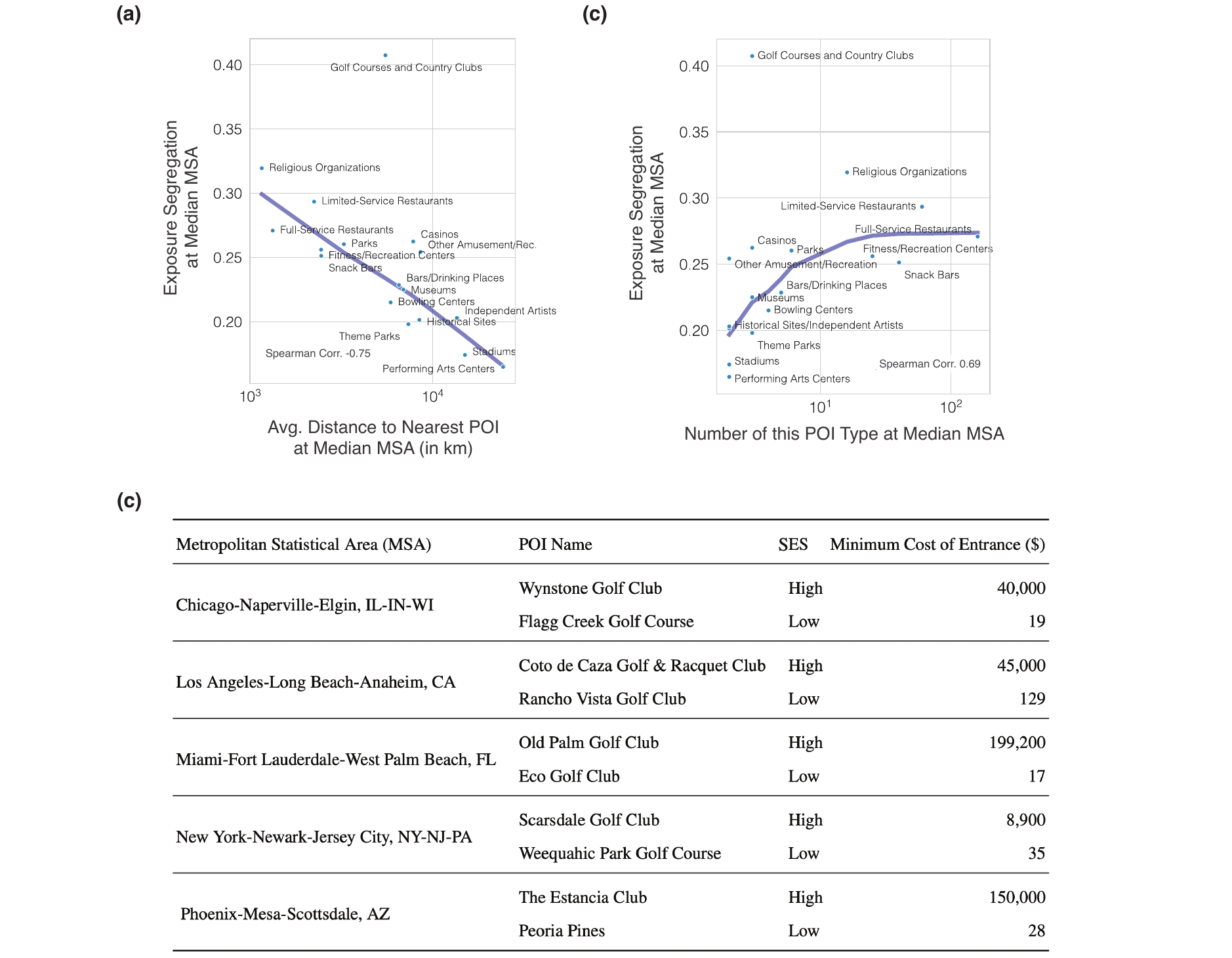}

\caption{\textbf{Understanding why \metric varies significantly across leisure sites.} We identify three primary facets of socioeconomic differentiation between POIs which explain the heterogeneous segregation levels of different leisure POIs (Figure \ref{fig:main_1}e): \textbf{(a)} localization, \textbf{(b)} quantity, and \textbf{(c)} stratification. \textbf{(a)} Localization strongly predicts segregation across all POI categories (Spearman Correlation -0.75, N=17, p$<$0.001). POIs which are more locally embedded into neighborhoods (e.g. religious organizations) are more segregated than activities in which POIs are serve multiple neighborhoods (e.g. stadiums). We operationalize localization as the average distance from each indvidual in the MSA to the nearest POI of that category. \textbf{(a)} The quantity of POIs also explains segregation (Spearman Correlation 0.69, N=17, p$<$0.01). Leisure activities with more options (e.g. restaurants) have differentiated venues catering to a specific economic standing (e.g. Michelin-star restaurants) compared to POIs which are small in number and cater to the overall city (e.g. stadiums) \textbf{(c)} Golf courses and country clubs (golf clubs) are an anomaly in that they have a small number of unlocalized POIs, but are highly segregated. We conduct a case study in which look at top and bottom golf clubs by mean visitor SES in 5 of the 10 largest MSAs. We find that the high segregation of golf clubs is due to extreme stratification between venues; for instance the minimum cost to play at the high-SES golf course in Miami, FL is 11717$\times$ higher than at the lowest-SES golf course. By contrast, the average cost of a MacDonalds Big Mac (\$5.65\cite{bigmac}) is only 63$\times$ higher than the average cost of a Michelin 3-star restaurant (\$357\cite{michelin}). Finally, these findings foreshadow Bridging Index (BI), which captures POI localization, quantity, and stratification (Extended Data Figure \ref{fig:ccbi_intuition}).}

 \label{fig:poi_disparity_explanation}
\end{figure}

\begin{figure}[htbp]
\vspace{20mm}
   \centering
   \hspace*{-17mm}
   \includegraphics[width=1.2\textwidth,keepaspectratio]{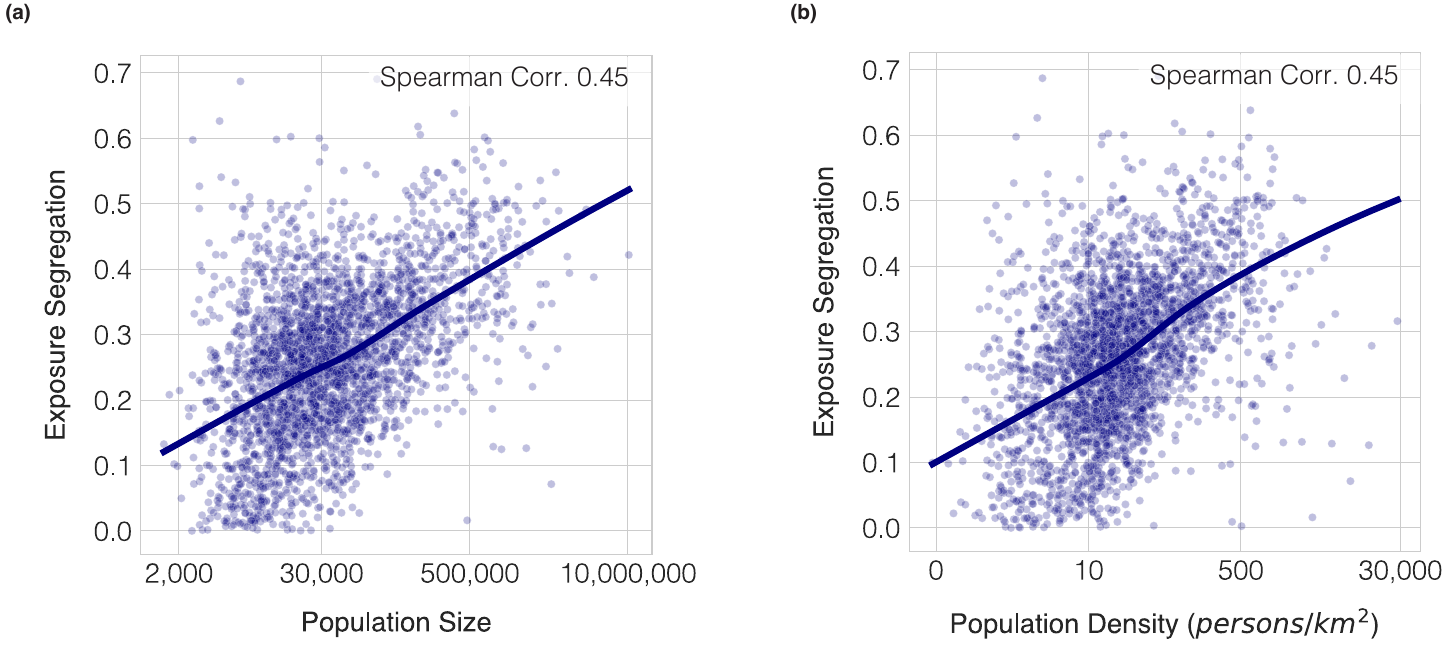}
   \caption{\textbf{Large, dense counties are more segregated}. We compute \metric across 2829 USA counties (94\% of the counties in the USA), excluding counties in which there are less than 50 individuals in our dataset. We find that at the county-level, \metric is also positively correlated with population size (Spearman Correlation 0.45, N=2829, p$<10^{-4}$) and population density (Spearman Correlation 0.45, N=2829, p$<10^{-4}$). These correlations reveal that the association between large, dense cities and \metric (Figure \ref{fig:main_2}a) is not an artifact of city boundaries, and may in fact be an emergent property from dynamics of individuals residing highly populated, dense geographic areas, which persists across multiple scales of granularity.}
   \label{fig:county_correlations}
\end{figure}

\begin{table}[ht]
\vspace*{-20mm}
\centering
\begin{scriptsize}
\setlength\tabcolsep{2pt} %
\begin{tabular}{@{\extracolsep{5pt}}lcccccc}
\\[-1.8ex]\hline
\hline \\[-1.8ex]
& \multicolumn{6}{c}{\textit{Dependent variable:}} \ Exposure Segregation
\cr \cline{6-7}
\\[-1.8ex] & (1) & (2) & (3) & (4) & (5) & (6) \\
\hline \\[-1.8ex]
 Intercept & 0.355$^{***}$ & 0.355$^{***}$ & 0.355$^{***}$ & 0.355$^{***}$ & 0.355$^{***}$ & 0.355$^{***}$ \\
  & (0.004) & (0.004) & (0.003) & (0.003) & (0.003) & (0.003) \\
 Log(Population Size) & 0.059$^{***}$ & & 0.041$^{***}$ & 0.044$^{***}$ & 0.026$^{***}$ & 0.028$^{***}$ \\
  & (0.004) & & (0.004) & (0.004) & (0.004) & (0.004) \\
 Gini Index (Estimated Rent) & & 0.064$^{***}$ & 0.050$^{***}$ & 0.051$^{***}$ & 0.045$^{***}$ & 0.047$^{***}$ \\
  & & (0.004) & (0.004) & (0.004) & (0.003) & (0.003) \\
 Political Alignment (\% Democrat in 2016 Election) & & & & 0.004$^{}$ & & 0.004$^{}$ \\
  & & & & (0.004) & & (0.004) \\
 Racial Demographics (\% non-Hispanic White) & & & & 0.001$^{}$ & & 0.006$^{*}$ \\
  & & & & (0.004) & & (0.003) \\
 Mean SES (Estimated Rent) & & & & -0.012$^{***}$ & & -0.005$^{}$ \\
  & & & & (0.004) & & (0.004) \\
 Walkability (Walkscore) & & & & & 0.002$^{}$ & 0.001$^{}$ \\
  & & & & & (0.003) & (0.003) \\
 Commutability (\% Commute to Work) & & & & & -0.011$^{***}$ & -0.010$^{***}$ \\
  & & & & & (0.003) & (0.004) \\
 Conventional Segregation (NSI) & & & & & 0.042$^{***}$ & 0.041$^{***}$ \\
  & & & & & (0.003) & (0.003) \\
\hline \\[-1.8ex]
 Observations & 382 & 382 & 382 & 376 & 382 & 376 \\
 $R^2$ & 0.350 & 0.419 & 0.567 & 0.578 & 0.704 & 0.705 \\
 Adjusted $R^2$ & 0.348 & 0.417 & 0.565 & 0.573 & 0.701 & 0.698 \\
\hline
\hline \\[-1.8ex]
\textit{} & \multicolumn{6}{r}{$^{*}$p$<$0.1; $^{**}$p$<$0.05; $^{***}$p$<$0.01} \\
\end{tabular}

\end{scriptsize}
\caption{\textbf{Population size is significantly associated with \metric, after controlling for MSA income inequality (Gini Index), political alignment (\% Democrat in 2016 election), racial demographics (\% non-Hispanic White), mean ES, walkability (Walkscore\cite{walkscore}), commutability (\% of residents commuting to work), and residential segregation (NSI)}. Here we show the coefficients (after normalizing via z-scoring to have mean 0 and variance 1) from the primary specifications estimating the effect of population size on \metric across all MSAs. Columns (1-5) are models specified with different subsets of covariates; Column 6 shows model specification with all covariates. Differences between sample size in models is due to missing data for several covariates in a small number of MSAs (Walkscores were not available for all MSAs). (*p $<$ 0.1; **p $<$ 0.05; *** p $<$ 0.01).}
\label{tab:segregation_population}
\end{table}

\begin{table}[ht]
\vspace*{-25mm}

\centering
\begin{scriptsize}
\setlength\tabcolsep{2pt} %

\begin{tabular}{@{\extracolsep{5pt}}lccccc}
\\[-1.8ex]\hline
\hline \\[-1.8ex]
& \multicolumn{5}{c}{\textit{Dependent variable:}} \ Exposure Segregation
\cr \cline{5-6}
\\[-1.8ex] & (1) & (2) & (3) & (4) & (5) \\
\hline \\[-1.8ex]
 Intercept & 0.355$^{***}$ & 0.355$^{***}$ & 0.355$^{***}$ & 0.355$^{***}$ & 0.355$^{***}$ \\
  & (0.003) & (0.003) & (0.003) & (0.003) & (0.003) \\
  Bridging Index & -0.078$^{***}$ & -0.059$^{***}$ & -0.058$^{***}$ & -0.035$^{***}$ & -0.036$^{***}$ \\
  & (0.003) & (0.005) & (0.005) & (0.006) & (0.006) \\
 Log(Population Size) & & 0.003$^{}$ & 0.008$^{*}$ & 0.010$^{**}$ & 0.017$^{***}$ \\
  & & (0.004) & (0.005) & (0.004) & (0.006) \\
 Gini Index (Estimated Rent) & & 0.031$^{***}$ & 0.032$^{***}$ & 0.035$^{***}$ & 0.036$^{***}$ \\
  & & (0.003) & (0.003) & (0.003) & (0.003) \\
 Political Alignment (\% Democrat in 2016 Election) & & & 0.001$^{}$ & & 0.002$^{}$ \\
  & & & (0.004) & & (0.004) \\
 Racial Demographics (\% non-Hispanic White) & & & 0.003$^{}$ & & 0.005$^{}$ \\
  & & & (0.003) & & (0.003) \\
 Mean SES (Estimated Rent) & & & -0.009$^{**}$ & & -0.005$^{}$ \\
  & & & (0.004) & & (0.003) \\
 Walkability (Walkscore) & & & & 0.002$^{}$ & 0.001$^{}$ \\
  & & & & (0.003) & (0.003) \\
 Commutability (\% Commute to Work) & & & & -0.011$^{***}$ & -0.009$^{**}$ \\
  & & & & (0.003) & (0.004) \\
 Conventional Segregation (NSI) & & & & 0.028$^{***}$ & 0.026$^{***}$ \\
  & & & & (0.004) & (0.004) \\
 \# of Exposure Hubs & & & & & -0.006$^{}$ \\
  & & & & & (0.005) \\
\hline \\[-1.8ex]
 Observations & 382 & 382 & 376 & 382 & 376 \\
 $R^2$ & 0.620 & 0.686 & 0.693 & 0.733 & 0.736 \\
 Adjusted $R^2$ & 0.619 & 0.684 & 0.688 & 0.729 & 0.729 \\
\hline
\hline \\[-1.8ex]
\textit{} & \multicolumn{5}{r}{$^{*}$p$<$0.1; $^{**}$p$<$0.05; $^{***}$p$<$0.01} \\
\end{tabular}

\end{scriptsize}
\caption{\textbf{Bridging Index (BI) is significantly associated with \metric, after controlling for population size, \# of hubs, MSA income inequality (Gini Index), political alignment (\% Democrat in 2016 election), racial demographics (\% non-Hispanic White), mean ES, walkability (Walkscore\cite{walkscore}), commutability (\% of residents commuting to work), and residential segregation (NSI)}. Here we show the coefficients (after normalizing via z-scoring to have mean 0 and variance 1) from the primary specifications estimating the effect of population size on \metric across all MSAs. Columns (1-4) are models specified with different subsets of covariates; Column 5 shows model specification with all covariates. Differences between sample size in models is due to missing data for several covariates in a small number of MSAs (Walkscores were not available for all MSAs). (*p $<$ 0.1; **p $<$ 0.05; *** p $<$ 0.01).}
\label{tab:segregation_ccbi}
\end{table}

\begin{table}[ht]
\vspace*{-25mm}

\centering
\begin{small}
\setlength\tabcolsep{2pt} %

\begin{tabular}{lrr}
\toprule
                     Measure &  Spearman $\rho^2$ &  Pearson $R^2$ \\
\midrule
                   \textbf{Bridging Index }&            \textbf{0.60} &           \textbf{0.62} \\
              Log(Population Size) &            0.39 &           0.35 \\
                        Gini Index (Estimated Rent) &            0.41 &           0.42 \\
               Political Alignment (\% Democrat in 2016 Election) &            0.06 &           0.05 \\
                  Racial Demographics (\% non-Hispanic White) &            0.09 &           0.05 \\
Mean SES (Estimated Rent)  &            0.09 &           0.05 \\
                   Walkability (Walkscore) &            0.01 &           0.02 \\
 Commutability (\% Commute to Work) &            0.04 &           0.03 \\
                         Conventional Segregation (NSI) &            0.44 &           0.42 \\
                   \# of Exposure Hubs  &            0.44 &           0.16 \\
\bottomrule
\end{tabular}

\end{small}
\caption{\textbf{Bridging Index (BI) strongly predicts \metric} and does so more accurately ($p$${<}$$10^{-4}$, Steiger's Z-test) than population size, \# of hubs, MSA income inequality (Gini Index), political alignment (\% Democrat in 2016 election), racial demographics (\% non-Hispanic White), mean ES, walkability (Walkscore\cite{walkscore}), commuthttps://www.overleaf.com/project/5fa8a3a6d3e9747a8dcef20dability (\% of residents commuting to work), and residential segregation (NSI)}.
\label{tab:segregation_corrs}
\end{table}

\begin{figure}[htbp]
\centering
\hspace*{-17mm}
\includegraphics[width=1.2\textwidth,keepaspectratio]{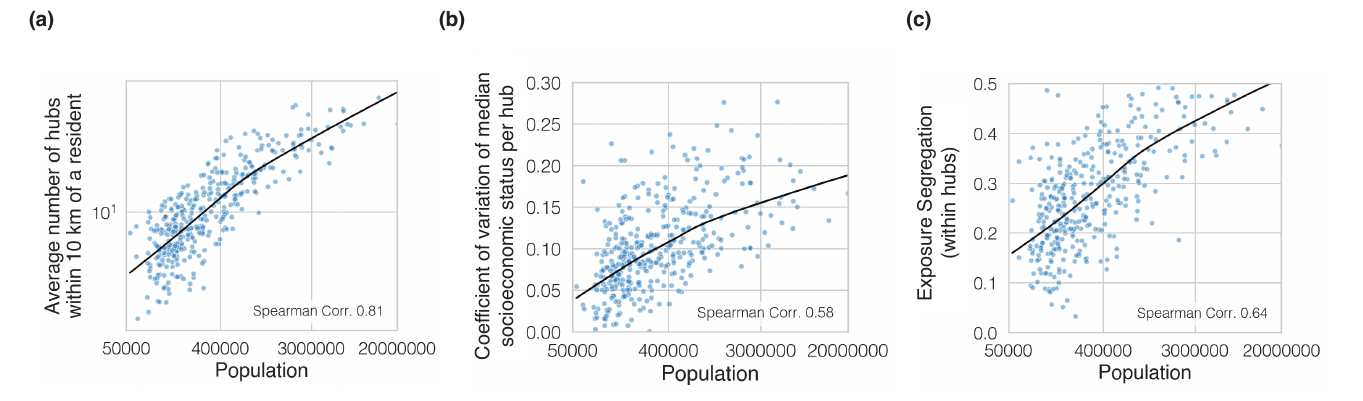}
\caption{\textbf{At higher levels of scale, spaces in large cities are more differentiated and consequently segregated: exposure hubs.} \textbf{(a-c)} Conducting an analogous analysis to that for restaurants in Figure~\ref{fig:main_3}c-e for exposure hubs (i.e. commercial centers which are higher-level clusters of restaurants, grocery stores, etc.). We find that higher segregation is driven by an increase in highly differentiated choice of exposure hubs in large cities: (\textbf{a}) Larger MSAs have more exposure hubs, giving residents more options to self-segregate (Spearman Correlation 0.81, N=382, p$<10^{-4}$). \textbf{(b)} Consequently, hubs in larger MSAs vary more in terms of the mean SES of their visitors (Spearman Correlation 0.58, N=382, p$<10^{-4}$) and as a result, (\textbf{c}) exposure segregation within hubs is higher in larger MSAs (Spearman Correlation 0.64, N=382, p$<10^{-4}$). Overall, this analysis suggests that across multiple levels of scale, large cities offer a greater choice of differentiated spaces targeted to specific socioeconomic groups, promoting everyday segregation in exposures. }
 \label{fig:hub_segregation}
\end{figure}

\begin{figure}[htbp]
\vspace{-20mm}

\centering

\begin{subfigure}[t]{.4\textwidth}
\includegraphics[width=\textwidth,keepaspectratio]{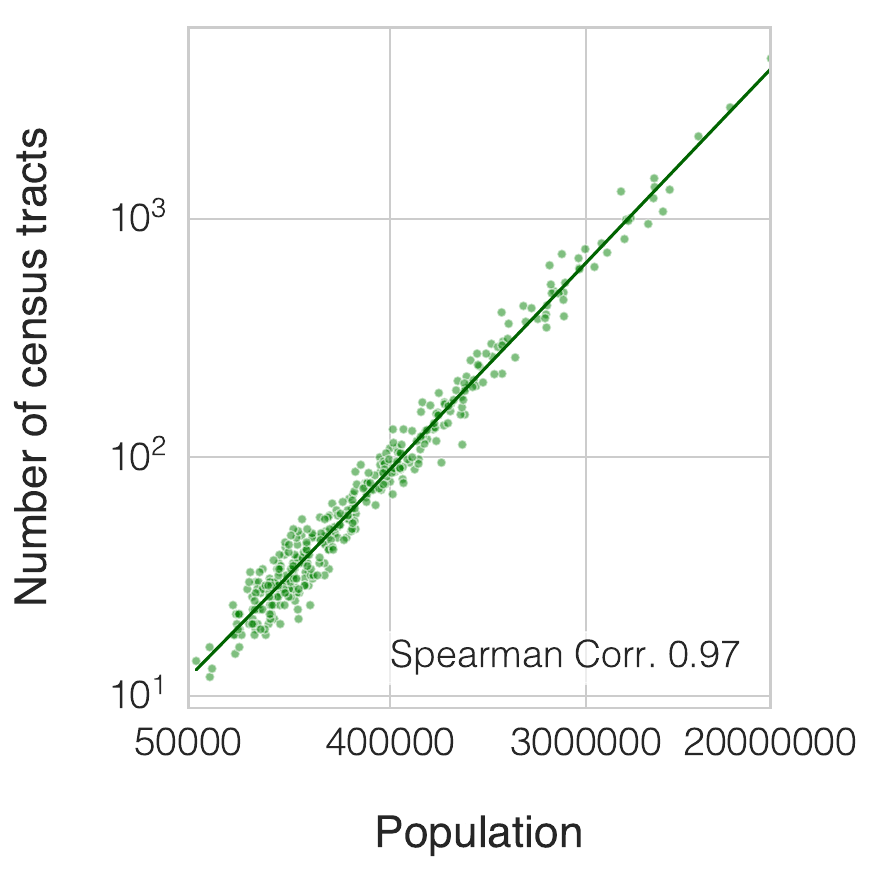}
   \caption{}
\end{subfigure}
\begin{subfigure}[t]{.4\textwidth}
\includegraphics[width=\textwidth,keepaspectratio]{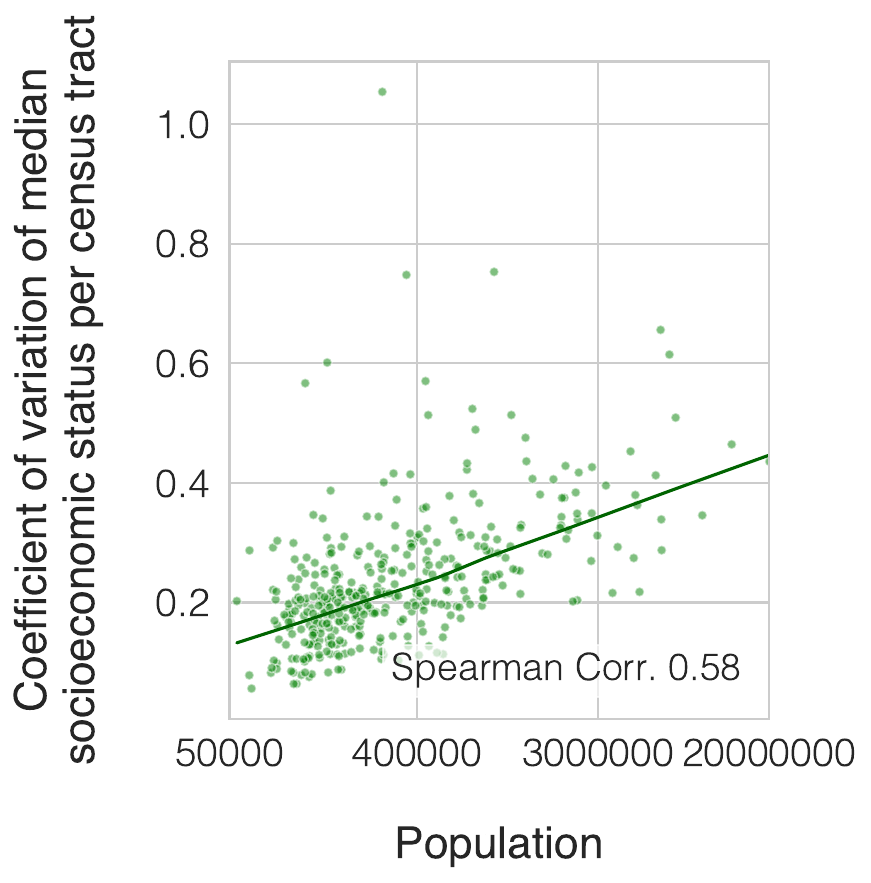}
   \caption{}
\end{subfigure}

\begin{subfigure}[t]{.5\textwidth}
  \includegraphics[width=\textwidth,keepaspectratio]{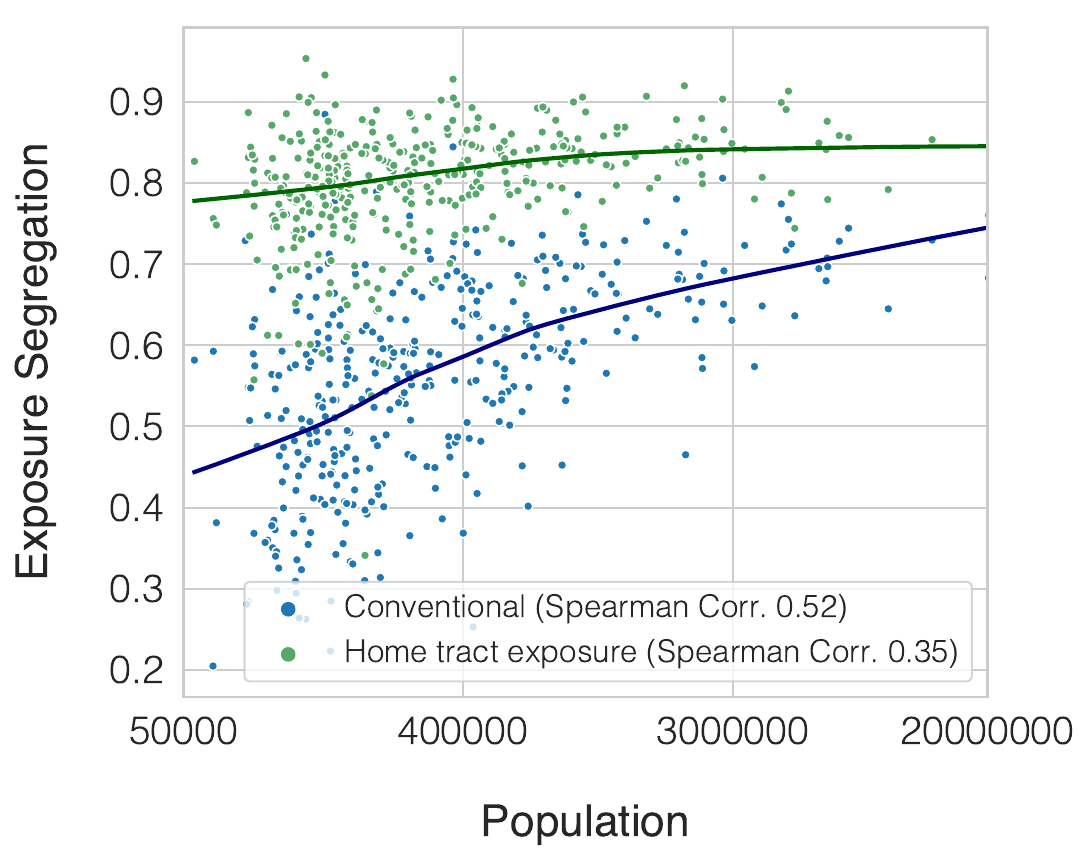}
  \caption{}
\end{subfigure}
\begin{subfigure}[t]{.4\textwidth}
  \includegraphics[width=\textwidth,keepaspectratio]{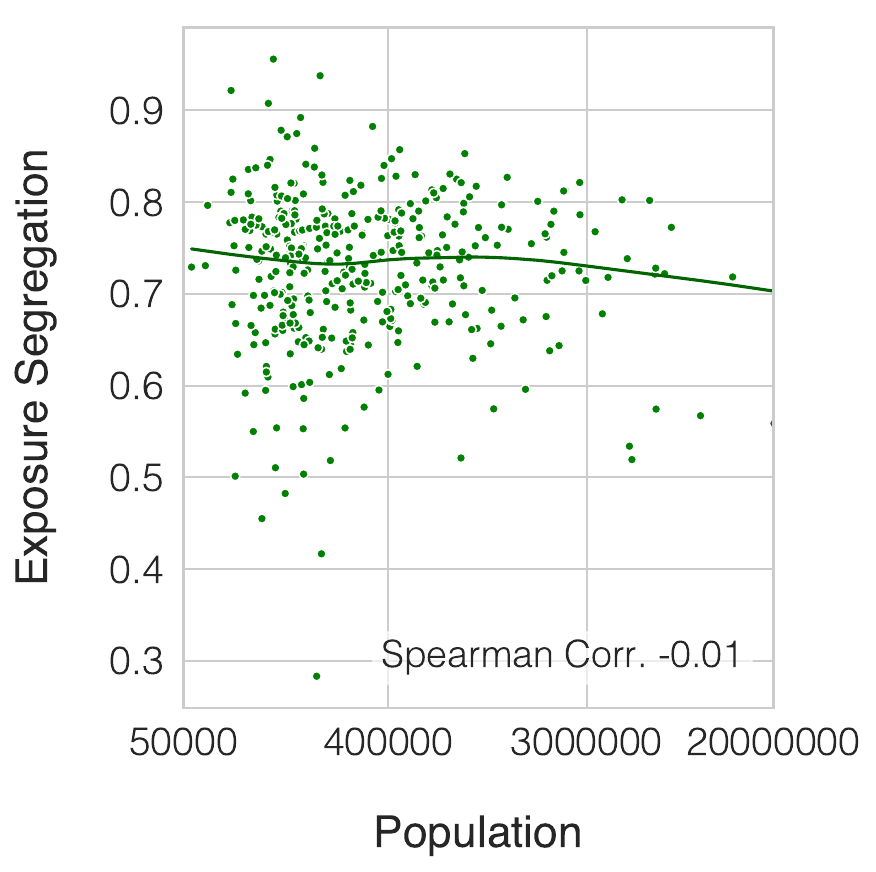}
  \caption{}
\end{subfigure}

\caption{\textbf{At higher levels of scale, spaces in large cities are more differentiated and consequently segregated: home neighborhoods.} \textbf{(a-c)} Conducting an analogous analysis to that for restaurants in Figure~\ref{fig:main_3}c-e, we find that higher segregation is driven by an increase in highly differentiated choice of neighborhoods in large cities: (\textbf{a}) Larger MSAs have more census tracts, giving residents more options to self-segregate (Spearman Correlation 0.97, N=382, p$<10^{-4}$). \textbf{(b)} Consequently, census tracts in larger MSAs vary more in terms of the mean SES of their residents (Spearman Correlation 0.58, N=382, p$<10^{-4}$) and as a result, (\textbf{c}) both conventional NSI and exposure segregation are higher (Spearman Correlations 0.52 and 0.35, N=382, p$<10^{-4}$ and p$<10^{-4}$). However, \textbf{(c)} also shows that exposure segregation (green series) rises more slowly with population than conventional segregation (blue series), suggesting that within-home-tract homophily, which increases exposure segregation but not conventional segregation, is \emph{not} more pronounced in large MSAs. Substantiating this, \textbf{(d)} shows that when home tract exposure segregation is computed using an alternate SES measure so it captures only within-home-tract-homophily, it is no higher in large MSAs (Spearman Correlation -0.01, N=382, p$>0.1$). (The alternative SES measure is computed by subtracting the mean SES in each Census tract; see Methods.) Overall, this analysis suggests that the higher home tract segregation in large MSAs is driven by people's greater choice of neighborhoods of varying SES in which to live, but not by a greater tendency to cross paths homophilously within their own neighborhood.}
 \label{fig:home_segregation}
\end{figure}

\begin{figure}[htbp]
   \centering
      \vspace*{-20mm}
   \includegraphics[width=1.0\textwidth,keepaspectratio]{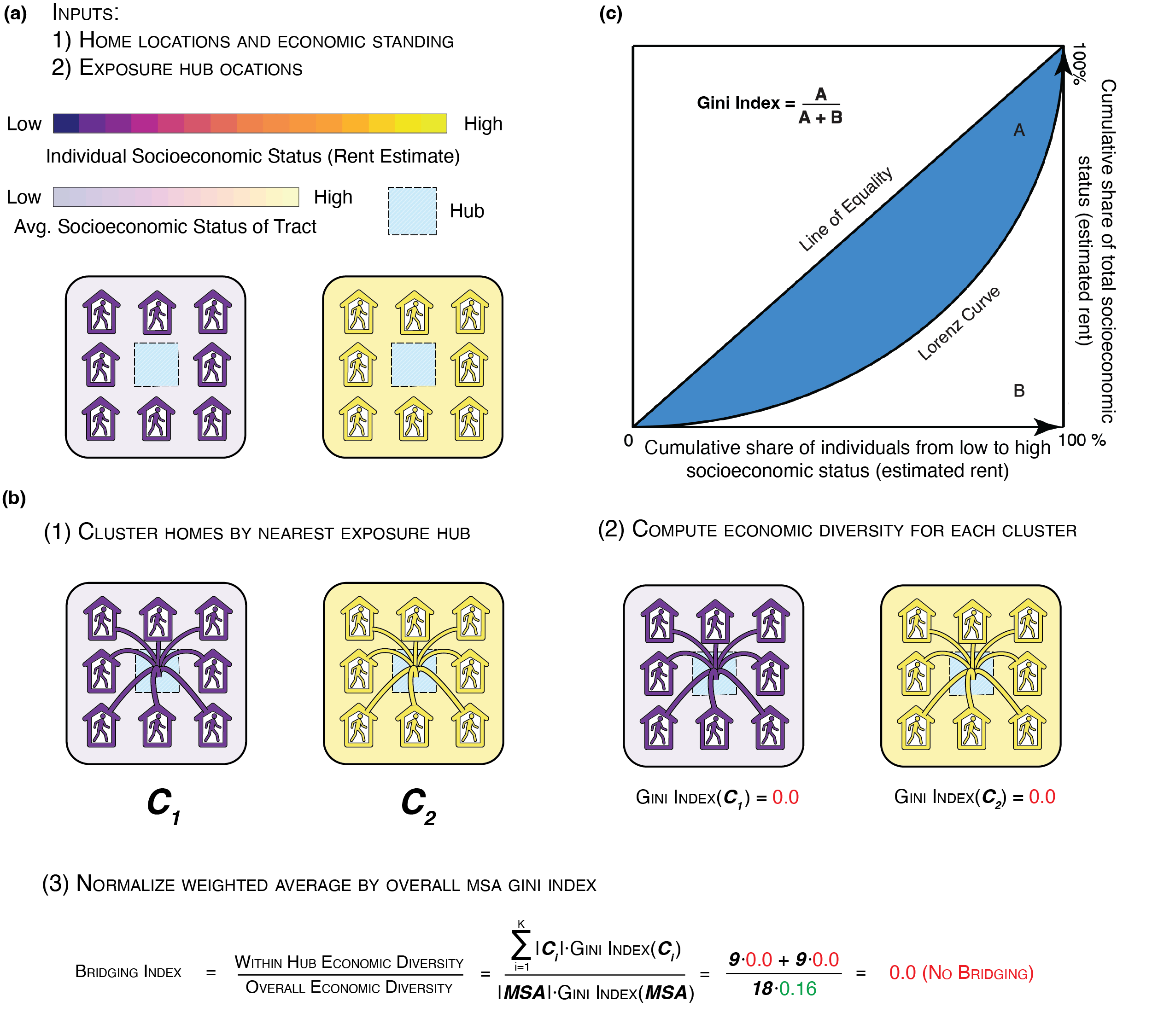}
   \caption{\textbf{Computing Bridging Index (BI)}. Illustration of our analytical pipeline for calculating BI. \textbf{(a)} BI is computed from the locations and number of POIs in the MSA which are expected to be hubs of exposure, as well the locations and economic standing values of all homes within MSA boundaries. We intentionally develop BI without using mobility data, with the intention of identifying a modifiable extrinsic aspect of an MSA that can be intervened on to \emph{impact} mobility patterns and decrease \metric  \textbf{(b)} In order, we (1) cluster all homes by nearest exposure hub (using straight line distance from home to hub), partitioning all homes into $K$ clusters, where $K$ is the number of hubs in the MSA  (2) compute the weighted average economic diversity (i.e. Gini Index) of the clusters, normalized by the overall economic diversity of the MSA to allow for comparisons between different MSAs of varying baseline levels of economic diversity (Extended Data Table~\ref{tab:segregation_population}) \textbf{(c)} The graphical definition of Gini Index is provided, which is a standard measure of economic dispersion\cite{dorfman1979formula}. Results are robust to the definition of economic diversity, and holds true when using variance in SES instead of Gini Index (Supplementary Figure \ref{fig:ccbi_variance}).}
   \label{fig:explain_metrics}
\end{figure}

\begin{figure}[htbp]
   \centering
   \vspace{-20mm}
   \includegraphics[width=1.0\textwidth,keepaspectratio]{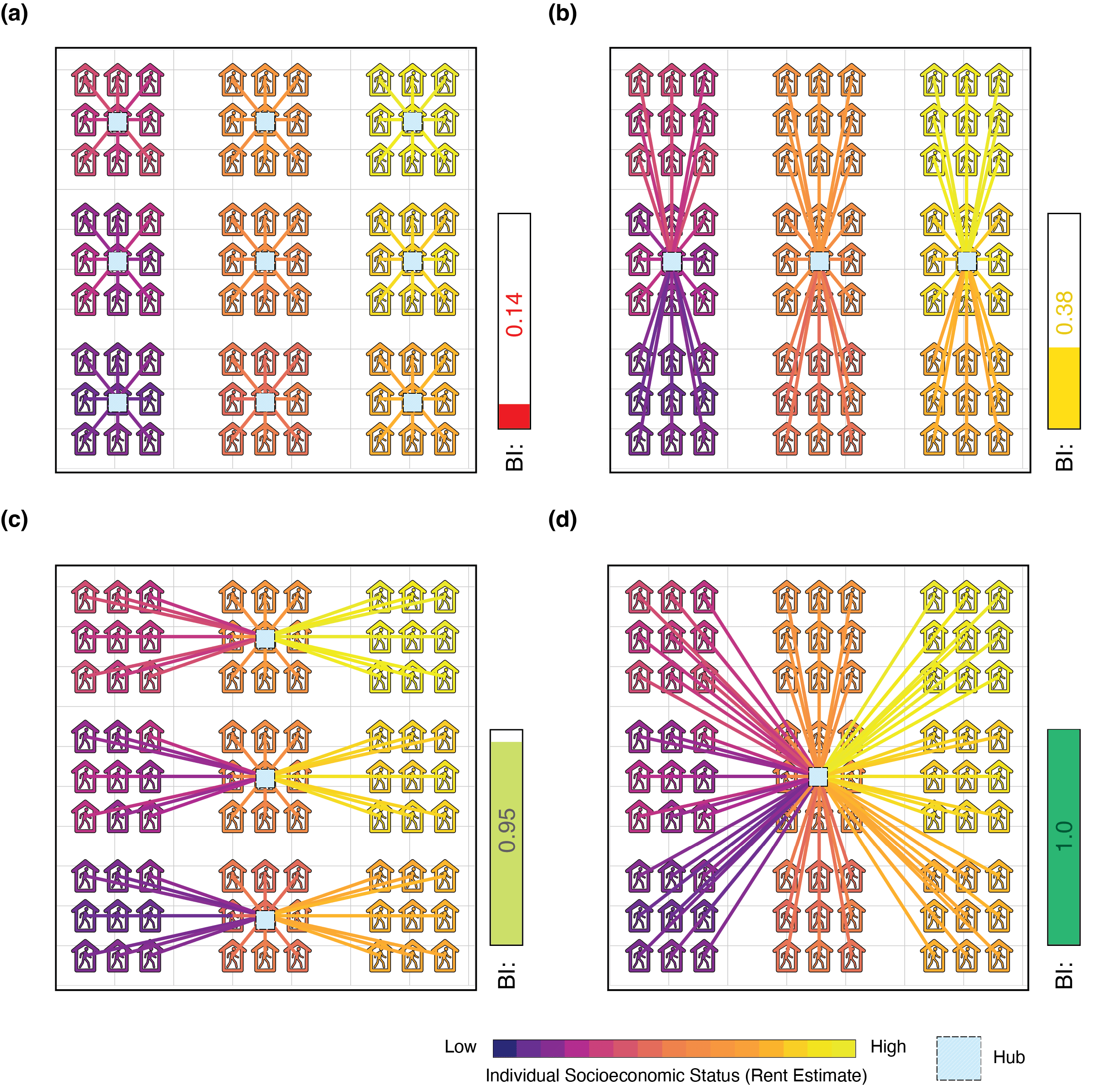}
\caption{\textbf{Understanding the determinants of BI.} The Bridging Index (BI) is a single metric which captures three important factors of built environment (see Supplementary Figure \ref{fig:ccbi_ablation} for contributions of these factors to explaining \metric): \\ (1) The locations of exposure hubs — If hubs are located in between diverse neighborhoods, BI will be high as hubs will bridge together diverse individuals.\\ (2) The \# of exposure hubs —  as \# of hubs decreases, BI increases (e.g if there is only 1 hub in a city, BI will be 1.0 as all individuals are unified by a single hub) \\(3) Residential segregation, i.e. the locations of homes and their associated economic standing — as residential segregation decreases we can expect that individuals residing near each hub will be more diverse.\\
This figure builds intuition for BI by showing how BI may vary for a single simulated city, consisting of highly segregated neighborhoods. We hold residential segregation (3) constant, and vary the location (1) and number (2) of exposure hubs across panels \textbf{(a)}, \textbf{(b)}, \textbf{(c)}, \textbf{(d)}, in order of increasing BI. Note that BI in \textbf{(c)} is substantially higher than BI in \textbf{(b)}, because hubs in \textbf{(c)} are better positioned to bridge diverse neighborhoods---even though the number of hubs remains constant.}
\label{fig:ccbi_intuition}
\end{figure}

\begin{figure}[htbp]

   \vspace{-20mm}
   \centering
   \hspace*{-17mm}
   \includegraphics[width=1.2\textwidth,keepaspectratio]{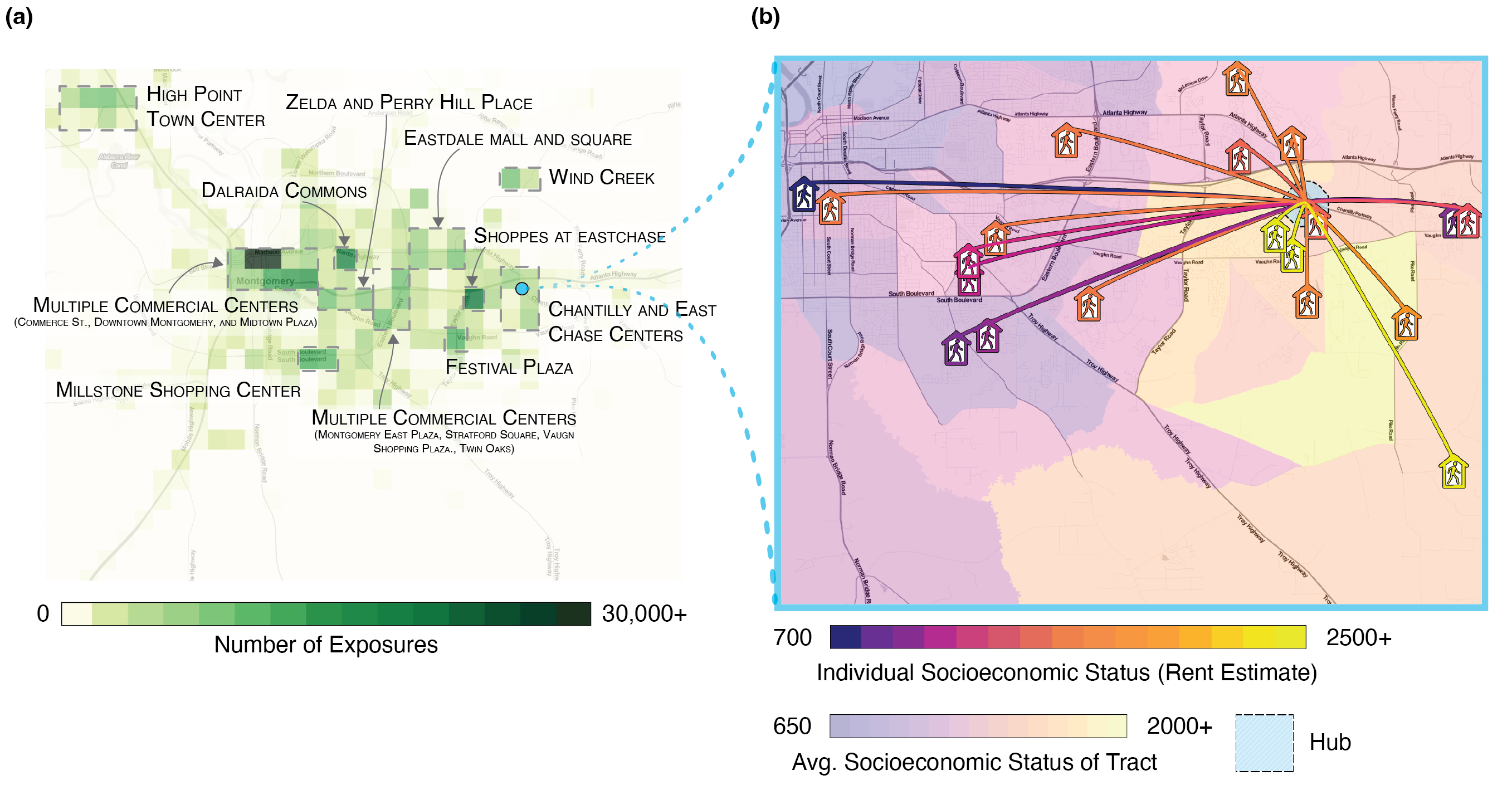}

   \caption{\textbf{Montgomery, AL}. We conduct an analogous analysis to Figure 3a,b but for Montgomery, AL, which has nearly identical population (374K vs 385K residents) and income inequality (55th vs 60th percentile Gini Index) to Fayetteville, NC but is 74\% more segregated (88th percentile vs. 21st percentile \metric). We find that the difference in segregation is explained by Montgomery, AL having a significantly higher BI compared to Fayetteville, NC (65th vs. 13th percentile). In Montgomery, AL exposure hubs (i.e. commercial centers) are differentiated by SES which results in high-SES individuals and low-SES individuals visiting separate hubs and prevents them from engaging in cross-SES exposures. (a) shows that, as with all MSAs, commercial centers  (e.g. shopping malls, plazas, etc.) are hubs of exposure. We illustrate that in Montgomery, AL all visually discernible hubs are associated with one or more commercial centers. (b) In Montgomery, AL, exposure hubs are located in different locations which cater separate to high and low SES residents, leading to segregated exposures. As an illustrative example, we show a zoomed-in map of one hub (Chantilly Center) in Montgomery, AL, and display a random sample of 10 exposures occurring inside of it. Chantilly Center in Montgomery, AL is located accessibly for high SES individuals but is far apart from low-SES tracts. As a result, the sample shows that the majority of exposures are middle-upper ES, and only a few low-SES individuals visit Chantilly Center and cross paths with these high-SES individuals. Home icons demarcate individual home location (up to 100 meters of random noise added to preserve anonymity); home colors denote individual ES; arcs indicate an exposure inside of the hub; background colors indicate mean census tract ES.}
   \label{fig:montgomery}
\end{figure}

\begin{figure}[htbp]

   \centering
   \includegraphics[width=1.0\textwidth,keepaspectratio]{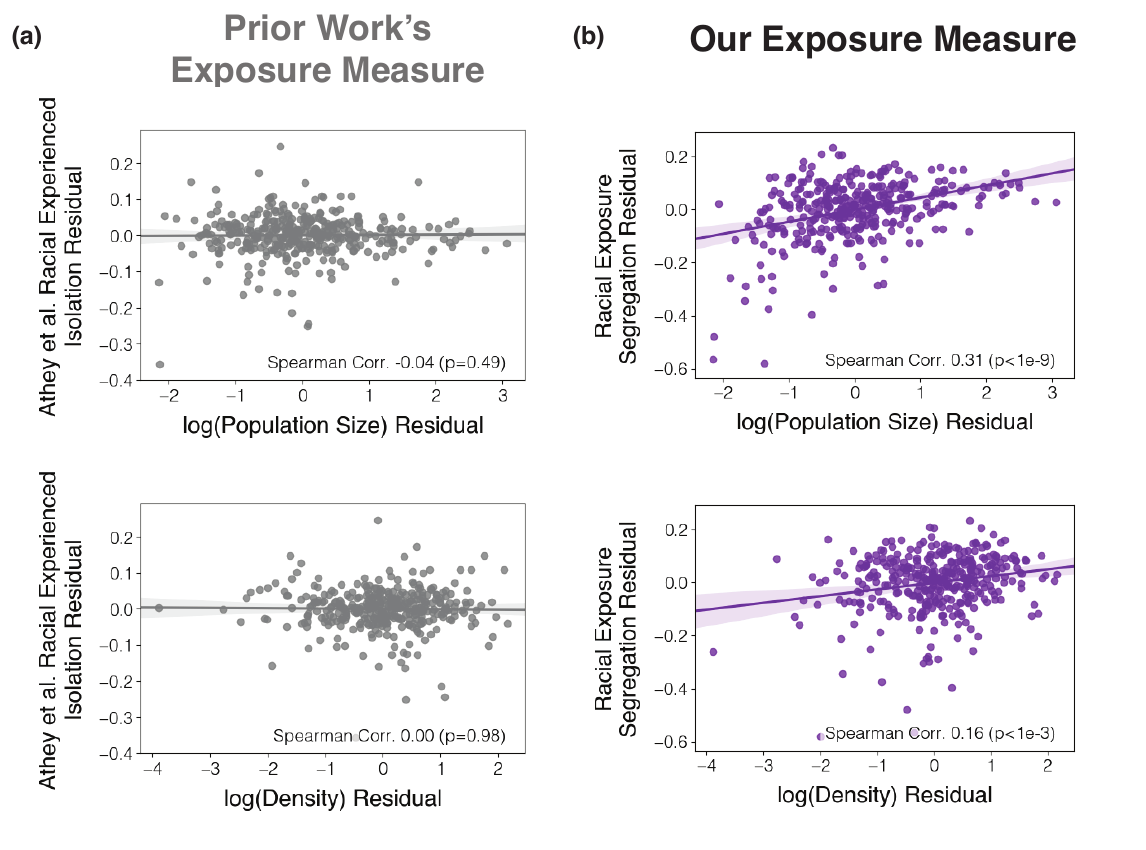}

   \caption{\new{\textbf{Our time-sensitive and high-resolution measure of exposure enables us to reject the cosmopolitan mixing hypothesis}. We analyze the consequences of  different measures of exposure by reproducing Athey et al.'s assessment of the cosmopolitan mixing hypothesis\cite{athey2020experienced}. Athey et al. estimate the correlation between population density and their measure of racial segregation, after controlling for residential (racial) segregation. They do so by correlating  the residuals of their racial segregation measure with the residuals of a population density, where the residuals are derived from regression on (racial) residential segregation controls.  We extend this analysis by using our own time-sensitive and high-resolution measure of racial exposure segregation (Supplementary Figure \ref{fig:robust_10}). \textbf{(a)} We find that when using Athey et al.'s measure (which consider two people exposed to each other if they ever visited the same 153 x 153m grid cell within 4 months), there is no significant correlation between population size and racial segregation (Spearman Corr.  -0.04, $p=0.49$), nor between population density and experienced segregation (Spearman Corr.  0.00, $p=0.98$). Athey et al.'s robustness checks further corroborate this point, showing no significant association between population density and segregation when MSAs are unweighted by population size (Athey et al. Supplementary Table S8). \textbf{(b)} When we use our own time-sensitive and  high-resolution  racial segregation measure, we find positive and statistically significant relationships between racial exposure segregation and population size (Spearman Corr.  0.31, $p<1\times 10^{-9}$
) as well as population density (Spearman Corr.  0.16, $p<0.001$). This comparison shows that precisely measuring exposure is necessary to reject the cosmopolitan mixing hypothesis.}
}
   \label{fig:atheyreproduce}
\end{figure}

\clearpage

\linespread{1.0}\selectfont
\FloatBarrier
\renewcommand\thesection{S\arabic{section}}
\renewcommand\thefigure{S\arabic{figure}}
\renewcommand\thetable{S\arabic{table}}
\renewcommand{\figurename}{Figure}
\renewcommand{\tablename}{Table}
\setcounter{section}{0}
\setcounter{figure}{0}
\setcounter{table}{0}
\captionsetup[table]{name=Supplementary Table}
\captionsetup[figure]{name=Supplementary Figure}
\clearpage

\section*{Supplementary Information}

\vspace{1em}

\textbf{Descriptive statistics.} We include high-level descriptive statistics of exposure network, users, and MSAs in Supplementary Table \ref{tab:combined_descriptive_statistics} and Supplementary Figure \ref{tab:combined_descriptive_statistics}. For detailed descriptive statistics of the exposure network see:

\begin{itemize}
\item Supplementary Table \ref{tab:interaction_distribution} Distribution of number of exposures
\item Supplementary Table \ref{tab:interaction_distribution2} Distribution of number of exposures (per active day)
\item Supplementary Table \ref{tab:interaction_distribution3} Distribution of tie strength
\item Supplementary Table \ref{tab:interaction_distribution4} Distribution of number of distinct exposures
\item Supplementary Table \ref{tab:interaction_distribution5} Distribution of number of distinct exposures (per active day)
\item Supplementary Figure \ref{fig:interactors_over_time} Average number of exposures over time
\item Supplementary Figure \ref{fig:active_over_time} Number of active individuals over time
\item Supplementary Figure \ref{fig:exposed_over_time} Total exposures over time

\end{itemize}

For descriptive statistics of robustness checks over time, distance, and length of exposure thresholds, see Supplementary Tables \ref{tab:poi_repetition1} and \ref{tab:poi_repetition2}, and Supplementary Figures \ref{fig:active_robust_1}-\ref{fig:robust_total_distinct_8}. 

For descriptive statistics of POIs (count, average SES, average exposure segregation, and average number of exposures) see Supplementary Tables \ref{tab:poi_desc_1}-\ref{tab:poi_desc_4}.

\vspace{1em}

\textbf{Robustness Checks}. For a high-level overview of all robustness checks, see Supplementary Table \ref{tab:correlations_between_segregation_measures}. For details of each robustness check, including nationwide correlations between exposure segregation, population size, and Bridging Index, see:

\begin{itemize}
\item Supplementary Figure \ref{fig:robust_1} Weighting exposures by repetition
\item Supplementary Figure \ref{fig:robust_2} Varying definitions of socioeconomic status 
\item Supplementary Figure \ref{fig:robust_3} Excluding exposures within roads, with residents of the same home, and in non-work/leisure contexts
\item Supplementary Figure \ref{fig:robust_4} Varying minimum distance between pings 
\item Supplementary Figure \ref{fig:robust_5} Varying minimum time between pings 
\item Supplementary Figures \ref{fig:robust_6}-\ref{fig:robust_7} Varying minimum tie strength 
\item Supplementary Figure \ref{fig:robust_8} Controlling for background density  
\item Supplementary Figure \ref{fig:robust_10} Racial segregation and economic segregation within race groups
\item Supplementary Figure \ref{fig:robust_11} Varying minimum stationary nights
\end{itemize}

\vspace{1em}

\textbf{Null models}. For null models of alternative homophily mechanisms, which do not explain segregation in large cities, see Supplementary Figures \ref{fig:leisure_null_models} and \ref{fig:leisure_null_models_robust}

\vspace{1em}

\textbf{Bridging Index}. For an explanation of why Bridging Index predicts exposure segregation, see \ref{fig:ccbi_ablation}. Supplementary Figure \ref{fig:ccbi_variance} shows that the hub bridging finding is robust to definition of SES diversity, and Supplementary Figure \ref{fig:ccbi_alternatives_2} shows null results for other alternatives to the hub bridging finding. Supplementary Figures \ref{fig:ccbi_example1}-\ref{fig:ccbi_example4} show illustrative examples of exposure hubs.  

\vspace{1em}

\textbf{Comparisons to conventional segregation measures}. Supplementary Figures \ref{fig:nsi_vs_mixed_model} and \ref{fig:low_correlation_figure_1} compared exposure segregation (ES) to conventional Neighborhood Sorting Index (NSI). 

\vspace{1em}

\textbf{Associations with downstream outcomes}. Supplementary Figure \ref{fig:external_validation_2} shows that exposure segregation predicts political polarization. Finally, Supplementary Figures \ref{fig:boroughs} and Supplementary Tables \ref{tab:friendship_distance1}-\ref{tab:friendship_distance2} show that our exposure network predicts friendship formation, even when controlling for distance.  

\vspace{1em}

\textbf{Robustness to noise}. We show that our network size is sufficient to draw statistical comparisons between large and small cities, and that our findings are robust to noise via bootstrapped confidence intervals (Supplementary Figure \ref{fig:bootstrap}) and by downsampling the network and reproducing our findings (Supplementary Figure \ref{fig:downsample_network}).
\vspace{1em}

\textbf{Types of exposures by tie strength}. We show how different venues vary by exposure reptition, length, and distance in Supplementary Tables \ref{tab:poi_unique_days}, \ref{tab:poi_repetition}, and \ref{tab:poi_length}. Supplementary Figure \ref{fig:external_validation_5} shows that exposure segregation predicts upward economic mobility, regardless of threshold on tie strength.

\vspace{1em}

\textbf{Race}. Supplementary Figure \ref{fig:poi_racial} shows how POIs segregation varies by racial and economic segregation. Supplementary Figure \ref{fig:robust_10} compares economic segregation to racial segregation, and splits economic segregation by race group. 

\vspace{1em}

\textbf{Temporal heterogeneity in segregation}. Supplementary Figure \ref{fig:segregation_by_hour} shows that exposure segregation varies by time, which is further illustrated by examples in Supplementary Figures \ref{fig:timestory1}-\ref{fig:timestory2}.

\vspace{1em}

\textbf{Miscellaneous}. For additional anlayses, see:
\begin{itemize}
\item Supplementary Figure \ref{fig:robustness_check_leisure_pois} POI differentiation in large cities is robust to POI category

\item Supplementary Table \ref{tab:longtable} Exposure Segregation and related variables (i.e. \# Exposures, Mean ES, NSI, Gini Index, Population Size, and Bridging Index (BI) for all 382 MSAs
\item Supplementary Figure \ref{tab:segregation_density} Population density finding robustness 
\item Supplementary Figure \ref{fig:county_rurality} Findings generalize to rural counties 

\end{itemize}

\pagebreak

\begin{table}[ht]
\centering
\begin{footnotesize}
\setlength\tabcolsep{2pt} %
\begin{tabular}{r r r r r || r r r}
      & \begin{tabular}[c]{@{}l@{}}\ Accurate pings \end{tabular} & \begin{tabular}[c]{@{}l@{}}\ Unique days \end{tabular} & \begin{tabular}[c]{@{}l@{}}\ Distinct user pairs\end{tabular} & \begin{tabular}[c]{@{}l@{}}\ \new{Exposures} \end{tabular} & \begin{tabular}[c]{@{}l@{}}\ Accurate pings\end{tabular}& \begin{tabular}[c]{@{}l@{}}\ Distinct user pairs \end{tabular} & \begin{tabular}[c]{@{}l@{}}\ \new{Exposures} \end{tabular} \\ \hline
count & 8,609,406 & \textcolor{white}{8,609,406}  &  \textcolor{white}{8,527,115} &  \textcolor{white}{8,527,115}  &            382 &   \textcolor{white}{382} &    \textcolor{white}{382}  \\
mean  &     3,273 &        35  &       184 &                      363  &     73,757,695 &   2,577,322 &                4,845,144  \\
std   &    16,507 &        20  &       374 &                    1,073  &    163,848,305 &   8,872,464 &               16,838,938  \\
min   &        11 &         2  &         1 &                        1  &      2,196,084 &      27,326 &                   53,350  \\
10\%  &       570 &        13  &          8 &                       17 &      8,398,875 &      140,251 &                  313,803 \\
50\%  &     1,471 &        30  &         76 &                      141 &     22,054,930 &      504,525 &                1,031,691 \\
90\%  &     5,857 &        63  &        436 &                      785 &    175,295,175 &    4,573,152 &                8,954,800 \\
max   & 4,755,081 &        95  &    42,323 &                  193,193  &  1,605,070,032 &  94,140,015 &              215,183,409  
\end{tabular}
\end{footnotesize}
\caption{Combined descriptive statistics for all individuals residing in 382 Metropolitan Statistical Areas (MSAs). 8,609,406 individuals reside in a Metropolitan Statistical Area (90\% of the overall 9,567,559 individuals in our study). The remaining 958,153 users live outside of MSAs, influencing the \new{exposure} segregation of an MSA by coming into contact with MSA residents. Descriptive statistics are grouped by individual (left) and MSA (right). At least one of two users in each \new{exposure} pair must live in an MSA to be included in this table.}
\label{tab:combined_descriptive_statistics}
\end{table}

\begin{figure}[htbp]
  \includegraphics[width=\textwidth,height=\textheight,keepaspectratio, trim=2 2 2 2,clip]{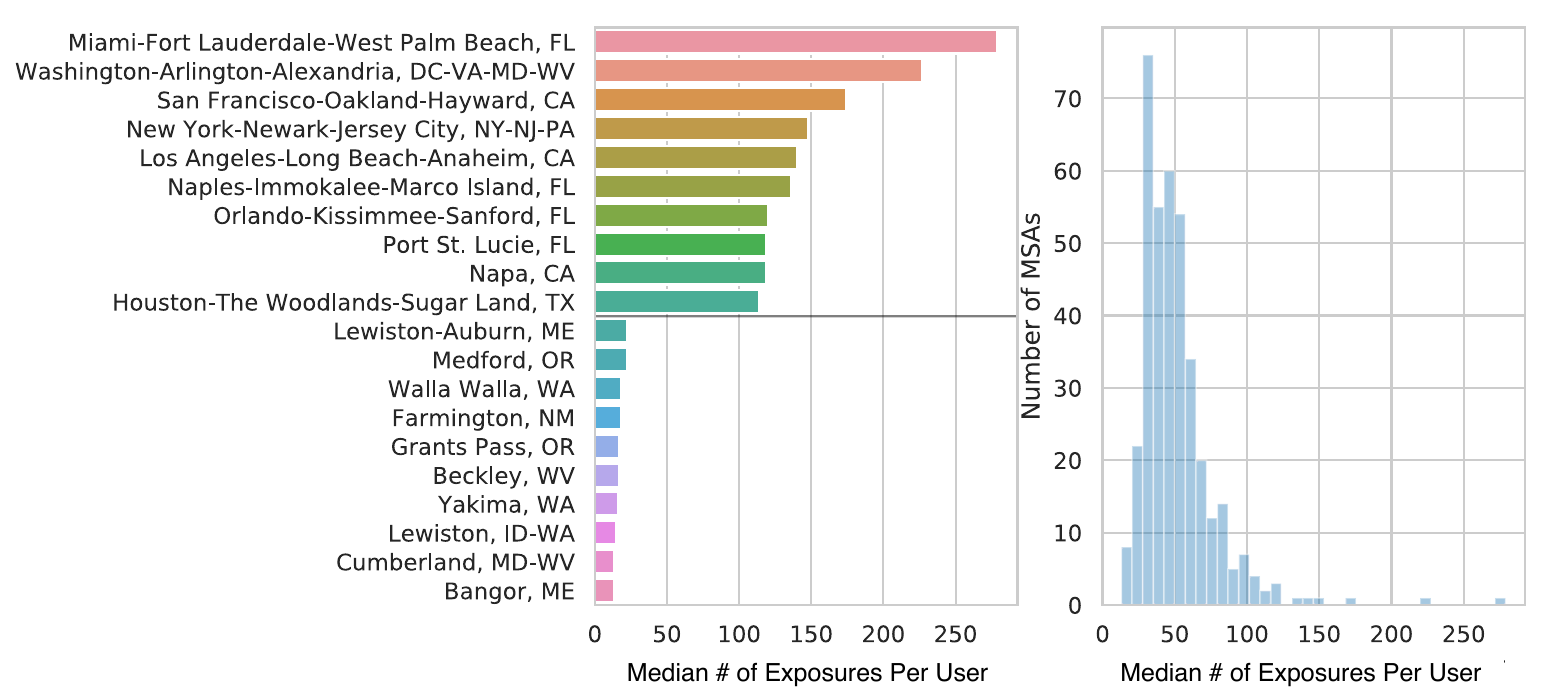}
  \caption{\textbf{Descriptive statistics of exposures.}\\
  (a) Ten Metropolitan Statistical Areas (MSAs) with the highest and lowest median number of exposures per user. \\
  (b) Overall distribution of median number of exposures per user over MSAs.}
\end{figure}

\begin{table}
\begin{scriptsize}

\begin{tabular}{lrrrrrrr}
\toprule
                   display\_name &  \# POIs (25\%) &  \# POIs (50\%) &  \# POIs (75\%) &  \# POIs (max) &  \# POIs (mean) &  \# POIs (min) &  \# POIs (std) \\
\midrule
       Full-Service Restaurants &          75.5 &         160.0 &         424.0 &      24,689.0 &          609.8 &          12.0 &      1,820.05 \\
                     Snack Bars &          18.0 &          40.0 &         110.0 &       6,266.0 &         169.76 &           1.0 &        511.17 \\
    Limited-Service Restaurants &          33.0 &          60.0 &         145.5 &       4,847.0 &         192.14 &           5.0 &        434.78 \\
                       Stadiums &           1.0 &           2.0 &           4.0 &          43.0 &           3.67 &           1.0 &          4.32 \\
        Performing Arts Centers &           1.0 &           2.0 &           4.0 &          28.0 &           3.25 &           1.0 &          3.41 \\
     Fitness/Recreation Centers &          10.0 &          25.0 &          72.0 &       4,877.0 &          126.6 &           1.0 &        414.26 \\
               Historical Sites &           1.0 &           2.0 &           7.0 &         206.0 &           9.16 &           1.0 &         21.65 \\
                    Theme Parks &           1.0 &           3.0 &           6.0 &         158.0 &           7.64 &           1.0 &         16.77 \\
           Bars/Drinking Places &           2.0 &           5.0 &          13.0 &         447.0 &          19.52 &           1.0 &         45.91 \\
                          Parks &           3.0 &           6.0 &          17.0 &         793.0 &          28.69 &           1.0 &         80.44 \\
        Religious Organizations &           7.0 &          16.0 &         41.25 &       2,644.0 &          63.28 &           1.0 &        196.97 \\
                Bowling Centers &           2.0 &           4.0 &           8.0 &         204.0 &           9.77 &           1.0 &         20.52 \\
                        Museums &           1.0 &           3.0 &           6.0 &         137.0 &           6.78 &           1.0 &         13.99 \\
                        Casinos &           1.0 &           3.0 &           7.0 &         188.0 &           8.05 &           1.0 &         17.22 \\
            Independent Artists &           1.0 &           2.0 &           5.0 &         130.0 &           7.55 &           1.0 &         17.42 \\
     Other Amusement/Recreation &           1.0 &           2.0 &           7.0 &         525.0 &          10.17 &           1.0 &         36.13 \\
 Golf Courses and Country Clubs &           2.0 &           3.0 &           7.0 &         101.0 &           8.07 &           1.0 &         13.77 \\
\bottomrule
\end{tabular}

\end{scriptsize}

\caption{POI descriptive statistics (\# of POIs in each MSA) for each of the fine-grained POI categories in Figure~\ref{fig:main_2}c.}
\label{tab:poi_desc_1}
\end{table}

\begin{table}
\begin{tiny}

\begin{tabular}{lrrrrrrr}
\toprule
                   display\_name &  POI SES (25\%) &  POI SES (50\%) &  POI SES (75\%) &  POI SES (max) &  POI SES (mean) &  POI SES (min) &  POI SES (std) \\
\midrule
       Full-Service Restaurants &      1,210.96 &       1,395.0 &      1,674.27 &      3,628.06 &       1,493.04 &         763.0 &        430.99 \\
                     Snack Bars &      1,229.73 &      1,412.35 &      1,684.69 &      3,621.34 &       1,513.61 &        788.12 &        433.86 \\
    Limited-Service Restaurants &      1,174.84 &      1,351.64 &       1,587.4 &      3,501.19 &       1,440.15 &        771.34 &        410.95 \\
                       Stadiums &       1,310.0 &       1,500.0 &       1,775.0 &      3,585.25 &       1,593.21 &         795.0 &        424.77 \\
        Performing Arts Centers &       1,395.0 &       1,583.1 &       1,832.4 &      3,632.78 &       1,659.56 &         875.0 &        431.06 \\
     Fitness/Recreation Centers &      1,230.03 &      1,431.79 &      1,703.94 &      3,749.05 &       1,528.73 &         700.0 &         453.4 \\
               Historical Sites &       1,325.0 &      1,527.94 &      1,793.75 &      3,618.58 &       1,627.62 &         757.5 &        452.96 \\
                    Theme Parks &       1,300.0 &      1,498.75 &       1,750.0 &       3,900.0 &       1,612.58 &         700.0 &        501.79 \\
           Bars/Drinking Places &      1,220.02 &      1,420.25 &       1,676.4 &      3,656.17 &       1,505.86 &         750.0 &        440.08 \\
                          Parks &      1,279.82 &      1,470.15 &      1,748.12 &      3,748.11 &       1,562.62 &         725.0 &        454.13 \\
        Religious Organizations &      1,269.27 &      1,459.86 &      1,677.08 &      3,670.38 &       1,529.02 &         754.0 &        428.42 \\
                Bowling Centers &      1,180.08 &      1,368.75 &      1,621.15 &      3,504.36 &       1,457.96 &         725.0 &        434.56 \\
                        Museums &       1,275.0 &      1,490.83 &      1,775.36 &      3,606.66 &       1,585.92 &         800.0 &        474.37 \\
                        Casinos &       1,200.0 &       1,400.0 &      1,655.54 &      3,606.17 &       1,503.88 &         725.0 &        469.68 \\
            Independent Artists &      1,374.38 &       1,611.5 &       1,904.6 &      3,691.68 &       1,725.42 &         850.0 &        528.33 \\
     Other Amusement/Recreation &       1,266.0 &       1,450.0 &      1,700.74 &      4,053.39 &       1,549.13 &         758.0 &        462.03 \\
 Golf Courses and Country Clubs &      1,399.06 &       1,648.4 &      1,964.19 &       4,248.5 &       1,765.92 &         900.0 &        542.13 \\
\bottomrule
\end{tabular}

\end{tiny}

\caption{POI descriptive statistics (average POI socioeconomic status in an MSA) for each of the fine-grained POI categories in Figure~\ref{fig:main_2}c. POI socioeconomic status is operationalized as the median visitor SES of the POI.}
\label{tab:poi_desc_2}
\end{table}

\begin{table}
\begin{scriptsize}

\begin{tabular}{lrrrrrrr}
\toprule
                   display\_name &  ES (25\%) &  ES (50\%) &  ES (75\%) &  ES (max) &  ES (mean) &  ES (min) &  ES (std) \\
\midrule
       Full-Service Restaurants &      0.22 &      0.27 &      0.32 &      0.48 &       0.27 &      0.08 &      0.07 \\
                     Snack Bars &       0.2 &      0.25 &      0.31 &       0.5 &       0.25 &      0.01 &      0.08 \\
    Limited-Service Restaurants &      0.24 &      0.29 &      0.34 &      0.47 &       0.29 &      0.04 &      0.08 \\
                       Stadiums &      0.14 &      0.17 &      0.22 &      0.36 &       0.18 &      0.02 &      0.06 \\
        Performing Arts Centers &      0.14 &      0.16 &      0.19 &      0.27 &       0.17 &      0.05 &      0.05 \\
     Fitness/Recreation Centers &       0.2 &      0.26 &      0.31 &      0.47 &       0.25 &      0.03 &      0.08 \\
               Historical Sites &      0.15 &       0.2 &      0.27 &      0.43 &       0.21 &       0.0 &      0.09 \\
                    Theme Parks &      0.16 &       0.2 &      0.25 &      0.42 &        0.2 &      0.02 &      0.08 \\
           Bars/Drinking Places &      0.18 &      0.23 &       0.3 &      0.42 &       0.23 &      0.06 &      0.08 \\
                          Parks &      0.19 &      0.26 &      0.33 &      0.47 &       0.26 &      0.05 &      0.09 \\
        Religious Organizations &      0.24 &      0.32 &      0.38 &      0.55 &       0.31 &      0.05 &       0.1 \\
                Bowling Centers &      0.16 &      0.21 &      0.26 &      0.44 &       0.22 &      0.03 &      0.08 \\
                        Museums &      0.18 &      0.22 &      0.28 &      0.45 &       0.24 &      0.06 &      0.08 \\
                        Casinos &       0.2 &      0.26 &      0.32 &      0.47 &       0.26 &      0.02 &      0.09 \\
            Independent Artists &      0.13 &       0.2 &      0.27 &      0.39 &       0.21 &      0.02 &      0.09 \\
     Other Amusement/Recreation &      0.18 &      0.25 &      0.31 &      0.71 &       0.25 &      0.02 &      0.12 \\
 Golf Courses and Country Clubs &      0.33 &      0.41 &       0.5 &      0.62 &        0.4 &       0.2 &      0.11 \\
\bottomrule
\end{tabular}

\end{scriptsize}

\caption{POI descriptive statistics (exposure segregation within-category) for each of the fine-grained POI categories in Figure~\ref{fig:main_2}c. Exposure segregation is calculated for each POI category by filtering for only exposures which occurred inside of the POI category, before estimating \metric (Methods).}
\label{tab:poi_desc_3}
\end{table}

\begin{table}
\begin{tiny}

\begin{tabular}{lrrrrrrr}
\toprule
                   display\_name &  \# Exposures (25\%) &  \# Exposures (50\%) &  \# Exposures (75\%) &  \# Exposures (max) &  \# Exposures (mean) &  \# Exposures (min) &  \# Exposures (std) \\
\midrule
       Full-Service Restaurants &             23,060.75 &              54,219.5 &             156,645.0 &          19,540,673.0 &             398,304.04 &               6,112.0 &          1,634,147.68 \\
                     Snack Bars &              15,582.0 &              38,954.0 &             120,225.0 &          14,128,466.0 &             291,523.01 &               5,233.0 &          1,205,873.07 \\
    Limited-Service Restaurants &              16,485.5 &              38,444.0 &             106,515.0 &          10,453,353.0 &              227,378.5 &               4,122.0 &             878,243.0 \\
                       Stadiums &              53,077.5 &              96,487.5 &            336,479.75 &           8,942,618.0 &             348,920.85 &              17,024.0 &            988,719.72 \\
        Performing Arts Centers &              66,712.0 &             120,256.0 &             384,770.0 &           6,972,326.0 &             403,378.22 &              27,589.0 &            932,589.88 \\
     Fitness/Recreation Centers &             11,165.25 &              21,541.5 &              62,380.5 &           5,630,299.0 &             158,025.53 &               3,740.0 &            573,788.93 \\
               Historical Sites &              18,351.5 &              47,793.0 &              98,385.5 &           6,362,665.0 &             187,978.09 &               5,147.0 &            684,470.65 \\
                    Theme Parks &              34,989.0 &              61,553.0 &             135,744.0 &           1,883,136.0 &             157,622.73 &              14,460.0 &            290,329.93 \\
           Bars/Drinking Places &              11,592.5 &              21,401.0 &              63,929.0 &           1,266,235.0 &              84,752.14 &               4,553.0 &            181,978.58 \\
                          Parks &             10,050.75 &              22,301.0 &             61,492.75 &           1,520,092.0 &              88,789.84 &               5,383.0 &            193,888.94 \\
        Religious Organizations &               6,014.0 &              13,002.0 &             35,157.25 &           2,206,316.0 &              60,683.34 &               2,739.0 &            212,948.46 \\
                Bowling Centers &              16,423.5 &              26,517.0 &              65,563.5 &           1,030,970.0 &              92,515.06 &               5,874.0 &            174,876.21 \\
                        Museums &              14,807.5 &              27,802.0 &             66,729.75 &             681,994.0 &              87,797.02 &               4,310.0 &            146,495.57 \\
                        Casinos &              13,844.0 &              23,109.0 &              60,222.0 &             826,676.0 &              68,012.52 &               7,474.0 &            124,063.14 \\
            Independent Artists &               8,795.0 &              23,789.5 &             56,736.75 &           1,106,402.0 &              87,662.95 &               3,951.0 &            198,394.53 \\
     Other Amusement/Recreation &               6,923.0 &              14,349.0 &              42,793.0 &             365,436.0 &              38,640.52 &               2,929.0 &             56,964.16 \\
 Golf Courses and Country Clubs &              4,765.75 &               7,836.5 &              15,555.0 &              58,348.0 &              13,047.42 &               2,636.0 &             13,348.76 \\
\bottomrule
\end{tabular}

\end{tiny}

\caption{POI descriptive statistics (number of exposures occurring inside POI category) for each of the fine-grained POI categories in Figure~\ref{fig:main_2}c.}
\label{tab:poi_desc_4}
\end{table}

\begin{table}
\begin{scriptsize}
\hspace*{-20mm}
\begin{tabular}{lrrrrrr}
\toprule
{} &  Pearson Corr. w/ Primary &  Spearman Corr. w/ Primary &  Median &  Mean &  \new{\% Pairs} &  \new{\% People} \\
Feature                                                        &                           &                            &         &       &          &           \\
\midrule
Primary Measure                                              &                       --- &                        --- &    0.35 &  0.35 &       \new{100.00} &   \new{ 100.00} \\
Primary Measure (+ Up-weight Multiple Exposures)          &                      0.89 &                       0.91 &    0.46 &  0.45 &       \new{100.00} &   \new{ 100.00} \\
\midrule    
SES Definition: Rent Zestimate Percentile                    &                      0.88 &                       0.89 &    0.42 &  0.42 &   100.00 &    100.00 \\
SES Definition: Within-MSA Rent Zestimate Percentile         &                      0.81 &                       0.83 &    0.54 &  0.53 &   100.00 &    100.00 \\
SES Definition: Census Median Household Income               &                      0.75 &                       0.77 &    0.47 &  0.46 &   100.00 &    100.00 \\
SES Definition: Educational Attainment (\% College or Higher) &                      0.70 &                       0.71 &    0.52 &  0.52 &   100.00 &    100.00 \\
\midrule   
Exclude Pri/Sec Roads                                        &                      0.99 &                       0.99 &    0.37 &  0.37 &       \new{ 74.30} &   \new{  99.87} \\
Exclude Roads                                                &                      0.98 &                       0.98 &    0.37 &  0.37 &       \new{ 38.66} &   \new{  98.56} \\
\new{Stationary Individuals (2 pings $<$ 10 meters in 1-10 min)}          &                      \new{0.86} &                       \new{0.87} &    \new{0.44} &  \new{0.43} &     \new{5.39} &     \new{79.46} \\
Exclude Same-home exposures                               &                      0.98 &                       0.98 &    0.34 &  0.34 &       \new{ 99.71} &   \new{  99.78} \\
Work/Leisure (Neither in Home Tract)                         &                      0.93 &                       0.93 &    0.31 &  0.31 &       \new{ 86.26} &   \new{  95.55} \\
Leisure (inside POI)                                         &                      0.85 &                       0.84 &    0.28 &  0.29 &       \new{ 16.15} &   \new{  76.32} \\
\midrule
Minimum Distance Between Pings: $<$ 25 meters                  &                      0.97 &                       0.97 &    0.34 &  0.34 &     \new{  49.53} &  \new{   97.71} \\
Minimum Distance Between Pings: $<$ 10 meters                  &                      0.95 &                       0.94 &    0.33 &  0.33 &     \new{  20.93} &  \new{   92.28} \\
\midrule   
Minimum Time Between Pings: $<$ 2 minutes                      &                      0.97 &                       0.97 &    0.34 &  0.34 &     \new{  53.59} &  \new{   98.92} \\
Minimum Time Between Pings: $<$ 60 seconds                     &                      0.97 &                       0.97 &    0.35 &  0.35 &     \new{  33.67} &  \new{   97.83} \\
\midrule   
Minimum Tie Strength: 2 consecutive exposures             &                      0.94 &                       0.95 &    0.35 &  0.35 &       \new{18.25} &    \new{ 94.80} \\
Minimum Tie Strength: 3 consecutive exposures             &                      0.83 &                       0.83 &    0.37 &  0.37 &       \new{ 3.06} &    \new{ 65.46} \\
Minimum Tie Strength: 2 unique days of exposure           &                      0.88 &                       0.90 &    0.47 &  0.46 &       \new{ 7.46} &    \new{ 83.56} \\
Minimum Tie Strength: 3 unique days of exposure           &                      0.73 &                       0.77 &    0.56 &  0.54 &       \new{ 2.32} &    \new{ 62.81} \\
\midrule
Dist. $<$ 25 meters, Time $<$ 2 min., $>=$ 2 consec. exposures  &                      0.92 &                       0.93 &    0.36 &  0.35 & \new{8.80} &     \new{86.64} \\
Dist. $<$ 25 meters, Time $<$ 2 min., $>=$ 2 unique days           &                      0.87 &                       0.89 &    0.46 &  0.44 & \new{3.84} &     \new{70.57} \\
Dist. $<$ 10 meters, Time $<$ 60 sec., $>= $3 consec. exposures &                      0.78 &                       0.79 &    0.39 &  0.38 & \new{0.94} &     \new{38.16} \\
Dist. $< $10 meters, Time $<$ 60 sec., $>=$ 3 unique days          &                      0.68 &                       0.68 &    0.52 &  0.51 & \new{0.58} &     \new{28.92} \\
\midrule
\new{Downweight Simultaneous Exposures}   &                      \new{0.97} &                       \new{0.97} &    \new{0.34} &  \new{0.34} &   \new{99.99} &     \new{99.99} \\
\new{Exclude Simultaneous Exposures}  &                      \new{0.94} &                       \new{0.95} &    \new{0.46} &  \new{0.46} &    \new{22.87} &     \new{99.61} \\
\midrule
\new{Tie Strength: 1 Exposure }                                  &                     \new{ 0.97} &                       \new{0.98} &    \new{0.31} &  \new{0.32} &    \new{74.24} &     \new{98.66} \\
\new{Tie Strength: 2 Exposures}                                  &                     \new{ 0.95} &                       \new{0.96} &    \new{0.33} &  \new{0.33} &    \new{16.14} &     \new{93.20} \\
\new{Tie Strength: 3 Exposures}                                  &                     \new{ 0.89} &                       \new{0.90} &    \new{0.36} &  \new{0.36} &    \new{ 4.09} &     \new{74.67} \\
\new{Tie Strength: 4 Exposures}                                  &                     \new{ 0.85} &                       \new{0.86} &    \new{0.37} &  \new{0.36} &    \new{ 1.82} &     \new{58.51} \\
\new{Tie Strength: 5+ Exposure}                                  &                     \new{ 0.83} &                       \new{0.85} &    \new{0.45} &  \new{0.45} &    \new{ 3.42} &     \new{70.62} \\
\midrule
\new{Minimum Stationary Nights: 6 nights}                          &                    \new{ 0.99} &                       \new{0.99} &    \new{0.35} &  \new{0.35} &    \new{77.45} &    \new{ 83.91} \\
\new{Minimum Stationary Nights: 9 nights  }                        &                    \new{ 0.99} &                       \new{0.99} &    \new{0.35} &  \new{0.36} &    \new{68.64} &    \new{ 73.42} \\
\new{Minimum Stationary Nights: 12 nights }                        &                    \new{ 0.98} &                       \new{0.98} &    \new{0.35} &  \new{0.36} &    \new{59.16} &    \new{ 63.66} \\
\midrule
\new{Racial Segregation (\% Non-White)                    }         &                   \new{   0.59} & \new{                      0.60} &   \new{ 0.46} & \new{ 0.46} & \new{    100.0} &\new{     100.0} \\
\new{Economic Segregation (White Overall)                 }        &                    \new{  0.94 }&  \new{                     0.95 }&    \new{0.34 }&  \new{0.34} &  \new{   55.97}& \new{    63.83} \\
\new{Economic Segregation (White Within-group)            }        &                    \new{  0.93 }&  \new{                     0.94 }&    \new{0.34 }&  \new{0.34} &  \new{   39.84 }& \new{    63.69} \\
\new{Economic Segregation (Non-White Overall)             }        &                    \new{  0.55 }&  \new{                     0.52 }&    \new{0.28 }&  \new{0.28} &  \new{   43.76 }& \new{    35.50} \\
\new{Economic Segregation (Non-White Within-group)        }        &                    \new{  0.47 }&  \new{                     0.44 }&    \new{0.28 }&  \new{0.30} &  \new{   27.50 }& \new{    35.31} \\
\bottomrule
\end{tabular}

\end{scriptsize}
\caption{\textbf{Robustness checks overview. We find that our definition of exposure segregation is robust to varying many parameters: weighting of repeated exposures between the same users, definition of socioeconomic status, inclusion/exclusion of roads and same-home exposures, filtering location of exposure, minimum distance, minimum time, and minimum tie strength (as well as the intersection of distance, time, and tie strength)}. The above variants all are strongly correlated to our primary measure (all have Spearman Corr. $>=$ 0.75). We also find that our primary findings that (1) large, dense cities \new{are more segregated} and (2) exposure hub locations accessible to diverse individuals may mitigate segregation are robust across all definitions of Exposure Segregation (Supplementary Figures \ref{fig:robust_1}-\ref{fig:robust_8}, \ref{fig:robust_10}, \ref{fig:robust_11}). Note that we exclude same-home exposures in robustness checks that vary minimum time, distance, or require repeated exposures, to ensure that results are not influenced by exposures with members of the same household (these exposures ordinarily have minimum influence on Exposure Segregation, as shown by the robustness check which excludes same-home exposures and results in virtually identical metric (Spearman Corr. 0.98); however, the influence of same-home exposures is higher after more conservative filters are applied to the definition of exposures, such as requiring a minimum tie strength of 3 consecutive exposure).} 
\label{tab:correlations_between_segregation_measures}
\end{table}

\begin{figure}[htbp]
\captionsetup[subfigure]{labelformat=empty}

\vspace{20mm}
   \centering
\begin{subfigure}[t]{1.0\textwidth}
\includegraphics[width=\textwidth,keepaspectratio]{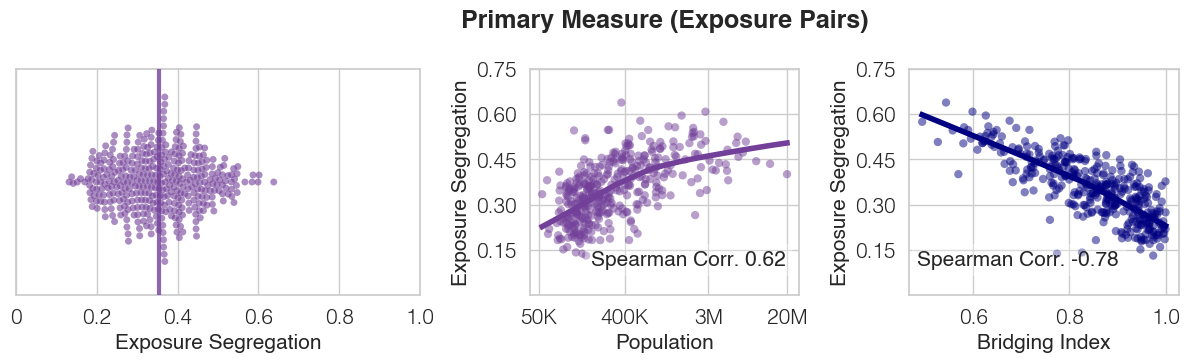}
   \caption{}
\end{subfigure}
\begin{subfigure}[t]{1.0\textwidth}
\includegraphics[width=\textwidth,keepaspectratio]{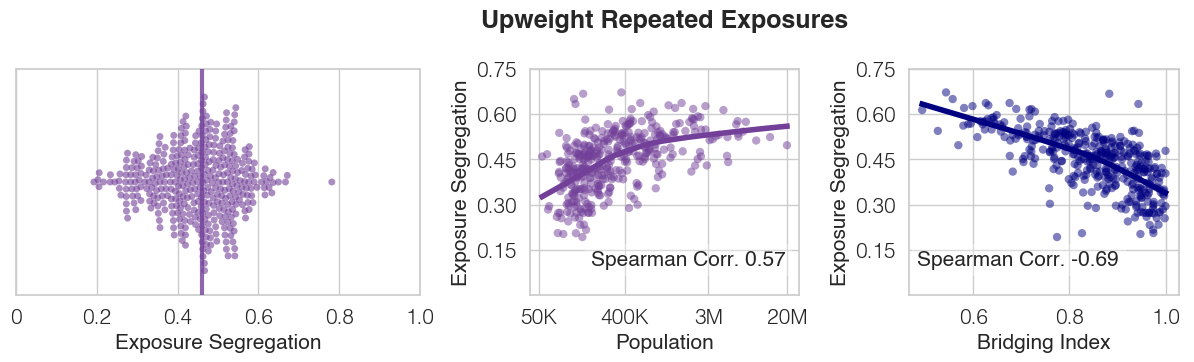}
   \caption{}
\end{subfigure}

\caption{\textbf{Robustness of primary study findings to weighting of repeated exposures.} We find that our primary study findings that, (1) large, dense cities \new{are more segregated} and (2) exposure hub locations accessible to diverse individuals may mitigate segregation, are robust to the choice of whether to upweight repeated exposures in our exposure network. We compare the results of:\\
\emph{Primary Measure:} Exposures are defined as pairs of users who have ever crossed paths within the study observation window (three months of 2017). We de-duplicate repeated exposures, as frequency of pings varies across smartphone users, to reduce bias from users with a higher frequency of pings. For instance, if an individual A with an individual B (ES \$1000) two times and individual C (ES \$2000) one, we compute the mean SES of individual A’s network as \$1500. \\
\emph{Upweight Repeated Exposures:} Repeated exposures are unweighted when calculating the mean SES of an individual’s exposure network.  For instance, if an individual A with an individual B (ES \$1000) two times and individual C (ES \$2000) once, we compute the mean SES of individual A’s network as \$1333.}
\label{fig:robust_1}
\end{figure}

\begin{figure}[htbp]
\captionsetup[subfigure]{labelformat=empty}

\vspace{-20mm}
   \centering
\begin{subfigure}[t]{0.7\textwidth}
\includegraphics[width=\textwidth,keepaspectratio]{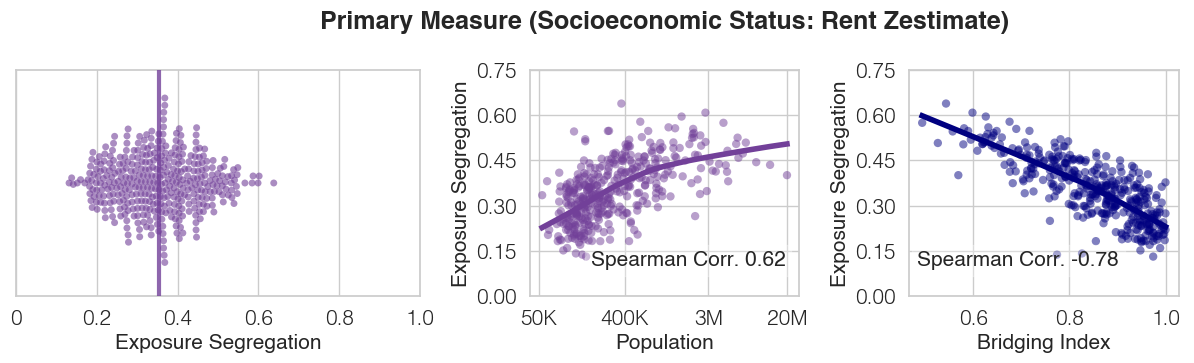}
   \caption{}
\end{subfigure}

\begin{subfigure}[t]{0.7\textwidth}
\includegraphics[width=\textwidth,keepaspectratio]{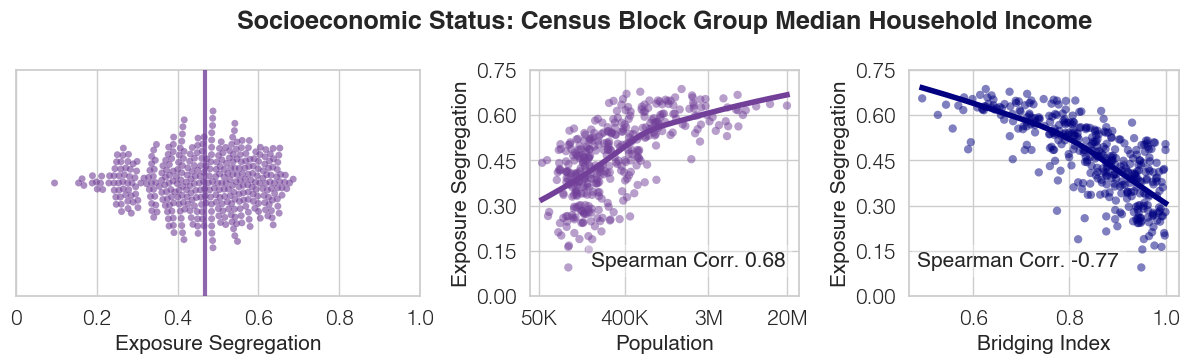}
   \caption{}
\end{subfigure}

\begin{subfigure}[t]{0.7\textwidth}
\includegraphics[width=\textwidth,keepaspectratio]{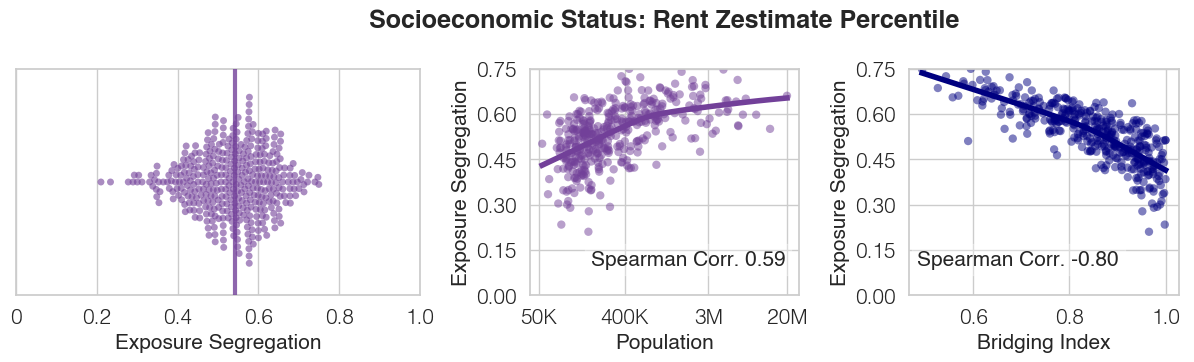}
   \caption{}
\end{subfigure}

\begin{subfigure}[t]{0.7\textwidth}
\includegraphics[width=\textwidth,keepaspectratio]{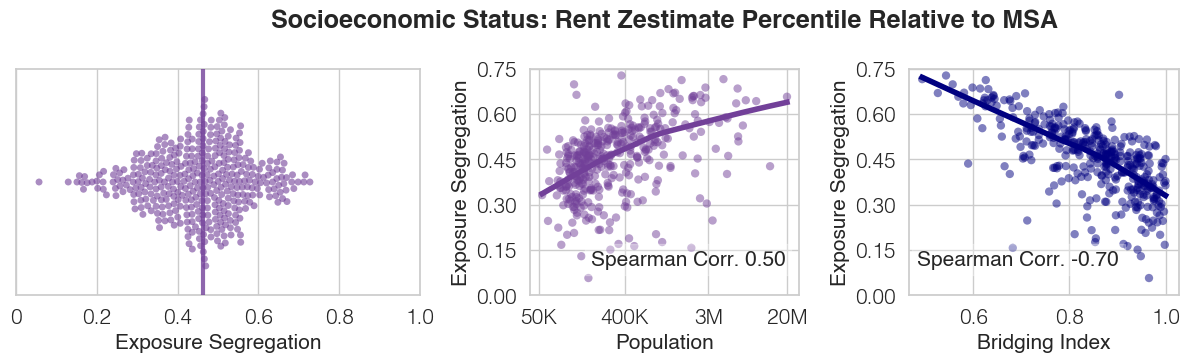}
   \caption{}
\end{subfigure}
\begin{subfigure}[t]{0.7\textwidth}
\includegraphics[width=\textwidth,keepaspectratio]{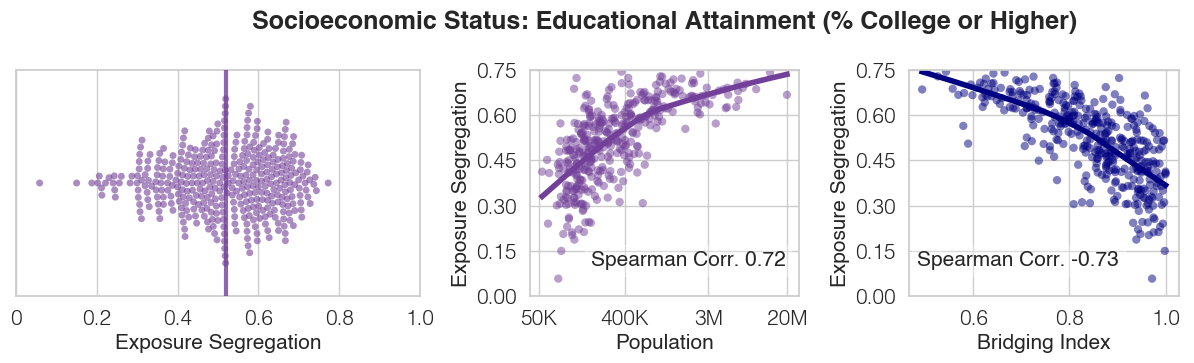}
   \caption{}
\end{subfigure}
\caption{\textbf{Robustness of primary study findings to definition of socioeconomic status.} We find that our primary study findings that, (1) large, dense cities \new{are more segregated} and (2) exposure hub locations accessible to diverse individuals may mitigate segregation, are robust to the definition of \new{socioeconomic status}. We compare the results of:\\
\emph{Primary Measure:} Our primary measure leverages estimated monthly rent value (Zillow Rent Zestimate). \\
\emph{Census Block Group (CBG) Median Household Income:} We define the SES of an individual is the median household income in the CBG in which they reside.\\
\emph{Rent Zestimate Percentile:} We normalize Rent Zestimate values across all individuals.\\
\emph{Primary measure Relative to MSA:} We normalize Rent Zestimate values across all individuals within an MSA, independent of other MSAs, to account for differences in cost of living across cities. \\
\new{\emph{Educational Attainment:} We define the SES of an individual as their educational attainment (\% with college education or higher)}.}
\label{fig:robust_2}
\end{figure}

\begin{figure}[htbp]
\captionsetup[subfigure]{labelformat=empty}
\vspace{-20mm}
   \centering

\begin{subfigure}[t]{0.5\textwidth}
\includegraphics[width=\textwidth,keepaspectratio]{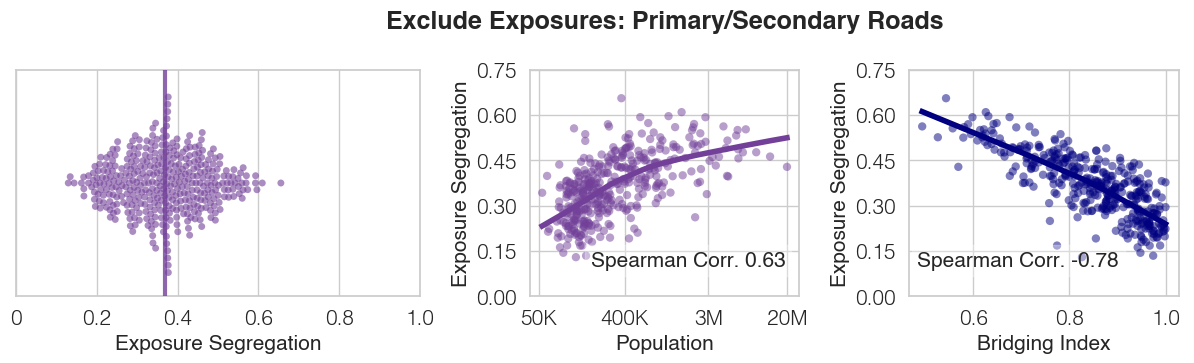}
   \caption{}
\end{subfigure}

\begin{subfigure}[t]{0.5\textwidth}
\includegraphics[width=\textwidth,keepaspectratio]{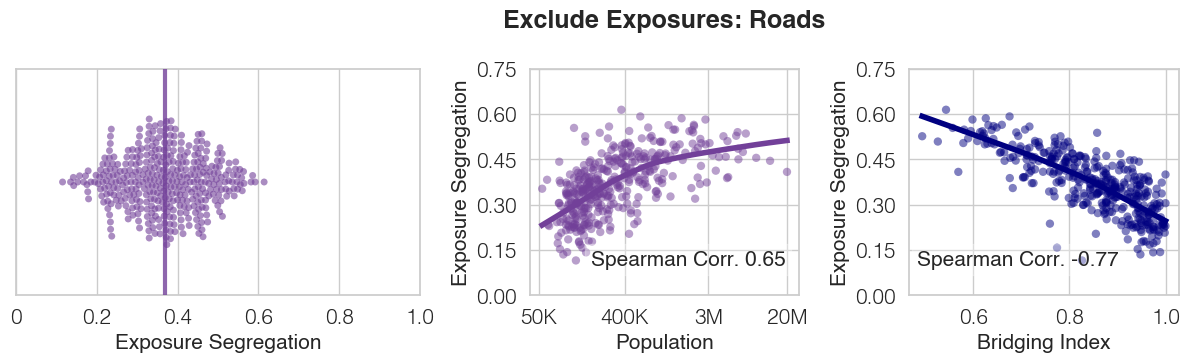}
   \caption{}
\end{subfigure}

\begin{subfigure}[t]{0.5\textwidth}
\includegraphics[width=\textwidth,keepaspectratio]{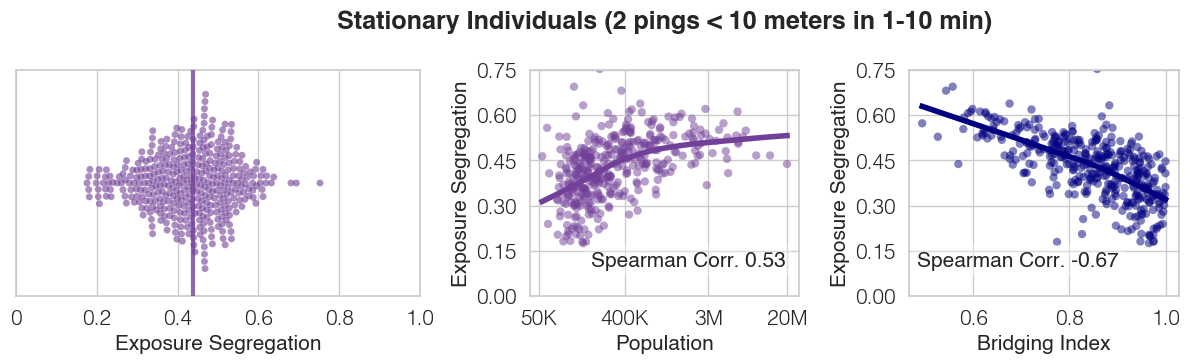}
   \caption{}
\end{subfigure}

\begin{subfigure}[t]{0.5\textwidth}
\includegraphics[width=\textwidth,keepaspectratio]{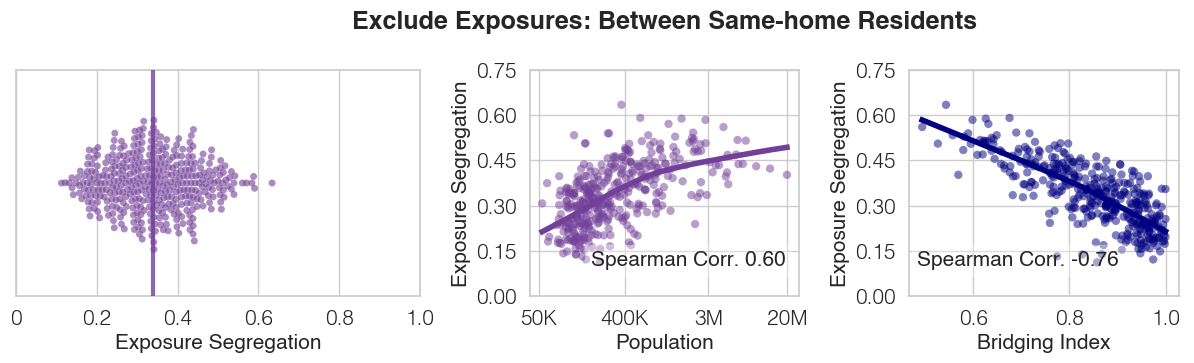}
   \caption{}
\end{subfigure}

\begin{subfigure}[t]{0.5\textwidth}
\includegraphics[width=\textwidth,keepaspectratio]{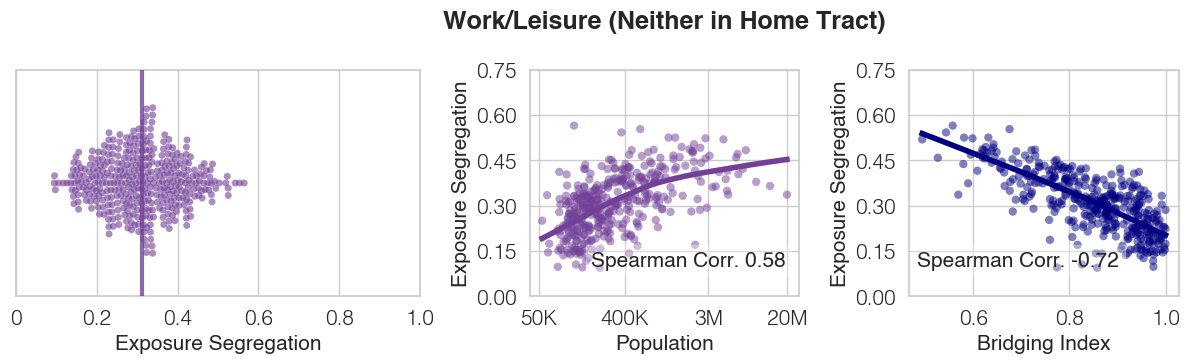}
   \caption{}
\end{subfigure}

\begin{subfigure}[t]{0.5\textwidth}
\includegraphics[width=\textwidth,keepaspectratio]{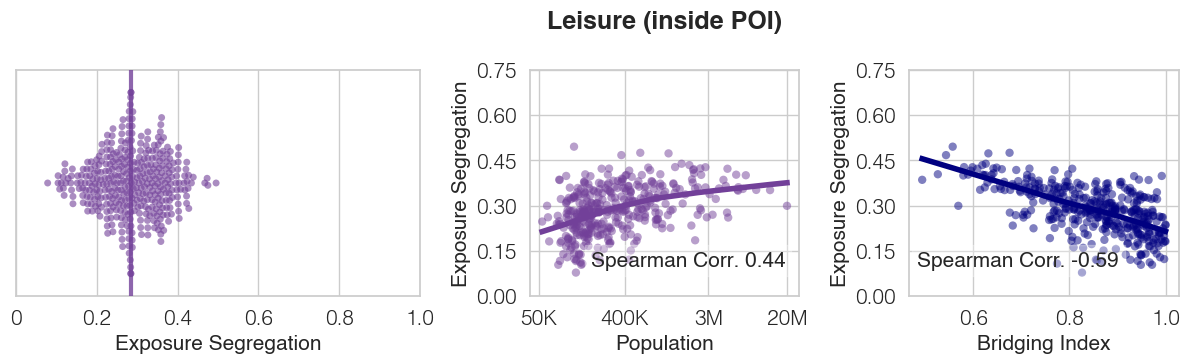}
   \caption{}
\end{subfigure}

\caption{\textbf{Robustness of primary study findings to exclusion of exposures within roads, exclusion exposures with residents of the same home, and exclusion of non-work/leisure exposures.} We find that our primary study findings that, (1) large, dense cities \new{are more segregated} and (2) exposure hub locations accessible to diverse individuals may mitigate segregation, are robust to filtering for a subset of exposures. We compare the results of:\\
\emph{Primary Measure:} Our primary measure includes all exposures, aiming to give a complete account of an individual’s exposure network including path crossings on roads as well as those they share a home with. \\
\emph{Excluding (primary/secondary) roads:} We filter to exclude exposures occurring on all roads, or only primary/secondary roads.\\
\emph{Only stationary pings:} We filter to include only exposures occurring when individuals are stationary (i.e. have two pings within 1-10 minutes and $<$ 10 meters apart).\\
\emph{Same home residents:} We filter to include only  exposures occurring between two people residing in different homes.\\
\emph{Work/Leisure:} We filter to include only  exposures likely to take place in the context of work or leisure, by excluding exposures which occurred when either individuals were located within their home tracts.\\
\emph{Leisure:} We filter for leisure exposures by including only exposures ocurring inside of the POIs categorized as related to leisure (Figure \ref{fig:main_2}c).
}

\label{fig:robust_3}
\end{figure}

\begin{figure}[htbp]
\captionsetup[subfigure]{labelformat=empty}

   \centering
\begin{subfigure}[t]{1.0\textwidth}
\includegraphics[width=\textwidth,keepaspectratio]{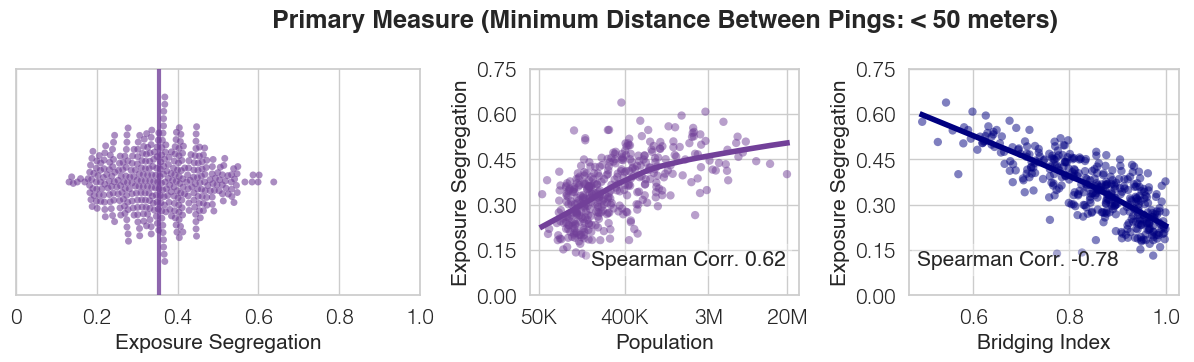}
   \caption{}
\end{subfigure}
\begin{subfigure}[t]{1.0\textwidth}
\includegraphics[width=\textwidth,keepaspectratio]{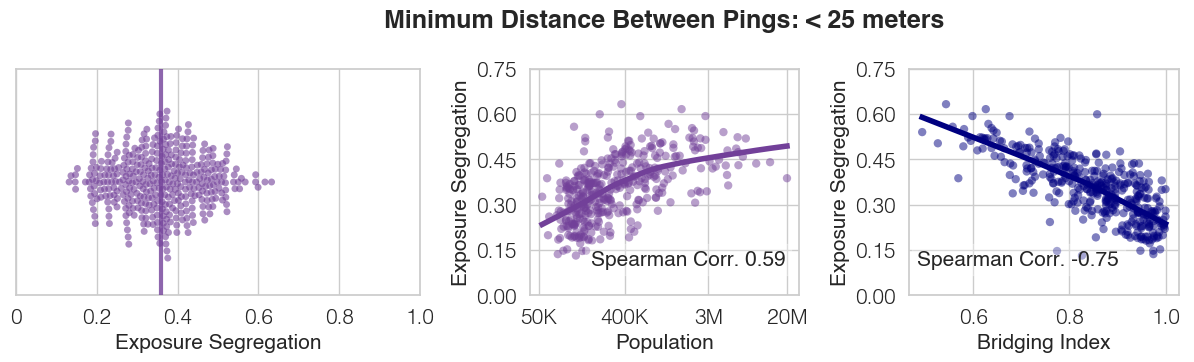}
   \caption{}
\end{subfigure}

\begin{subfigure}[t]{1.0\textwidth}
\includegraphics[width=\textwidth,keepaspectratio]{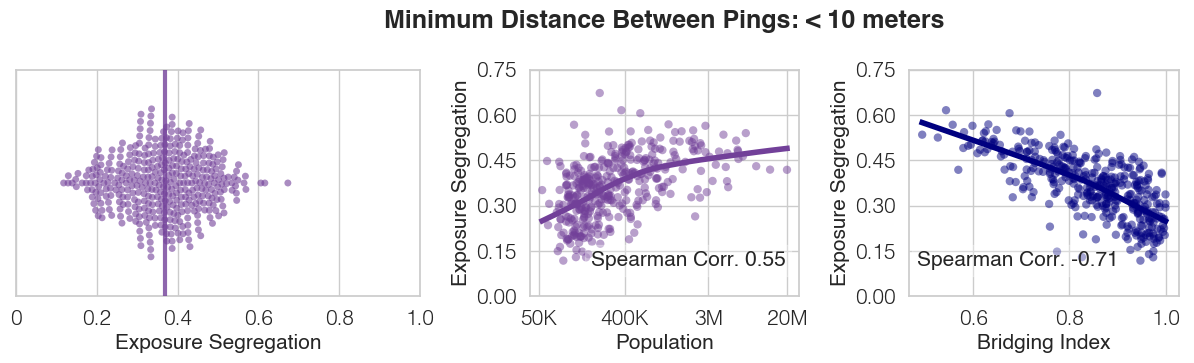}
   \caption{}
\end{subfigure}

\caption{\textbf{Robustness of primary study findings to minimum distance required between two GPS pings for individuals to be considered crossing paths.} We find that our primary study findings that, (1) large, dense cities \new{are more segregated} and (2) exposure hub locations accessible to diverse individuals may mitigate segregation, are robust to the time threshold used in our definition of exposure:\\
\emph{Primary Measure:} Our primary measure uses a threshold of 50 meters, based on prior literature which shows that even distant exposure to diverse individuals is predictive of long-term behaviors\cite{brown2021childhood}. \\
\emph{Alternative measures:} We alternatively consider more conservative thresholds of 25 meters and 10 meters, with 10 meters being the lowest threshold due to limitations of GPS ping accuracy\cite{merry2019smartphone, menard2011comparing}.}
\label{fig:robust_4}
\end{figure}

\begin{figure}[htbp]
\captionsetup[subfigure]{labelformat=empty}

   \centering
\begin{subfigure}[t]{1.0\textwidth}
\includegraphics[width=\textwidth,keepaspectratio]{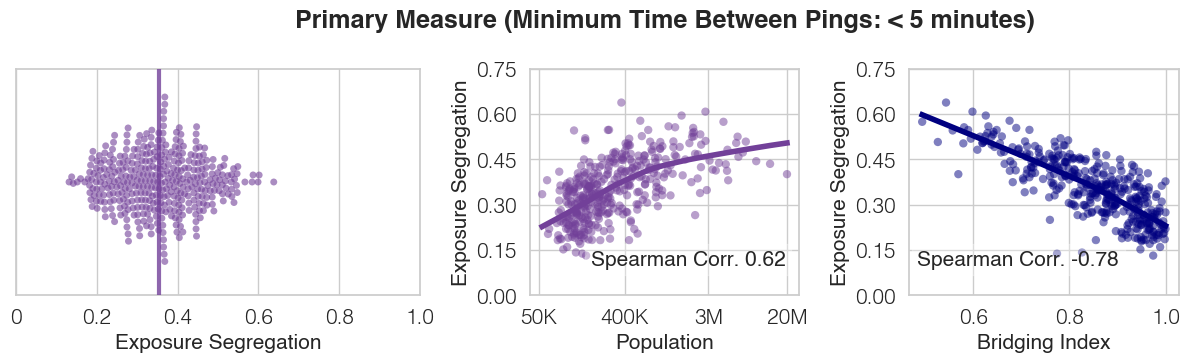}
   \caption{}
\end{subfigure}
\begin{subfigure}[t]{1.0\textwidth}
\includegraphics[width=\textwidth,keepaspectratio]{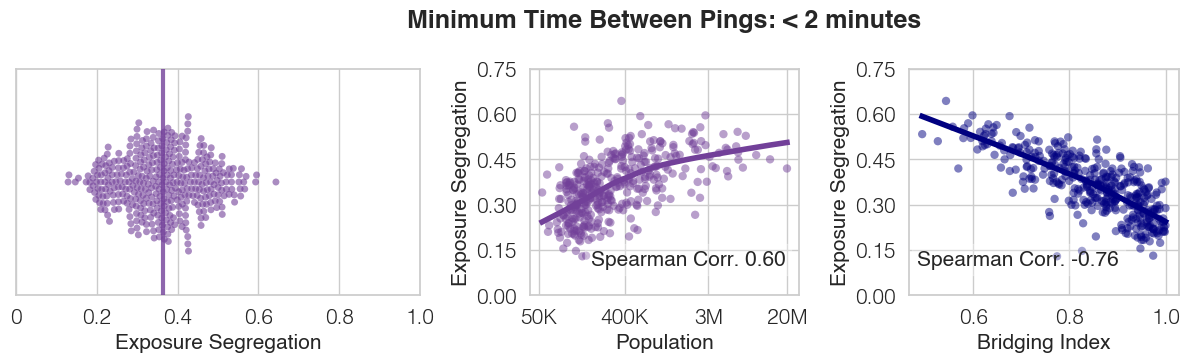}
   \caption{}
\end{subfigure}

\begin{subfigure}[t]{1.0\textwidth}
\includegraphics[width=\textwidth,keepaspectratio]{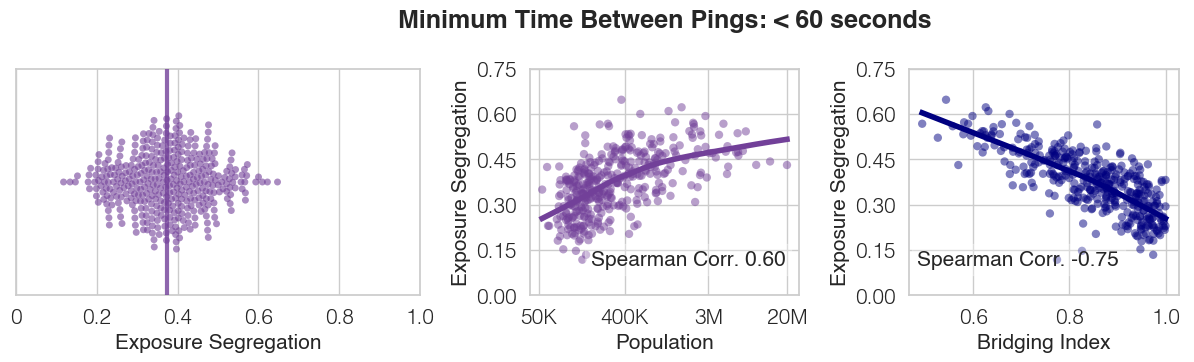}
   \caption{}
\end{subfigure}

\caption{\textbf{Robustness of primary study findings to minimum time elapsed between two pings to constitute an exposure.} We find that our primary study findings that, (1) large, dense cities \new{are more segregated} and (2) exposure hub locations accessible to diverse individuals may mitigate segregation, are robust to the time threshold used in our definition of exposure:\\
\emph{Primary Measure:} Our primary measure uses a threshold of 5 minutes, to be inclusive of users with sparse pings (e.g., for a subset of users, we only have 1 ping per day, while for others we have 100+ pings per day) while maintaining a reasonable confidence that an exposure may have occurred. \\
\emph{Alternative measures:} We alternatively consider more conservative thresholds of 2 minutes and 1 minute.
}
\label{fig:robust_5}
\end{figure}

\begin{figure}[htbp]
\captionsetup[subfigure]{labelformat=empty}

\vspace{-20mm}
   \centering

\begin{subfigure}[t]{0.6\textwidth}
\includegraphics[width=\textwidth,keepaspectratio]{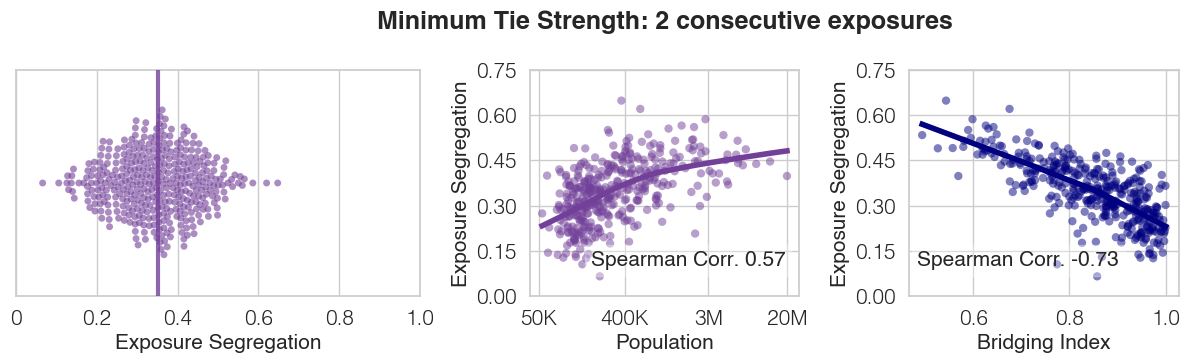}
   \caption{}
\end{subfigure}

\begin{subfigure}[t]{0.6\textwidth}
\includegraphics[width=\textwidth,keepaspectratio]{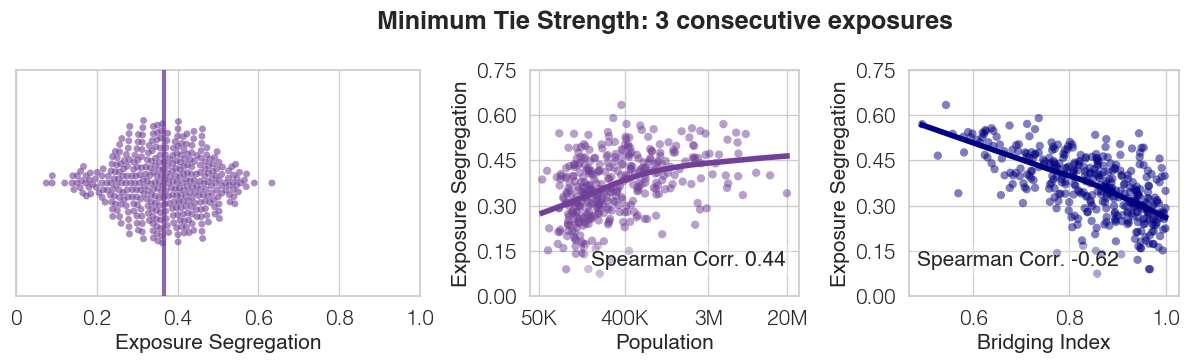}
   \caption{}
\end{subfigure}

\begin{subfigure}[t]{0.6\textwidth}
\includegraphics[width=\textwidth,keepaspectratio]{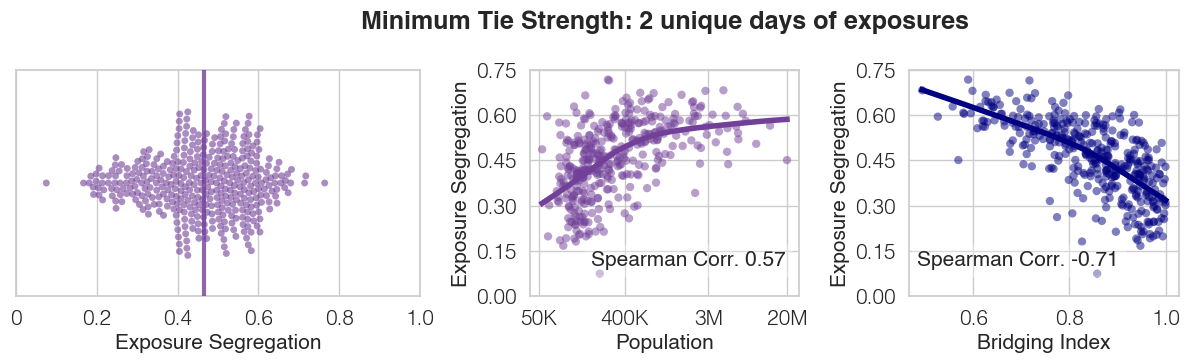}
   \caption{}
\end{subfigure}

\begin{subfigure}[t]{0.6\textwidth}
\includegraphics[width=\textwidth,keepaspectratio]{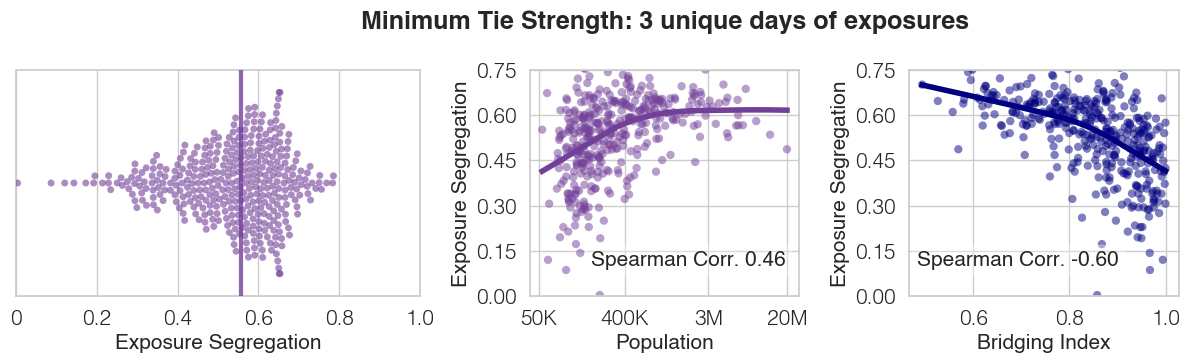}
   \caption{}
\end{subfigure}

\caption{\textbf{Robustness of primary study findings to minimum tie strength required to constitute an exposure.} We find that our primary study findings that, (1) large, dense cities \new{are more segregated} and (2) exposure hub locations accessible to diverse individuals may mitigate segregation, are robust regardless of the minimum tie strength threshold between two individuals to be constitute an exposure:\\
\emph{Primary Measure:}  Our primary measure only requires a single pair of pings between users to constitute an exposure, to be inclusive of users with sparse pings (e.g., for a subset of users, we only have 1 ping per day, while for others we have 100+ pings per day). \\
\emph{Alternative measures:} We alternatively consider more conservative thresholds of 2 or 3 consecutive exposures, as well as 2 or 3 exposures across unique days. Requiring consecutive exposures increases the likelihood that individuals actually came into contact together; exposures across unique days increases the likelihood that exposures are not merely path crossings, but social exposures between individuals who are familiar with each other. \textbf{(continued on next page)}}
\label{fig:robust_6}
\end{figure}

\begin{figure}[htbp]
\captionsetup[subfigure]{labelformat=empty}

\vspace{-20mm}
   \centering
\begin{subfigure}[t]{0.8\textwidth}
\includegraphics[width=\textwidth,keepaspectratio]{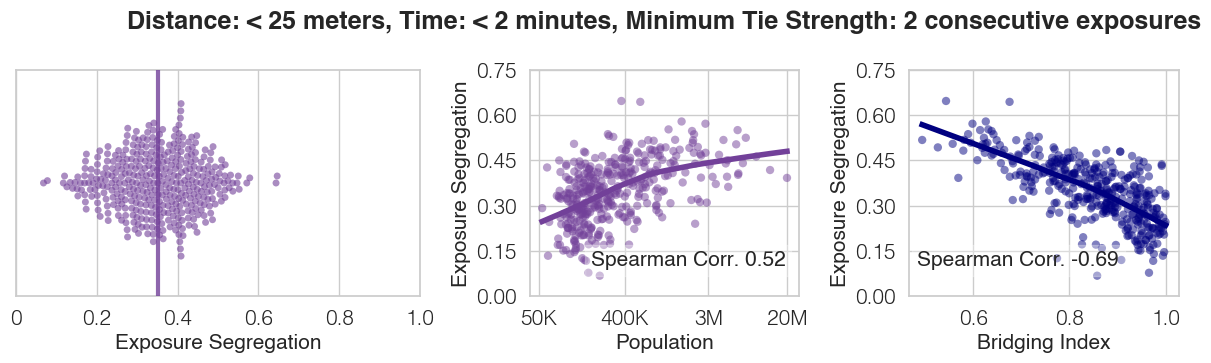}
   \caption{}
\end{subfigure}

\begin{subfigure}[t]{0.8\textwidth}
\includegraphics[width=\textwidth,keepaspectratio]{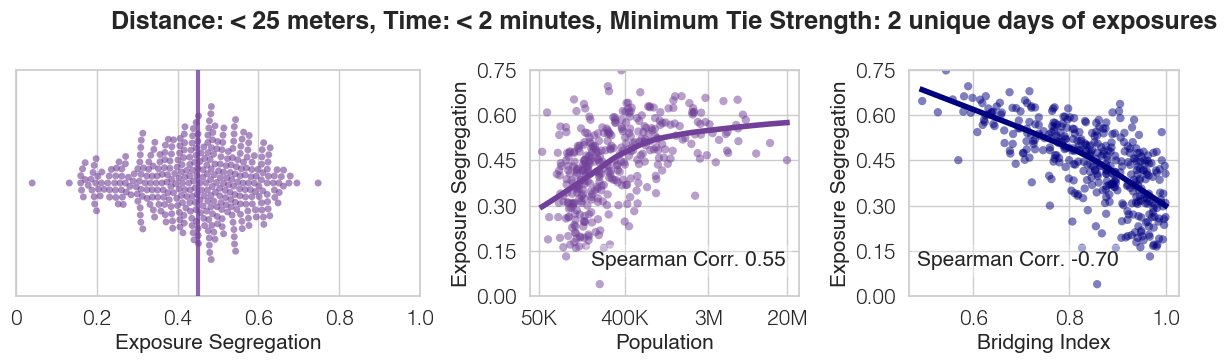}
   \caption{}
\end{subfigure}

\begin{subfigure}[t]{0.8\textwidth}
\includegraphics[width=\textwidth,keepaspectratio]{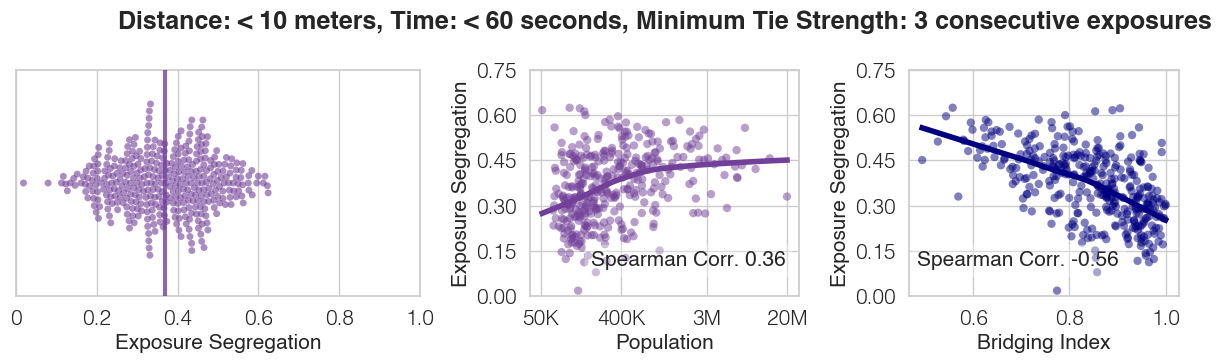}
   \caption{}
\end{subfigure}

\begin{subfigure}[t]{0.8\textwidth}
\includegraphics[width=\textwidth,keepaspectratio]{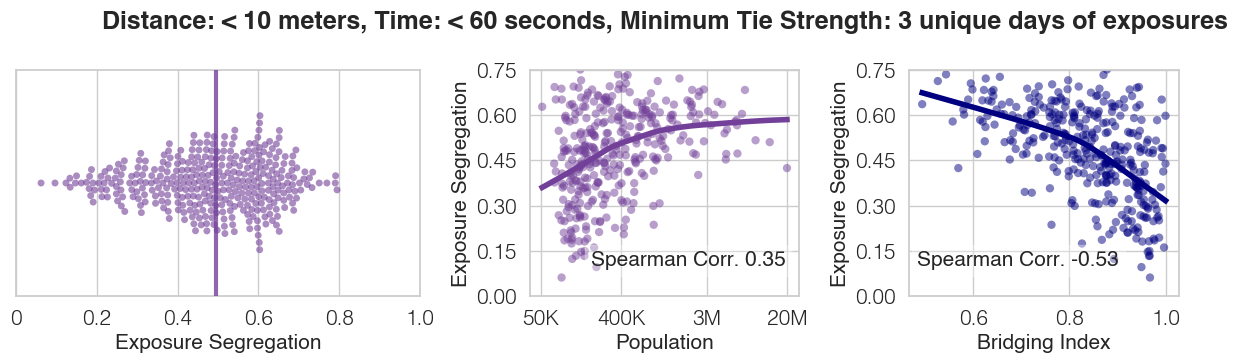}
   \caption{}
\end{subfigure}

\caption{\textbf{(continued from previous page).} We find that our primary study findings that, (1) large, dense cities \new{are more segregated} and (2) exposure hub locations accessible to diverse individuals may mitigate segregation, are \emph{robust to the combination of the minimum time, minimum distance, and minimum tie strength threshold parameters}. To account for exposures between threshold parameters, we also consider combinations of parameter variants. For instance, the most conservative robustness check defines an exposure as two individuals being $<$ 10 meters apart within a $<$ 60 second window, and for this to have occurred either for either 3 consecutive minutes (second figure from the bottom) or across 3 unique days (bottom figure). }
\label{fig:robust_7}
\end{figure}

\begin{figure}[htbp]
\captionsetup[subfigure]{labelformat=empty}

\vspace{20mm}
   \centering
\begin{subfigure}[t]{1.0\textwidth}
\includegraphics[width=\textwidth,keepaspectratio]{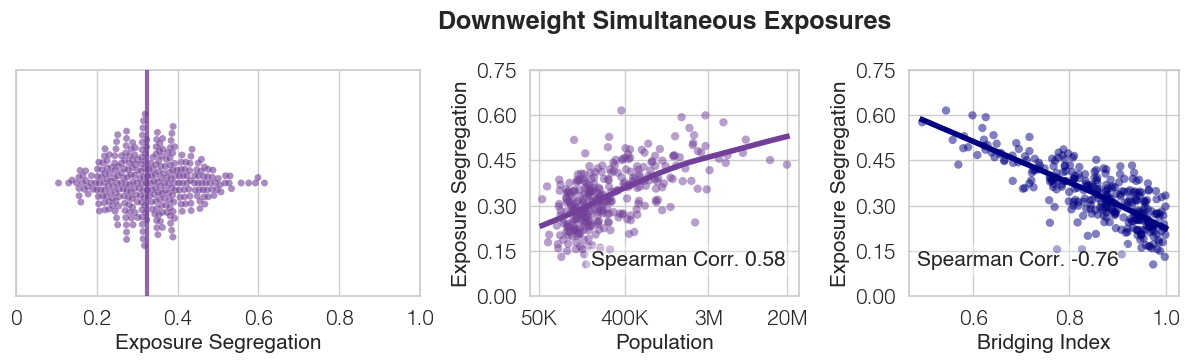}
   \caption{}
\end{subfigure}
\begin{subfigure}[t]{1.0\textwidth}
\includegraphics[width=\textwidth,keepaspectratio]{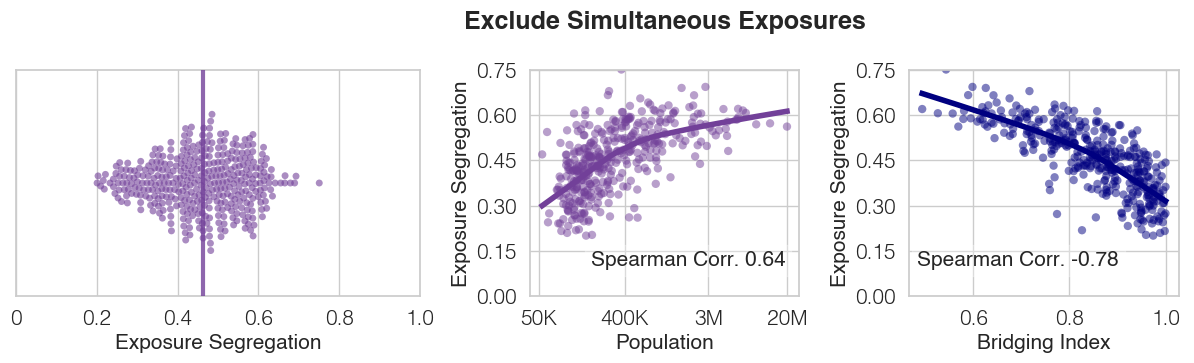}
   \caption{}
\end{subfigure}

\caption{\new{\textbf{Robustness of primary study findings to background density of exposures.} We find that our primary study findings that, (1) large, dense cities \new{are more segregated} and (2) exposure hub locations accessible to diverse individuals may mitigate segregation, are robust to the choice of whether to downweight or exclude simultaneous exposures in our exposure network. We conduct these robustness checks to ensure that our results are not overly influenced by background density of exposures\cite{crandall2010inferring}, and that our results generalize to exposures in areas with low background density:\\
\emph{Exclude Simultaneous Exposures:} For all exposures between two people (Person A and Person B), we exposures in which either person was simultaneously exposed to an additional individual (i.e. Person C was observed within 50 meters and 5 minutes of either Person A or Person B). \\ 
\emph{Downweight Simultaneous Exposures:} For each exposure between Person A and Person B, we compute $S$ the total number of additional people that Person A and Person B are simultaneously exposed to (within a 5 minute window of their pings). Exposures are then weighted by $min(25 - S, 0)$, i.e. exposure weight decreases as there are more simultaneous individuals exposed, and exposures with more than 25 people are excluded entirely\\}
}
\label{fig:robust_8}
\end{figure}

\begin{figure}[htbp]
\vspace*{-20mm}
   \centering

  \hspace*{-15mm}
\begin{subfigure}[t]{.5\textwidth}
\includegraphics[width=\textwidth,keepaspectratio]{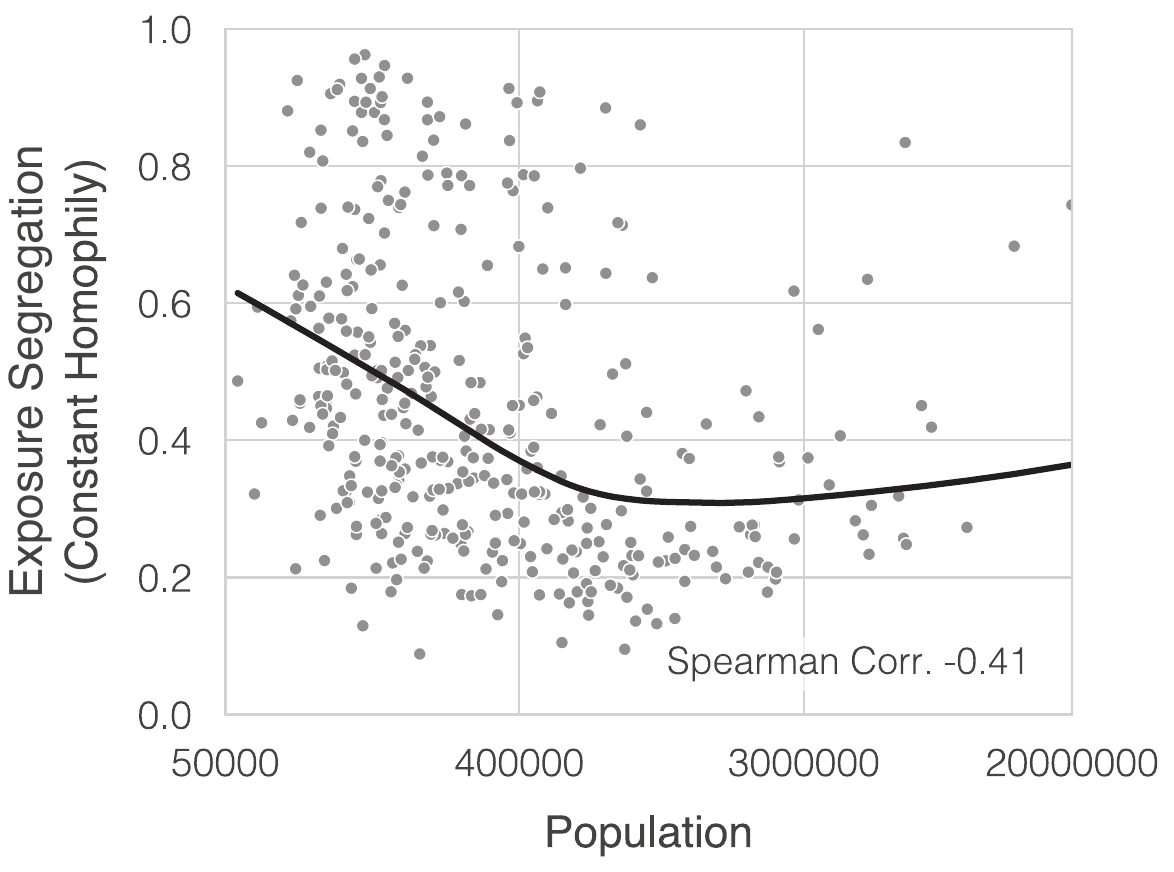}
   \caption{}
\end{subfigure}
\begin{subfigure}[t]{.5\textwidth}
\includegraphics[width=\textwidth,keepaspectratio]{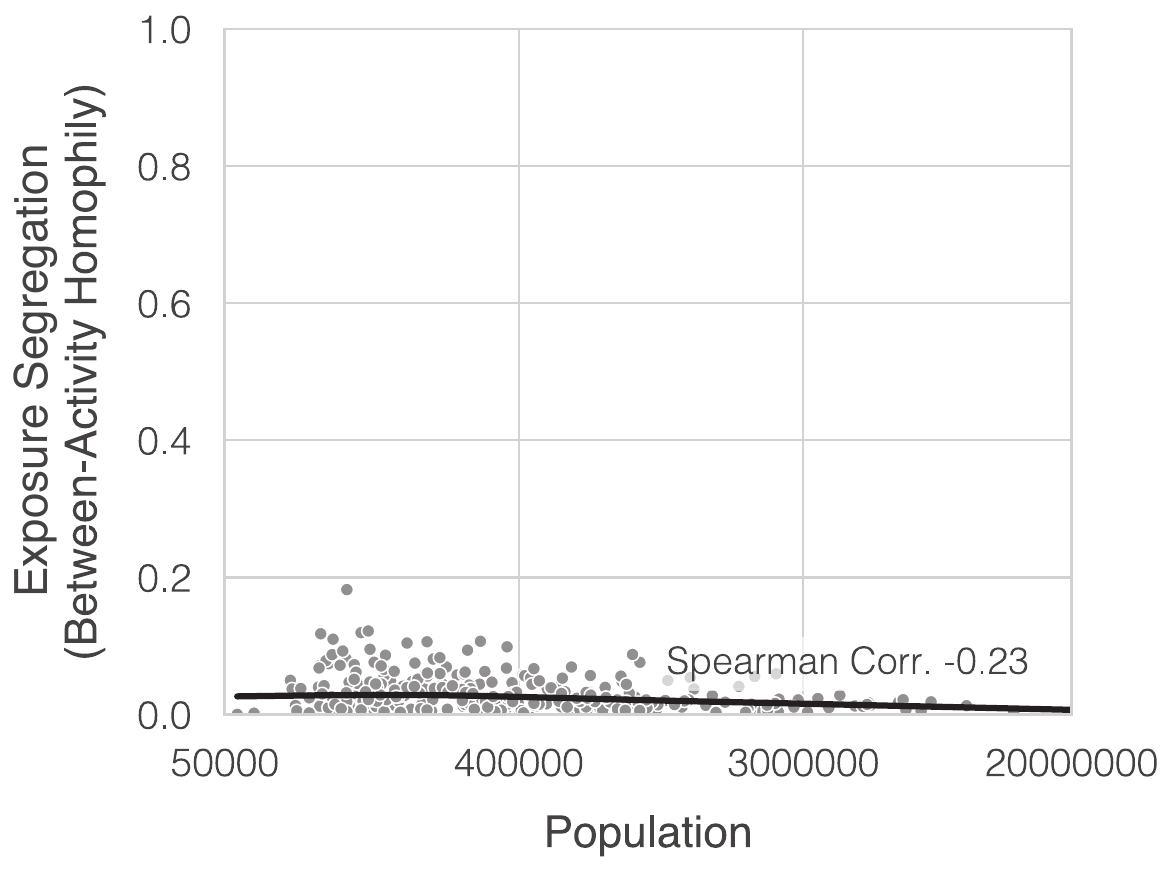}
   \caption{}
\end{subfigure}
\caption{\textbf{Alternative homophily mechanisms do not explain segregation in large cities}. We consider the possibility of two alternative hypotheses which may explain the trend towards high segregation in large cities\\
\textbf{(a)} \emph{Constant Homophily}: i.e. individuals have the same proclivity for crossing paths with individuals of similar SES regardless of if they live in large or small cities, and it is instead change in distribution of socioeconomic status that drives segregation in large cities (e.g. in large cities there may be a greater supply of people in the same economic class available to cross paths with). We test this hypothesis via a null network model, in which which we preserve network nodes (individuals and their SES values) but randomize edges\cite{barabasi2013network, maslov2002specificity}. We randomly assign exposures between pairs of people, weighting the likelihood of exposure between people of similar SES higher according to a constant homophily function. Specifically, the probability of exposure ($p_{i,j}$) between two individuals of $ES_i$ and $ES_j$ is weighted by their similarity in ES, defined as the complement of the normalized Euclidean distance in ES: $p_{i,j} \propto Similarity(ES_i,ES_j)= 1 - \frac{|ES_i - ES_j|}{max(ES)-min(ES)}$. We choose 75 exposures per person such that the mean number of exposures per person is 150, which corresponds to Dunbar’s number\cite{dunbar2010many}. We find that under this null model, there is no positive association between exposure segregation and population size; in fact, larger cities are less segregated on average, as there is an increase in supply of diverse individuals in socioeconomic status in larger cities. These findings are also robust to a variety of null model specifications (Supplementary Figure \ref{fig:leisure_null_models_robust}).\\
\textbf{(b)} \emph{Between Activity Homophily}: i.e. it is not the differentiation of individual venues that drives segregation, but rather that in large cities individuals choose different categories of activities which results in segregation (e.g. in small cities, there are less country clubs so everybody visits restaurants to socialize, whereas in large cities high-SES individuals segregate by spending a higher proportion of time in exclusive venues such as country clubs). We test this hypothesis via a configuration model \cite{barabasi2013network, maslov2002specificity}, a prominent null network model in which node degree is preserved. Specifically, by applying a configuration model to reconfigure network edges for each leisure category separately, we preserve network nodes (individuals and their SES values) as well as the number of exposures they had in each category of POI (node degree), but randomize the specific venue in which each exposure occurred. For instance, if an individual crosses paths with 5 people inside of restaurants and 100 people inside of a fitness center, they will be randomly assigned to cross paths with 5 people from all of those who visited restaurants, and 100 people from all of those who visited fitness centers. This null model preserves between-activity homophily which results from activity choices (e.g. whether to visit a country club or restaurant), but erases  within-activity homophily (e.g. individuals who visit any restaurant are equally likely to cross paths). We find that under this null model, there is no positive association between exposure segregation and population size; in fact, there is minimal segregation across all cities as variation between activity categories is insufficient to retain segregation. This is further supported by Supplementary Table \ref{tab:poi_desc_2}, which shows relatively small differences in SES between participants in different categories of leisure activity (e.g. the lowest SES activity, limited service restaurants has a median visitor SES of \$1,352, the highest SES activity, golf courses and country clubs has a median visitor SES of \$1,648.4). }
\label{fig:leisure_null_models}
\end{figure}

\begin{figure}[htbp]
\vspace*{-20mm}
  \centering
  \includegraphics[width=0.95\textwidth]{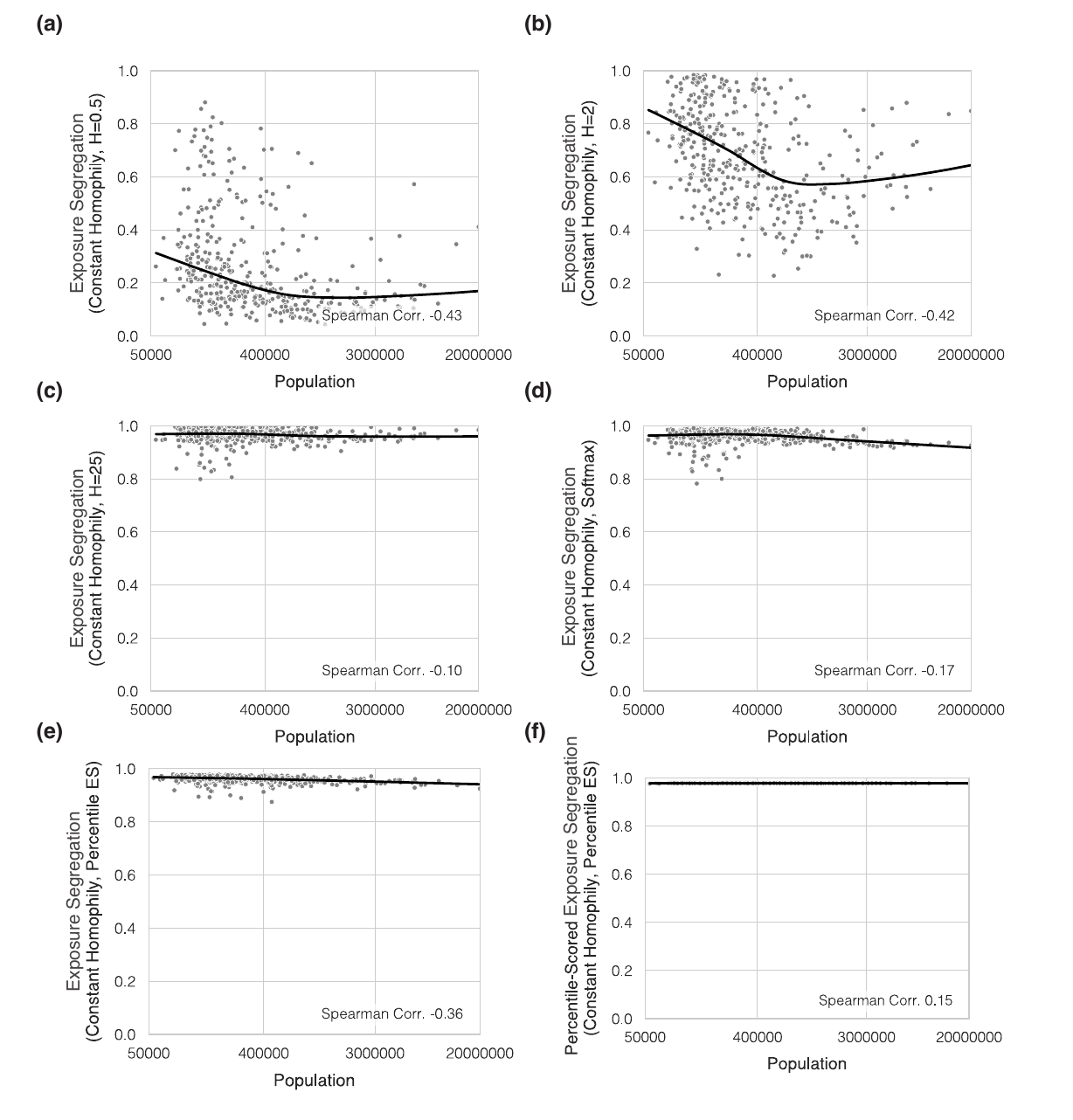}
  \caption{\textbf{Baseline Homophily null model results are robust to varying null model specifications}. We re-run the analysis in Supplementary Figure \ref{fig:leisure_null_models}a under a variety of null model specifications, and find that \emph{in all cases there is no evidence to suggest that the Constant Homophily hypothesis explains the high segregation observed in large cities}. \textbf{(a-c)} We first consider varying the extent of homophily by adding a constant parameter $H$ to the homophilous weighting of edges, to exponentially increase/decrease the extent of homophily in our null model for the probability of individuals $i$ and $j$ crossing paths: $p_{i,j} \propto Similarity(ES_i,ES_j)^H$. We find that regardless of if we \textbf{(a)} decrease homophily (H=0.5) or \textbf{(b-c)} increase homophily mildly (H=2) or strongly (H=25), there is no positive association between population size and segregation in our simulations. In fact, larger cities are less segregated on average, as there is an increase in supply of diverse individuals in socioeconomic status in larger cities. We also consider alternative null model specifications such as \textbf{(d)} a softmax homophily function $p_{i,j} \propto \frac{e^{Similarity(ES_i,ES_j)}}{\sum_{k=1}^{N} e^{Similarity(ES_i,ES_k)}}$ 
  , \textbf{(e)} applying the original null model to percentile-scored values socioeconomic status \textbf{(f)} applying the original null model to percentile-scored values socioeconomic status, and calculating Exposure Segregation using percentile-scored values socioeconomic status. This suggests that the high segregation in large cities is due to a change in resident behavior, facilitated by the built environment of large cities, and not an artifact the socioeconomic status distribution in large cities. }
  \label{fig:leisure_null_models_robust}

\end{figure}

\begin{figure}[htbp]
   \centering
   \vspace{-20mm}
   \includegraphics[width=0.5\textwidth,keepaspectratio]{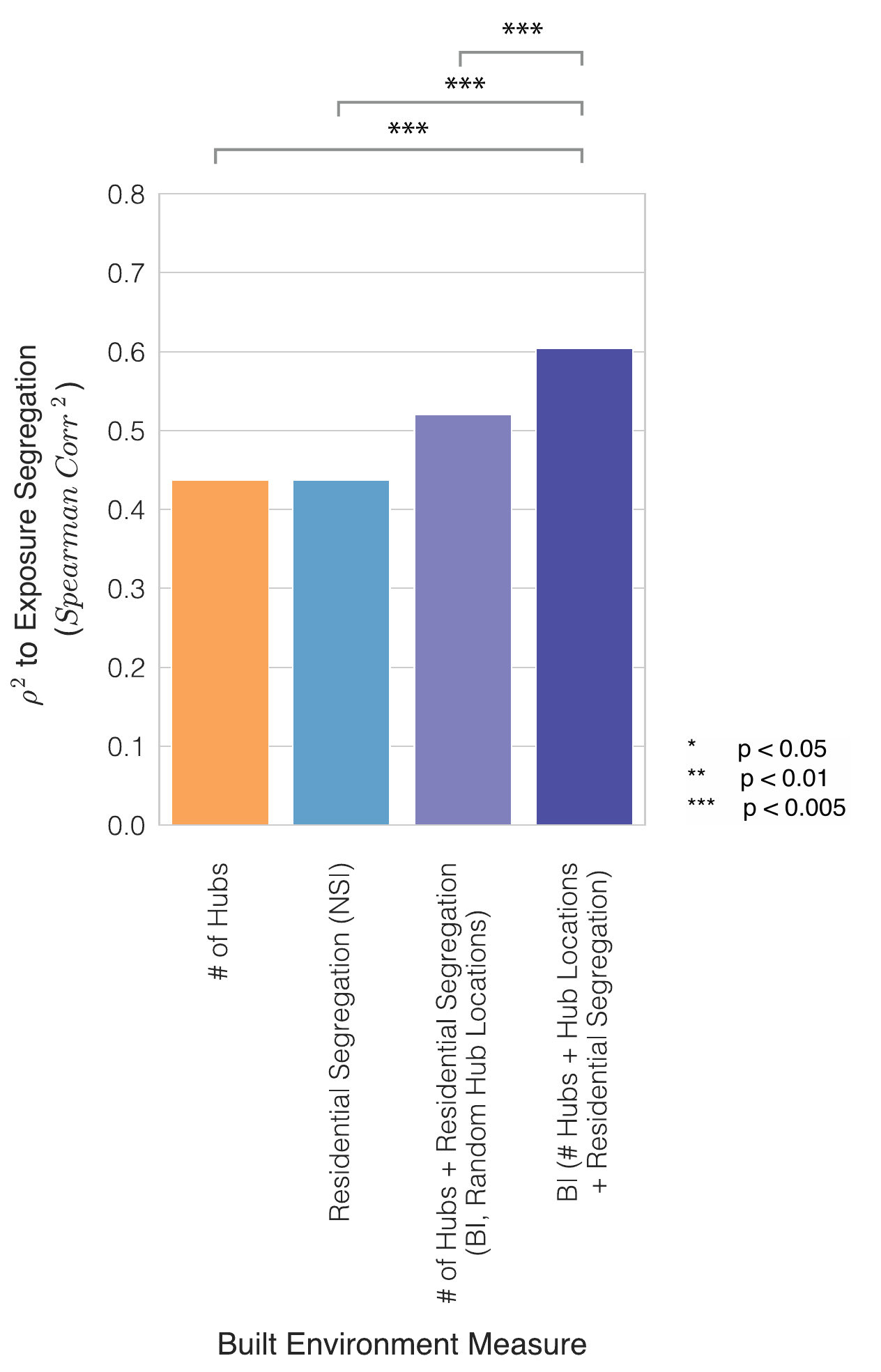}
\caption{\textbf{Understanding why BI explains Exposure Segregation.} We show via an ablation study that exposure hub locations, in addition to number of hubs and residential segregation, contributes to the explanatory power of BI. As illustrated in Extended Data Figure \ref{fig:ccbi_intuition}, BI captures three factors of built environment: (1) locations of exposure hubs (2) number of exposure hubs and (3) residential segregation. In this analysis, we aim to disentangle how these three factors contribute to the ability of BI to explain \metric (as measured by $\rho^2$, the squared Spearman correlation with \metric). We find that number of hubs (orange, $\rho^2=0.436$) and residential segregation (blue, $\rho^2=0.437$) are each correlated with \metric. To measure the combined explanatory power of these two factors within BI, independent of exposure hub locations, we conduct an ablation study in which we calculate Bridging Index for each MSA, using the actual home location data and number of exposure hubs for each MSA, but randomize hub locations (light purple, $\rho^2=0.523$). For each MSA, we estimate this value over 1000 random trials. We find that calculating BI using randomized hub locations is a significantly weaker predictor (p=0.0006$<$0.01, Steiger's Z-test) compared to BI values computed using actual hub locations (dark purple, $\rho^2=0.604$). This demonstrates that hub locations contribute to the explanatory power of BI, i.e. BI explains \metric because it captures the extent to which the locations of hubs in different cities facilitate the exposure of diverse individuals.}
\label{fig:ccbi_ablation}
\end{figure}

\begin{figure}[htbp]
\vspace{20mm}
   \centering
   \hspace*{-15mm}
\includegraphics[width=0.5\textwidth,keepaspectratio]{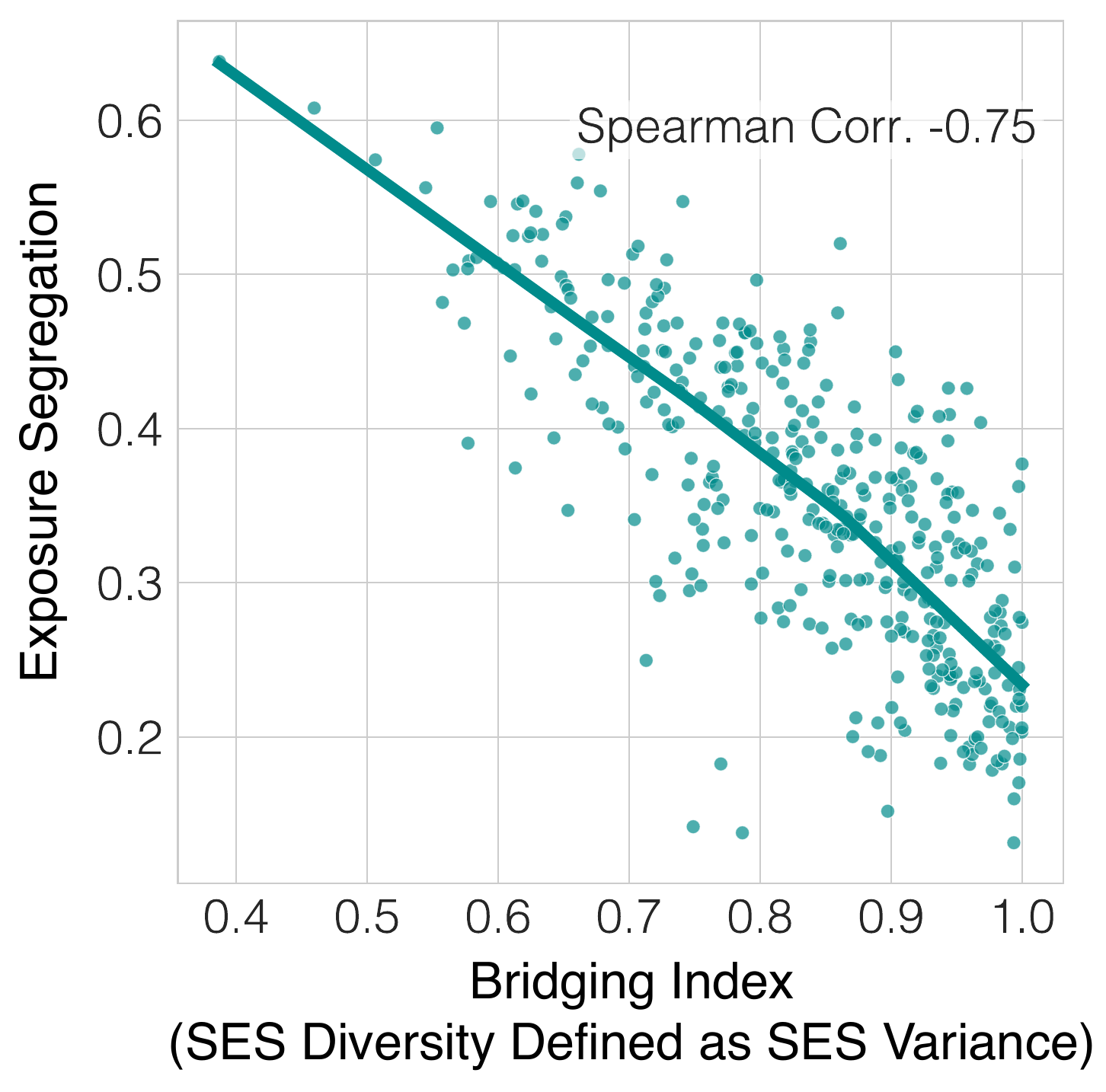}
\caption{Robustness of BI to definition of SES diversity. We calculate a version of Bridging Index which uses variance in to operationalize economic diversity: $Bridging\:Index\:(BI) = \frac{\sum_{i=1}^{K}|\mathcal{C}_i| \cdot Var(\mathcal{C}_i)}{|\mathcal{V}_{MSA}| \cdot Var(\mathcal{V}_{MSA})}$. This variant of BI explains \metric comparably (Spearman Corr. -0.75 vs. -0.78, both N=382, both p$<10^{-4}$) to our primary measure of BI which uses Gini Index to operationalize economic diversity. Thus, we find that the ability of Bridging Index to explain \metric is robust to the definition of SES diversity.}
\label{fig:ccbi_variance}
\end{figure}

\begin{figure}[htbp]
   \centering

   \centering
   \hspace*{-15mm}
\includegraphics[width=1.2\textwidth,keepaspectratio]{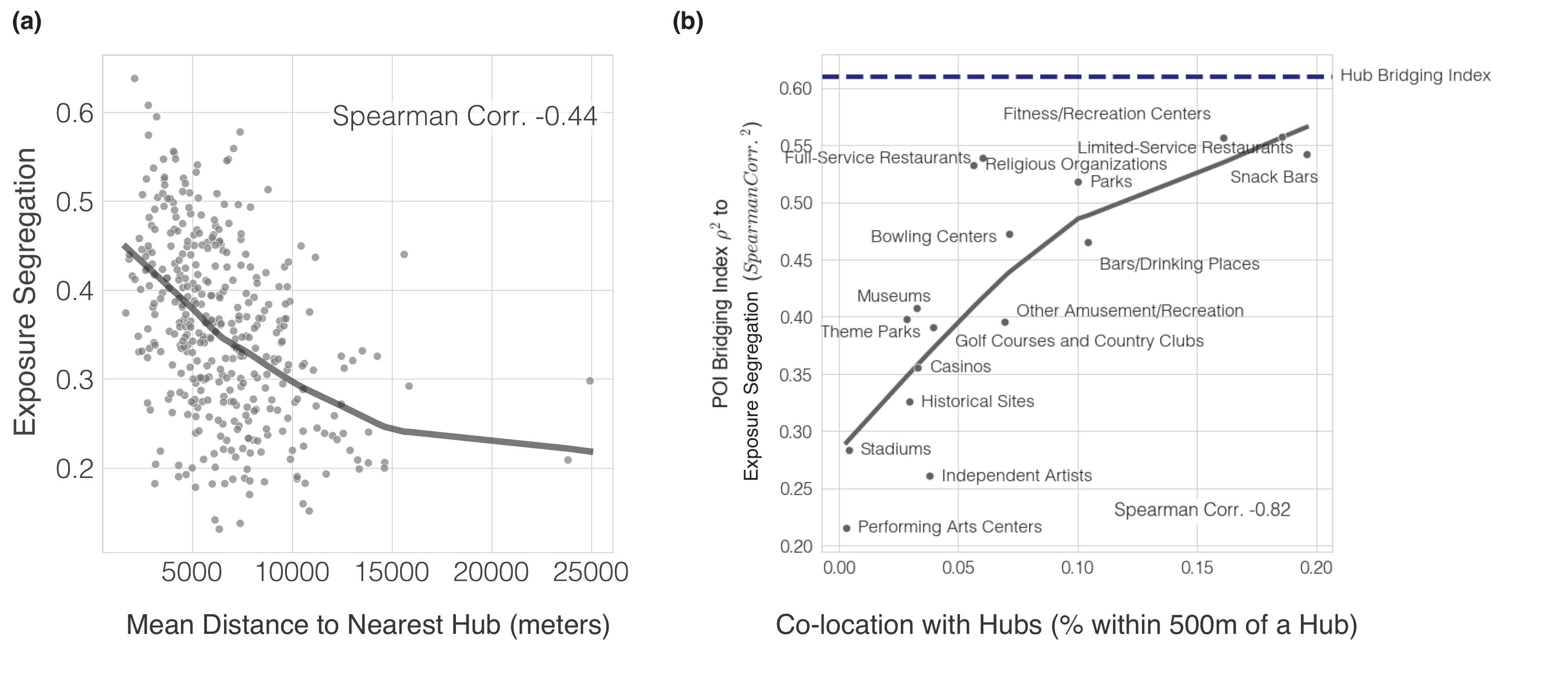}

\caption{\textbf{We consider alternative \new{processes} through which built environment may  mitigate \metric.} \textbf{(a)} Following the inverse relationship between POI localization and segregation established in Extended Data Figure \ref{fig:poi_disparity_explanation}a, we consider whether the de-localization of exposure hubs alone can provide an alternative to BI. We compute mean distance to nearest exposure hub for each MSA, which is the same measure in Extended Data Figure \ref{fig:poi_disparity_explanation}a but calculated for exposure hubs. We find that while hub localization is inversely correlated with \metric (Spearman Corr -0.44, N=382, p$<10^{-4}$), this correlation is significantly less ($p<10^{-4}$) than the correlation between BI and \metric (Spearman Corr -0.78, N=382, p$<10^{-4}$). This suggests that hub bridging, as quantified by BI may be a more promising direction to investigate as a potential mitigator of segregation. \textbf{(b)} We also consider whether fine-grained POIs may function as bridges between diverse individuals. For each of the fine-grained leisure POI categories in Figure \ref{fig:main_2}c, we calculate a Bridging Index across all MSAs (using the same procedure to calculate BI as shown in Extended Data Figure \ref{fig:ccbi_variance}, except using fine-grained POI locations instead of exposure hub locations). For instance, to calculate the Bridging Index for restaurants, we cluster all homes by the nearest restaurant location, and then calculate: $(Restaurant)\:Bridging\:Index\: = \frac{\sum_{i=1}^{K}|\mathcal{R}_i| \cdot Gini\:Index(\mathcal{R}_i)}{|\mathcal{V}_{MSA}| \cdot Gini\:Index(\mathcal{V}_{MSA})}$. After calculating the bridging index for all fine-grained POI categories and for each of the 382 MSAs, we then measure the correlation between each bridging index and \metric across all MSAs (as measured by $\rho^2$, the squared Spearman correlation). We find that BI for hubs provides a stronger correlation ($\rho^2=0.604$, horizontal line), than all other bridging indices which are plotted as points on the scatter-plot in \textbf{(b)}. Further, we find that POI categories which are often located inside or near exposure hubs (co-location, X-axis) have bridging indices which are stronger predictors of \metric (e.g. for fitness/recreation centers, snack bars etc.). The high correlation between (Spearman Correlation -0.82, N=17, $p < 0.001$) between co-location of POIs and bridging index predictive ability demonstrates asymptotic convergence between all other predictive bridging index metrics  and our primary BI measure. This further suggests that bridging of exposure hubs should be the primary metric of interest for mitigators of segregation, because other bridging indexes computed for fine-grained POI locations are at best proxies for BI which leverages higher-level exposure hub locations. Supplementary Figures \ref{fig:ccbi_example1}-\ref{fig:ccbi_example4} illustrate the frequent co-location between hubs and other fine-grained POIs.}
\label{fig:ccbi_alternatives_2}
\end{figure}

\begin{figure}[htbp]
   \centering
   \hspace*{-15mm}
   \includegraphics[width=0.9\textwidth,keepaspectratio]{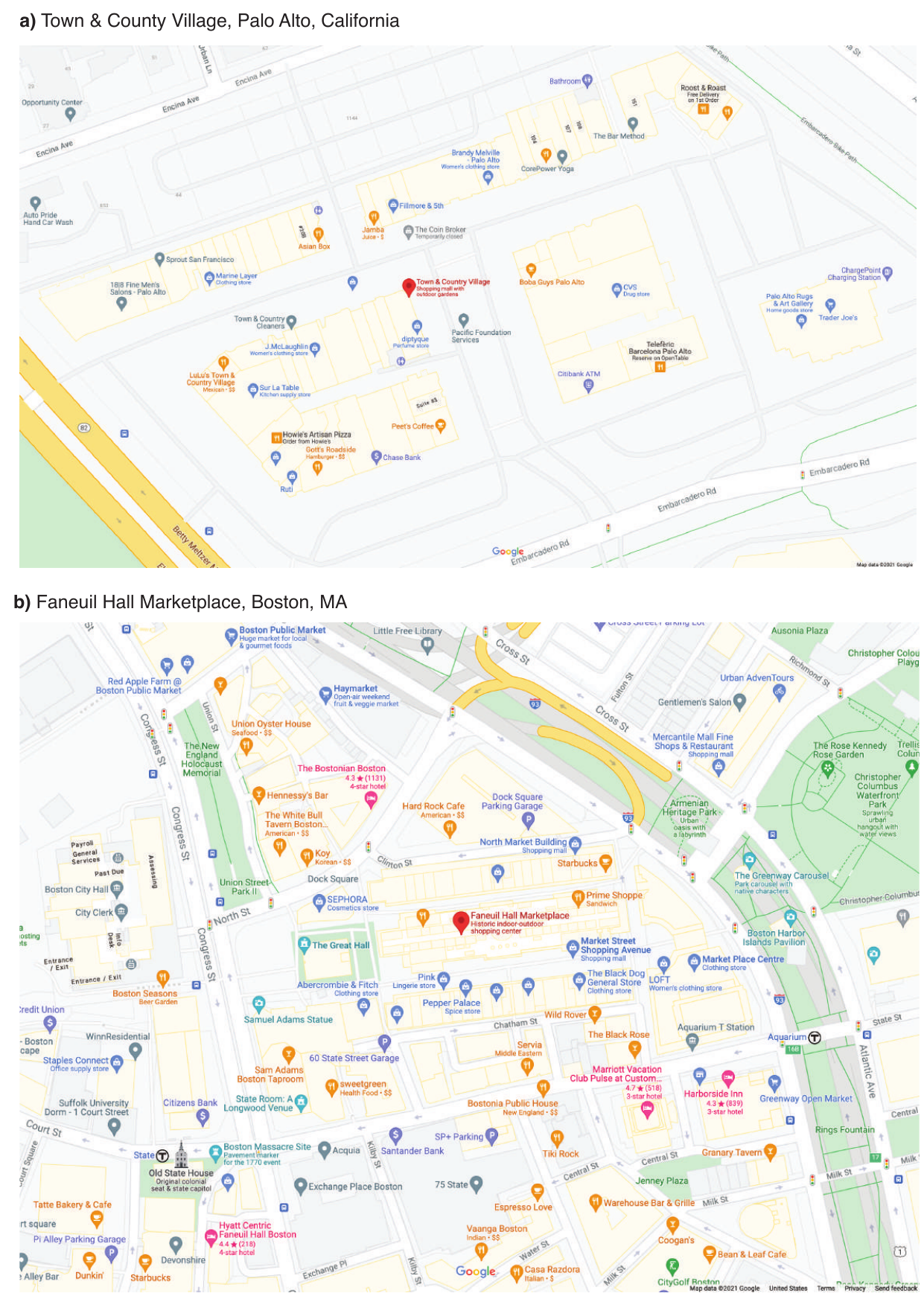}
   \caption{\textbf{Examples of exposure hubs in coastal cities of (a) San Francisco Bay Area and (b) Boston, MA.} Hubs frequently contain a diverse assortment of POIs including  restaurants, fitness centers/gyms, grocery stores, etc. and are also frequently hubs around which other POIs are located nearby.}
   \label{fig:ccbi_example1}
\end{figure}

\begin{figure}[htbp]
   \centering
   \hspace*{-15mm}
   \includegraphics[width=1.0\textwidth,keepaspectratio]{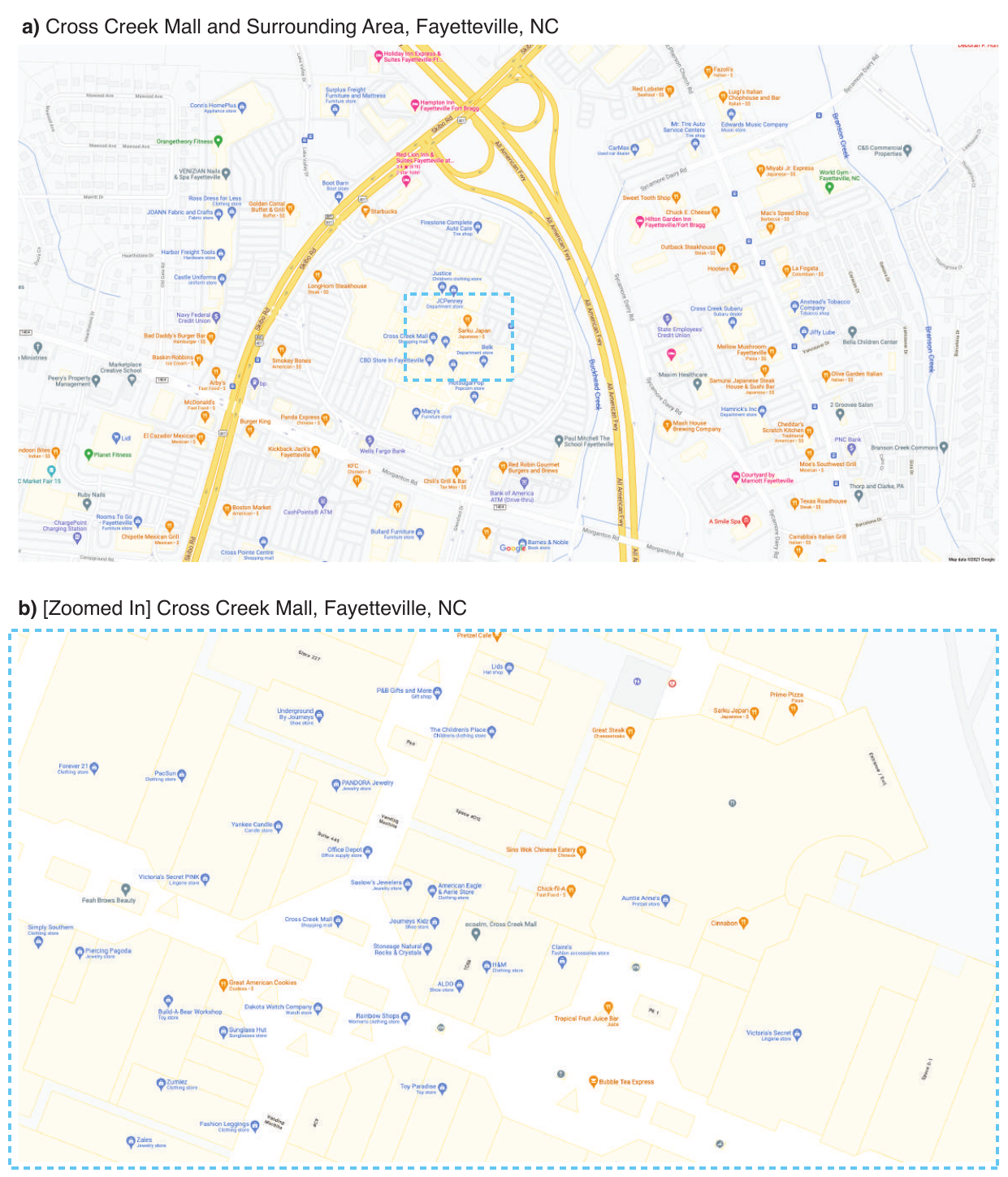}
   \caption{\textbf{Example of a major hubs in Fayetteville, NC (a) zoomed-out view of hub and surrounding co-located POIs  (b) zoomed-in view of the hub core and businesses contained inside.} We find that in Fayetteville, a city with a high Bridging Index, large hubs contain a variety of POIs which cater to diverse individuals of both high and low-ES.}
   \label{fig:ccbi_example2}
\end{figure}

\begin{figure}[htbp]
   \centering
   \hspace*{-15mm}
   \includegraphics[width=0.7\textwidth,keepaspectratio]{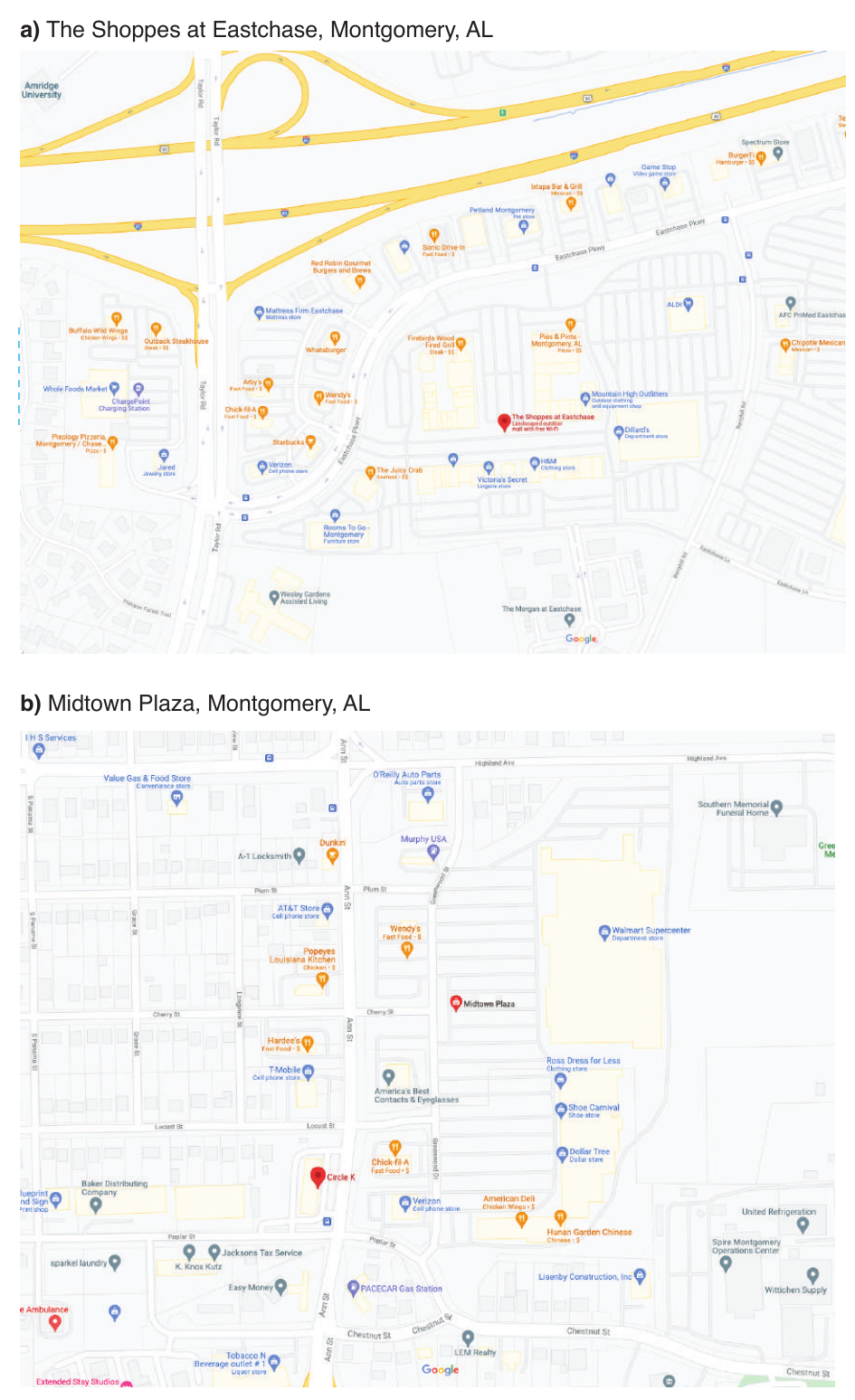}
   \caption{\textbf{Examples two hubs in Montgomery, AL which have visitors of predominantly (a) high socioeconomic status  (b) low socioeconomic status.} We find that in Montgomery, AL a city with a low Bridging Index, smaller hubs exist which contain POIs which cater to a narrow band of individuals in a specific economic stratum. For instance, we find that the nearby grocery store \textbf{(a)} is a Whole Foods Market in the high-SES hub, in contrast to the \textbf{(b)} Walmart Supercenter in the low-SES hub.}
   \label{fig:ccbi_example4}
\end{figure}

\begin{figure}[htbp]
  \centering
  \includegraphics[width=0.5\textwidth]{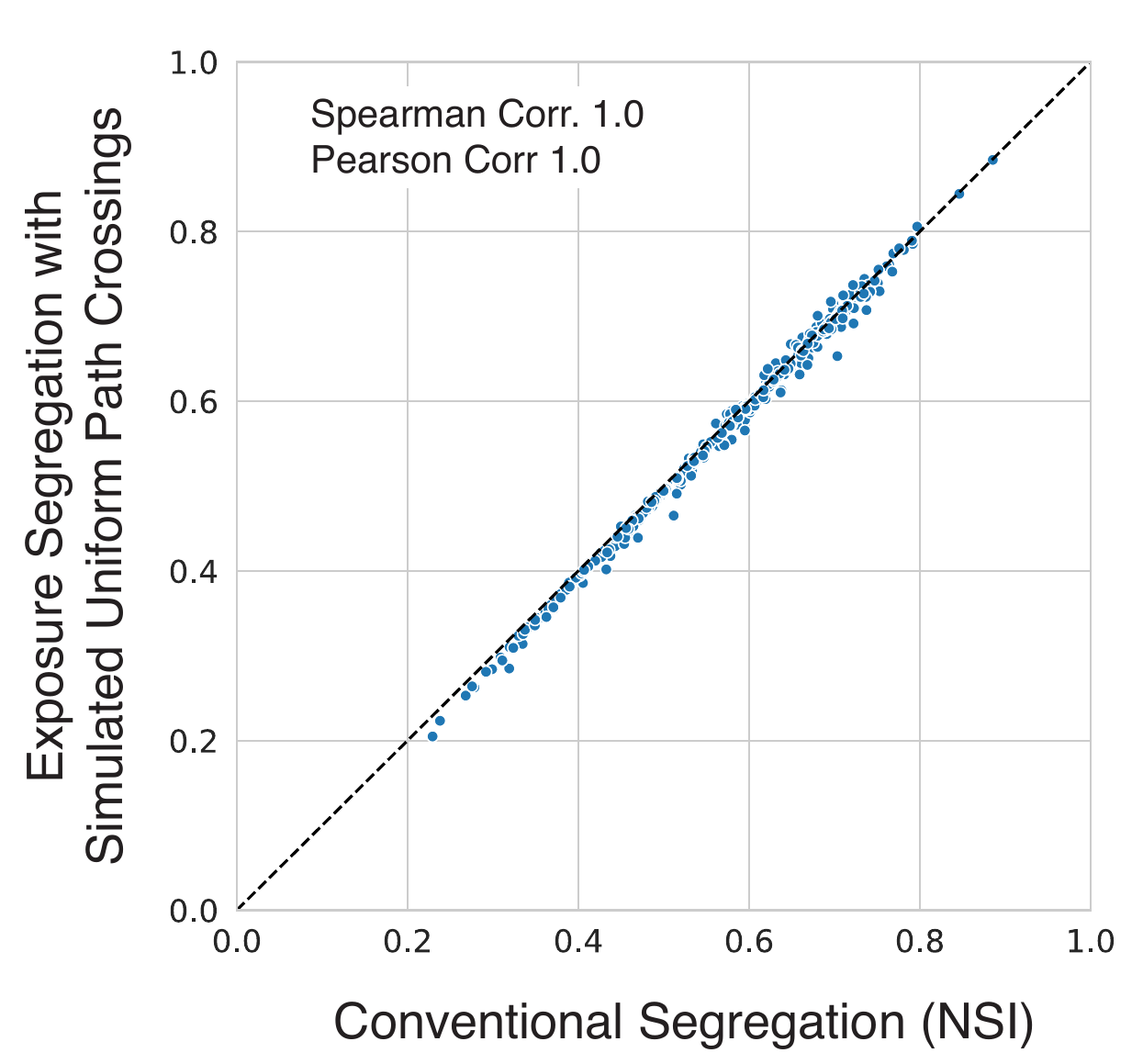}
  \caption{\textbf{Our model with simulated uniform within-tract crossings is equivalent to the conventional neighbourhood sorting index (NSI).} Each point is a Metropolitan Statistical Area (MSA). The y-axis shows the \metric estimate from the mixed model with a simulated path crossing between every person in a tract (in our dataset). The x-axis shows the correlation between a person's SES and the average SES of people in their tract, which is the neighbourhood sorting index (NSI). As these measures are equivalent, Spearman Corr = 1.0 and Pearson Corr. = 1.0.}
  \label{fig:nsi_vs_mixed_model}
\end{figure}

\begin{figure}[htbp]
  \centering
  \includegraphics[width=0.6\textwidth]{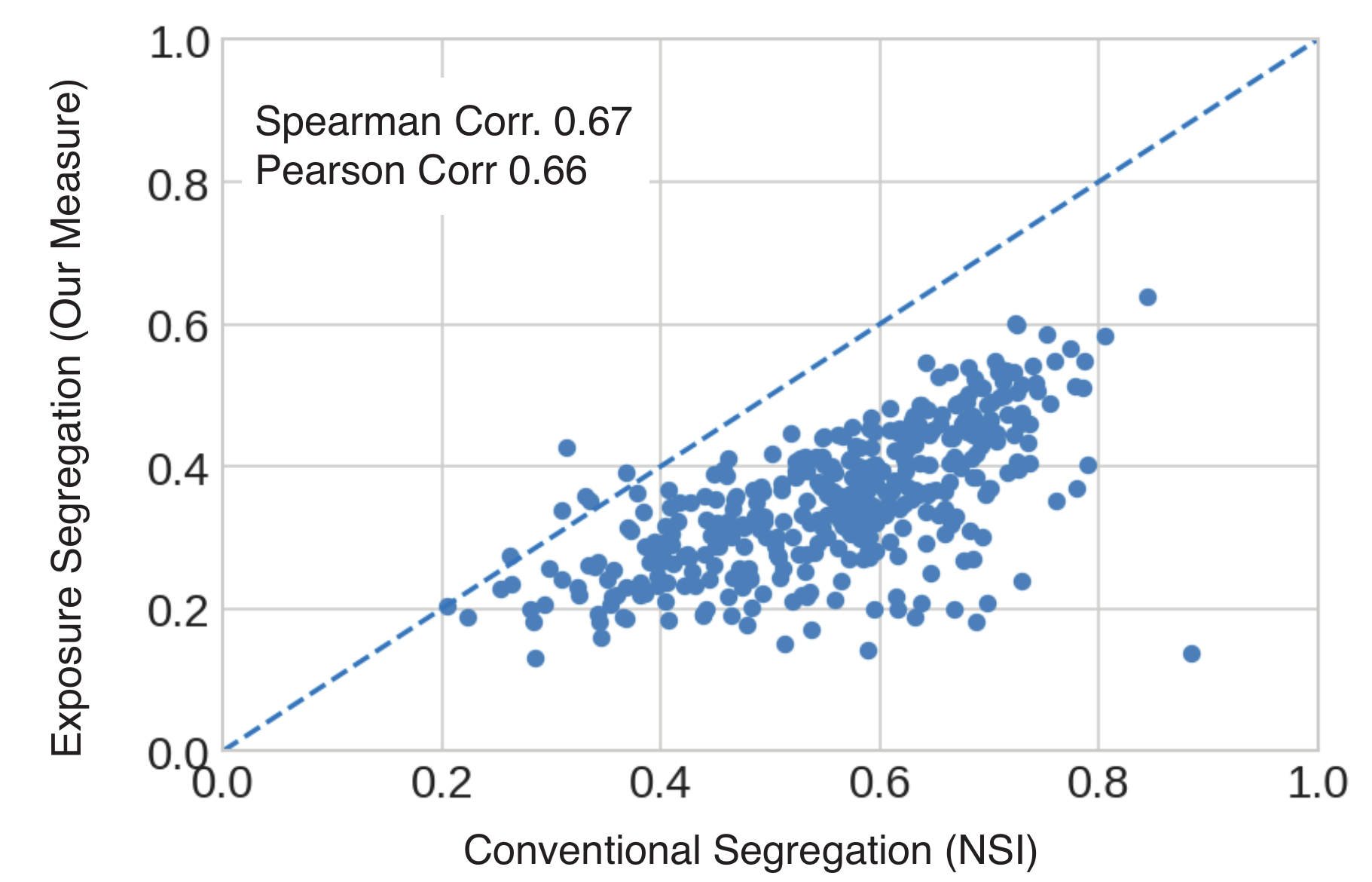}
  \caption{\textbf{Our segregation measure versus a conventional residential segregation, neighborhood sorting index (NSI).} Each point is a Metropolitan Statistical Area (MSA). Regardless of whether we compare the numerical segregation values (Pearson Correlation 0.67) or the MSA ranking (Spearman Correlation 0.66), only moderate correlation indicates that our measure is different in kind from residential segregaiton as measured conventionally by NSI.}
  \label{fig:low_correlation_figure_1}
\end{figure}

\begin{figure}[htbp]
  \centering
  \includegraphics[width=0.5\textwidth]{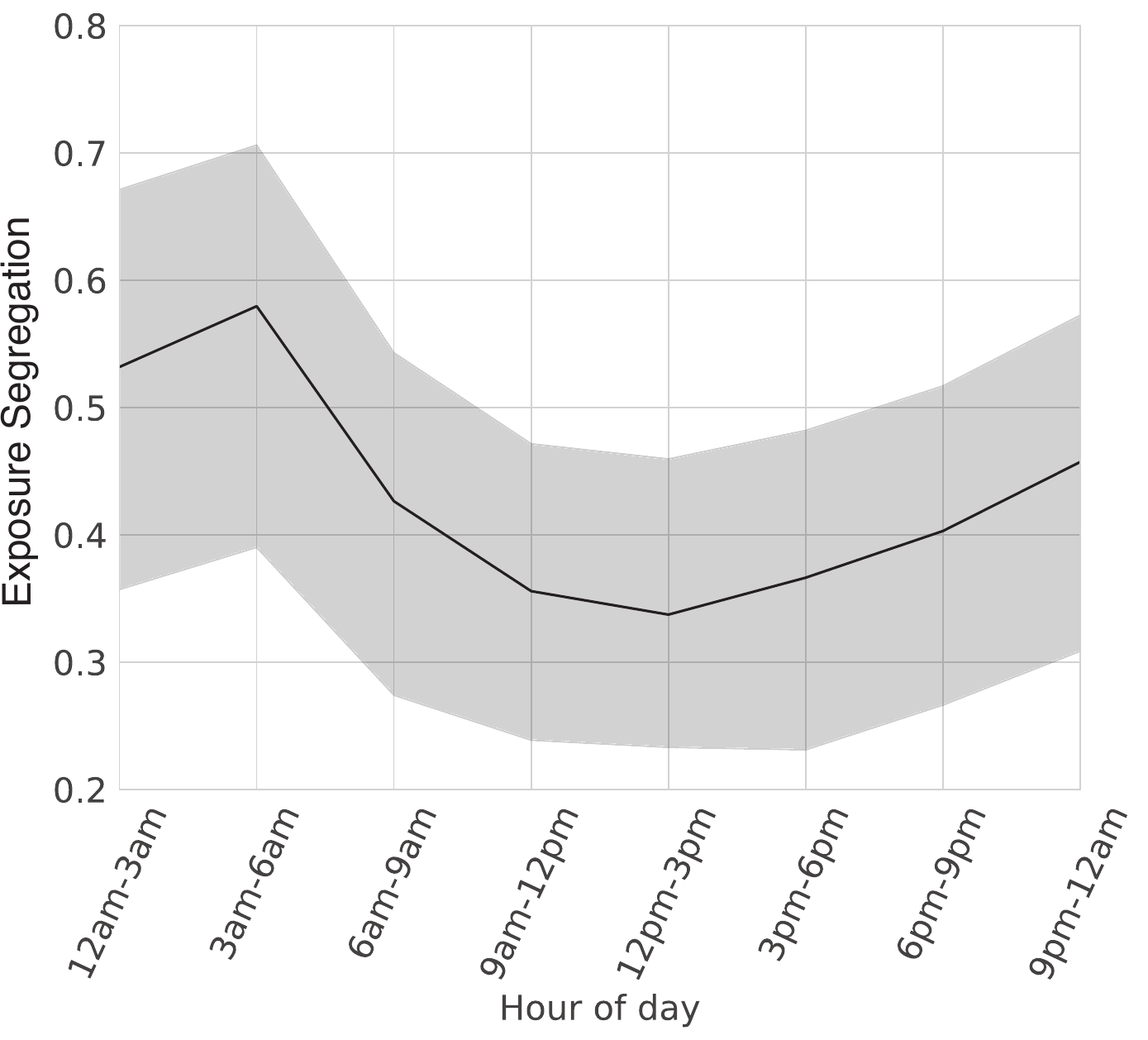}
  \caption{\textbf{Segregation decomposed by time.} As described in Methods \ref{sec:segregation_by_time}, our fine-grained exposure network allows us to decompose our overall \metric into  estimates of segregation during different hours of the day, by filtering for exposures that occurred within a specific hour. In Supplementary Figure \ref{fig:segregation_by_hour}, we partition estimates of segregation by 3 hour windows to illustrate how segregation varies throughout the day (see Supplementary Information).  We observe that segregation increases by 61\% between the afternoon and early morning hours. Segregation is lowest during commute and work hours, indicating higher levels of exposure with people of different SES while at work or otherwise away from home. Segregation is higher during nighttime hours. This is driven by individuals returning to their home neighborhoods, which are more homogeneous in socioeconomic status, as we well as mechanically a result of SES being is defined by rent value, such that people who live in the same household will have the same SES (and thus will be highly segregated).}
  \label{fig:segregation_by_hour}
\end{figure}

\begin{figure}[htbp]

\begin{subfigure}{\textwidth}
  \centering
  \includegraphics[width=.3\textwidth]{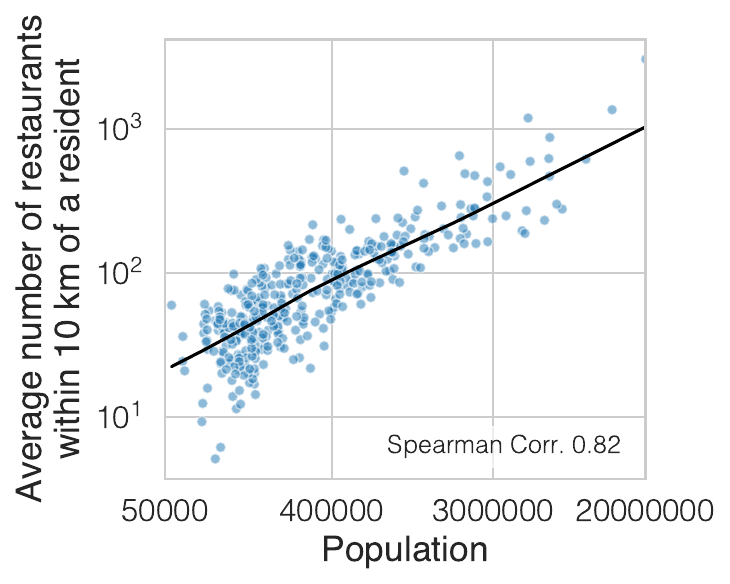}
  \includegraphics[width=.3\textwidth]{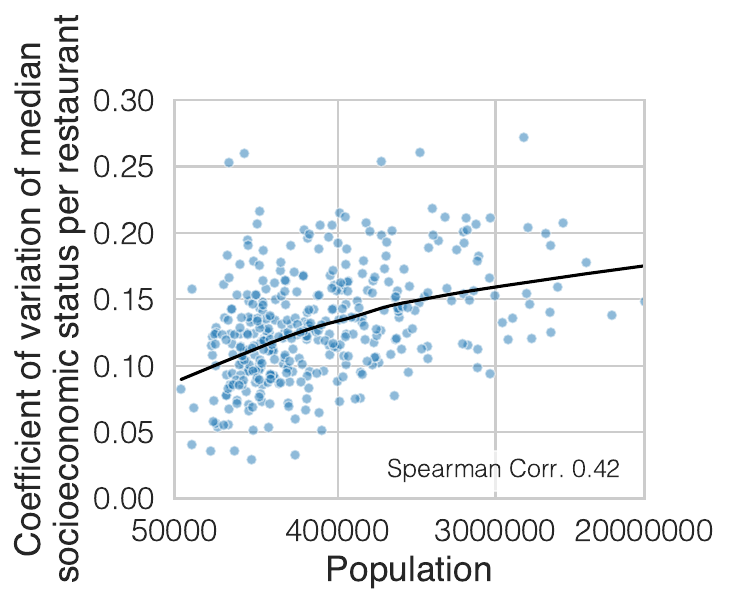}
  \includegraphics[width=.3\textwidth]{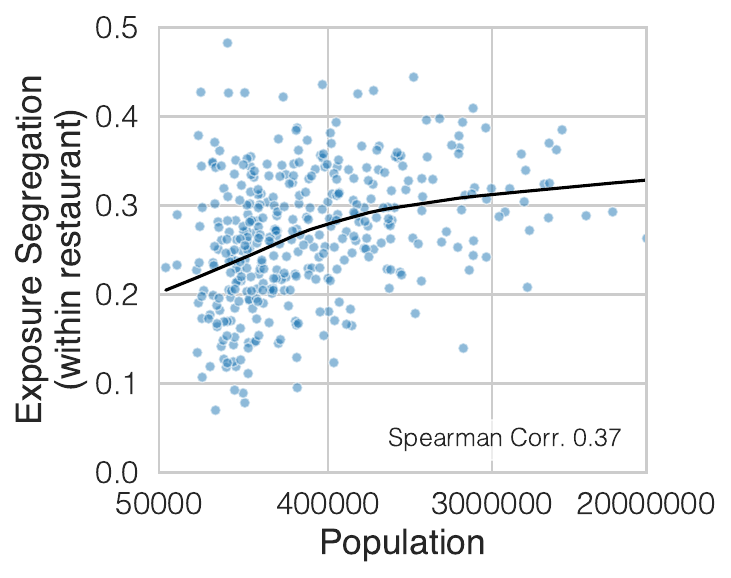}
  \caption{Full-Service Restaurants}
\end{subfigure}

\begin{subfigure}{\textwidth}
  \centering
  \includegraphics[width=.3\textwidth]{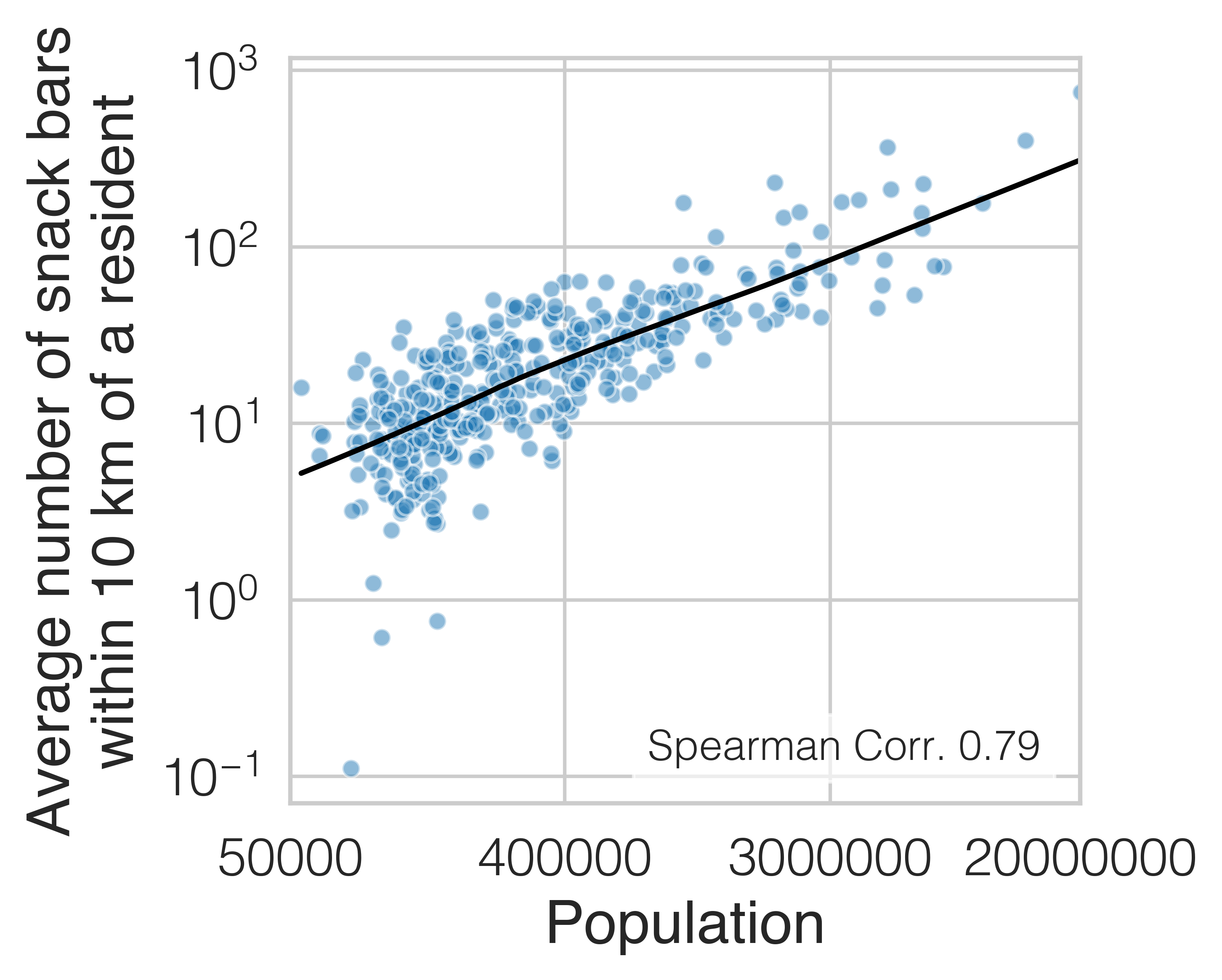}
  \includegraphics[width=.3\textwidth]{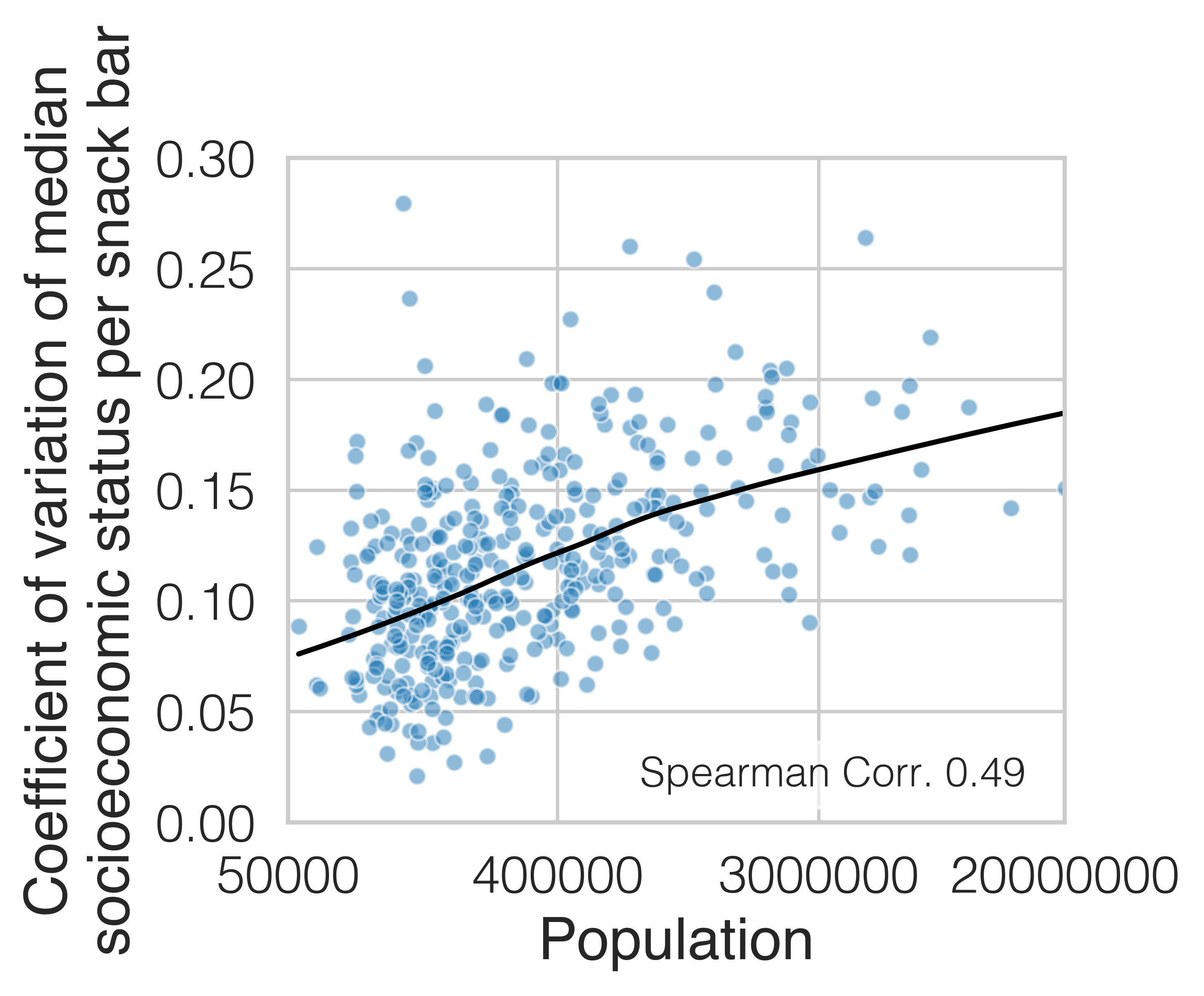}
  \includegraphics[width=.3\textwidth]{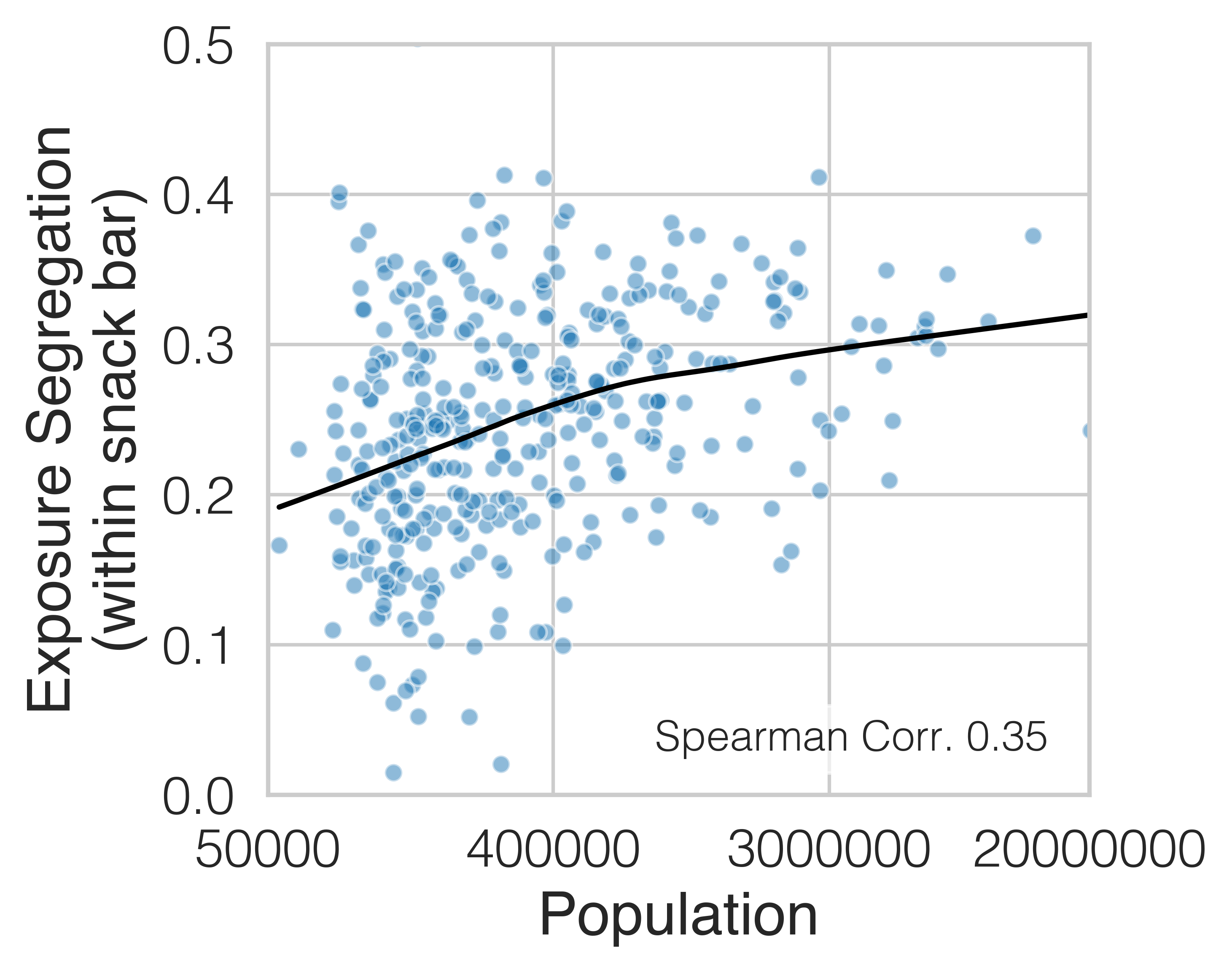}
  \caption{Snack Bars}
\end{subfigure}

\begin{subfigure}{\textwidth}
  \centering
  \includegraphics[width=.3\textwidth]{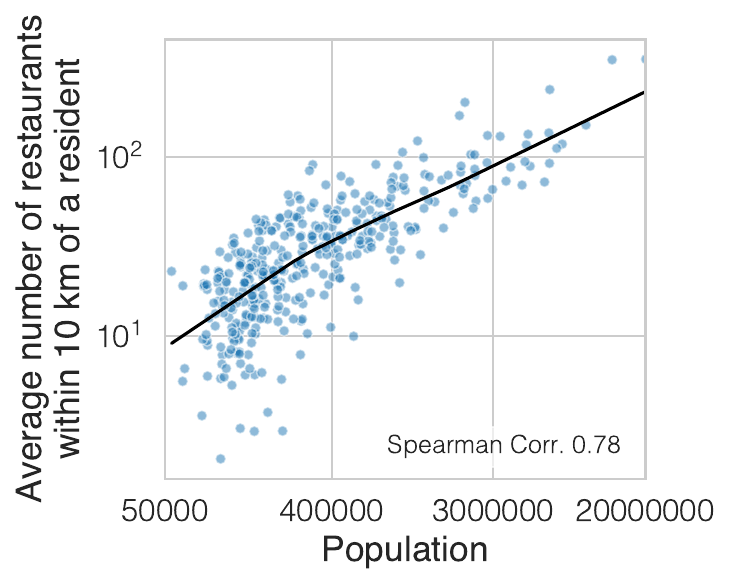}
  \includegraphics[width=.3\textwidth]{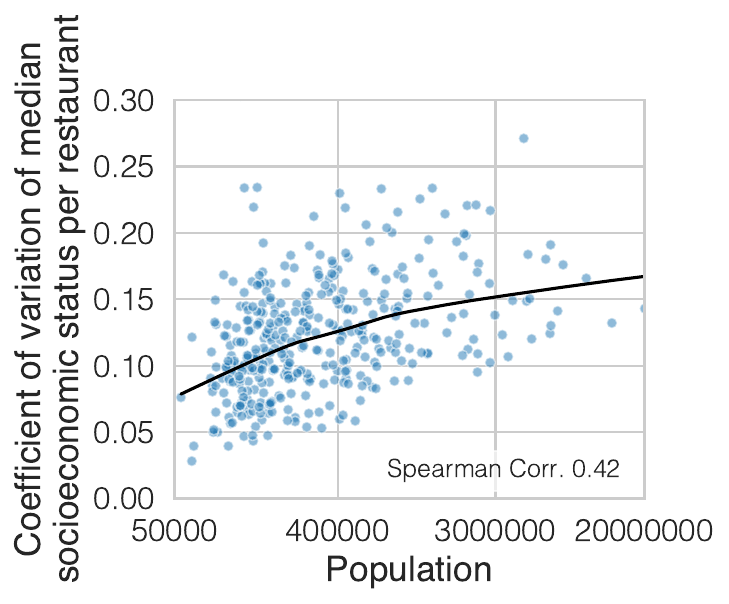}
  \includegraphics[width=.3\textwidth]{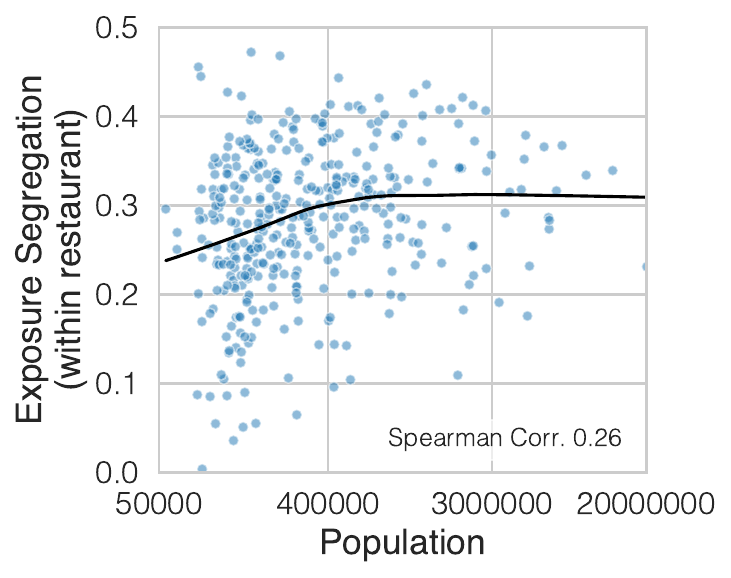}
  \caption{Limited-Service Restaurants}
\end{subfigure}

\end{figure}

\begin{figure}[htbp]\ContinuedFloat

\begin{subfigure}{\textwidth}
  \centering
  \includegraphics[width=.3\textwidth]{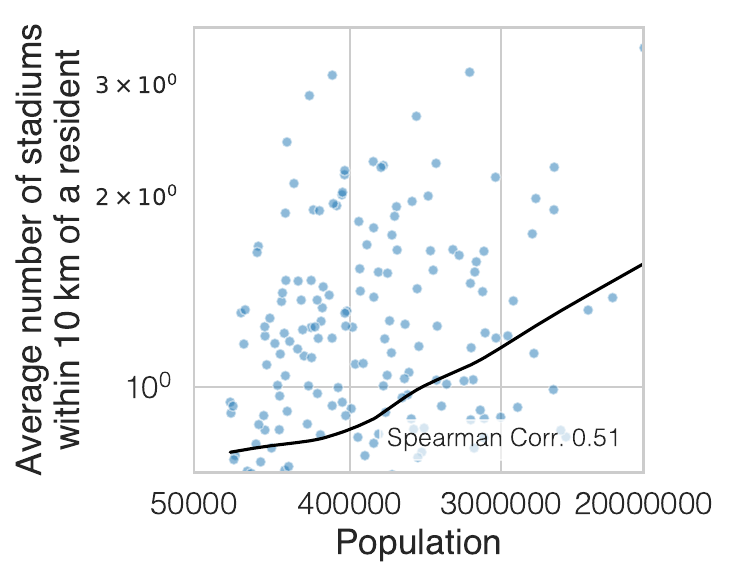}
  \includegraphics[width=.3\textwidth]{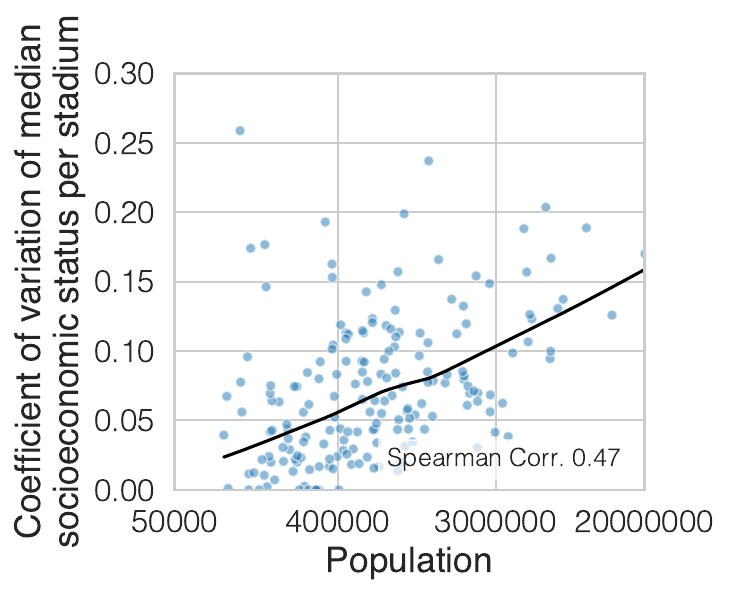}
  \includegraphics[width=.3\textwidth]{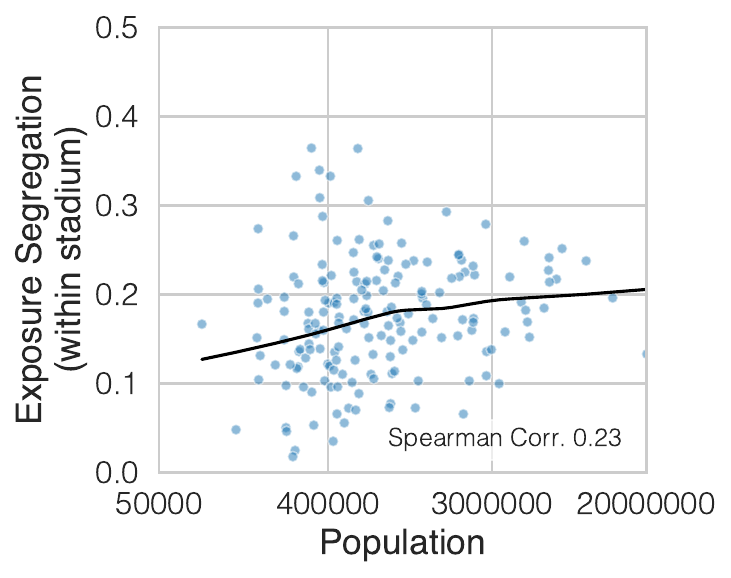}
  \caption{Stadiums}
\end{subfigure}

\begin{subfigure}{\textwidth}
  \centering
  \includegraphics[width=.3\textwidth]{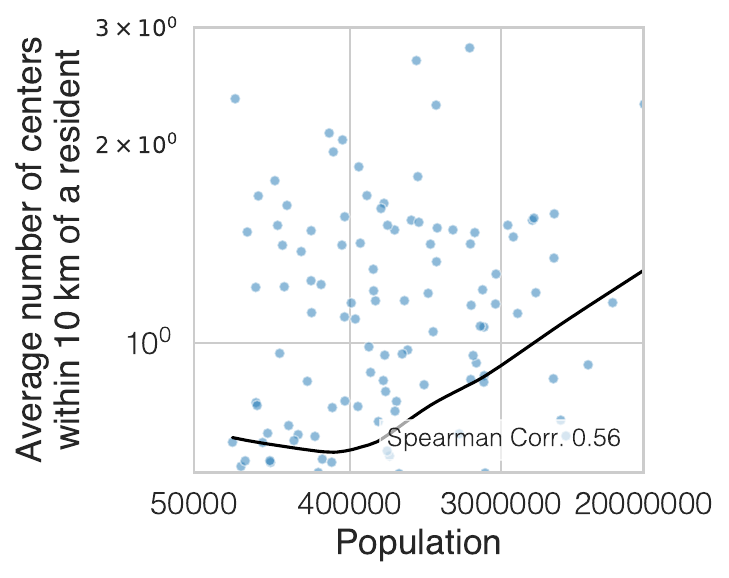}
  \includegraphics[width=.3\textwidth]{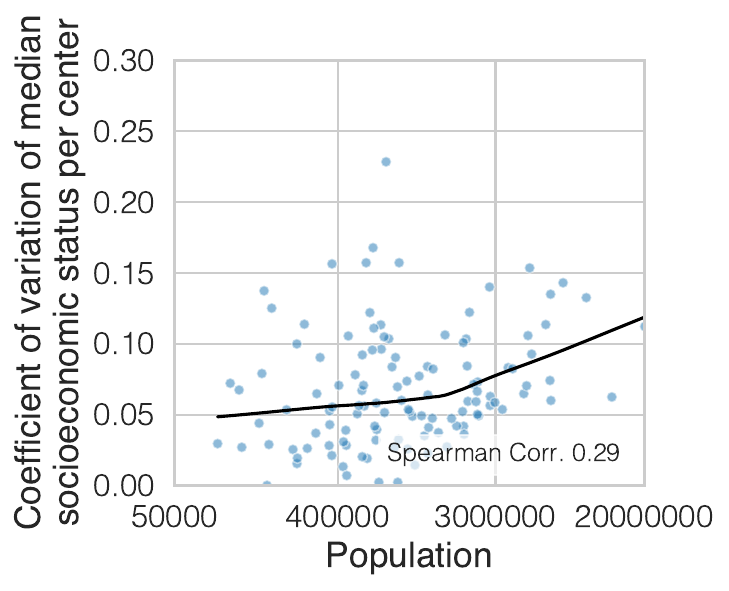}
  \includegraphics[width=.3\textwidth]{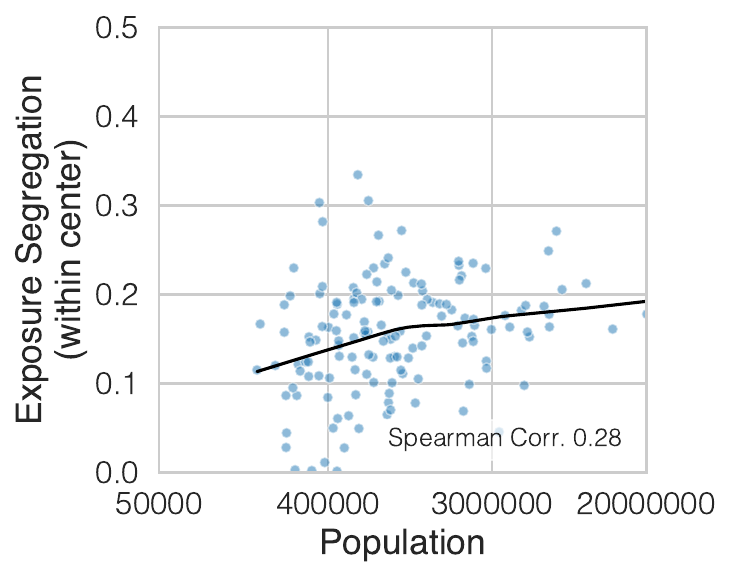}
  \caption{Performing Arts Centers}
\end{subfigure}

 \caption{\textbf{Across many activities, POIs in large cities are more differentiated and consequently more segregated}. This figure shows that the trend towards more options, increased differentiation, and consequently higher segregation is consistent across many prominent POI categories. Here we find similar results for the 5 most frequently visited fine-grained Safegraph place features. The analyses for full-service correspond to Figures \ref{fig:main_2}c-e, and we additionally show the same trend for snack bars, limited-service restaurants, stadiums, and performing arts centers (ranked 2-5 after full-service restaurants in terms of most frequently visited POIs among Safegraph places). Across the board, large, densely populated metropolitan areas are associated with increased options and economic differentiation of POIs, which may facilitate higher self-segregation.}

\label{fig:robustness_check_leisure_pois}

\end{figure}

\begin{table}[ht]
\centering
\begin{footnotesize}
\setlength\tabcolsep{2pt} %

\begin{tabular}{@{\extracolsep{5pt}}lcccccc}
\\[-1.8ex]\hline
\hline \\[-1.8ex]
& \multicolumn{6}{c}{\textit{Dependent variable:}} \ Exposure Segregation
\cr \cline{6-7}
\\[-1.8ex] & (1) & (2) & (3) & (4) & (5) & (6) \\
\hline \\[-1.8ex]
 Intercept & 0.355$^{***}$ & 0.355$^{***}$ & 0.355$^{***}$ & 0.356$^{***}$ & 0.355$^{***}$ & 0.355$^{***}$ \\
  & (0.005) & (0.004) & (0.004) & (0.004) & (0.003) & (0.003) \\
 Population Density & 0.039$^{***}$ & & 0.024$^{***}$ & 0.022$^{***}$ & 0.017$^{***}$ & 0.017$^{***}$ \\
  & (0.005) & & (0.004) & (0.004) & (0.004) & (0.004) \\
 Gini Index (Estimated Rent) & & 0.064$^{***}$ & 0.058$^{***}$ & 0.059$^{***}$ & 0.049$^{***}$ & 0.050$^{***}$ \\
  & & (0.004) & (0.004) & (0.004) & (0.003) & (0.003) \\
 Political Alignment (\% Democrat in 2016 Election) & & & & 0.009$^{*}$ & & 0.006$^{}$ \\
  & & & & (0.005) & & (0.004) \\
 Racial Demographics (\% non-Hispanic White) & & & & -0.005$^{}$ & & 0.003$^{}$ \\
  & & & & (0.004) & & (0.003) \\
 Mean SES (Estimated Rent) & & & & -0.009$^{*}$ & & -0.003$^{}$ \\
  & & & & (0.005) & & (0.004) \\
 Walkability (Walkscore) & & & & & 0.002$^{}$ & 0.001$^{}$ \\
  & & & & & (0.003) & (0.004) \\
 Commutability (\% Commute to Work) & & & & & -0.012$^{***}$ & -0.013$^{***}$ \\
  & & & & & (0.004) & (0.004) \\
 Conventional Segregation (NSI) & & & & & 0.048$^{***}$ & 0.047$^{***}$ \\
  & & & & & (0.003) & (0.003) \\
\hline \\[-1.8ex]
 Observations & 382 & 382 & 382 & 376 & 382 & 376 \\
 $R^2$ & 0.151 & 0.419 & 0.475 & 0.490 & 0.682 & 0.680 \\
 Adjusted $R^2$ & 0.149 & 0.417 & 0.472 & 0.483 & 0.678 & 0.673 \\
\hline
\hline \\[-1.8ex]
\textit{} & \multicolumn{6}{r}{$^{*}$p$<$0.1; $^{**}$p$<$0.05; $^{***}$p$<$0.01} \\
\end{tabular}

\end{footnotesize}
\caption{\textbf{Population density is significantly associated with \metric, after controlling for MSA income inequality (Gini Index), political alignment (\% Democrat in 2016 election), racial demographics (\% non-Hispanic White), mean ES, walkability (Walkscore\cite{walkscore}), commutability (\% of residents commuting to work), and residential segregation (NSI)}. This table is from an analogous regression to the regression shown in Extended Data Table \ref{tab:segregation_population}, using population density instead of population size (we look at each separately due to co-linearity between population size and density). Here we show the coefficients (after normalizing via z-scoring to have mean 0 and variance 1) from the primary specifications estimating the effect of population density on \metric across all MSAs. Columns (1-5) are models specified with different subsets of covariates; Column 6 shows model specification with all covariates. Differences between sample size in models is due to missing data for several covariates in a small number of MSAs (Walkscores were not available for all MSAs). (*p $<$ 0.1; **p $<$ 0.05; *** p $<$ 0.01).}
\label{tab:segregation_density}
\end{table}

\afterpage{
\begin{tiny}

\begin{longtable}{lrrrrrrr}

\caption{Exposure Segregation and related variables (i.e. \# Exposures, Mean ES, NSI, Gini Index, Population Size, and Bridging Index (BI) by MSA}\\
\toprule
                                            MSA & Exposure Segregation & \# Exposures &  Mean SES &  NSI & Gini &     Pop. Size & BI \\
\midrule
\endfirsthead
\caption[]{(cont'd) Exposure Segregation and related variables (i.e. \# Exposures, Mean ES, NSI, Gini Index, Population Size, and Bridging Index (BI) by MSA} \\
\toprule
                                            MSA & Exposure Segregation & \# Exposures &  Mean SES &  NSI & Gini &     Pop. Size & BI \\
\midrule
\endhead
\midrule
\multicolumn{8}{r}{{Continued on next page}} \\
\midrule
\endfoot

\bottomrule
\endlastfoot

                                    Abilene, TX &                    0.44 &     561,896.00 & 1,245.47 & 0.63 & 0.21 &    170,516.00 & 0.79 \\
                                      Akron, OH &                    0.55 &   2,211,810.00 & 1,338.21 & 0.71 & 0.27 &    704,367.00 & 0.65 \\
                                     Albany, GA &                    0.40 &     355,999.00 & 1,077.89 & 0.47 & 0.26 &    151,293.00 & 0.79 \\
                                     Albany, OR &                    0.27 &     153,057.00 & 1,425.95 & 0.40 & 0.11 &    124,977.00 & 0.95 \\
                    Albany-Schenectady-Troy, NY &                    0.40 &   2,058,079.00 & 1,651.16 & 0.62 & 0.18 &    882,130.00 & 0.81 \\
                                Albuquerque, NM &                    0.35 &   1,979,325.00 & 1,316.79 & 0.59 & 0.19 &    912,897.00 & 0.78 \\
                                 Alexandria, LA &                    0.37 &     267,228.00 & 1,011.98 & 0.52 & 0.23 &    153,604.00 & 0.88 \\
              Allentown-Bethlehem-Easton, PA-NJ &                    0.44 &   2,629,311.00 & 1,647.92 & 0.62 & 0.19 &    838,081.00 & 0.79 \\
\end{longtable}
\end{tiny}
}

\begin{table}[ht]
\centering
\begin{scriptsize}
\setlength\tabcolsep{2pt} %
\begin{tabular}{rrrr}
\toprule
 Percentile &  \# of Exposures &  Percentile &  \# of Exposures \\
\midrule
        0.0 &                 1.0 &        51.0 &               146.0 \\
        1.0 &                 2.0 &        52.0 &               152.0 \\
        2.0 &                 4.0 &        53.0 &               157.0 \\
        3.0 &                 5.0 &        54.0 &               163.0 \\
        4.0 &                 7.0 &        55.0 &               168.0 \\
        5.0 &                 8.0 &        56.0 &               174.0 \\
        6.0 &                10.0 &        57.0 &               181.0 \\
\bottomrule
\end{tabular}
\end{scriptsize}
\caption{\new{Distribution of number of exposures for all individuals residing in 382 Metropolitan Statistical Areas (MSAs). The median individual had 141 exposures overall. 8,609,406 individuals reside in a Metropolitan Statistical Area (90\% of the overall 9,567,559 individuals in our study). The remaining 958,153 users live outside of MSAs, influencing the \metric of an MSA by coming into contact with MSA residents.}}
\label{tab:interaction_distribution}
\end{table}

\begin{figure}[htbp]
\vspace*{-20mm}
   \centering
   \hspace*{-15mm}
\begin{subfigure}[t]{.5\textwidth}
\includegraphics[width=\textwidth,keepaspectratio]{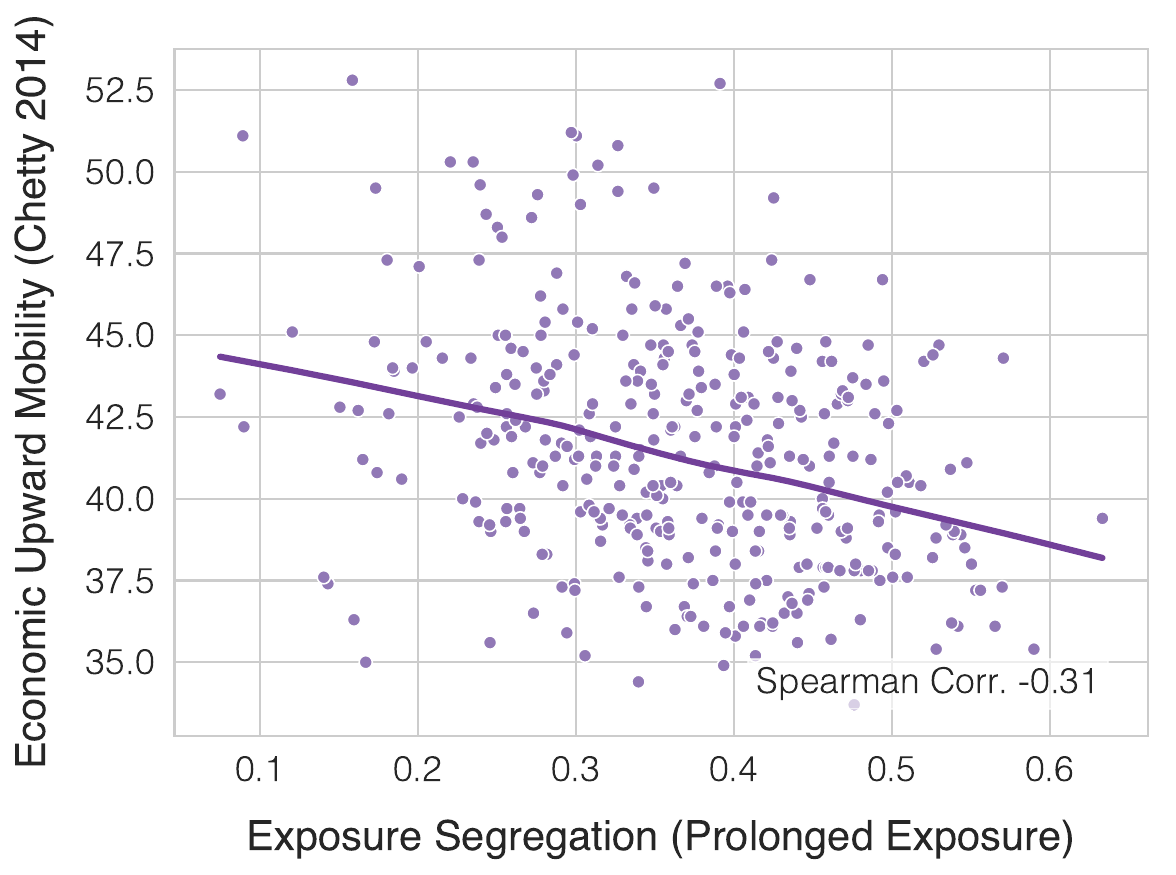}
   \caption{}
\end{subfigure}
\begin{subfigure}[t]{.5\textwidth}
\includegraphics[width=\textwidth,keepaspectratio]{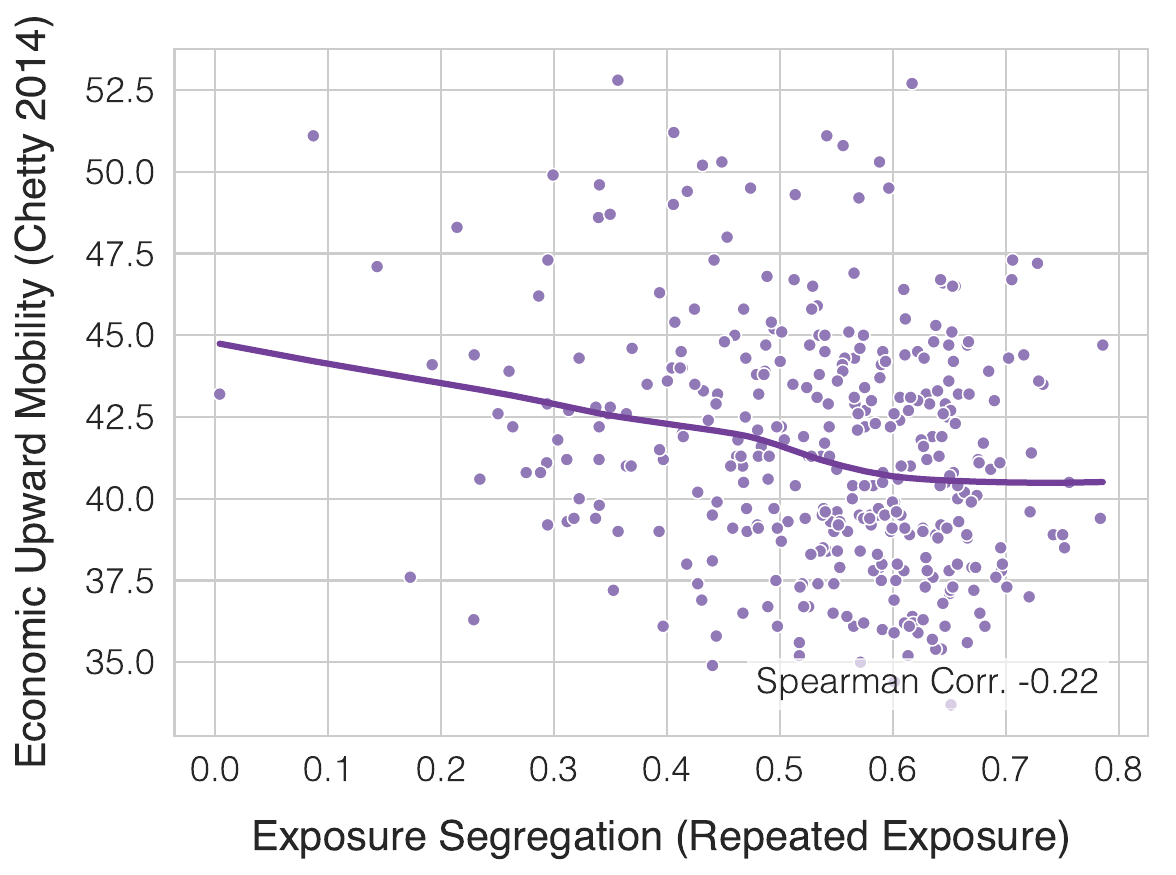}
   \caption{}
\end{subfigure}
\begin{subfigure}[b]{.5\textwidth}
\includegraphics[width=\textwidth,keepaspectratio]{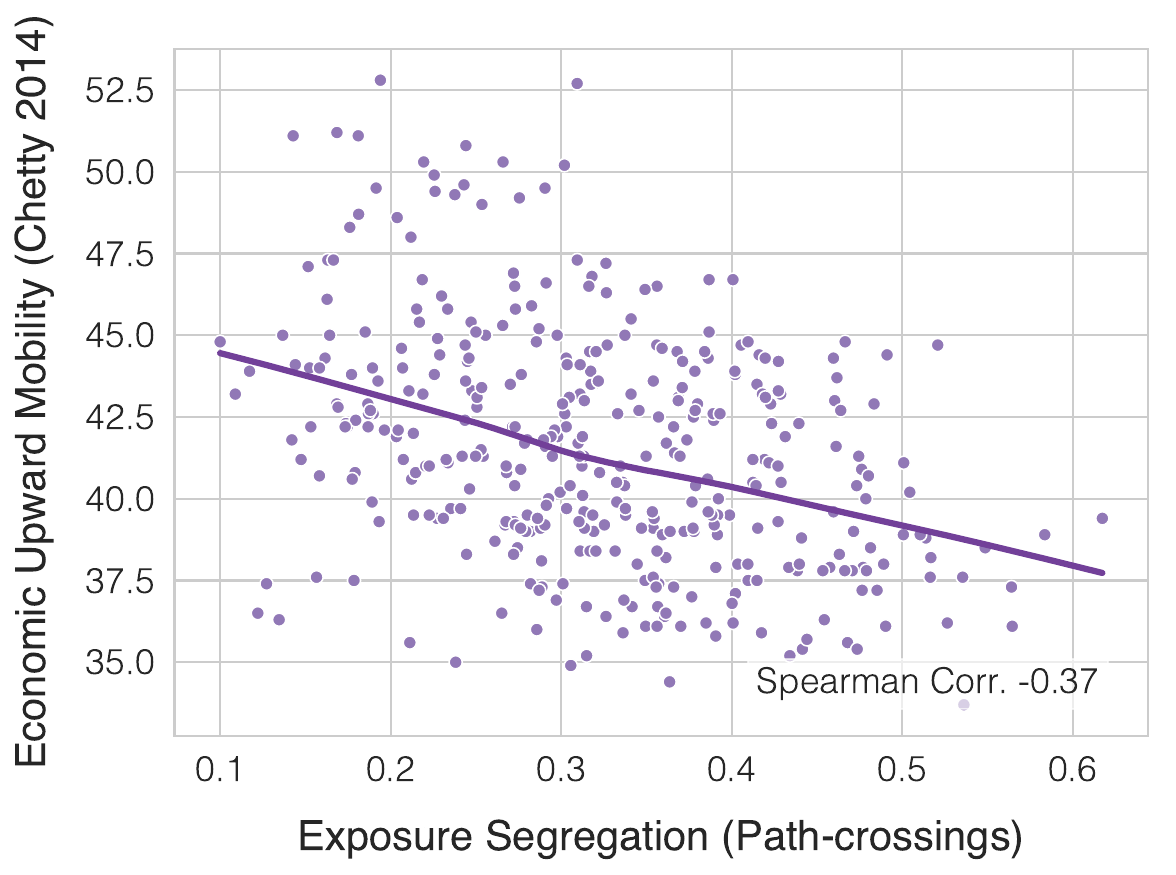}
   \caption{}
\end{subfigure}
\hspace*{-15mm}   
   \caption{\new{\textbf{We measure the external validity of alternative measures in which the strictness of our exposure definition is varied}. We compute exposure segregation using only prolonged exposure of 3+ consecutive intervals of exposure on the same day,  repeated exposure of 3+ consecutive intervals of exposure on different days, and path crossings (i.e. pairs of users that had only one instance of being within proximity of each other). We find all measures of \metric correlate to (absolute) upward economic mobility (Spearman Correlation -0.31, $N=365$, $p  <10^{-4}$), (Spearman Correlation -0.22, $N=364$, $p  <10^{-4}$), and (Spearman Correlation -0.37, $N=382$, $p  <10^{-4}$) respectively. Associations are significant even for the weakest definition of exposure (path-crossings), which may reflect the strength of weak ties in shaping upward economic mobility\cite{granovetter1973strength,rajkumar2022causal}.}}
   \label{fig:external_validation_5}
\end{figure}

\pagebreak

\begin{figure}[htbp]
   \centering
   \hspace*{-15mm}
\begin{subfigure}[t]{.5\textwidth}
\includegraphics[width=\textwidth,keepaspectratio]{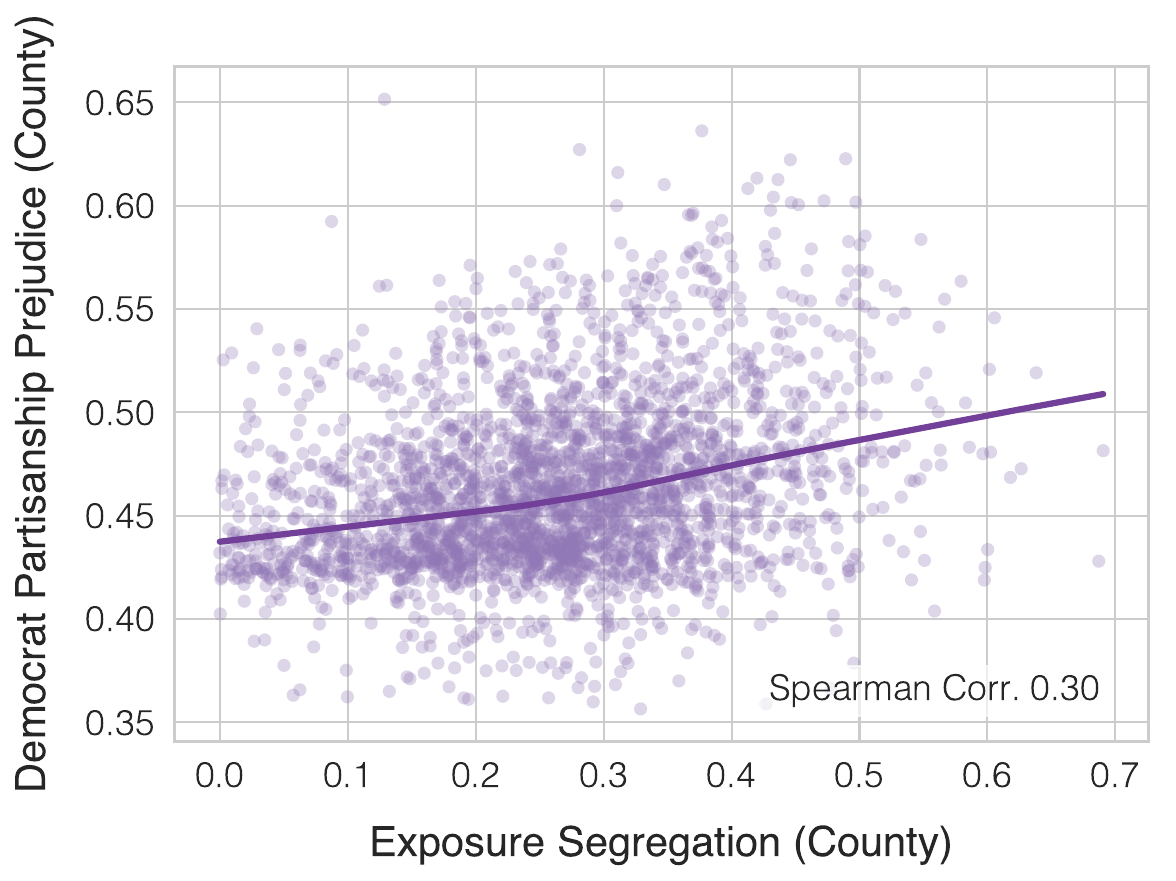}
   \caption{}
\end{subfigure}
\begin{subfigure}[t]{.5\textwidth}
\includegraphics[width=\textwidth,keepaspectratio]{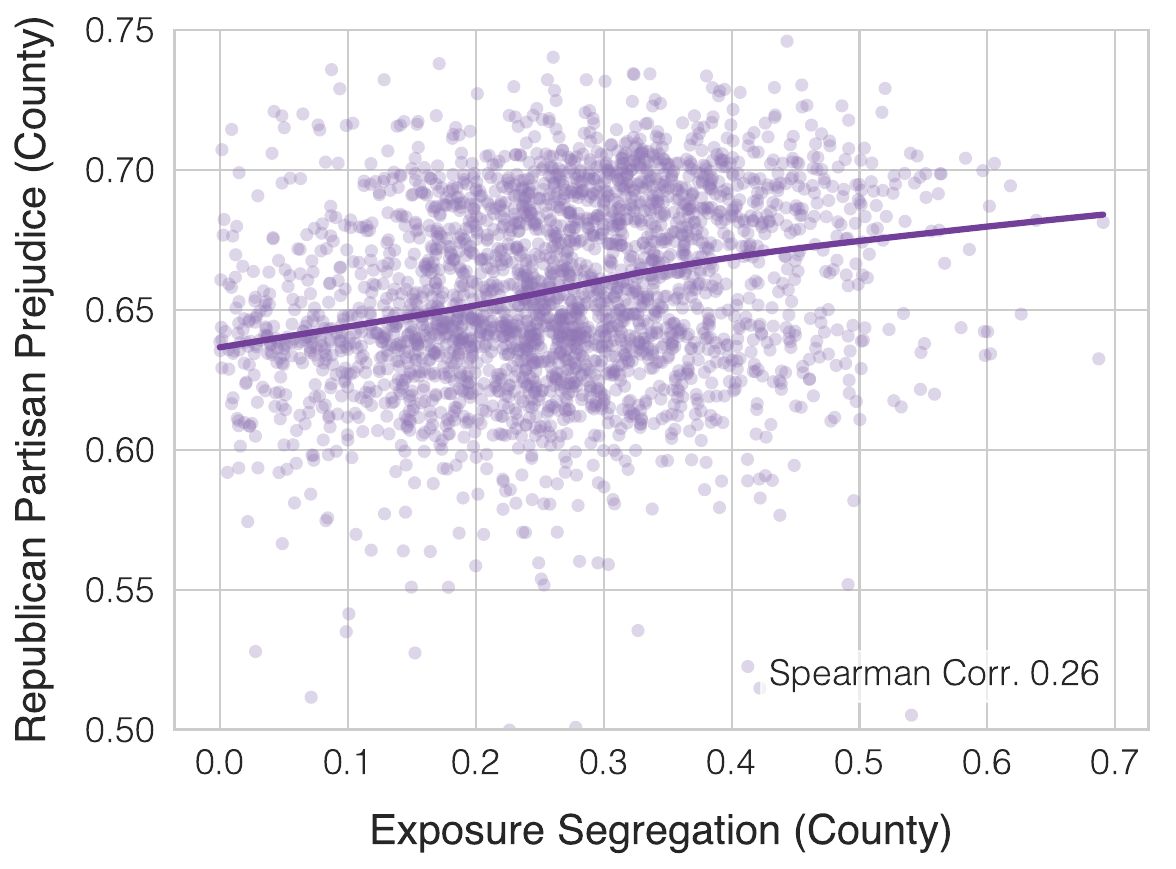}
   \caption{}
\end{subfigure}
\hspace*{-15mm}   
   \caption{\new{\textbf{Exposure segregation predicts political polarization outcomes}. We measure the external validity of our definition of exposure segregation, by linking our measure to outcomes from a large-scale survey of political polarization\cite{ripley_2021}. We find that county-level \metric correlates to political prejudice among both (a) Democrats (Spearman Correlation 0.30, $N=2828$, $p  <10^{-4}$) and (b) Republicans (Spearman Correlation 0.26, $N=2828$, $p  <10^{-4}$). These findings suggest that exposure to diverse others may lead to increased tolerance of inter-group differences, following following prior work\cite{brown2021childhood}.}}.
   \label{fig:external_validation_2}
\end{figure}

\pagebreak

\begin{figure}[htbp]
   \centering
\includegraphics[width=\textwidth,keepaspectratio]{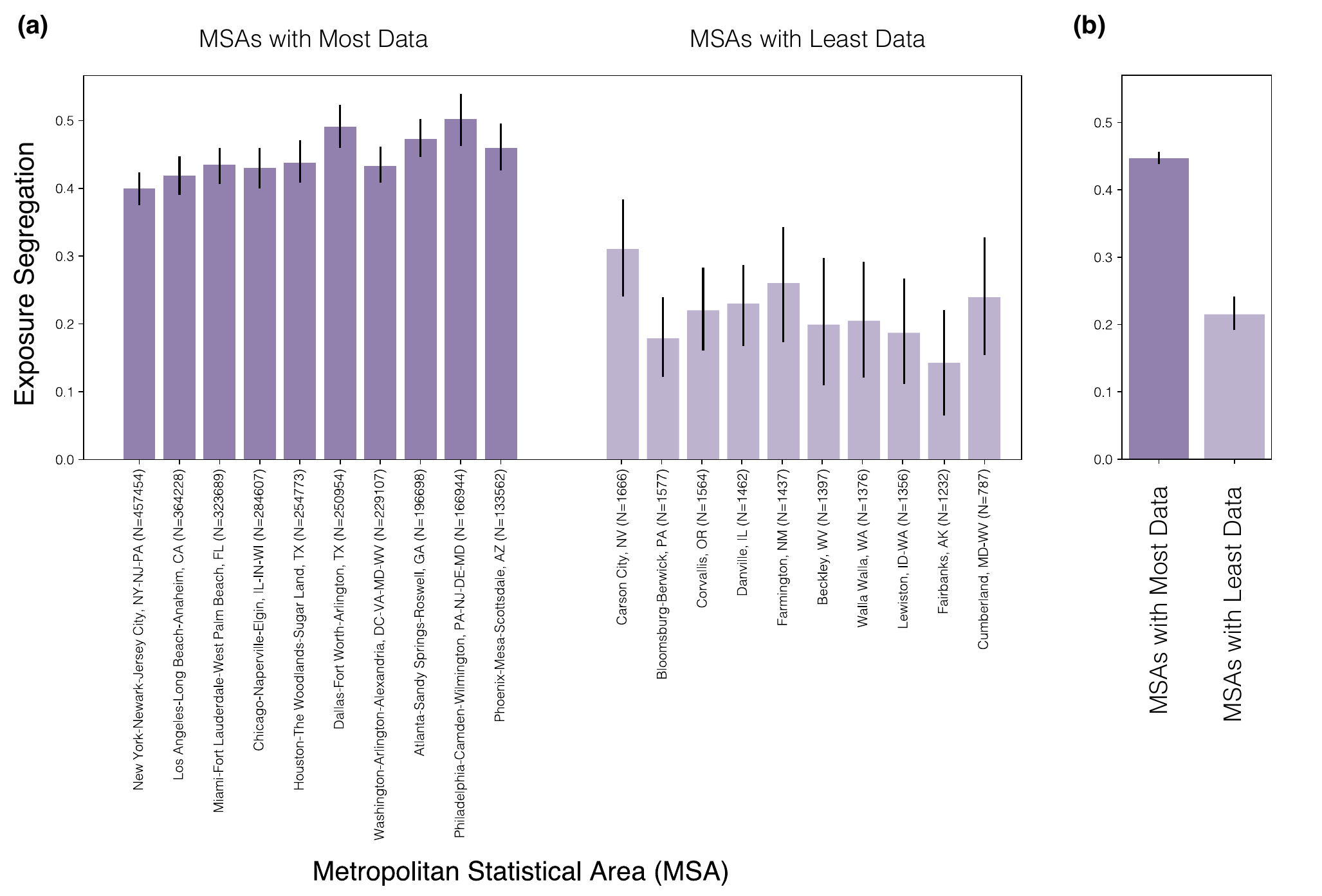}
   \caption{\new{\textbf{Precision of exposure segregation estimates}. (a) We quantify precision of our exposure segregation estimates by computing 95\% bootstrap confidence intervals in each MSA. Nonparametric bootstrap confidence intervals indicate resampling nodes in the network with replacement and recomputing exposure segregation in each MSA using our mixed model (with N=1000 replications). To illustrate the precision of our estimates, we show the 10 MSAs with the most individuals in our dataset compared to the 10 MSAs with the least individuals in our dataset. Nearly all pairs of small and large MSAs have non-overlapping confidence intervals. Even in smallest cities with little data, we can estimate exposure segregation with sufficient confidence to be able to compare to larger cities, finding that larger cities are more segregated. (b) We compare the mean ES of large and small MSAs and find that the mean of the top 10 is higher across all 1000 replicates ($p < 0.001$). 
}}.
   \label{fig:bootstrap}
\end{figure}

\begin{figure}[htbp]
   \centering
\begin{subfigure}[b]{1.0\textwidth}
\includegraphics[width=\textwidth,keepaspectratio]{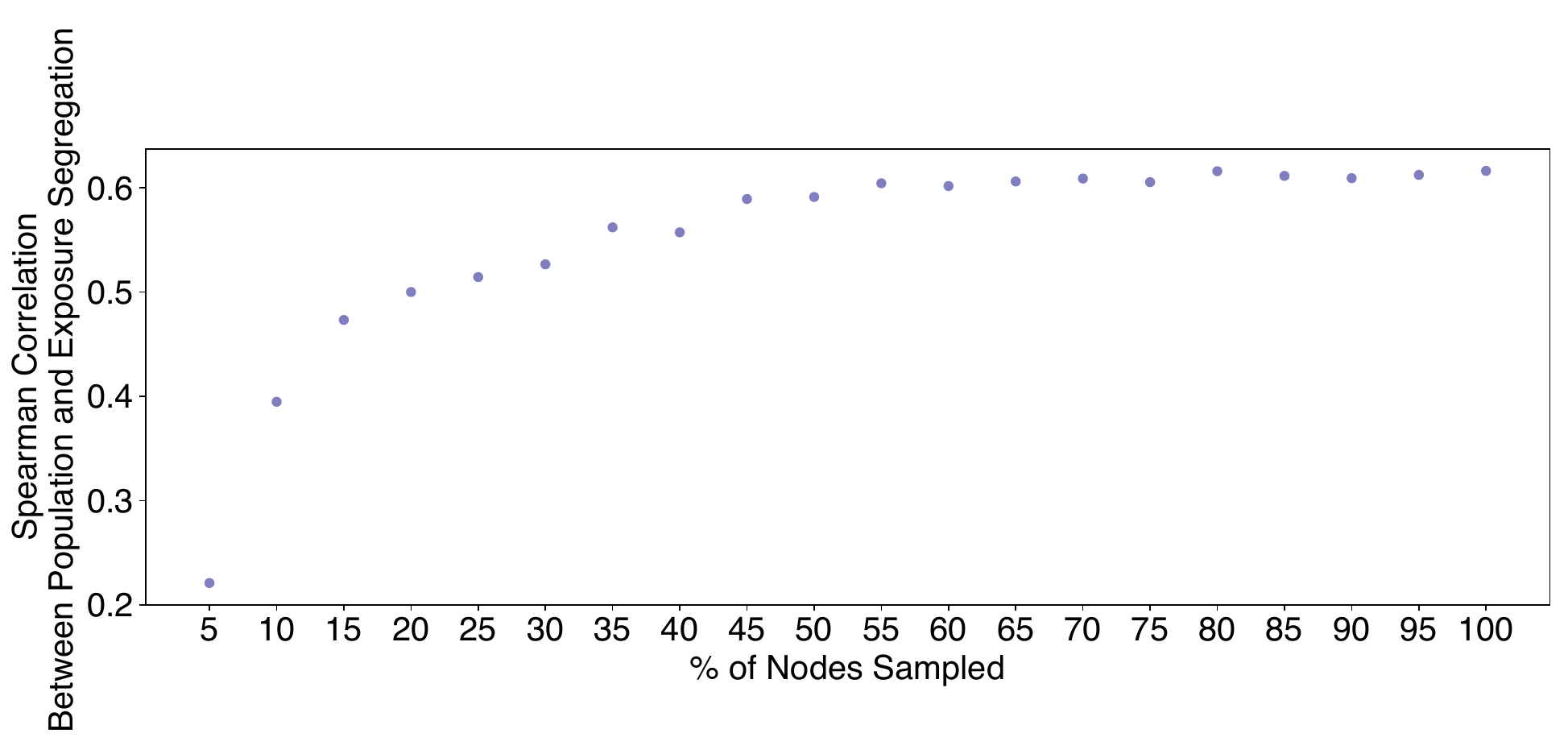}
   \caption{}
\end{subfigure}
\begin{subfigure}[b]{1.0\textwidth}
\includegraphics[width=\textwidth,keepaspectratio]{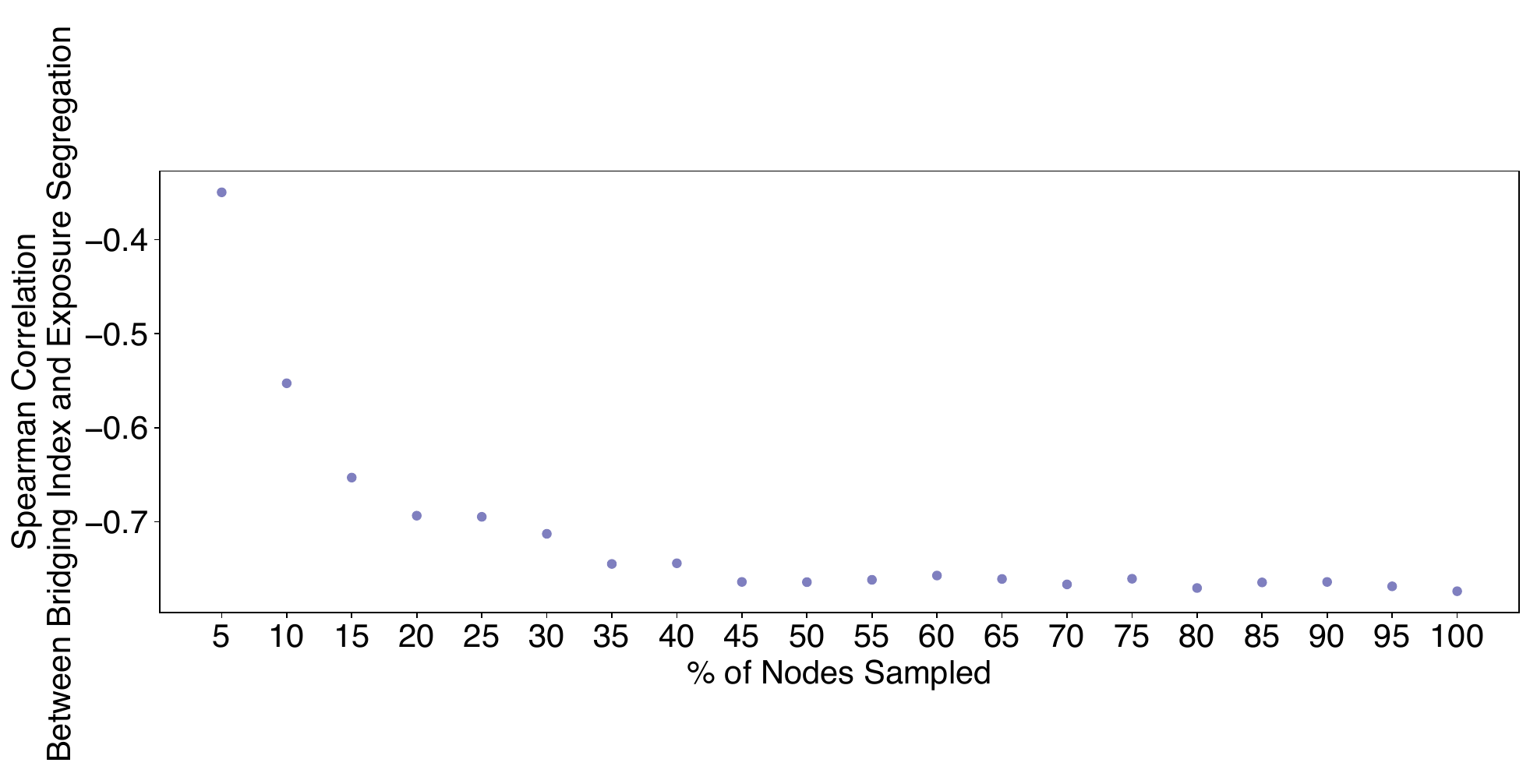}
   \caption{}
\end{subfigure}
   \caption{\new{\textbf{Robustness of primary findings to network size.} We further assess whether our key findings remain robust to smaller and sparser networks than the one we have amassed. To this end we downsample the nodes in the network and recompute our key findings using only the downsampled network. Statistically significant correlations ($p < 0.0001$) across all samples, as well as minimal variation in Spearman Corr. (0.60 to 0.62 for population size, and -0.75 to -0.78 for Bridging Index) after over 50\% of the network is sampled reveals that we have abundant data to support our claims. Diminishing returns are to be expected from amassing further network data.}}
   \label{fig:downsample_network}
\end{figure}

\begin{table}[ht]
\centering
\begin{scriptsize}
\setlength\tabcolsep{2pt} %
\begin{tabular}{rrrr}
\toprule
 Percentile &  \# of Exposures &  Percentile &  \# of Exposures \\
\midrule
        0.0 &        1.00 &        51.0 &        7.10 \\
        1.0 &        1.00 &        52.0 &        7.27 \\
        2.0 &        1.29 &        53.0 &        7.44 \\
        3.0 &        1.50 &        54.0 &        7.62 \\
        4.0 &        1.60 &        55.0 &        7.81 \\
        5.0 &        1.73 &        56.0 &        8.00 \\
        6.0 &        1.86 &        57.0 &        8.20 \\
        7.0 &        2.00 &        58.0 &        8.40 \\
        8.0 &        2.00 &        59.0 &        8.62 \\
        9.0 &        2.17 &        60.0 &        8.83 \\
       10.0 &        2.27 &        61.0 &        9.06 \\
       11.0 &        2.38 &        62.0 &        9.30 \\
       12.0 &        2.50 &        63.0 &        9.54 \\
       13.0 &        2.57 &        64.0 &        9.80 \\
       14.0 &        2.67 &        65.0 &       10.06 \\
       15.0 &        2.77 &        66.0 &       10.33 \\
       16.0 &        2.88 &        67.0 &       10.62 \\
       17.0 &        3.00 &        68.0 &       10.93 \\
       18.0 &        3.05 &        69.0 &       11.25 \\
       19.0 &        3.17 &        70.0 &       11.57 \\
       20.0 &        3.26 &        71.0 &       11.93 \\
       21.0 &        3.36 &        72.0 &       12.29 \\
       22.0 &        3.47 &        73.0 &       12.68 \\
       23.0 &        3.57 &        74.0 &       13.09 \\
       24.0 &        3.67 &        75.0 &       13.52 \\
       25.0 &        3.77 &        76.0 &       14.00 \\
       26.0 &        3.88 &        77.0 &       14.49 \\
       27.0 &        4.00 &        78.0 &       15.00 \\
       28.0 &        4.08 &        79.0 &       15.58 \\
       29.0 &        4.19 &        80.0 &       16.19 \\
       30.0 &        4.30 &        81.0 &       16.83 \\
       31.0 &        4.41 &        82.0 &       17.55 \\
       32.0 &        4.52 &        83.0 &       18.32 \\
       33.0 &        4.64 &        84.0 &       19.17 \\
       34.0 &        4.75 &        85.0 &       20.09 \\
       35.0 &        4.87 &        86.0 &       21.12 \\
       36.0 &        5.00 &        87.0 &       22.27 \\
       37.0 &        5.11 &        88.0 &       23.57 \\
       38.0 &        5.23 &        89.0 &       25.04 \\
       39.0 &        5.36 &        90.0 &       26.74 \\
       40.0 &        5.50 &        91.0 &       28.71 \\
       41.0 &        5.62 &        92.0 &       31.06 \\
       42.0 &        5.75 &        93.0 &       33.92 \\
       43.0 &        5.89 &        94.0 &       37.48 \\
       44.0 &        6.00 &        95.0 &       42.10 \\
       45.0 &        6.17 &        96.0 &       48.46 \\
       46.0 &        6.31 &        97.0 &       57.88 \\
       47.0 &        6.46 &        98.0 &       74.00 \\
       48.0 &        6.62 &        99.0 &      112.19 \\
       49.0 &        6.77 &       100.0 &     5351.00 \\
       50.0 &        6.94 &          &          \\
\bottomrule
\end{tabular}
\end{scriptsize}
\caption{\new{Distribution of average number of exposures (per active day) for all individuals residing in 382 Metropolitan Statistical Areas (MSAs). The median individual had 6.94 exposures on the average day of activity. 8,609,406 individuals reside in a Metropolitan Statistical Area (90\% of the overall 9,567,559 individuals in our study). The remaining 958,153 users live outside of MSAs, influencing the \metric of an MSA by coming into contact with MSA residents. Activity is defined as one or more exposures occurring on a given day. For details on activity over time, see Supplementary Figure \ref{fig:active_over_time}}.}
\label{tab:interaction_distribution2}
\end{table}

\begin{figure}[htbp]
   \centering
   \vspace{-5em}
\includegraphics[width=0.4\textwidth,keepaspectratio]{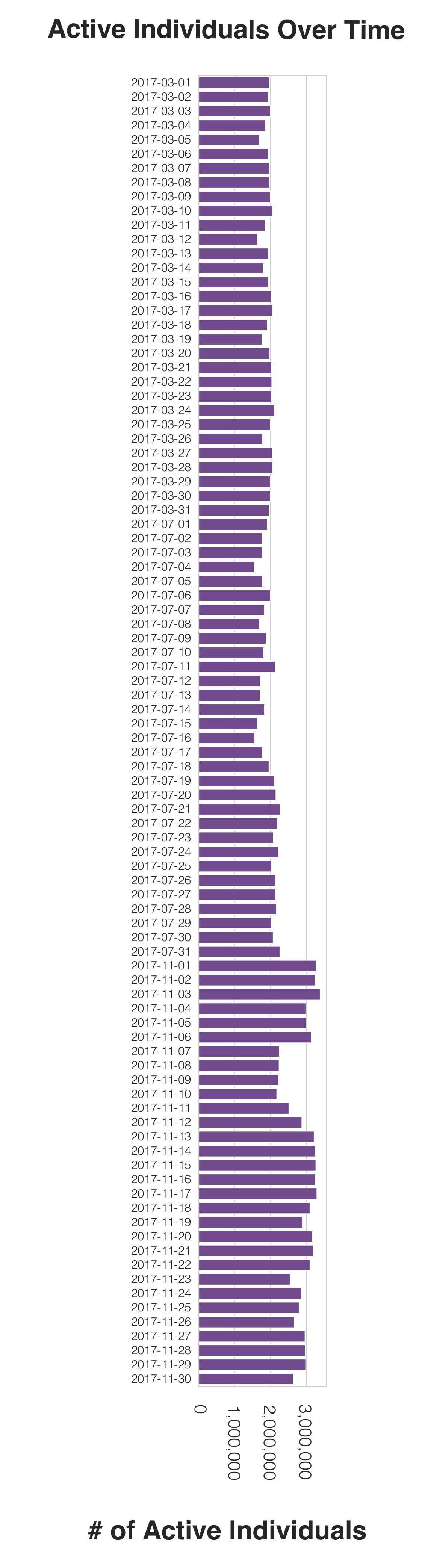}
   \vspace{-0.75em}
   \caption{\new{\textbf{Active individuals over time}. Number of active individuals (i.e. nodes in the network) over the study observation period. Activity is defined as one or more exposures occurring on a given day. 
}}
   \label{fig:active_over_time}
\end{figure}

\begin{figure}[htbp]
   \centering
   \vspace{-5em}
\includegraphics[width=0.65\textwidth,keepaspectratio]{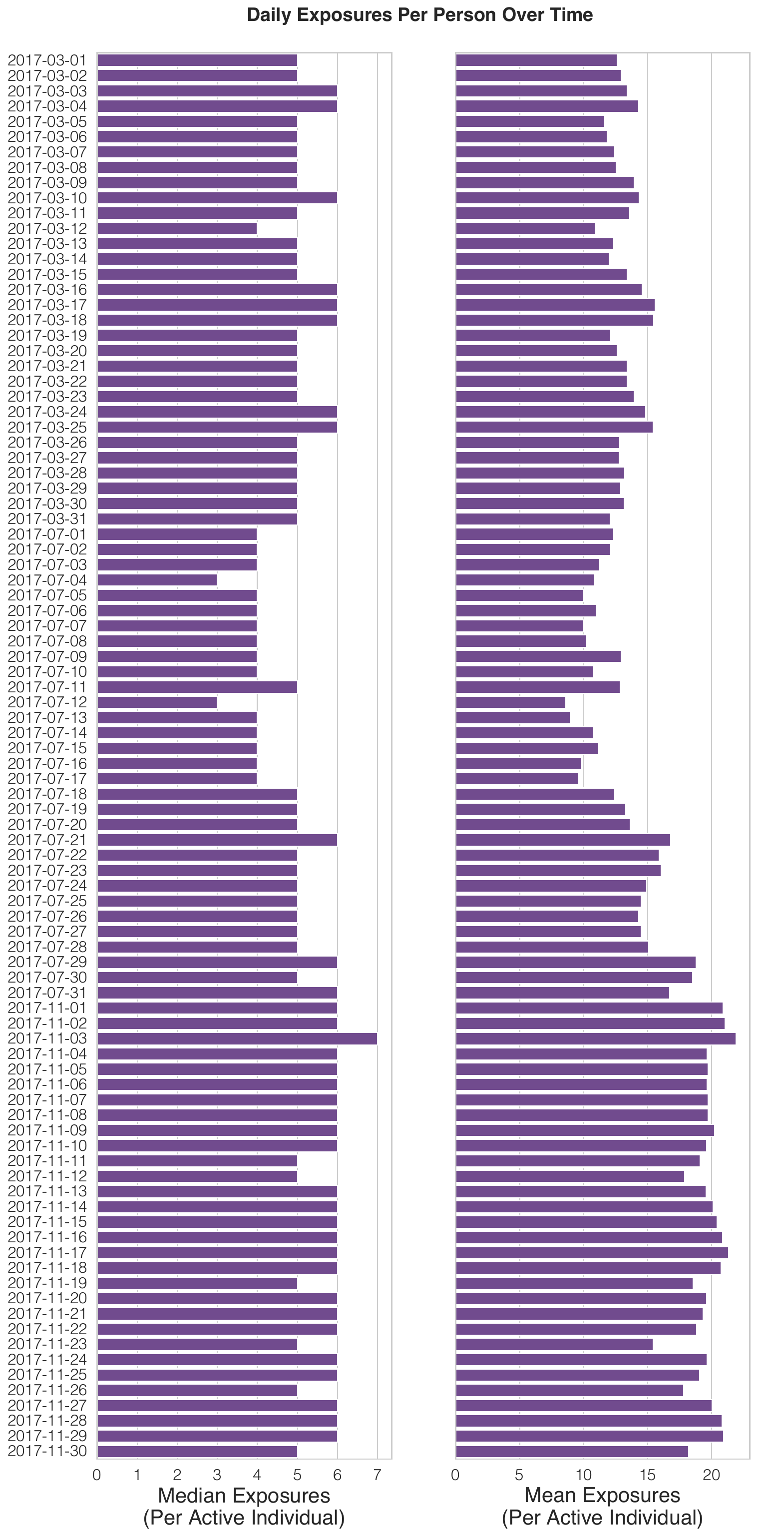}
   \vspace{-0.75em}
   \caption{\new{\textbf{Average number of exposures over time}. Mean/median exposures per active individuals (i.e. nodes in the network) over the study observation period. Activity is defined as one or more exposures occurring on a given day. 
}}
   \label{fig:interactors_over_time}
\end{figure}

\begin{table}[ht]
\centering
\begin{scriptsize}
\setlength\tabcolsep{2pt} %
\begin{tabular}{rrrr}
\toprule
 Percentile &  Tie Strength &  Percentile &  Tie Strength \\
\midrule
        0.0 &           1 &        51.0 &           1 \\
        1.0 &           1 &        52.0 &           1 \\
        2.0 &           1 &        53.0 &           1 \\
        3.0 &           1 &        54.0 &           1 \\
        4.0 &           1 &        55.0 &           1 \\
        5.0 &           1 &        56.0 &           1 \\
        6.0 &           1 &        57.0 &           1 \\
        7.0 &           1 &        58.0 &           1 \\
        8.0 &           1 &        59.0 &           1 \\
        9.0 &           1 &        60.0 &           1 \\
       10.0 &           1 &        61.0 &           1 \\
       11.0 &           1 &        62.0 &           1 \\
       12.0 &           1 &        63.0 &           1 \\
       13.0 &           1 &        64.0 &           1 \\
       14.0 &           1 &        65.0 &           1 \\
       15.0 &           1 &        66.0 &           1 \\
       16.0 &           1 &        67.0 &           1 \\
       17.0 &           1 &        68.0 &           1 \\
       18.0 &           1 &        69.0 &           1 \\
       19.0 &           1 &        70.0 &           1 \\
       20.0 &           1 &        71.0 &           1 \\
       21.0 &           1 &        72.0 &           1 \\
       22.0 &           1 &        73.0 &           1 \\
       23.0 &           1 &        74.0 &           1 \\
       24.0 &           1 &        75.0 &           2 \\
       25.0 &           1 &        76.0 &           2 \\
       26.0 &           1 &        77.0 &           2 \\
       27.0 &           1 &        78.0 &           2 \\
       28.0 &           1 &        79.0 &           2 \\
       29.0 &           1 &        80.0 &           2 \\
       30.0 &           1 &        81.0 &           2 \\
       31.0 &           1 &        82.0 &           2 \\
       32.0 &           1 &        83.0 &           2 \\
       33.0 &           1 &        84.0 &           2 \\
       34.0 &           1 &        85.0 &           2 \\
       35.0 &           1 &        86.0 &           2 \\
       36.0 &           1 &        87.0 &           2 \\
       37.0 &           1 &        88.0 &           2 \\
       38.0 &           1 &        89.0 &           2 \\
       39.0 &           1 &        90.0 &           2 \\
       40.0 &           1 &        91.0 &           3 \\
       41.0 &           1 &        92.0 &           3 \\
       42.0 &           1 &        93.0 &           3 \\
       43.0 &           1 &        94.0 &           3 \\
       44.0 &           1 &        95.0 &           4 \\
       45.0 &           1 &        96.0 &           4 \\
       46.0 &           1 &        97.0 &           5 \\
       47.0 &           1 &        98.0 &           7 \\
       48.0 &           1 &        99.0 &          11 \\
       49.0 &           1 &       100.0 &       11644 \\
       50.0 &           1 &          &           \\
\bottomrule
\end{tabular}
\end{scriptsize}
\caption{\new{Distribution of tie strength (\# of exposures) for all pairs of individuals residing in 382 Metropolitan Statistical Areas (MSAs).}
}
\label{tab:interaction_distribution3}
\end{table}

\begin{table}[ht]
\centering
\begin{scriptsize}
\setlength\tabcolsep{2pt} %
\begin{tabular}{rrrrrr}
\toprule
 Percentile &  Accurate Pings &  Raw Pings &  Percentile &  Accurate Pings &  Raw Pings \\
\midrule
       0 &           11 &     500 &       51 &        1,507 &     1,668 \\
       1 &          370 &     513 &       52 &        1,544 &     1,706 \\
       2 &          424 &     526 &       53 &        1,582 &     1,745 \\
       3 &          458 &     541 &       54 &        1,621 &     1,786 \\
       4 &          483 &     555 &       55 &        1,661 &     1,827 \\
       5 &          501 &     571 &       56 &        1,702 &     1,869 \\
       6 &          514 &     586 &       57 &        1,745 &     1,913 \\
       7 &          528 &     602 &       58 &        1,789 &     1,958 \\
       8 &          542 &     618 &       59 &        1,834 &     2,004 \\
       9 &          556 &     634 &       60 &        1,880 &     2,052 \\
      10 &          570 &     651 &       61 &        1,927 &     2,101 \\
      11 &          585 &     668 &       62 &        1,976 &     2,152 \\
      12 &          600 &     686 &       63 &        2,027 &     2,204 \\
      13 &          616 &     703 &       64 &        2,080 &     2,259 \\
      14 &          631 &     721 &       65 &        2,134 &     2,315 \\
      15 &          647 &     739 &       66 &        2,190 &     2,374 \\
      16 &          664 &     757 &       67 &        2,249 &     2,434 \\
      17 &          680 &     776 &       68 &        2,310 &     2,498 \\
      18 &          697 &     795 &       69 &        2,373 &     2,565 \\
      19 &          714 &     814 &       70 &        2,440 &     2,634 \\
      20 &          731 &     833 &       71 &        2,509 &     2,708 \\
      21 &          749 &     852 &       72 &        2,582 &     2,785 \\
      22 &          767 &     872 &       73 &        2,659 &     2,867 \\
      23 &          785 &     892 &       74 &        2,740 &     2,955 \\
      24 &          803 &     913 &       75 &        2,827 &     3,048 \\
      25 &          822 &     933 &       76 &        2,919 &     3,148 \\
      26 &          842 &     955 &       77 &        3,019 &     3,256 \\
      27 &          861 &     976 &       78 &        3,125 &     3,372 \\
      28 &          881 &     998 &       79 &        3,241 &     3,498 \\
      29 &          901 &   1,021 &       80 &        3,367 &     3,636 \\
      30 &          922 &   1,043 &       81 &        3,504 &     3,788 \\
      31 &          944 &   1,067 &       82 &        3,656 &     3,954 \\
      32 &          966 &   1,090 &       83 &        3,824 &     4,139 \\
      33 &          988 &   1,114 &       84 &        4,011 &     4,344 \\
      34 &        1,011 &   1,139 &       85 &        4,220 &     4,576 \\
      35 &        1,034 &   1,164 &       86 &        4,460 &     4,837 \\
      36 &        1,058 &   1,190 &       87 &        4,733 &     5,137 \\
      37 &        1,083 &   1,216 &       88 &        5,051 &     5,479 \\
      38 &        1,108 &   1,243 &       89 &        5,420 &     5,873 \\
      39 &        1,134 &   1,271 &       90 &        5,857 &     6,327 \\
      40 &        1,160 &   1,300 &       91 &        6,370 &     6,862 \\
      41 &        1,187 &   1,329 &       92 &        6,987 &     7,492 \\
      42 &        1,215 &   1,360 &       93 &        7,735 &     8,258 \\
      43 &        1,244 &   1,391 &       94 &        8,669 &     9,211 \\
      44 &        1,274 &   1,423 &       95 &        9,885 &    10,444 \\
      45 &        1,304 &   1,456 &       96 &       11,543 &    12,116 \\
      46 &        1,336 &   1,489 &       97 &       14,011 &    14,602 \\
      47 &        1,369 &   1,523 &       98 &       18,150 &    18,735 \\
      48 &        1,402 &   1,558 &       99 &       27,407 &    27,938 \\
      49 &        1,436 &   1,594 &      100 &    4,755,081 & 4,777,213 \\
      50 &        1,471 &   1,630 &             &                 &              \\
\bottomrule
\end{tabular}
\end{scriptsize}
\caption{\new{Distribution of total pings for all included individuals residing in 382 Metropolitan Statistical Areas (MSAs). The median individual has 1,471 accurate pings. Accurate pings are those with $< 100$ meters error. }}
\label{tab:pings_distribution}
\end{table}

\begin{table}[ht]
\centering
\begin{scriptsize}
\setlength\tabcolsep{2pt} %
\begin{tabular}{rrrr}
\toprule
 Percentile &  Unique Days &  Percentile &  Unique Days \\
\midrule
       0 &  2 &       51 & 30 \\
       1 &  5 &       52 & 31 \\
       2 &  7 &       53 & 31 \\
       3 &  8 &       54 & 31 \\
       4 &  9 &       55 & 31 \\
       5 &  9 &       56 & 31 \\
       6 & 10 &       57 & 31 \\
       7 & 11 &       58 & 32 \\
       8 & 11 &       59 & 32 \\
       9 & 12 &       60 & 32 \\
      10 & 13 &       61 & 32 \\
      11 & 13 &       62 & 32 \\
      12 & 14 &       63 & 33 \\
      13 & 15 &       64 & 34 \\
      14 & 15 &       65 & 34 \\
      15 & 16 &       66 & 35 \\
      16 & 16 &       67 & 37 \\
      17 & 17 &       68 & 38 \\
      18 & 17 &       69 & 39 \\
      19 & 18 &       70 & 40 \\
      20 & 18 &       71 & 41 \\
      21 & 19 &       72 & 43 \\
      22 & 19 &       73 & 44 \\
      23 & 20 &       74 & 45 \\
      24 & 20 &       75 & 47 \\
      25 & 21 &       76 & 48 \\
      26 & 21 &       77 & 49 \\
      27 & 22 &       78 & 51 \\
      28 & 22 &       79 & 52 \\
      29 & 23 &       80 & 53 \\
      30 & 23 &       81 & 54 \\
      31 & 24 &       82 & 56 \\
      32 & 24 &       83 & 57 \\
      33 & 25 &       84 & 58 \\
      34 & 25 &       85 & 59 \\
      35 & 25 &       86 & 60 \\
      36 & 26 &       87 & 61 \\
      37 & 26 &       88 & 62 \\
      38 & 26 &       89 & 62 \\
      39 & 27 &       90 & 63 \\
      40 & 27 &       91 & 64 \\
      41 & 27 &       92 & 67 \\
      42 & 27 &       93 & 70 \\
      43 & 28 &       94 & 74 \\
      44 & 28 &       95 & 78 \\
      45 & 28 &       96 & 82 \\
      46 & 29 &       97 & 86 \\
      47 & 29 &       98 & 90 \\
      48 & 29 &       99 & 93 \\
      49 & 30 &      100 & 95 \\
      50 & 30 &          &    \\
\bottomrule
\end{tabular}
\end{scriptsize}
\caption{\new{Distribution of unique days of ping data coverage for all included individuals residing in 382 Metropolitan Statistical Areas (MSAs). The median individual has 30 days of data coverage.}}
\label{tab:pings_distribution_unique_days}
\end{table}

\begin{table}[ht]
\centering
\begin{scriptsize}
\setlength\tabcolsep{2pt} %
\begin{tabular}{rrrr}
\toprule
 Percentile &  Time Elapsed Between Pings (seconds) &  Percentile & Time Elapsed Between Pings (seconds) \\
\midrule
       0 &       1 &       51  &          6  \\
       1 &       1 &       52  &          7  \\
       2 &       1 &       53  &          8  \\
       3 &       1 &       54  &          9  \\
       4 &       1 &       55  &          9  \\
       5 &       1 &       56  &         11  \\
       6 &       1 &       57  &         14  \\
       7 &       1 &       58  &         18  \\
       8 &       1 &       59  &         24  \\
       9 &       1 &       60  &         31  \\
      10 &       1 &       61  &         40  \\
      11 &       1 &       62  &         50  \\
      12 &       1 &       63  &         60  \\
      13 &       1 &       64  &         65  \\
      14 &       1 &       65  &         73  \\
      15 &       1 &       66  &         83  \\
      16 &       1 &       67  &         96  \\
      17 &       2 &       68  &        112  \\
      18 &       2 &       69  &        124  \\
      19 &       2 &       70  &        138  \\
      20 &       2 &       71  &        154  \\
      21 &       2 &       72  &        172  \\
      22 &       2 &       73  &        182  \\
      23 &       3 &       74  &        190  \\
      24 &       3 &       75  &        203  \\
      25 &       3 &       76  &        221  \\
      26 &       3 &       77  &        241  \\
      27 &       4 &       78  &        261  \\
      28 &       4 &       79  &        288  \\
      29 &       4 &       80  &        300  \\
      30 &       4 &       81  &        304  \\
      31 &       5 &       82  &        322  \\
      32 &       5 &       83  &        357  \\
      33 &       5 &       84  &        402  \\
      34 &       5 &       85  &        473  \\
      35 &       5 &       86  &        561  \\
      36 &       5 &       87  &        603  \\
      37 &       5 &       88  &        640  \\
      38 &       5 &       89  &        742  \\
      39 &       5 &       90  &        897  \\
      40 &       5 &       91  &      1,081  \\
      41 &       5 &       92  &      1,355  \\
      42 &       5 &       93  &      1,679  \\
      43 &       5 &       94  &      1,888  \\
      44 &       5 &       95  &      2,418  \\
      45 &       5 &       96  &      3,482  \\
      46 &       5 &       97  &      4,732  \\
      47 &       5 &       98  &      8,185  \\
      48 &       5 &       99  &     21,454  \\
      49 &       6 &      100  & 23,053,986  \\
      50 &       6 &           &             \\
\bottomrule
\end{tabular}
\end{scriptsize}
\caption{\new{Distribution of interevent times (seconds elapsed between pings)  for all included individuals residing in 382 Metropolitan Statistical Areas (MSAs). The median interevent time is 6 seconds, and 96\% of intervent times are $<$ 60 minutes. Distribution is estimated using a random sample of 1\% of users, corresponding to 326,039,078 pings}}
\label{tab:pings_distribution_unique_days2}
\end{table}

\begin{figure}[htbp]
   \centering
   \vspace{-5em}
\includegraphics[width=0.5\textwidth,keepaspectratio]{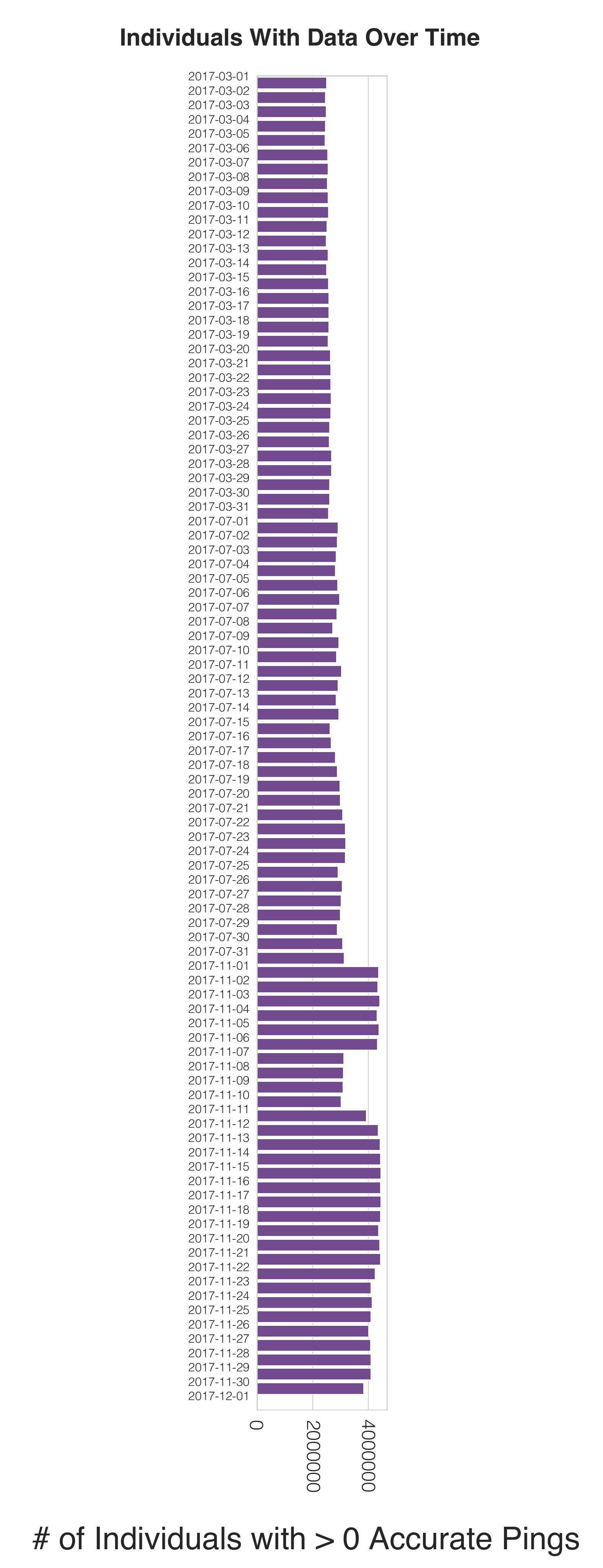}
   \vspace{-0.75em}
   \caption{\new{\textbf{Individuals with data coverage over time}. Number of individuals with data (i.e. $> 0$ accurate pings) over the study observation period. Accurate pings are those with $< 100$ meters error. 
}}
   \label{fig:active_over_time_2}
\end{figure}

\begin{figure}[htbp]
   \centering
   \vspace{-5em}
\includegraphics[width=0.35\textwidth,keepaspectratio]{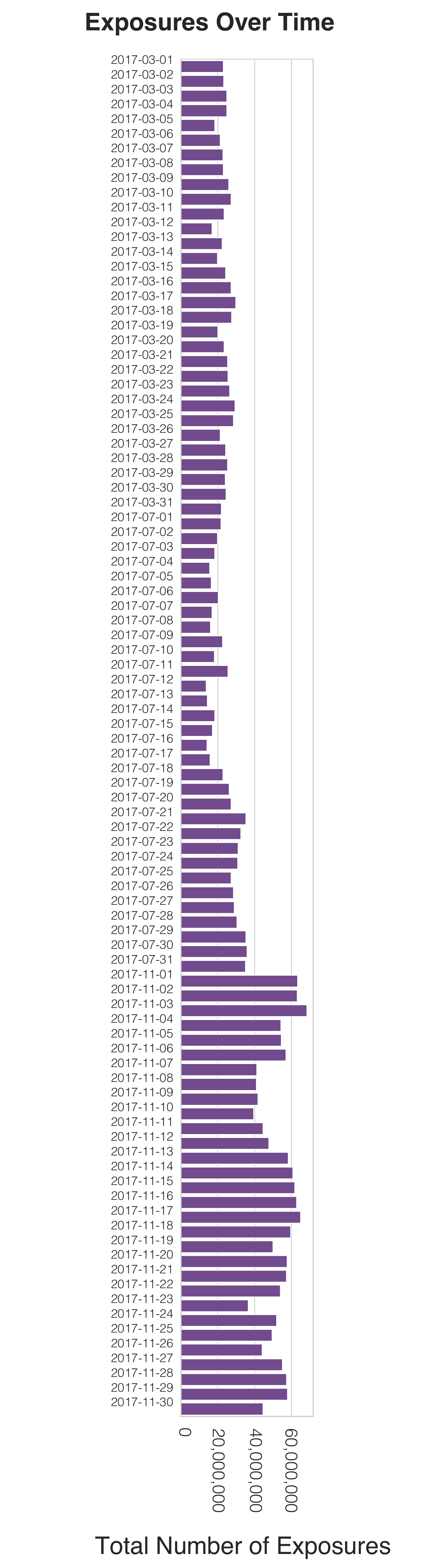}
   \vspace{-0.75em}
   \caption{\new{\textbf{Exposures} over time across all individuals residing in the 382 MSAs.}}
   \label{fig:exposed_over_time}
\end{figure}

\begin{figure}[htbp]
   \centering
   \vspace{-5em}
\includegraphics[width=1.0\textwidth,keepaspectratio]{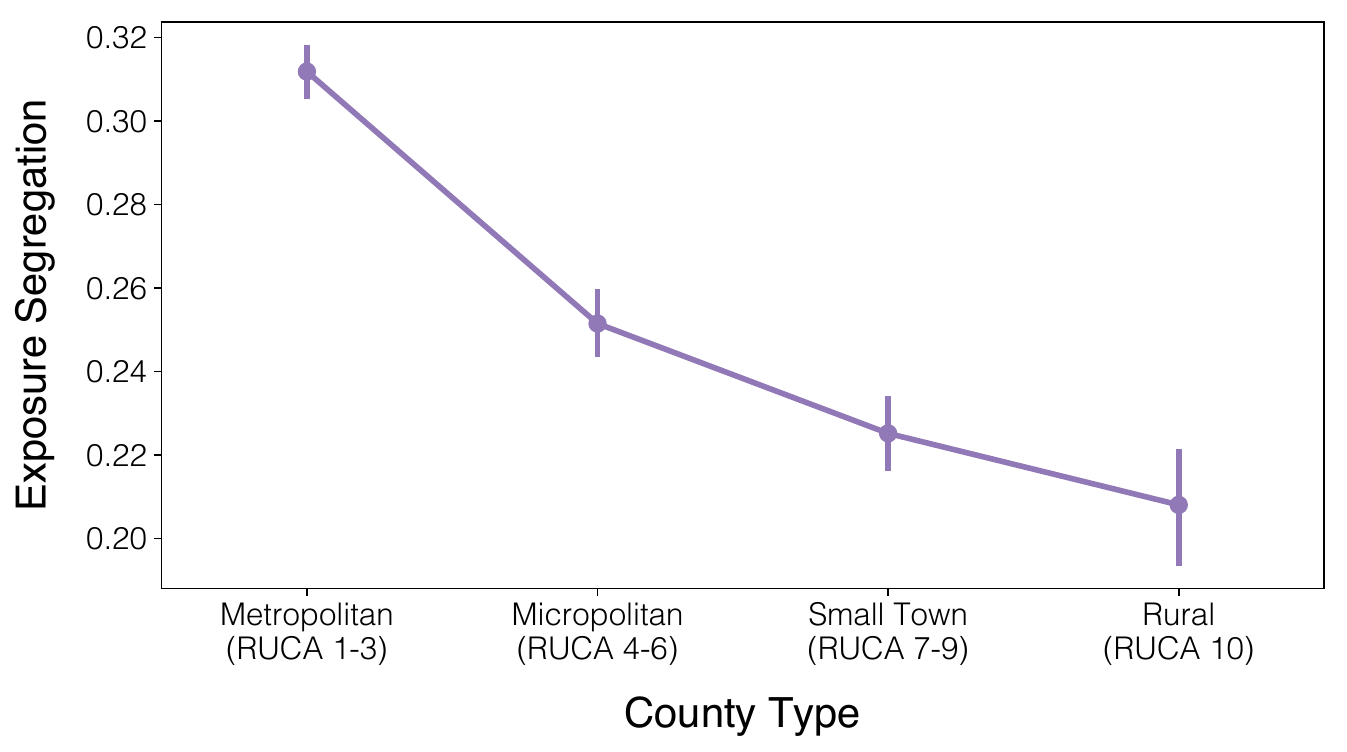}
   \vspace{-0.75em}
   \caption{\new{\textbf{Counties grouped by extent of county urbanization.} Urbanized counties are more segregated than rural counties. To operationalize county urbanization, we use the median rural urban continuum (RUCA code) of the tracts in a county.\cite{cromartie2005rural}}}
   \label{fig:county_rurality}
\end{figure}

\begin{figure}[htbp]
   \centering
   \vspace{-5em}
\includegraphics[width=1.0\textwidth,keepaspectratio]{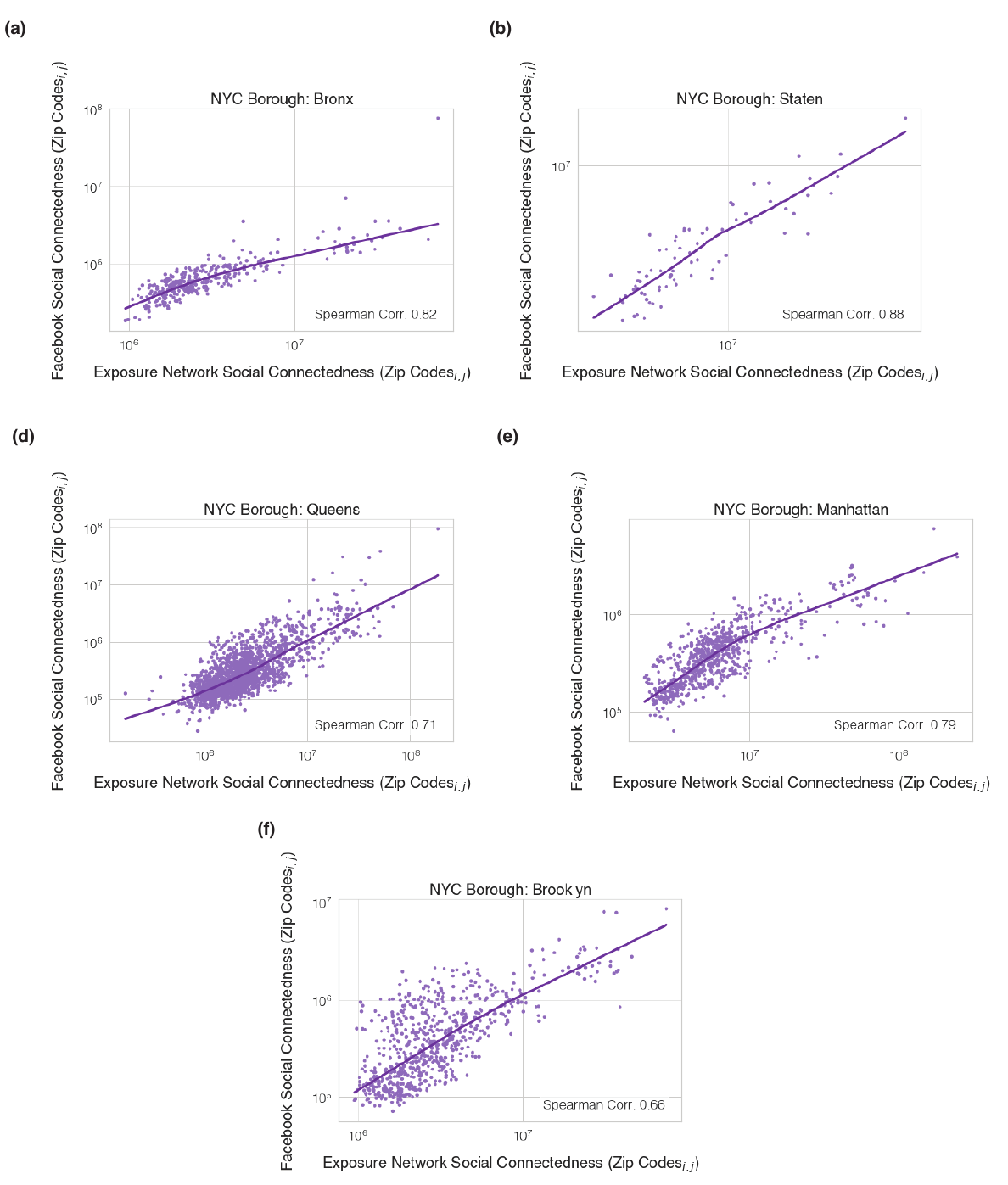}
   \vspace{-0.75em}
   \caption{\new{\textbf{Exposure network connections strongly correlate to friendship formation even within fine-grained geographical areas.} We reproduce the the Facebook Social Connectedness Index\cite{bailey2018social} at zip code-level (Extended Data Figure \ref{fig:external_validation}) for each of the five boroughs of New York City, and find strong correlations to online friendships in all five boroughs (Spearman Correlations 0.66-0.88, all $p  <10^{-4}$). Strong correlations suggest that exposure segregation is likely related to segregation of friendships and other strong social ties.}}
   \label{fig:boroughs}
\end{figure}

\newpage
\renewcommand\floatpagefraction{0.1}

\begin{table}[ht]
\centering
\begin{scriptsize}
\setlength\tabcolsep{2pt} %
\begin{tabular}{llrl}
\toprule
Mean \# of Unique Days of Exposure &  POI Type \\
\midrule
 1.027096 &         Performing Arts Centers \\
 1.029548 &                        Stadiums \\
 1.036697 &                     Theme Parks \\
 1.044692 &                 Bowling Centers \\
 1.050582 &      Other Amusement/Recreation \\
 1.054688 &            Bars/Drinking Places \\
 1.068597 &                         Museums \\
 1.068676 &                Historical Sites \\
 1.073173 &             Independent Artists \\
 1.087597 &                         Casinos \\
 1.089572 &     Limited-Service Restaurants \\
 1.092399 &                           Parks \\
 1.097890 &                      Snack Bars \\
 1.102125 &        Full-Service Restaurants \\
 1.117464 &      Fitness/Recreation Centers \\
 1.147761 &  Golf Courses and Country Clubs \\
 1.153269 &         Religious Organizations \\
\bottomrule
\end{tabular}
\end{scriptsize}
\caption{\new{\textbf{Exposure repetition by setting.} For each leisure POI category, we compute the mean number of unique days of exposure over all pairs of individuals in the exposure network. POIs associated with most repeated exposures (religious organizations, golf courses and country clubs, and fitness centers) are all venues with relatively static membership structures (e.g. religious affiliation, annual gym membership) and in which frequent attendance is a norm (e.g. visiting church every Sunday, weekly workout). By contrast, the POIs with least repetition (performing arts centers and stadiums) are those which are typically attended only special occasions and typically without a commitments (e.g. buying a single ticket to see a sports game).}}
\label{tab:poi_unique_days}
\end{table}

\begin{table}[ht]
\centering
\begin{scriptsize}
\setlength\tabcolsep{2pt} %
\begin{tabular}{llrl}
\toprule
Mean Exposure Distance (meters) &  POI Type \\
\midrule
  14.148335 &                         Museums \\
  20.783428 &                Historical Sites \\
  23.343675 &  Golf Courses and Country Clubs \\
  24.539538 &         Religious Organizations \\
  24.954691 &                        Stadiums \\
  25.063481 &         Performing Arts Centers \\
  25.168213 &                 Bowling Centers \\
  25.765411 &        Full-Service Restaurants \\
  25.845487 &                           Parks \\
  25.973838 &      Other Amusement/Recreation \\
  26.080283 &            Bars/Drinking Places \\
  26.102137 &      Fitness/Recreation Centers \\
  26.214838 &                     Theme Parks \\
  26.305469 &     Limited-Service Restaurants \\
  26.423661 &                      Snack Bars \\
  26.775229 &             Independent Artists \\
  27.048224 &                         Casinos \\
\bottomrule
\end{tabular}
\end{scriptsize}
\caption{\new{\textbf{Exposure distance by setting.} For each leisure POI category, we compute the mean distance during exposure over all pairs of individuals in the exposure network. POIs associated with furthest exposures (casinos, independent artists) are those in which mobile phone usage is typically restricted (e.g. many casinos do not allow mobile phone usage to ensure fair  play), which is likely to lead to sparse GPS pings. By contrast, the POIs with least distance (museums, historical sites) are those which phones may be actively used to enhance the experience (e.g. to take photos, look up information, or use a virtually guided tour).}}
\label{tab:poi_repetition}
\end{table}

\begin{table}[ht]
\centering
\begin{scriptsize}
\setlength\tabcolsep{2pt} %
\begin{tabular}{llrl}
\toprule
Mean Length (\# of consecutive five minute intervals) &  POI Type \\
\midrule
  1.617740 &                         Museums \\
  1.620565 &                     Theme Parks \\
  1.693283 &      Other Amusement/Recreation \\
  1.748765 &            Bars/Drinking Places \\
  1.776146 &             Independent Artists \\
  1.782905 &     Limited-Service Restaurants \\
  1.840704 &        Full-Service Restaurants \\
  1.851464 &                      Snack Bars \\
  1.871852 &                         Casinos \\
  1.882052 &      Fitness/Recreation Centers \\
  1.910205 &                Historical Sites \\
  2.039442 &                           Parks \\
  2.061254 &         Religious Organizations \\
  2.082550 &  Golf Courses and Country Clubs \\
  2.209579 &                 Bowling Centers \\
  2.436712 &                        Stadiums \\
  2.464210 &         Performing Arts Centers \\
\bottomrule
\end{tabular}
\end{scriptsize}
\caption{\new{\textbf{Exposure length by setting.} For each leisure POI category, we compute the mean length during exposure over all pairs of individuals in the exposure network (in number of consecutive 5 minute intervals). POIs associated with longest exposures (performing arts centers, stadiums) are those in which attendance is typically prolonged and mobility is restricted (e.g. sitting in the same seat for multiple hours to watch a game). By contrast, the POIs with shortest exposure (museums, theme parks) are those which mobility is a necessary part of the experience (e.g. walking to different exhibits or attractions).}}
\label{tab:poi_length}
\end{table}

\begin{figure}[htbp]
\vspace*{-20mm}
   \centering
   \hspace*{-15mm}
\begin{subfigure}[t]{.5\textwidth}
\includegraphics[width=\textwidth,keepaspectratio]{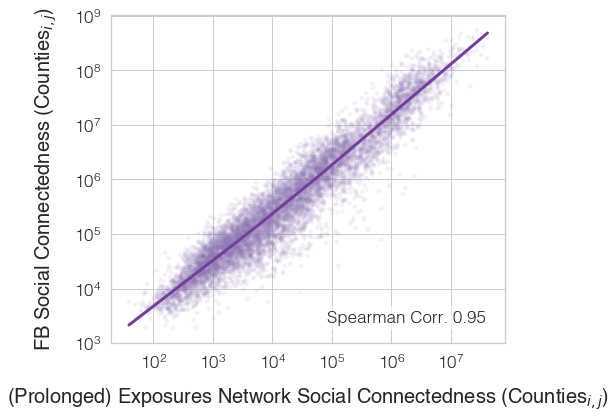}
   \caption{}
\end{subfigure}
\begin{subfigure}[t]{.5\textwidth}
\includegraphics[width=\textwidth,keepaspectratio]{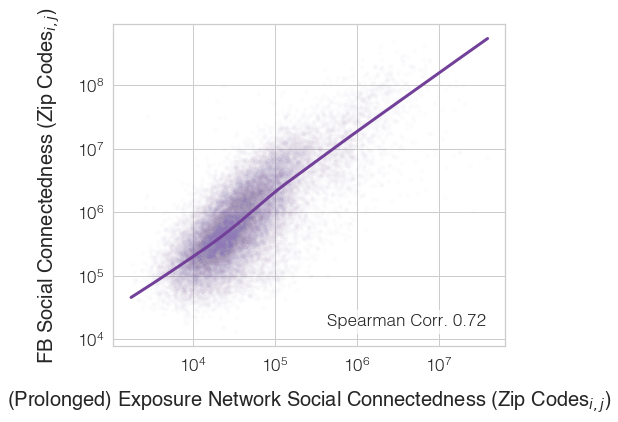}
   \caption{}
\end{subfigure}
\hspace*{-15mm}   
\begin{subfigure}[b]{.5\textwidth}
\includegraphics[width=\textwidth,keepaspectratio]{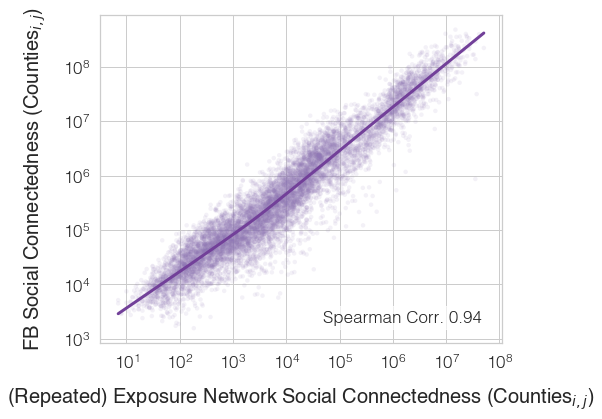}
   \caption{}
\end{subfigure}
\begin{subfigure}[b]{.5\textwidth}
\includegraphics[width=\textwidth,keepaspectratio]{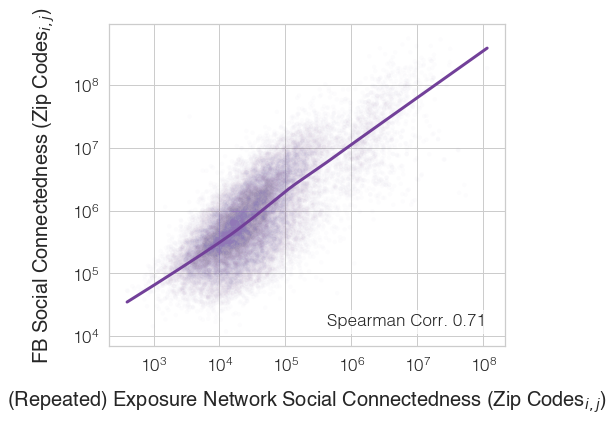}
   \caption{}
\end{subfigure}
   \hspace*{-15mm}
\begin{subfigure}[t]{.5\textwidth}
\includegraphics[width=\textwidth,keepaspectratio]{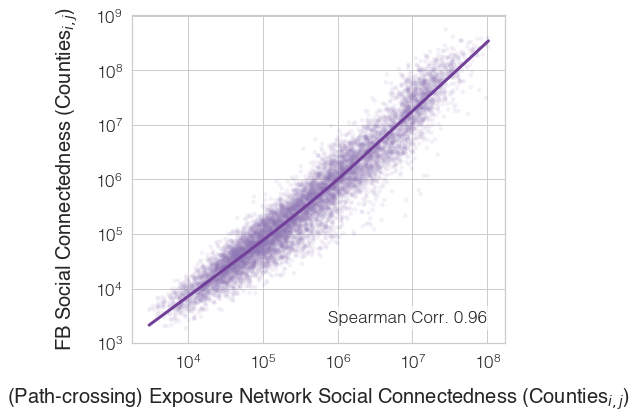}
   \caption{}
\end{subfigure}
\begin{subfigure}[t]{.5\textwidth}
\includegraphics[width=\textwidth,keepaspectratio]{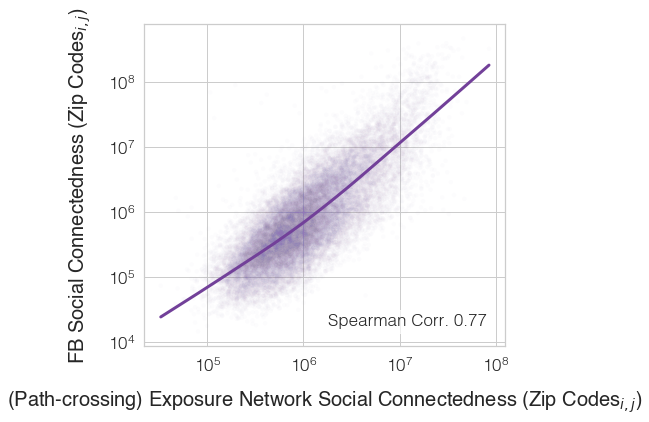}
   \caption{}
\end{subfigure}
   \caption{\new{\textbf{We measure the external validity of alternative measures in which the strictness of our exposure definition is varied}. We filter the exposure network to include (a-b) only prolonged exposure of 3+ consecutive intervals of exposure on the same day, (c-d)  repeated exposure of 3+ consecutive intervals of exposure on different days, and (e-f) path crossings (i.e. pairs of users that had only one instance of being within proximity of each other). We find that social connectedness across all three definitions of exposure correlates strongly to social connectedness measured by online friendship linked (detailed in Extended Data Figure \ref{fig:external_validation})}}
   \label{fig:external_validation_5_social}
\end{figure}

\begin{figure}[htbp]
\captionsetup[subfigure]{labelformat=empty}

\vspace{-20mm}
   \centering
\begin{subfigure}[t]{0.65\textwidth}
\includegraphics[width=\textwidth,keepaspectratio]{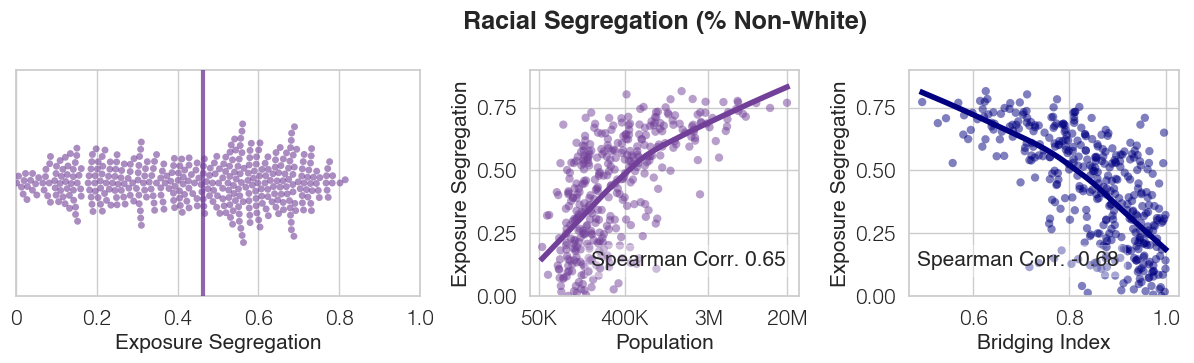}
   \caption{}
\end{subfigure}

\begin{subfigure}[t]{0.65\textwidth}
\includegraphics[width=\textwidth,keepaspectratio]{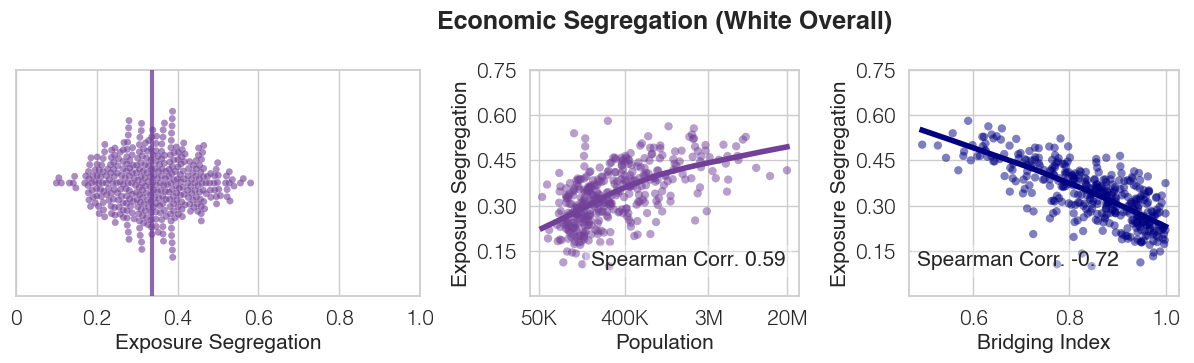}
   \caption{}
\end{subfigure}

\begin{subfigure}[t]{0.65\textwidth}
\includegraphics[width=\textwidth,keepaspectratio]{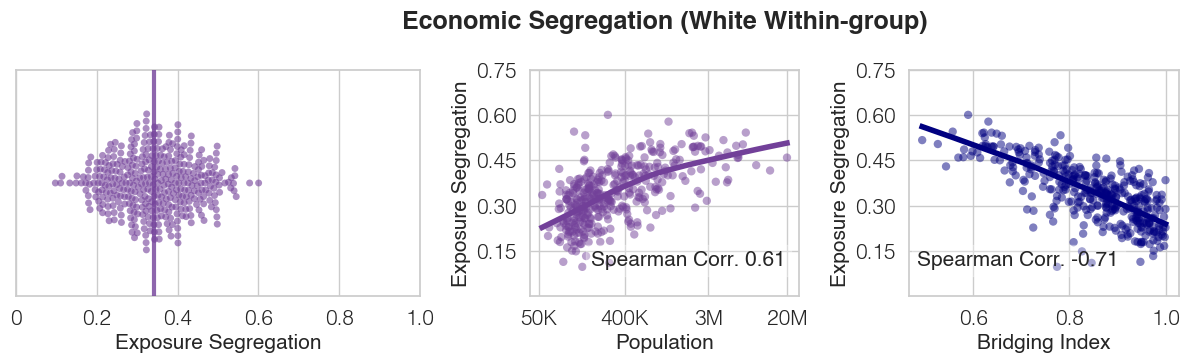}
   \caption{}
\end{subfigure}

\begin{subfigure}[t]{0.65\textwidth}
\includegraphics[width=\textwidth,keepaspectratio]{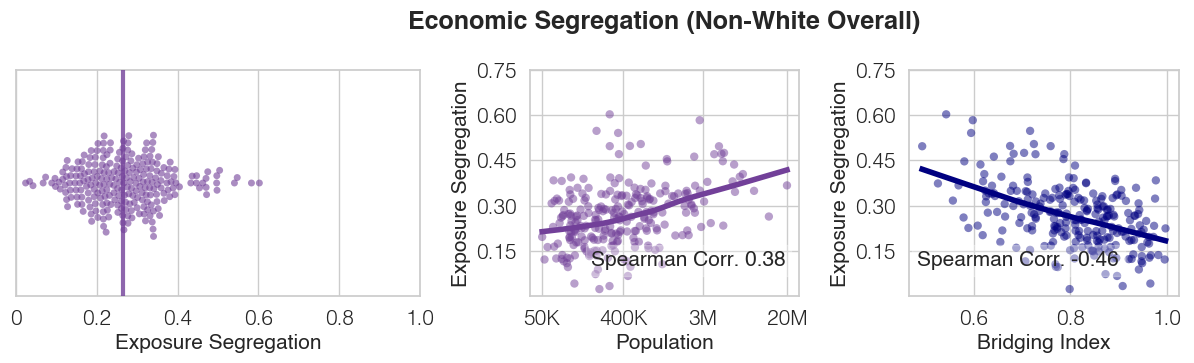}
   \caption{}
\end{subfigure}
\begin{subfigure}[t]{0.65\textwidth}
\includegraphics[width=\textwidth,keepaspectratio]{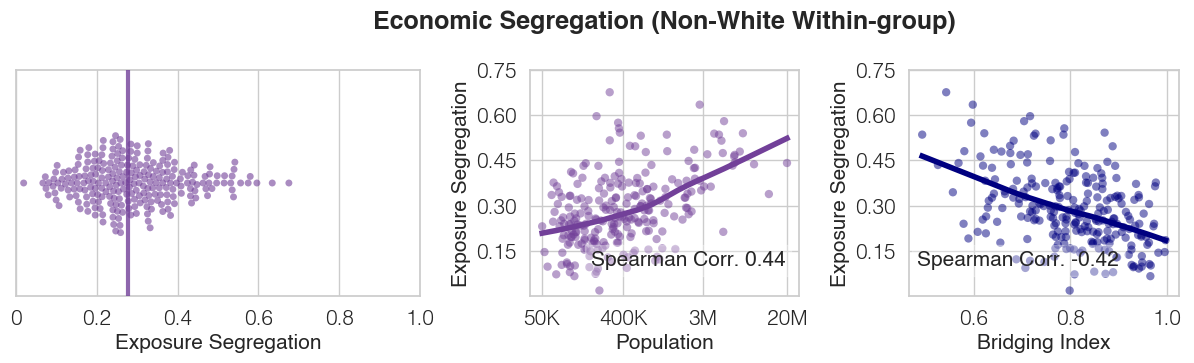}
   \caption{}
\end{subfigure}
\caption{\new{\textbf{Racial exposure segregation, and economic exposure segregation broken down by race group.} We find that our primary study findings that, (1) large, dense cities are more segregated and (2) exposure hub locations accessible to diverse individuals may mitigate segregation, are robust to replacing socioeconomic status with race (using \% non-White in home neighborhoods), and to isolated exposures of individuals residing in predominantly White and non-White neighborhoods. Neighborhood is operationalized as census block group. Overall, racial exposure segregation (top row) is 31\% higher in the median MSA compared to the economic exposure segregation in the median MSA. Moreover, there is much higher variance in racial exposure segregation: the standard deviation of racial exposure segregation is 114\% higher than the standard deviation of racial exposure segregation. Additionally, economic exposure segregation is 27\% higher for White individuals. However, for both groups, economic segregation is similar when comparing within-race exposures to overall exposures.}
}
\label{fig:robust_10}
\end{figure}

\begin{figure}[htbp]
   \centering
\begin{subfigure}[b]{0.5\textwidth}
\includegraphics[width=\textwidth,keepaspectratio]{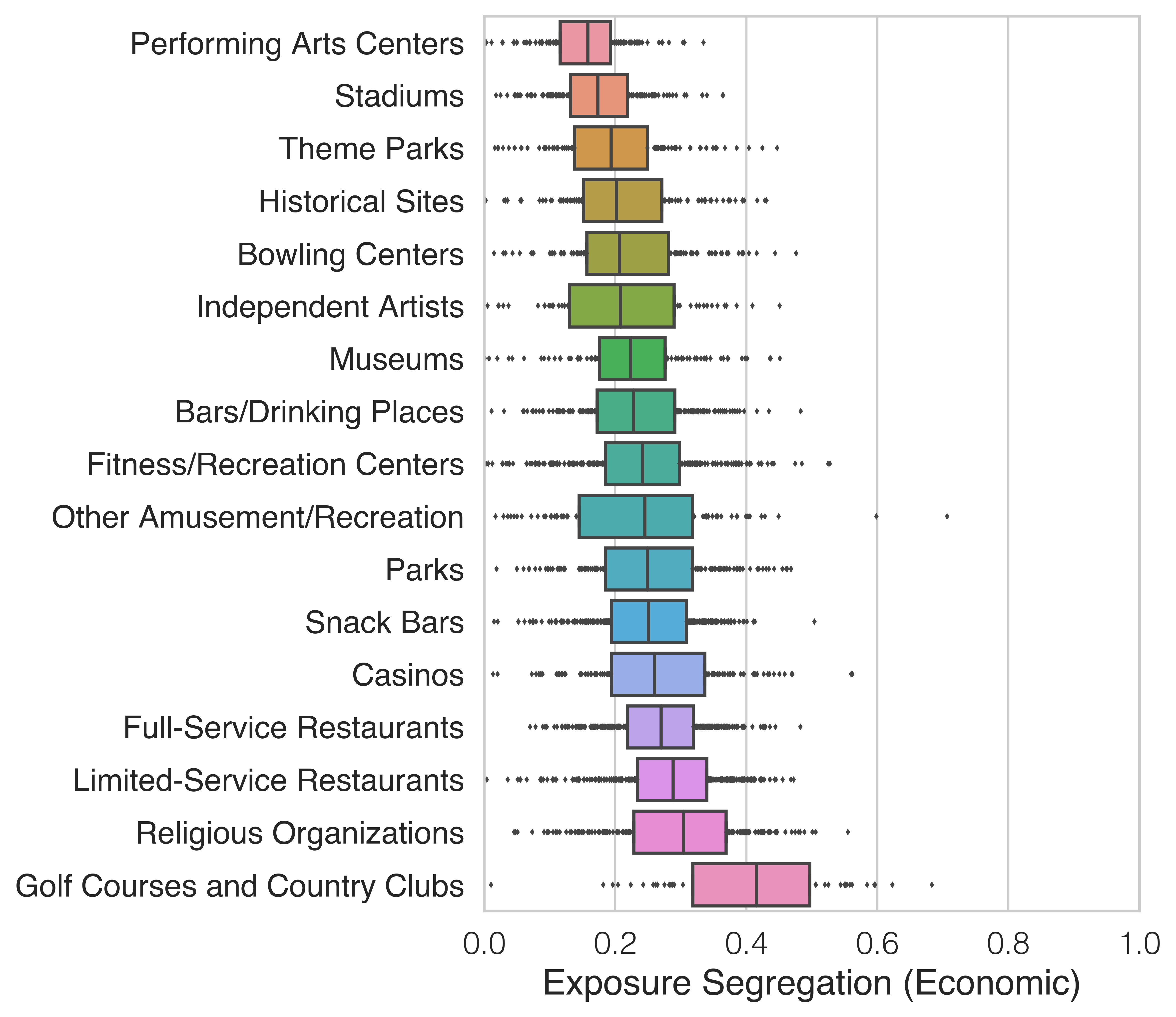}
   \caption{}
\end{subfigure}
\begin{subfigure}[b]{0.5\textwidth}
\includegraphics[width=\textwidth,keepaspectratio]{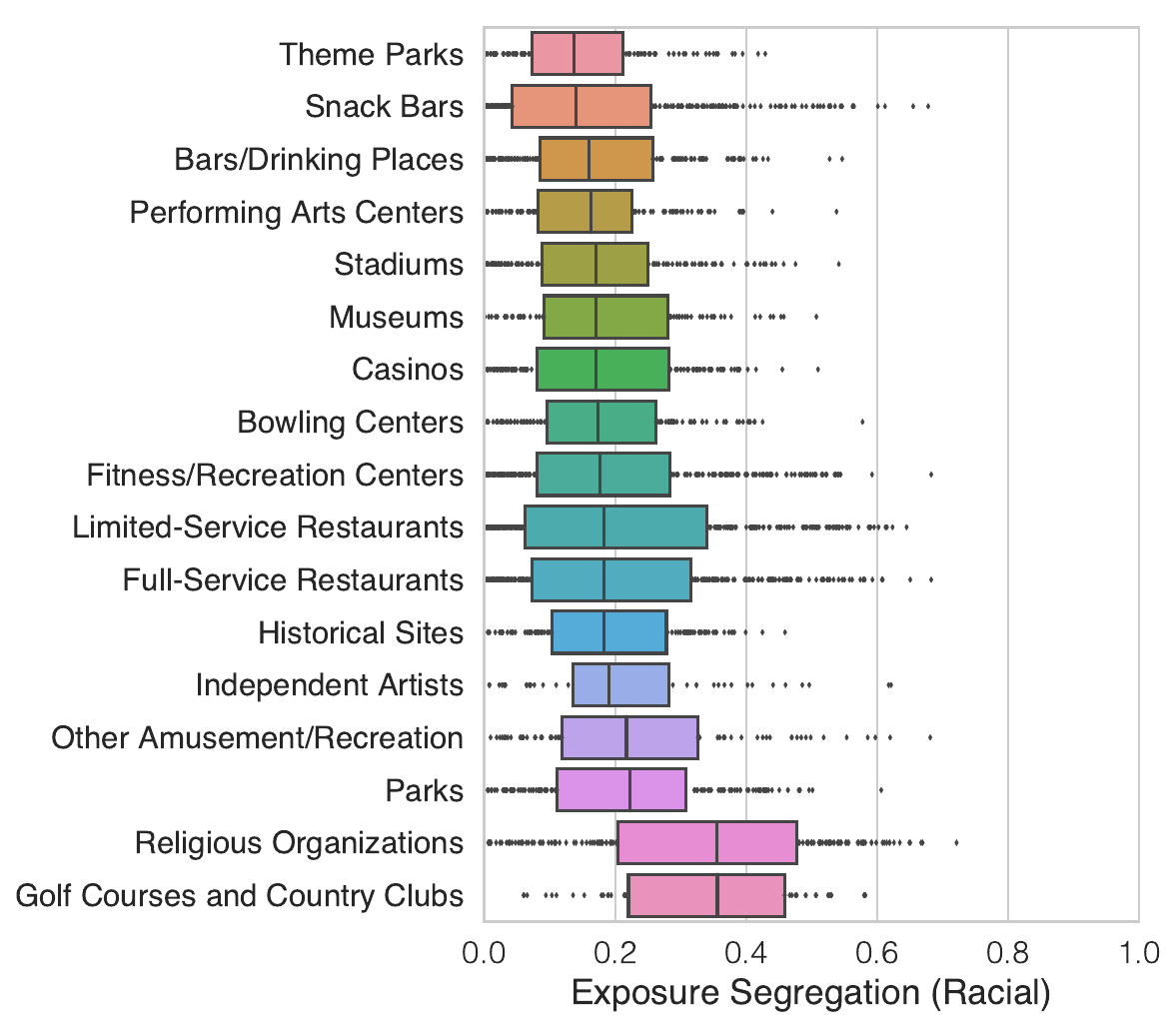}
   \caption{}
\end{subfigure}
   \caption{\new{\textbf{Comparing economic and racial exposure segregation by POI.} We observe significant differences between how POIs are segregated by race vs. economic status. For instance, are 47\% more segregated by economic status than by race. At the same time, golf courses and country clubs, as well as religious organizations, are the most segregated across both socioeconomic status and race. The least racially segregated POI category is theme parks, whereas the least economically segreagted POI category is performing arts centers.}}
   \label{fig:poi_racial}
\end{figure}

\begin{table}[ht]
\centering
\begin{scriptsize}
\setlength\tabcolsep{2pt} %
\begin{tabular}{rrrr}
\toprule
 Percentile &  \# Distinct Exposures &  Percentile &  \# Distinct Exposures \\
\midrule
        0.0 &                 1.0 &        51.0 &                79.0 \\
        1.0 &                 1.0 &        52.0 &                81.0 \\
        2.0 &                 2.0 &        53.0 &                85.0 \\
        3.0 &                 3.0 &        54.0 &                88.0 \\
        4.0 &                 3.0 &        55.0 &                91.0 \\
        5.0 &                 4.0 &        56.0 &                94.0 \\
        6.0 &                 5.0 &        57.0 &                98.0 \\
        7.0 &                 6.0 &        58.0 &               102.0 \\
        8.0 &                 7.0 &        59.0 &               106.0 \\
        9.0 &                 8.0 &        60.0 &               109.0 \\
       10.0 &                 8.0 &        61.0 &               114.0 \\
       11.0 &                 9.0 &        62.0 &               118.0 \\
       12.0 &                10.0 &        63.0 &               122.0 \\
       13.0 &                11.0 &        64.0 &               127.0 \\
       14.0 &                12.0 &        65.0 &               132.0 \\
       15.0 &                13.0 &        66.0 &               137.0 \\
       16.0 &                14.0 &        67.0 &               142.0 \\
       17.0 &                15.0 &        68.0 &               148.0 \\
       18.0 &                17.0 &        69.0 &               154.0 \\
       19.0 &                18.0 &        70.0 &               160.0 \\
       20.0 &                19.0 &        71.0 &               167.0 \\
       21.0 &                20.0 &        72.0 &               173.0 \\
       22.0 &                21.0 &        73.0 &               181.0 \\
       23.0 &                23.0 &        74.0 &               188.0 \\
       24.0 &                24.0 &        75.0 &               196.0 \\
       25.0 &                25.0 &        76.0 &               205.0 \\
       26.0 &                27.0 &        77.0 &               214.0 \\
       27.0 &                28.0 &        78.0 &               224.0 \\
       28.0 &                30.0 &        79.0 &               234.0 \\
       29.0 &                31.0 &        80.0 &               245.0 \\
       30.0 &                33.0 &        81.0 &               257.0 \\
       31.0 &                34.0 &        82.0 &               271.0 \\
       32.0 &                36.0 &        83.0 &               285.0 \\
       33.0 &                38.0 &        84.0 &               300.0 \\
       34.0 &                40.0 &        85.0 &               317.0 \\
       35.0 &                41.0 &        86.0 &               336.0 \\
       36.0 &                43.0 &        87.0 &               357.0 \\
       37.0 &                45.0 &        88.0 &               380.0 \\
       38.0 &                47.0 &        89.0 &               406.0 \\
       39.0 &                49.0 &        90.0 &               436.0 \\
       40.0 &                51.0 &        91.0 &               471.0 \\
       41.0 &                53.0 &        92.0 &               511.0 \\
       42.0 &                55.0 &        93.0 &               560.0 \\
       43.0 &                58.0 &        94.0 &               620.0 \\
       44.0 &                60.0 &        95.0 &               695.0 \\
       45.0 &                62.0 &        96.0 &               796.0 \\
       46.0 &                65.0 &        97.0 &               939.0 \\
       47.0 &                67.0 &        98.0 &              1171.0 \\
       48.0 &                70.0 &        99.0 &              1655.0 \\
       49.0 &                73.0 &       100.0 &             42323.0 \\
       50.0 &                76.0 &                &                  \\
\bottomrule
\end{tabular}
\end{scriptsize}
\caption{\new{Distribution of number of distinct exposures for all individuals residing in 382 Metropolitan Statistical Areas (MSAs). The median individual had exposures to 76 distinct people. 8,609,406 individuals reside in a Metropolitan Statistical Area (90\% of the overall 9,567,559 individuals in our study). The remaining 958,153 users live outside of MSAs, influencing the \metric of an MSA by coming into contact with MSA residents.}}
\label{tab:interaction_distribution4}
\end{table}

\begin{table}[ht]
\centering
\begin{scriptsize}
\setlength\tabcolsep{2pt} %
\begin{tabular}{rrrr}
\toprule
 Percentile &  \# of Distinct Exposures &  Percentile &  \# of Distinct Exposures \\
\midrule
        0.0 &        1.00 &        51.0 &        4.55 \\
        1.0 &        1.00 &        52.0 &        4.65 \\
        2.0 &        1.00 &        53.0 &        4.76 \\
        3.0 &        1.00 &        54.0 &        4.87 \\
        4.0 &        1.14 &        55.0 &        5.00 \\
        5.0 &        1.23 &        56.0 &        5.09 \\
        6.0 &        1.30 &        57.0 &        5.21 \\
        7.0 &        1.36 &        58.0 &        5.33 \\
        8.0 &        1.43 &        59.0 &        5.46 \\
        9.0 &        1.50 &        60.0 &        5.60 \\
       10.0 &        1.56 &        61.0 &        5.73 \\
       11.0 &        1.62 &        62.0 &        5.87 \\
       12.0 &        1.67 &        63.0 &        6.00 \\
       13.0 &        1.75 &        64.0 &        6.17 \\
       14.0 &        1.80 &        65.0 &        6.32 \\
       15.0 &        1.87 &        66.0 &        6.49 \\
       16.0 &        1.94 &        67.0 &        6.66 \\
       17.0 &        2.00 &        68.0 &        6.83 \\
       18.0 &        2.00 &        69.0 &        7.00 \\
       19.0 &        2.10 &        70.0 &        7.21 \\
       20.0 &        2.17 &        71.0 &        7.41 \\
       21.0 &        2.23 &        72.0 &        7.62 \\
       22.0 &        2.29 &        73.0 &        7.86 \\
       23.0 &        2.36 &        74.0 &        8.09 \\
       24.0 &        2.42 &        75.0 &        8.33 \\
       25.0 &        2.50 &        76.0 &        8.61 \\
       26.0 &        2.55 &        77.0 &        8.89 \\
       27.0 &        2.62 &        78.0 &        9.20 \\
       28.0 &        2.68 &        79.0 &        9.52 \\
       29.0 &        2.75 &        80.0 &        9.87 \\
       30.0 &        2.82 &        81.0 &       10.24 \\
       31.0 &        2.89 &        82.0 &       10.65 \\
       32.0 &        3.00 &        83.0 &       11.08 \\
       33.0 &        3.00 &        84.0 &       11.57 \\
       34.0 &        3.10 &        85.0 &       12.09 \\
       35.0 &        3.17 &        86.0 &       12.67 \\
       36.0 &        3.25 &        87.0 &       13.33 \\
       37.0 &        3.33 &        88.0 &       14.05 \\
       38.0 &        3.40 &        89.0 &       14.88 \\
       39.0 &        3.48 &        90.0 &       15.83 \\
       40.0 &        3.56 &        91.0 &       16.94 \\
       41.0 &        3.64 &        92.0 &       18.25 \\
       42.0 &        3.72 &        93.0 &       19.83 \\
       43.0 &        3.81 &        94.0 &       21.78 \\
       44.0 &        3.89 &        95.0 &       24.31 \\
       45.0 &        4.00 &        96.0 &       27.71 \\
       46.0 &        4.07 &        97.0 &       32.71 \\
       47.0 &        4.16 &        98.0 &       41.00 \\
       48.0 &        4.25 &        99.0 &       59.00 \\
       49.0 &        4.35 &       100.0 &     1740.25 \\
       50.0 &        4.45 &             &             \\
\bottomrule
\end{tabular}
\end{scriptsize}
\caption{\new{Distribution of average number of distinct exposures (per active day) for all individuals residing in 382 Metropolitan Statistical Areas (MSAs). The median individual had 4.45 unique exposures on the average day of activity. 8,609,406 individuals reside in a Metropolitan Statistical Area (90\% of the overall 9,567,559 individuals in our study). The remaining 958,153 users live outside of MSAs, influencing the \metric of an MSA by coming into contact with MSA residents. Activity is defined as one or more exposures occurring on a given day. For details on activity over time, see Supplementary Figure \ref{fig:active_over_time}}.}
\label{tab:interaction_distribution5}
\end{table}

\begin{figure}[htbp]
   \centering
   \vspace{-5em}
\includegraphics[width=0.65\textwidth,keepaspectratio]{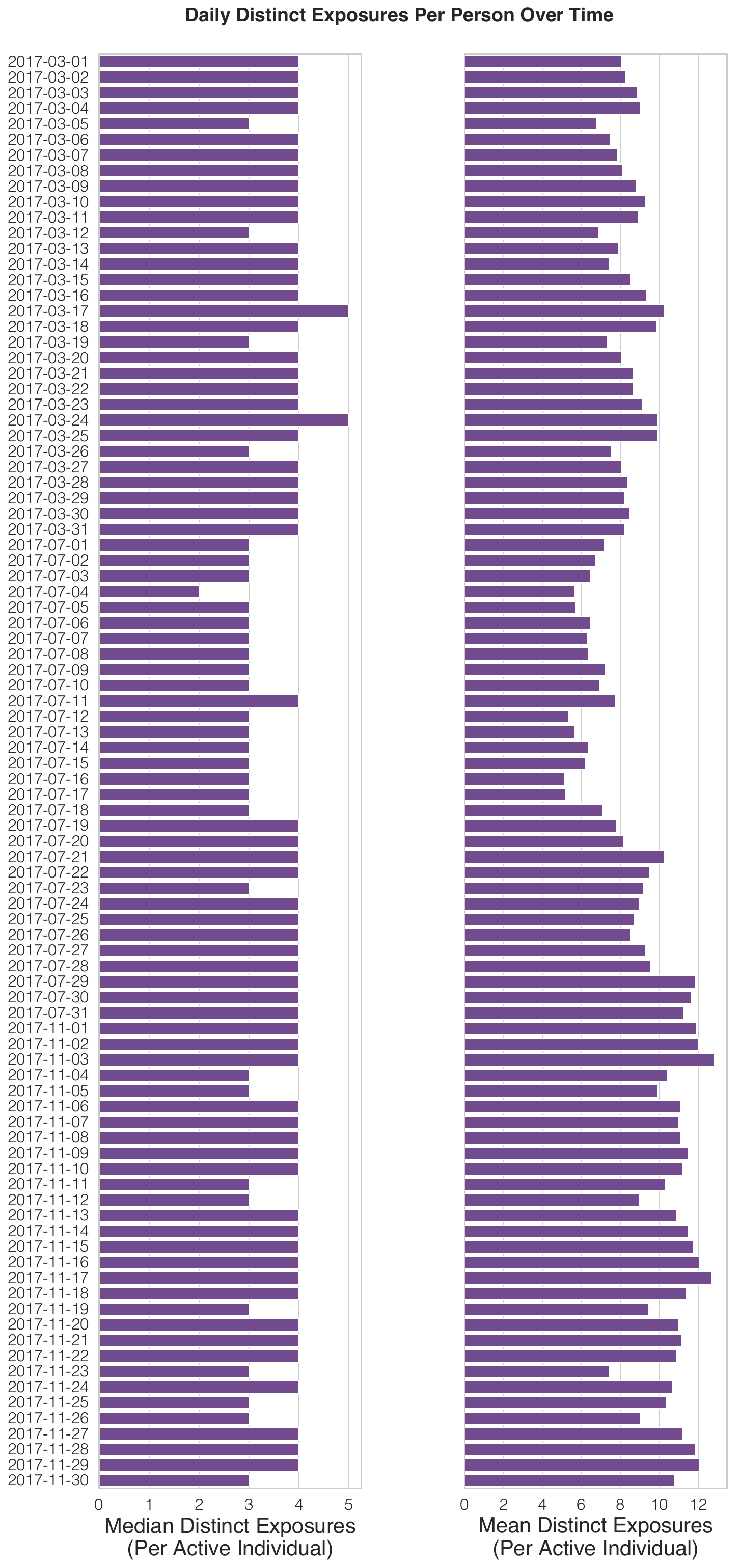}
   \vspace{-0.75em}
   \caption{\new{\textbf{Average number of distinct exposures over time}. Mean/median distinct exposures per active individuals (i.e. nodes in the network) over the study observation period. Activity is defined as one or more exposures occurring on a given day. 
}}
   \label{fig:interactors_over_time_distinct}
\end{figure}

\begin{figure}[htbp]
   \centering
   \vspace{-5em}
\includegraphics[width=0.5\textwidth,keepaspectratio]{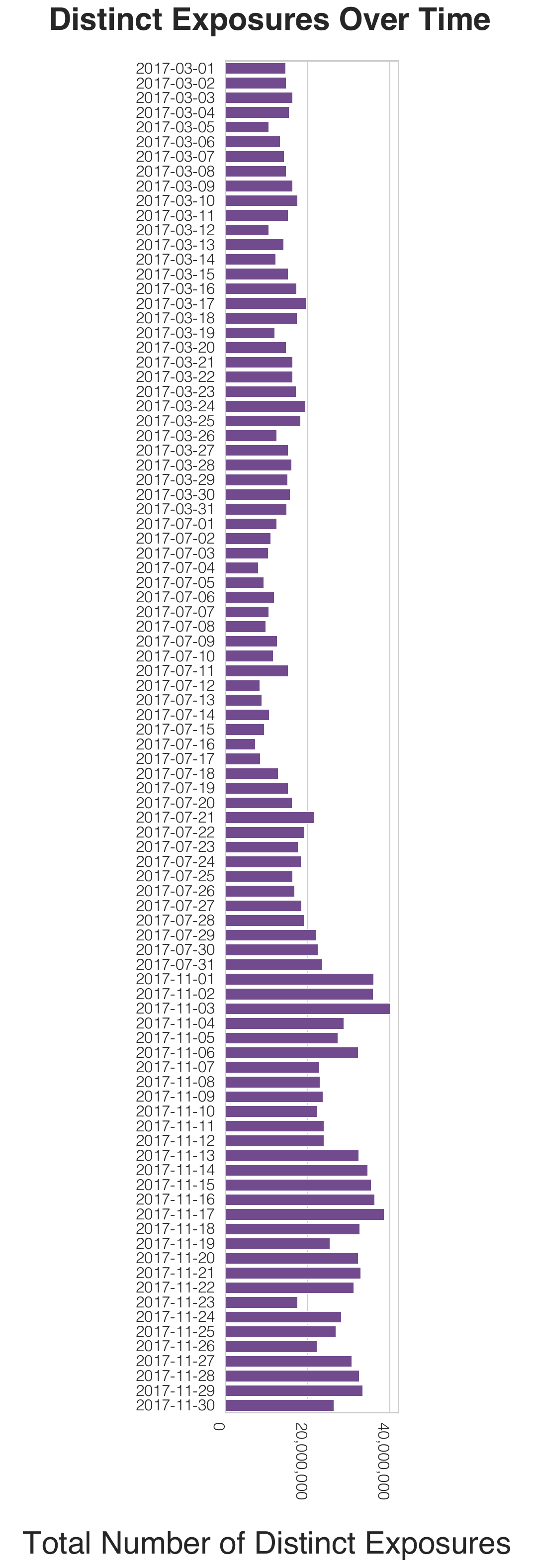}
   \vspace{-0.75em}
   \caption{\new{\textbf{Distinct exposures} over time across all individuals residing in the 382 MSAs.}}
   \label{fig:active_over_time_distinct}
\end{figure}

\begin{table}[ht]
\centering
\begin{scriptsize}
\setlength\tabcolsep{2pt} %
\begin{tabular}{lrrrrrrrrrr}
\toprule
                                       Measure &     Mean &     std &  min &  10\% &   25\% &   50\% &    75\% &    90\% &      max \\
\midrule
                               Primary Measure &    184.21 &  373.98 &  1.0 &  8.0 &  25.0 &  76.0 &  196.0 &  436.0 &  42323.0 \\
\midrule
         Minimum Distance Between Pings: $<$ 25m &   93.23 &  200.13 &  1.0 &  4.0 &  12.0 &  37.0 &   97.0 &  219.0 &  26778.0 \\
         Minimum Distance Between Pings: $<$  10m &   41.47 &   99.99 &  1.0 &  2.0 &   5.0 &  14.0 &   38.0 &   94.0 &  12907.0 \\
\midrule
       Minimum Time Between Pings: $<$  2 minutes &   99.92 &  212.03 &  1.0 &  5.0 &  14.0 &  41.0 &  104.0 &  232.0 &  26592.0 \\
      Minimum Time Between Pings: $<$  60 seconds &   63.50 &  141.10 &  1.0 &  4.0 &   9.0 &  26.0 &   65.0 &  144.0 &  18912.0 \\
\midrule
 Minimum Tie Strength: 2 consecutive exposures &   35.51 &   91.98 &  1.0 &  2.0 &   5.0 &  13.0 &   33.0 &   78.0 &  13513.0 \\
 Minimum Tie Strength: 3 consecutive exposures &    8.53 &   23.31 &  1.0 &  1.0 &   1.0 &   3.0 &    7.0 &   18.0 &   3107.0 \\
           Minimum Tie Strength: 2 unique days &   16.39 &   54.78 &  1.0 &  1.0 &   2.0 &   5.0 &   13.0 &   33.0 &  15454.0 \\
           Minimum Tie Strength: 3 unique days &    6.74 &   22.04 &  1.0 &  1.0 &   1.0 &   3.0 &    6.0 &   13.0 &   7685.0 \\
\midrule
           Dist. $<$  25m, Time $<$  2 minutes, Length 2+ unique days  &   9.83 &   33.09 &  1.0 &  1.0 &   2.0 &   3.0 &    8.0 &   19.0 &   9951.0 \\
           Dist. $<$ 25m, Time $<$  2 minutes, Length 2+ consecutive exposures  &  18.62 &   49.25 &  1.0 &  1.0 &   3.0 &   7.0 &   17.0 &   41.0 &   8447.0 \\
          Dist. $<$  10m, Time $<$ 1 minutes, Length 3+ unique days  &   3.30 &   10.72 &  1.0 &  1.0 &   1.0 &   1.0 &    3.0 &    6.0 &   2019.0 \\
           Dist. $<$  10m, Time $<$  1 minutes, Length 3+ consecutive exposures  &   4.31 &   12.40 &  1.0 &  1.0 &   1.0 &   2.0 &    4.0 &    9.0 &   2893.0 \\
\bottomrule
\end{tabular}
\end{scriptsize}
\caption{\new{\textbf{Distribution of number of distinct exposures per person (by time, distance, and length threshold).} We compute the number of distinct pairs of exposures for all residents of the 382 Metropolitan Statistical Areas (MSAs), for each robustness check which varies the time, distance, length thresholds for the definition of exposure. Summary of the per person number of distinct exposures.}}
\label{tab:poi_repetition1}
\end{table}

\begin{table}[ht]
\centering
\begin{scriptsize}
\setlength\tabcolsep{2pt} %
\begin{tabular}{lrrrrrrrrrr}
\toprule
                                       Measure &     Mean &     std &  min &  10\% &   25\% &   50\% &    75\% &    90\% &      max \\
\midrule
                                    Primary Metric &   7.90 &  13.90 &  1.0 &  1.56 &  2.50 &  4.45 &  8.33 &  15.83 &  1740.25 \\
\midrule
       Minimum Distance Between Pings: $<$ 25 meters &   3.25 &   5.39 &  1.0 &  1.00 &  1.25 &  1.90 &  3.20 &   6.06 &   907.75 \\
       Minimum Distance Between Pings: $<$ 10 meters &   2.54 &   4.58 &  1.0 &  1.00 &  1.00 &  1.40 &  2.17 &   4.33 &   510.14 \\
\midrule
           Minimum Time Between Pings: $<$ 2 minutes &   3.17 &   5.25 &  1.0 &  1.00 &  1.29 &  1.95 &  3.23 &   5.86 &   973.75 \\
          Minimum Time Between Pings: $<$ 60 seconds &   2.49 &   3.79 &  1.0 &  1.00 &  1.13 &  1.62 &  2.50 &   4.33 &   674.50 \\
\midrule
  Minimum Tie Strength: 2 consecutive exposures &   2.19 &   3.59 &  1.0 &  1.00 &  1.00 &  1.40 &  2.11 &   3.67 &   622.00 \\
  Minimum Tie Strength: 3 consecutive exposures &   1.61 &   2.18 &  1.0 &  1.00 &  1.00 &  1.00 &  1.50 &   2.45 &   344.00 \\
\midrule
 Dist. $<$ 25 meters, Time $<$ 2 minutes, Length 2+ consecutive exposures &   1.76 &   2.30 &  1.0 &  1.00 &  1.00 &  1.18 &  1.75 &   2.89 &  1148.00 \\
 Dist. $<$ 10 meters, Time $<$ 60 minutes, Length 3+ consecutive exposures &   1.43 &   2.03 &  1.0 &  1.00 &  1.00 &  1.00 &  1.30 &   2.00 &   342.00 \\
\bottomrule
\end{tabular}
\end{scriptsize}
\caption{\new{\textbf{Distribution of average number of distinct exposures (per active day) per person across all days of activity (by time, distance, and length threshold).} Summary statistics are shown for  for all residents of the 382 Metropolitan Statistical Areas (MSAs), for each robustness check which varies the time, distance, length thresholds for the definition of exposure. Activity is defined as one or more exposure occurring on a given day. Summary of the per person number of distinct exposures.}}
\label{tab:poi_repetition2}
\end{table}

\begin{figure}[htbp]
   \centering
   \hspace*{-15mm}
\begin{subfigure}[t]{.35\textwidth}
\includegraphics[width=\textwidth,keepaspectratio]{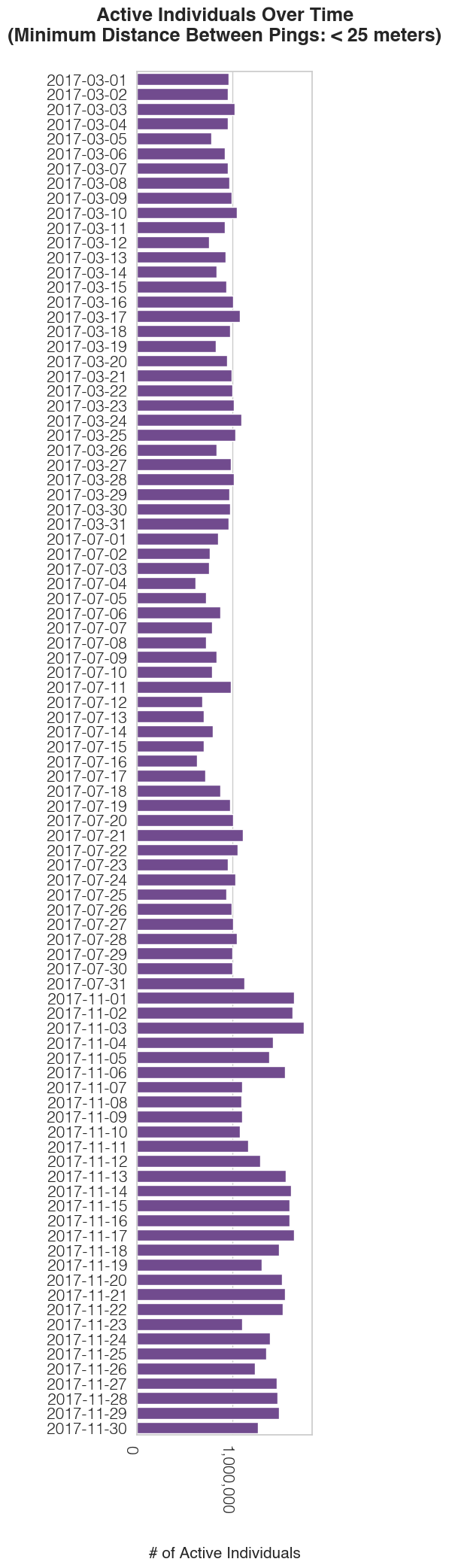}
   \caption{}
\end{subfigure}
   \caption{\new{\textbf{Active individuals over time (exposure distance threshold: 25 meters)}. Number of active individuals (i.e. nodes in the network) over the study observation period. Activity is defined as one or more exposures occurring on a given day. }}
   \label{fig:active_robust_1}
\end{figure}

\begin{figure}[htbp]
   \centering
   \hspace*{-15mm}
\begin{subfigure}[t]{.35\textwidth}
\includegraphics[width=\textwidth,keepaspectratio]{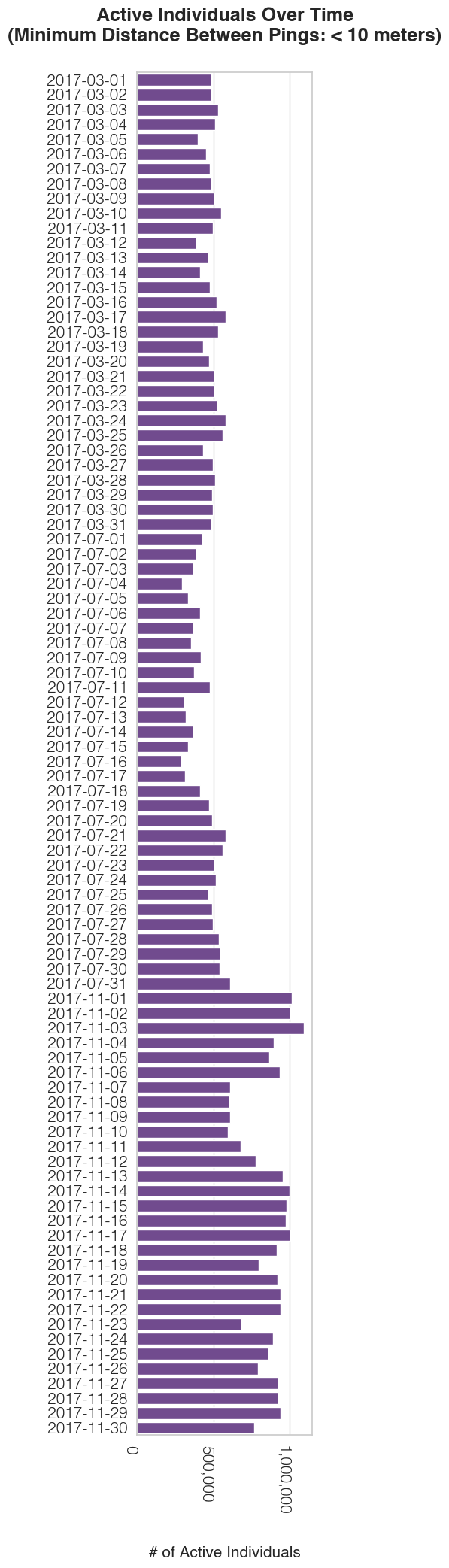}
   \caption{}
\end{subfigure}
   \caption{\new{\textbf{Active individuals over time (exposure distance threshold: 10 meters)}. Number of active individuals (i.e. nodes in the network) over the study observation period. Activity is defined as one or more exposures occurring on a given day. }}
   \label{fig:active_robust_2}
\end{figure}

\begin{figure}[htbp]
   \centering
   \hspace*{-15mm}
\begin{subfigure}[t]{.35\textwidth}
\includegraphics[width=\textwidth,keepaspectratio]{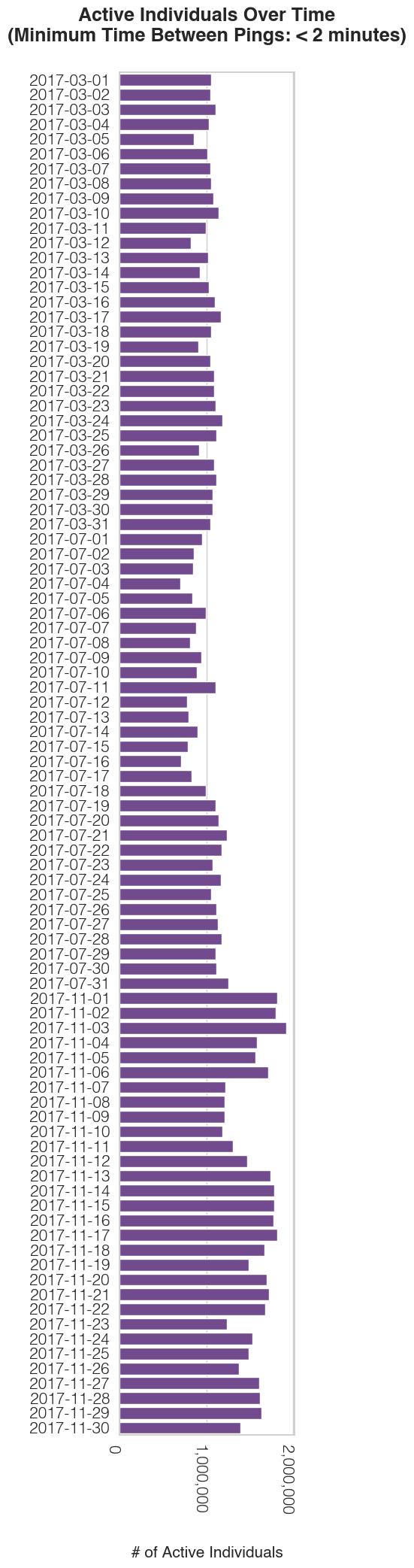}
   \caption{}
\end{subfigure}
   \caption{\new{\textbf{Active individuals over time (exposure time threshold: 2 minutes)}. Number of active individuals (i.e. nodes in the network) over the study observation period. Activity is defined as one or more exposures occurring on a given day. }}
   \label{fig:active_robust_3}
\end{figure}

\begin{figure}[htbp]
   \centering
   \hspace*{-15mm}
\begin{subfigure}[t]{.35\textwidth}
\includegraphics[width=\textwidth,keepaspectratio]{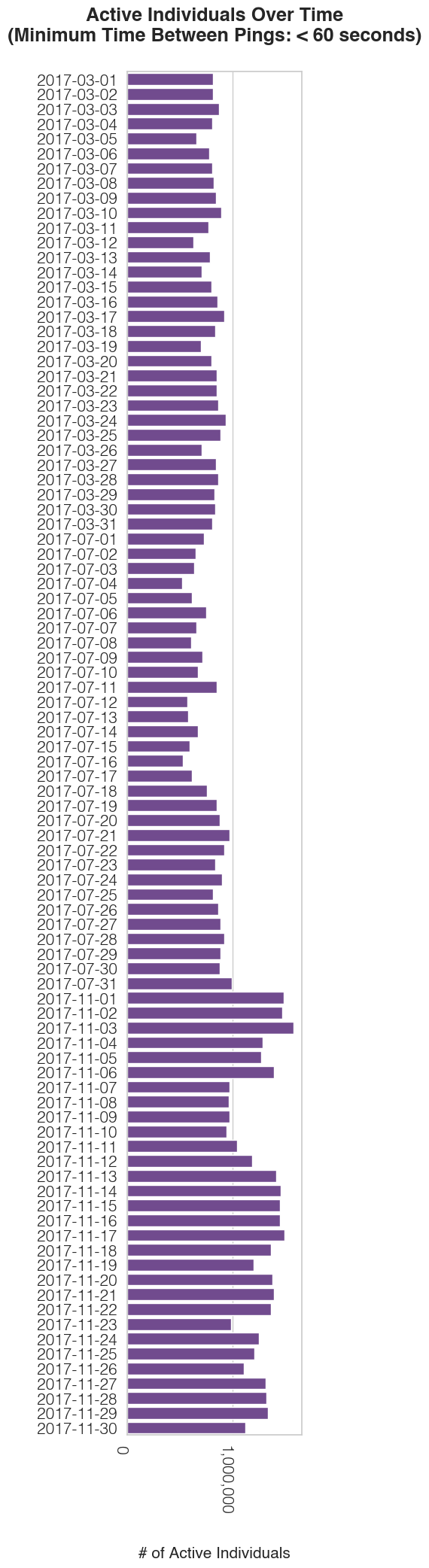}
   \caption{}
\end{subfigure}
   \caption{\new{\textbf{Active individuals over time (exposure time threshold: 60 seconds)}. Number of active individuals (i.e. nodes in the network) over the study observation period. Activity is defined as one or more exposures occurring on a given day. }}
   \label{fig:active_robust_4}
\end{figure}

\begin{figure}[htbp]
   \centering
   \hspace*{-15mm}
\begin{subfigure}[t]{.35\textwidth}
\includegraphics[width=\textwidth,keepaspectratio]{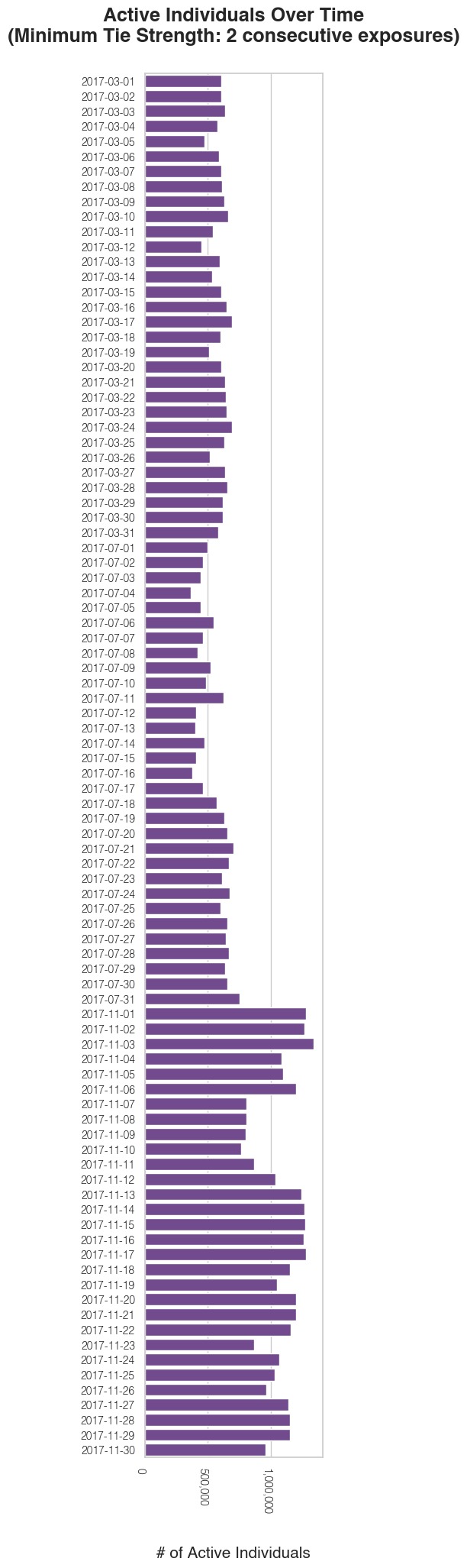}
   \caption{}
\end{subfigure}
   \caption{\new{\textbf{Active individuals over time (exposure length threshold: 2 consecutive exposures over five minute intervals)}. Number of active individuals (i.e. nodes in the network) over the study observation period. Activity is defined as one or more exposures occurring on a given day. }}
   \label{fig:active_robust_5}
\end{figure}

\begin{figure}[htbp]
   \centering
   \hspace*{-15mm}
\begin{subfigure}[t]{.35\textwidth}
\includegraphics[width=\textwidth,keepaspectratio]{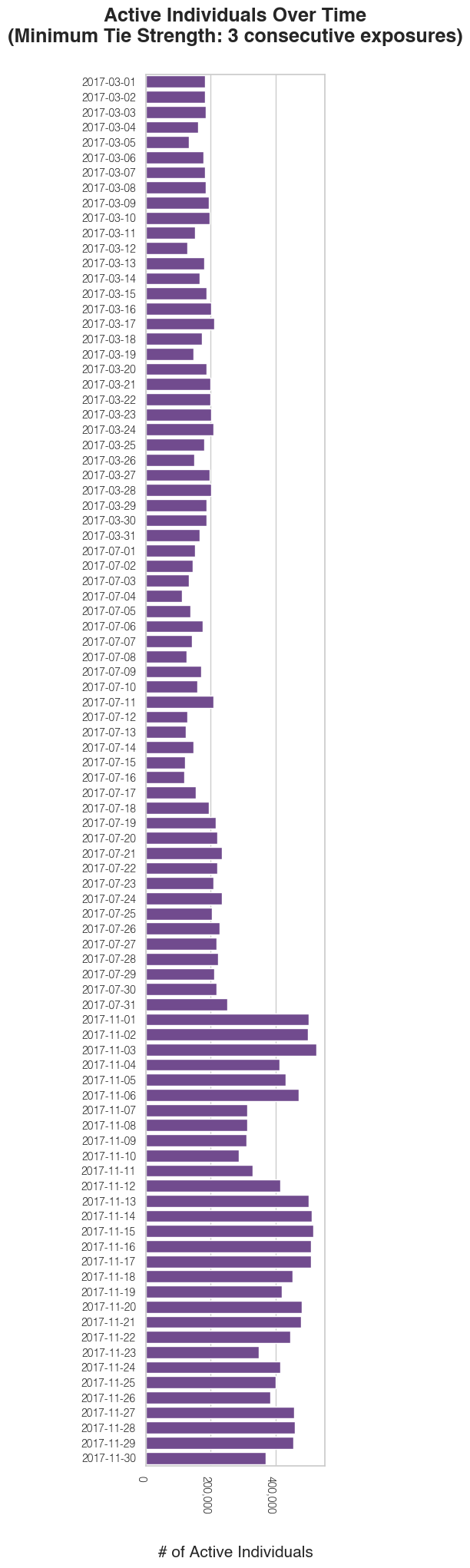}
   \caption{}
\end{subfigure}
   \caption{\new{\textbf{Active individuals over time (exposure length threshold: 3 consecutive exposures over five minute intervals)}. Number of active individuals (i.e. nodes in the network) over the study observation period. Activity is defined as one or more exposures occurring on a given day. }}
   \label{fig:active_robust_6}
\end{figure}

\begin{figure}[htbp]
   \centering
   \hspace*{-15mm}
\begin{subfigure}[t]{.35\textwidth}
\includegraphics[width=\textwidth,keepaspectratio]{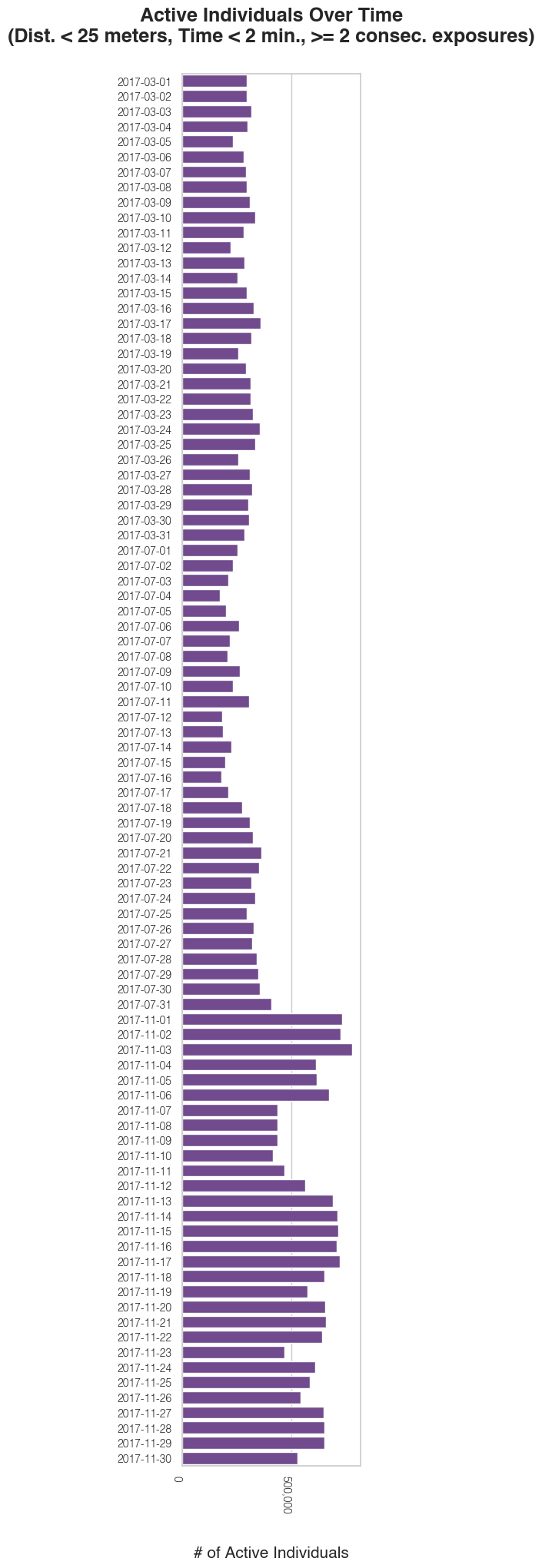}
   \caption{}
\end{subfigure}
   \caption{\new{\textbf{Active individuals over time (distance threshold: 25 meters, time threshold: 2 minutes, length threshold: 2 consecutive exposures of five minute intervals)}. Number of active individuals (i.e. nodes in the network) over the study observation period. Activity is defined as one or more exposures occurring on a given day. }}
   \label{fig:active_robust_7}
\end{figure}

\begin{figure}[htbp]
   \centering
   \hspace*{-15mm}
\begin{subfigure}[t]{.35\textwidth}
\includegraphics[width=\textwidth,keepaspectratio]{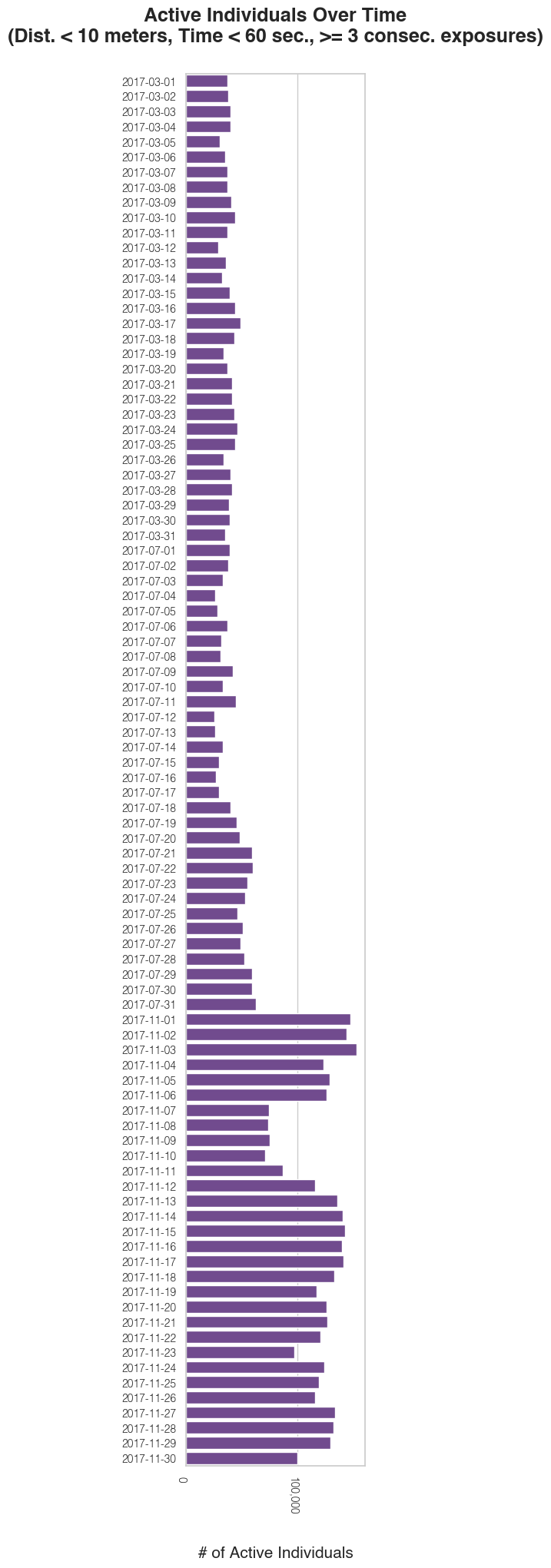}
   \caption{}
\end{subfigure}
   \caption{\new{\textbf{Active individuals over time (distance threshold: 10 meters, time threshold: 60 seconds, length threshold: 3 consecutive exposures of five minute intervals)}. Number of active individuals (i.e. nodes in the network) over the study observation period. Activity is defined as one or more exposures occurring on a given day. }}
   \label{fig:active_robust_8}
\end{figure}

\begin{figure}[htbp]
   \centering
   \vspace{-5em}
\includegraphics[width=0.5\textwidth,keepaspectratio]{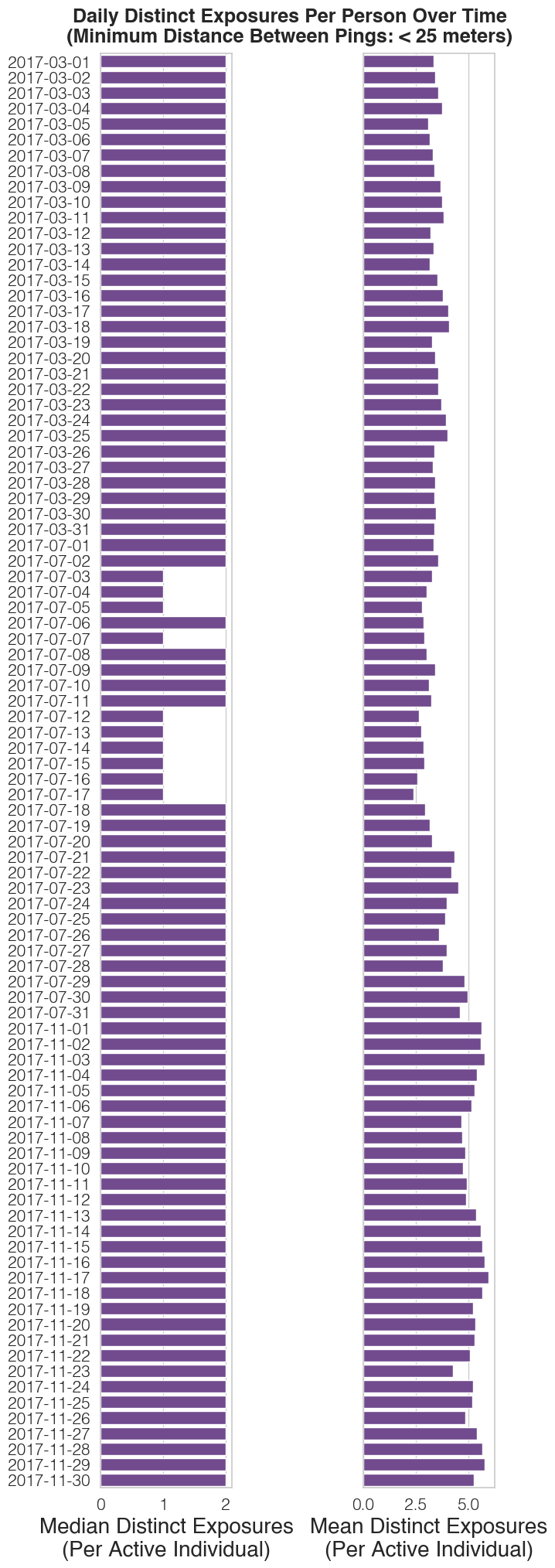}
   \vspace{-0.75em}
   \caption{\new{\textbf{Average number of distinct exposures over time (distance threshold: 25 meters)}. Mean/median distinct exposures per active individuals (i.e. nodes in the network) over the study observation period. Activity is defined as one or more exposures occurring on a given day. 
}}
   \label{fig:interactors_over_time_robust_1}
\end{figure}

\begin{figure}[htbp]
   \centering
   \vspace{-5em}
\includegraphics[width=0.5\textwidth,keepaspectratio]{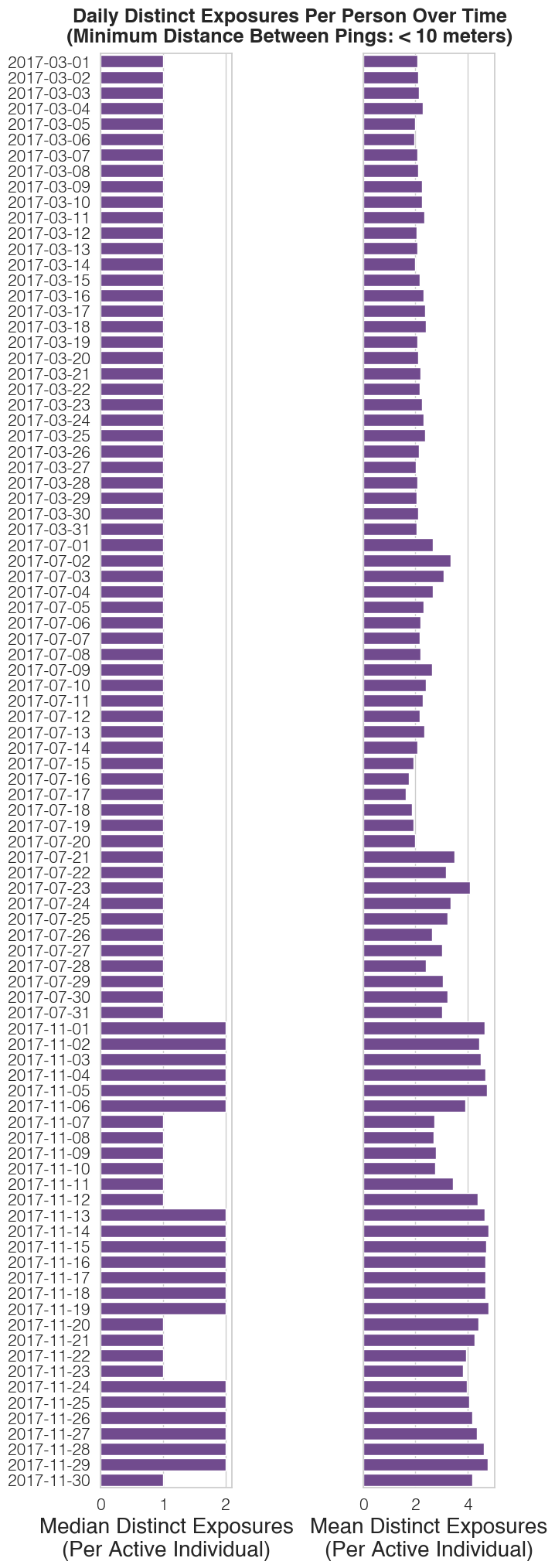}
   \vspace{-0.75em}
   \caption{\new{\textbf{Average number of distinct exposures over time (distance threshold: 10 meters)}. Mean/median distinct exposures per active individuals (i.e. nodes in the network) over the study observation period. Activity is defined as one or more exposures occurring on a given day. 
}}
   \label{fig:interactors_over_time_robust_2}
\end{figure}
\begin{figure}[htbp]
   \centering
   \vspace{-5em}
\includegraphics[width=0.5\textwidth,keepaspectratio]{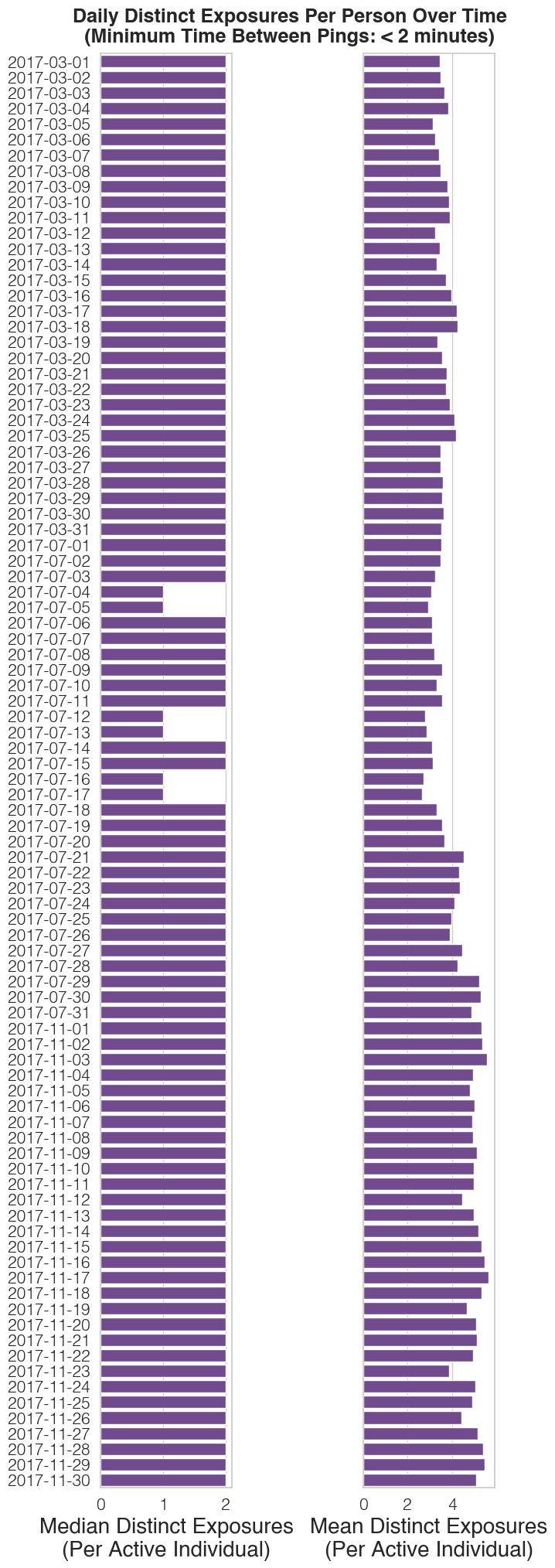}
   \vspace{-0.75em}
   \caption{\new{\textbf{Average number of distinct exposures over time (time threshold: 2 minutes)}. Mean/median distinct exposures per active individuals (i.e. nodes in the network) over the study observation period. Activity is defined as one or more exposures occurring on a given day. 
}}
   \label{fig:interactors_over_time_robust_3}
\end{figure}
\begin{figure}[htbp]
   \centering
   \vspace{-5em}
\includegraphics[width=0.5\textwidth,keepaspectratio]{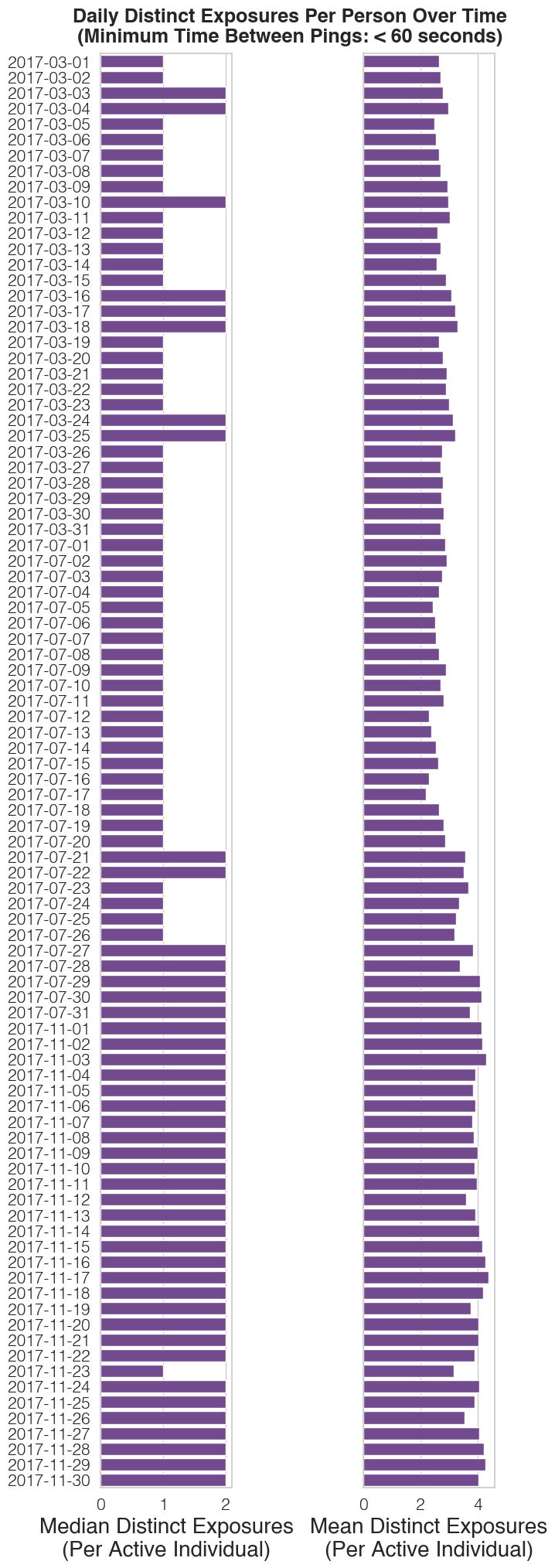}
   \vspace{-0.75em}
   \caption{\new{\textbf{Average number of distinct exposures over time (time threshold: 60 seconds)}. Mean/median distinct exposures per active individuals (i.e. nodes in the network) over the study observation period. Activity is defined as one or more exposures occurring on a given day. 
}}
   \label{fig:interactors_over_time_robust_4}
\end{figure}
\begin{figure}[htbp]
   \centering
   \vspace{-5em}
\includegraphics[width=0.5\textwidth,keepaspectratio]{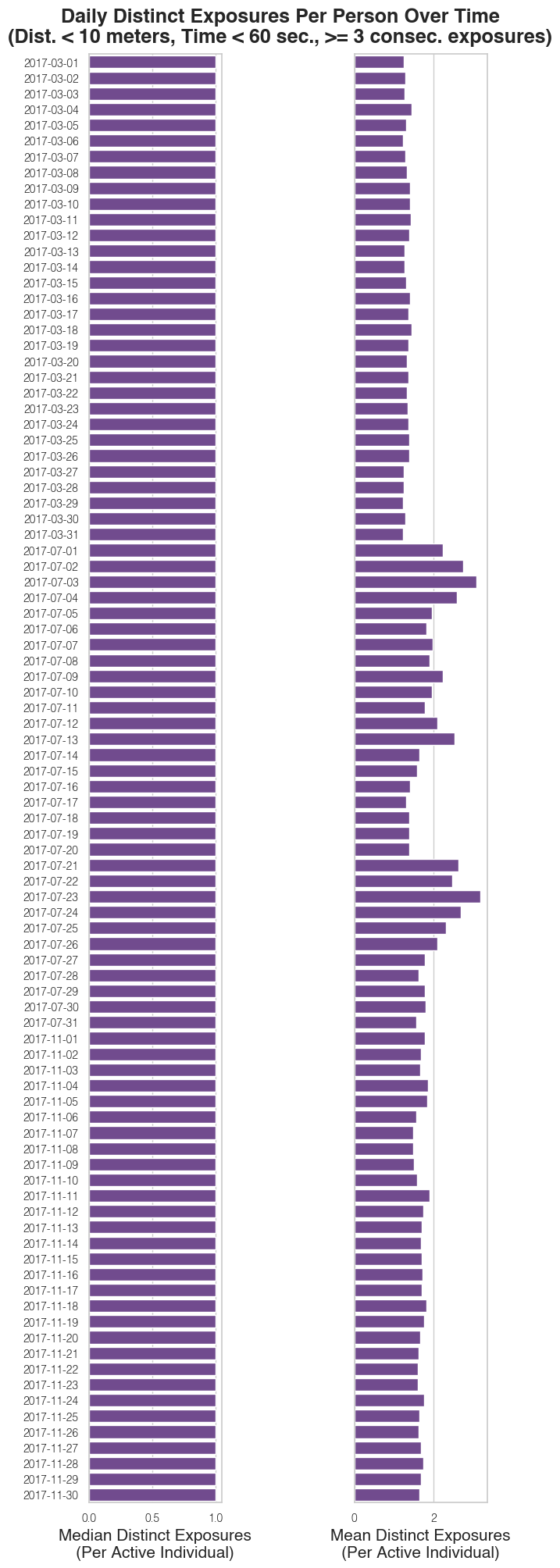}
   \vspace{-0.75em}
   \caption{\new{\textbf{Average number of distinct exposures over time (exposure length threshold: 2 consecutive exposures over five minute intervals)}. Mean/median distinct exposures per active individuals (i.e. nodes in the network) over the study observation period. Activity is defined as one or more exposures occurring on a given day. 
}}
   \label{fig:interactors_over_time_robust_5}
\end{figure}
\begin{figure}[htbp]
   \centering
   \vspace{-5em}
\includegraphics[width=0.5\textwidth,keepaspectratio]{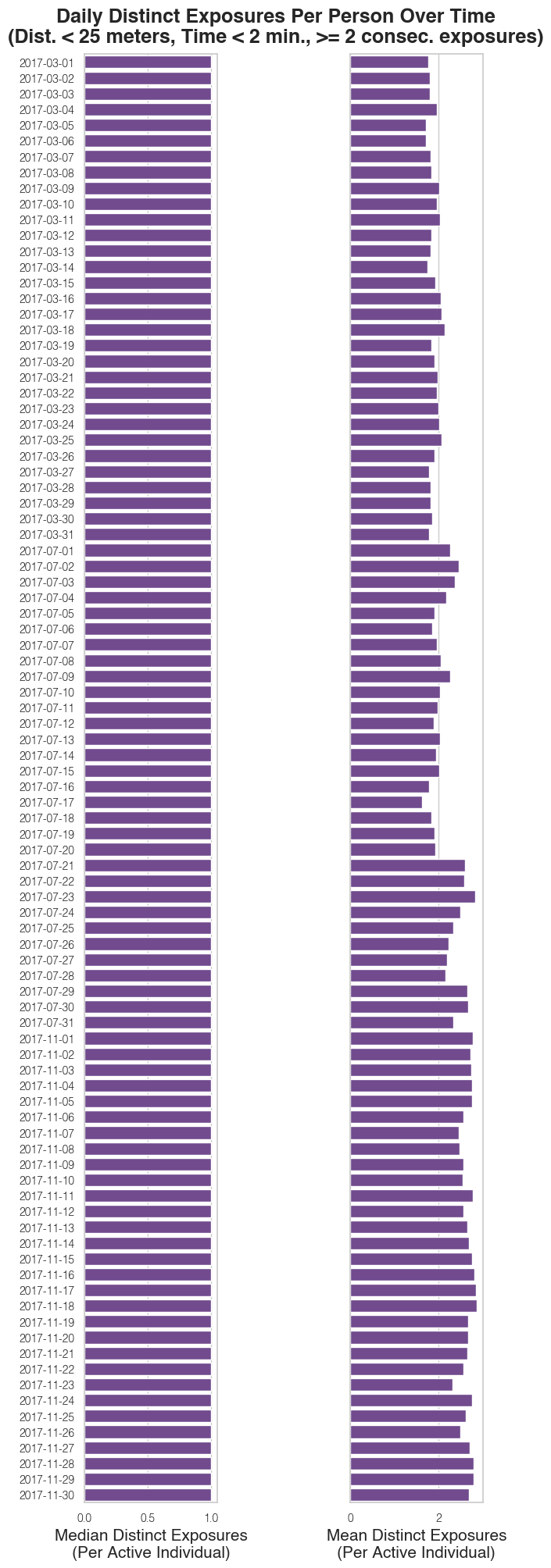}
   \vspace{-0.75em}
   \caption{\new{\textbf{Average number of distinct exposures over time (exposure length threshold: 3 consecutive exposures over five minute intervals)}. Mean/median distinct exposures per active individuals (i.e. nodes in the network) over the study observation period. Activity is defined as one or more exposures occurring on a given day. 
}}
   \label{fig:interactors_over_time_robust_6}
\end{figure}
\begin{figure}[htbp]
   \centering
   \vspace{-5em}
\includegraphics[width=0.5\textwidth,keepaspectratio]{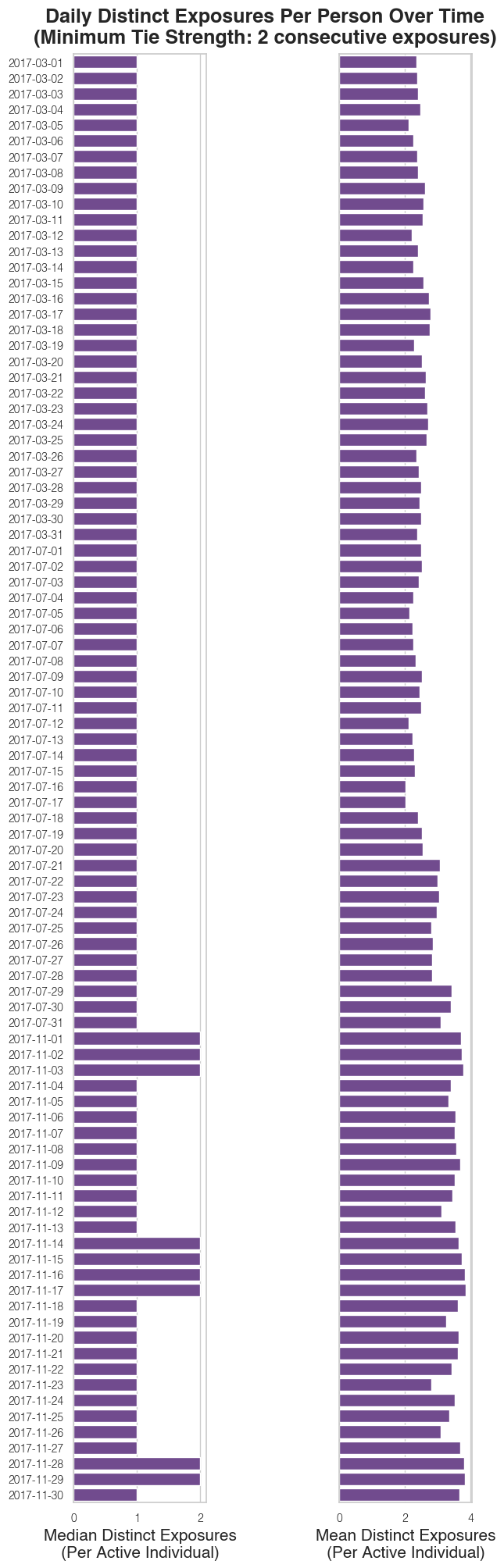}
   \vspace{-0.75em}
   \caption{\new{\textbf{Average number of distinct exposures over time (distance threshold: 25 meters, time threshold: 2 minutes, length threshold: 2 consecutive exposures of five minute intervals)}. Mean/median distinct exposures per active individuals (i.e. nodes in the network) over the study observation period. Activity is defined as one or more exposures occurring on a given day. 
}}
   \label{fig:interactors_over_time_robust_7}
\end{figure}
\begin{figure}[htbp]
   \centering
   \vspace{-5em}
\includegraphics[width=0.5\textwidth,keepaspectratio]{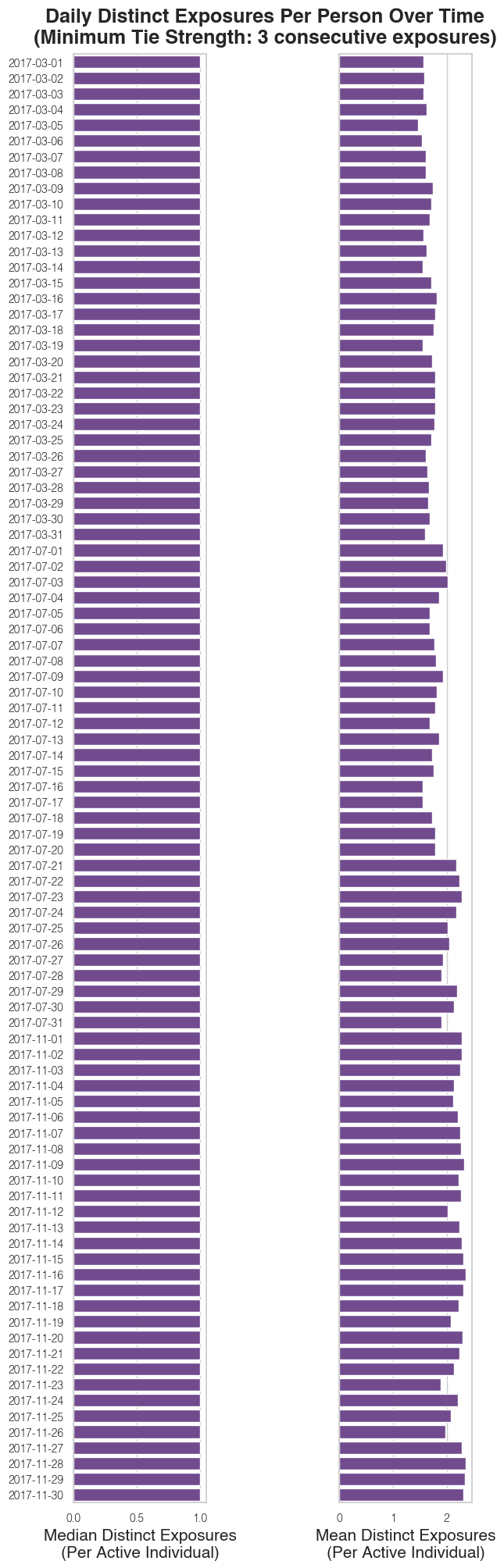}
   \vspace{-0.75em}
   \caption{\new{\textbf{Average number of distinct exposures over time (distance threshold: 10 meters, time threshold: 60 seconds, length threshold: 3 consecutive exposures of five minute intervals)}. Mean/median distinct exposures per active individuals (i.e. nodes in the network) over the study observation period. Activity is defined as one or more exposures occurring on a given day. 
}}
   \label{fig:interactors_over_time_robust_8}
\end{figure}

\begin{figure}[htbp]
   \centering
   \vspace{-5em}
\includegraphics[width=0.4\textwidth,keepaspectratio]{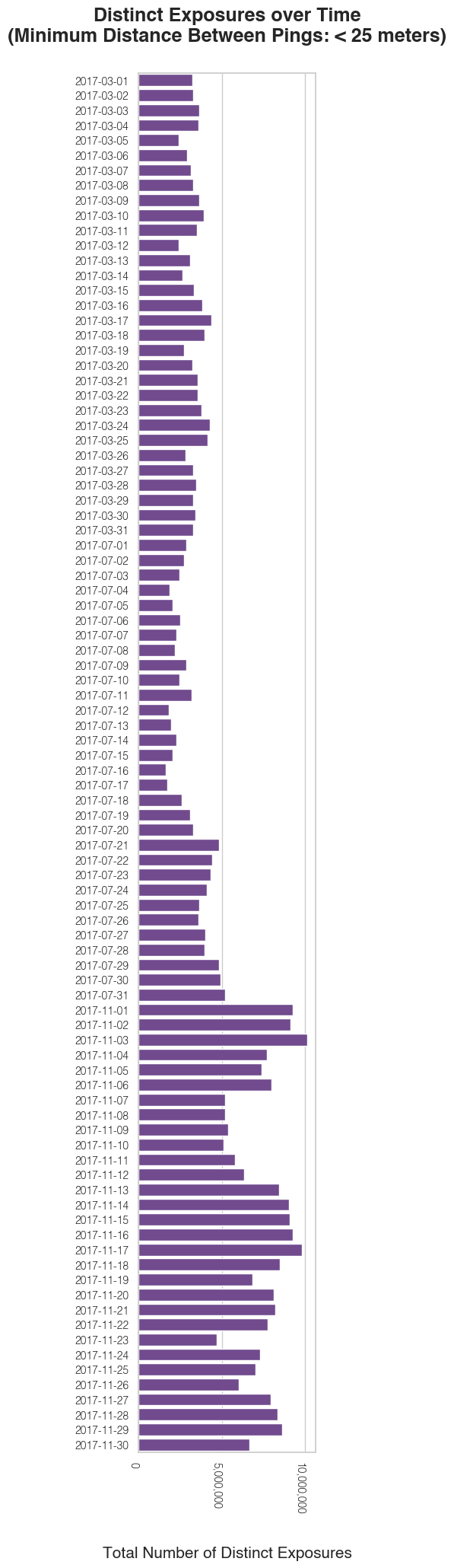}
   \vspace{-0.75em}
   \caption{\new{\textbf{Number of distinct exposures (distance threshold: 25 meters)} over time across all individuals residing in the 382 MSAs.}}
   \label{fig:robust_total_distinct_1}

\end{figure}

\begin{figure}[htbp]
   \centering
   \vspace{-5em}
\includegraphics[width=0.4\textwidth,keepaspectratio]{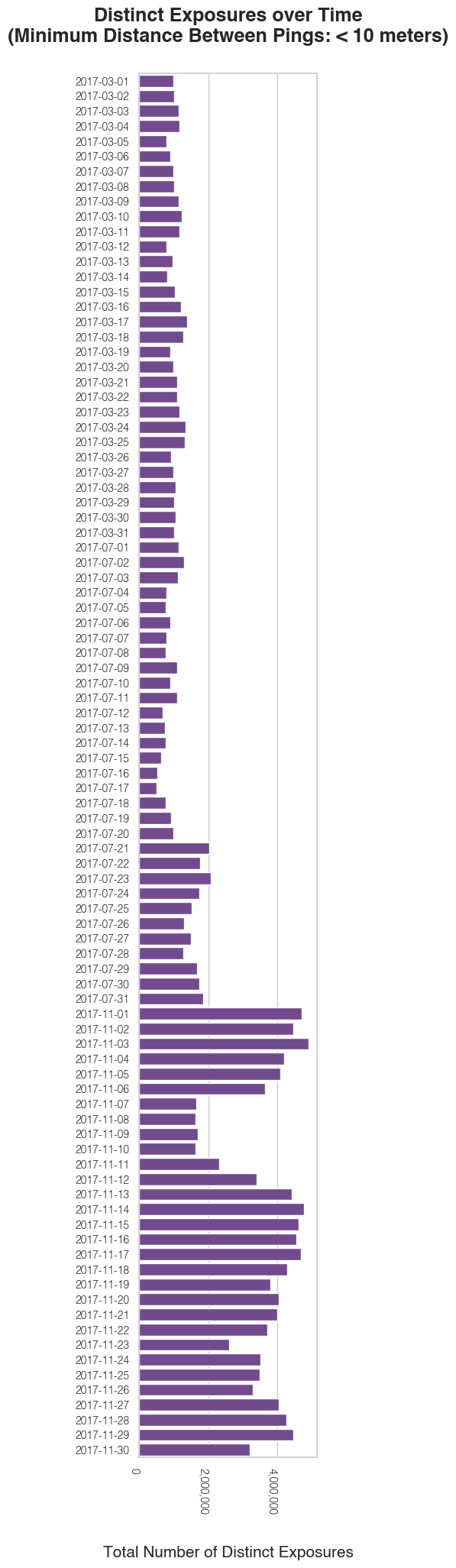}
   \vspace{-0.75em}
   \caption{\new{\textbf{Number of distinct exposures (distance threshold: 10 meters)} over time across all individuals residing in the 382 MSAs.}}
   \label{fig:robust_total_distinct_2}

\end{figure}

\begin{figure}[htbp]
   \centering
   \vspace{-5em}
\includegraphics[width=0.4\textwidth,keepaspectratio]{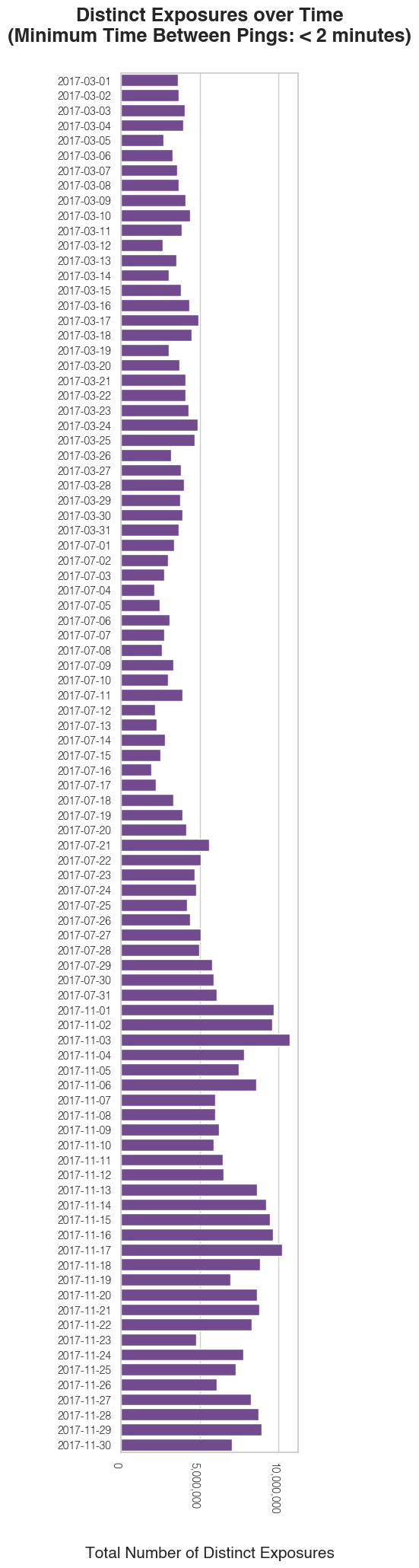}
   \vspace{-0.75em}
   \caption{\new{\textbf{Number of distinct exposures (time threshold: 2 minutes)} over time across all individuals residing in the 382 MSAs.}}
   \label{fig:robust_total_distinct_3}

\end{figure}

\begin{figure}[htbp]
   \centering
   \vspace{-5em}
\includegraphics[width=0.4\textwidth,keepaspectratio]{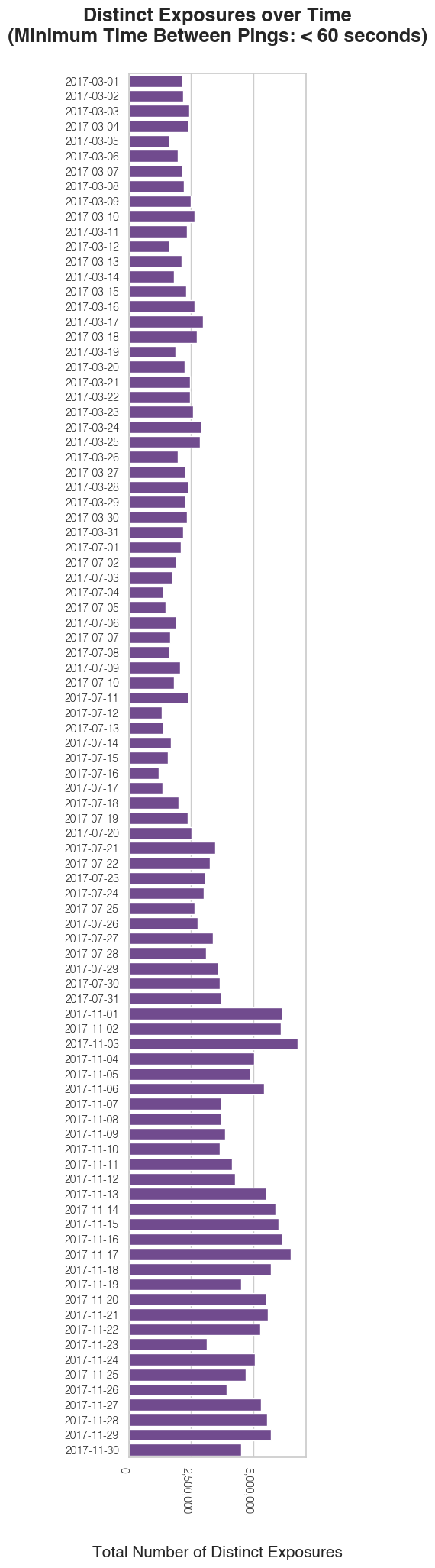}
   \vspace{-0.75em}
   \caption{\new{\textbf{Number of distinct exposures (time threshold: 60 seconds)} over time across all individuals residing in the 382 MSAs.}}
   \label{fig:robust_total_distinct_4}

\end{figure}

\begin{figure}[htbp]
   \centering
   \vspace{-5em}
\includegraphics[width=0.4\textwidth,keepaspectratio]{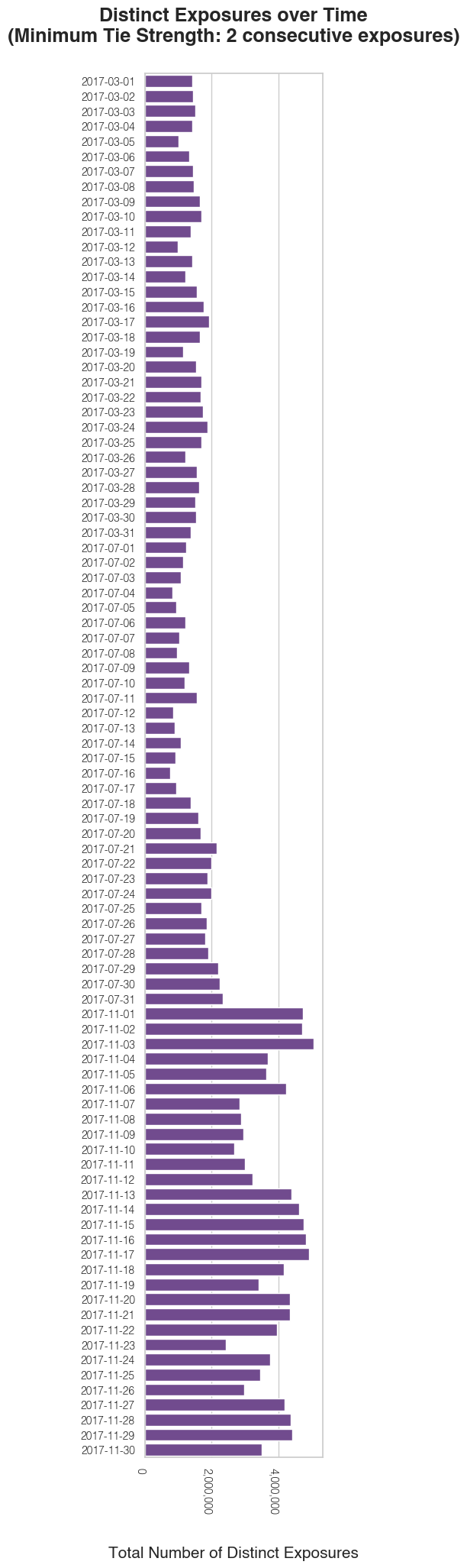}
   \vspace{-0.75em}
   \caption{\new{\textbf{Number of distinct exposures (exposure length threshold: 2 consecutive exposures over five minute intervals)} over time across all individuals residing in the 382 MSAs.}}
   \label{fig:robust_total_distinct_5}

\end{figure}

\begin{figure}[htbp]
   \centering
   \vspace{-5em}
\includegraphics[width=0.4\textwidth,keepaspectratio]{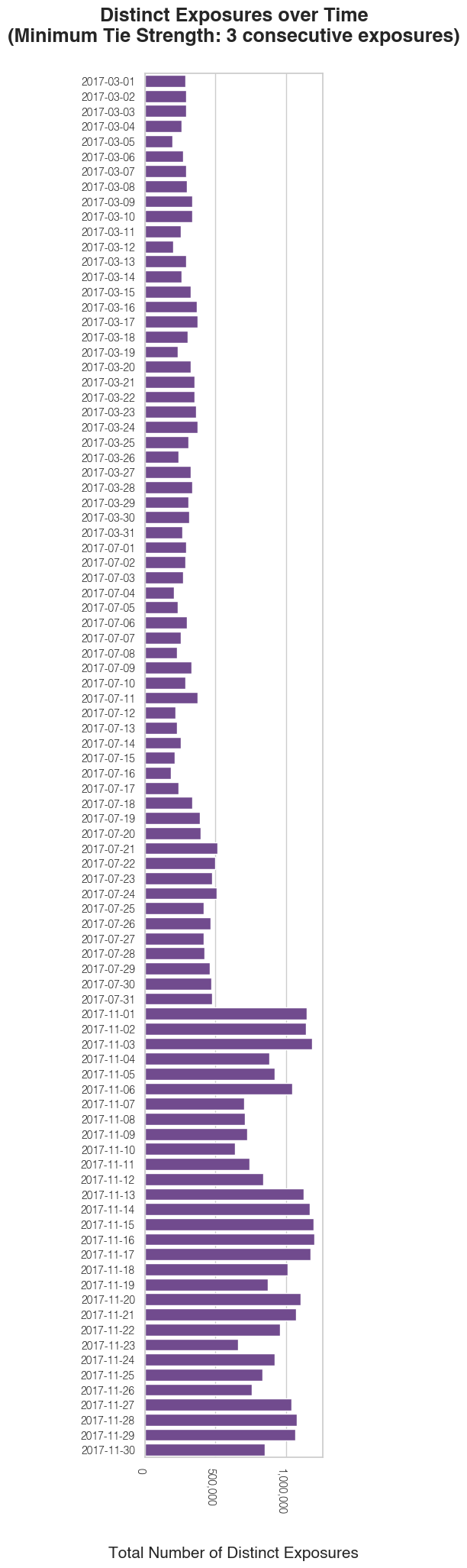}
   \vspace{-0.75em}
   \caption{\new{\textbf{Number of distinct exposures (exposure length threshold: 3 consecutive exposures over five minute intervals)} over time across all individuals residing in the 382 MSAs.}}
   \label{fig:robust_total_distinct_6}

\end{figure}

\FloatBarrier

\begin{figure}[htbp]
   \centering
   \vspace{-5em}
\includegraphics[width=0.4\textwidth,keepaspectratio]{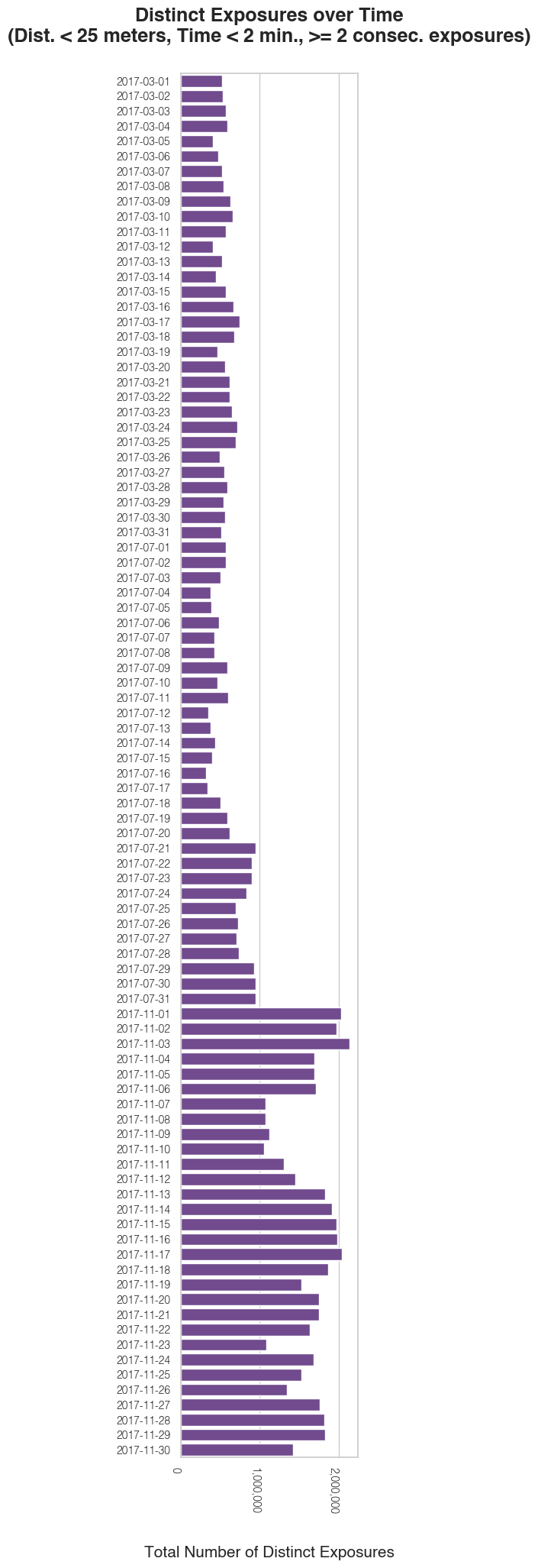}
   \vspace{-0.75em}
   \caption{\new{\textbf{Number of distinct exposures (distance threshold: 25 meters, time threshold: 2  minutes, length threshold: 2 consecutive exposures of five minute intervals)} over time across all individuals residing in the 382 MSAs.}}
   \label{fig:robust_total_distinct_7}
\end{figure}

\FloatBarrier

\begin{figure}[htbp]
   \centering
   \vspace{-5em}
\includegraphics[width=0.4\textwidth,keepaspectratio]{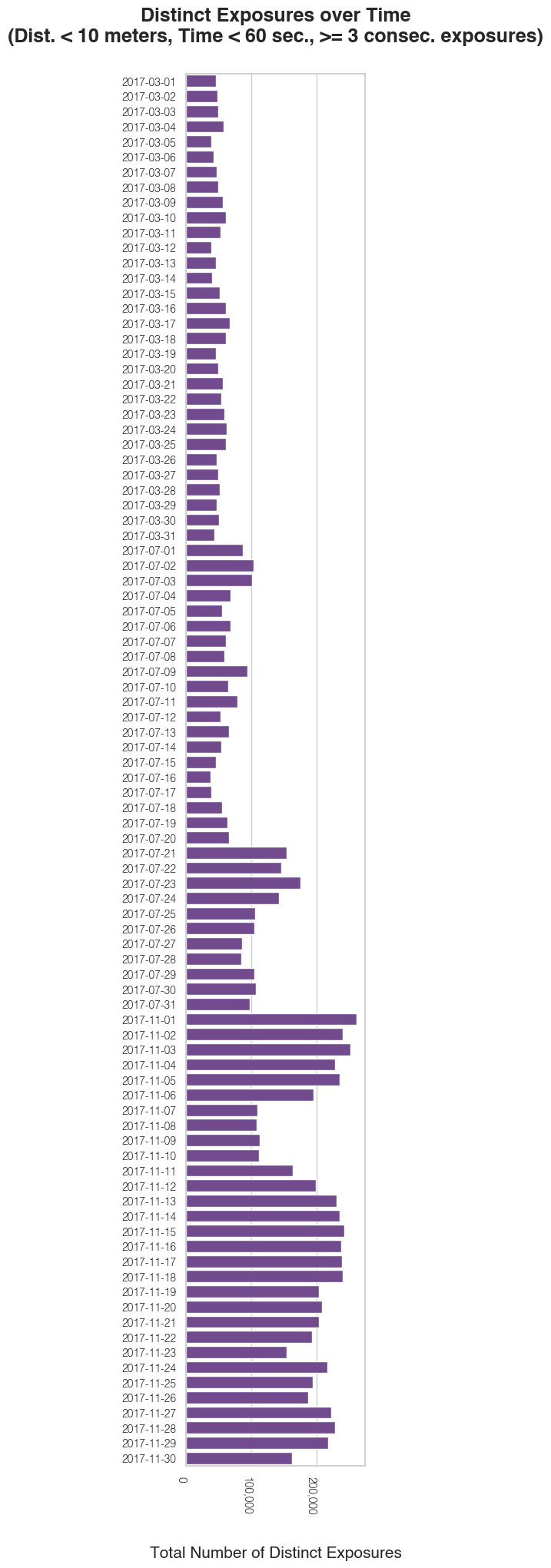}
   \vspace{-0.75em}
   \caption{\new{\textbf{Number of distinct exposures (distance threshold: 10 meters, time threshold: 60 seconds, length threshold: 3 consecutive exposures of five minute intervals)} over time across all individuals residing in the 382 MSAs.}}
   \label{fig:robust_total_distinct_8}
\end{figure}

\FloatBarrier

\begin{figure}[htbp]
\captionsetup[subfigure]{labelformat=empty}

\vspace{-20mm}
   \centering
\begin{subfigure}[t]{0.9\textwidth}
\includegraphics[width=\textwidth,keepaspectratio]{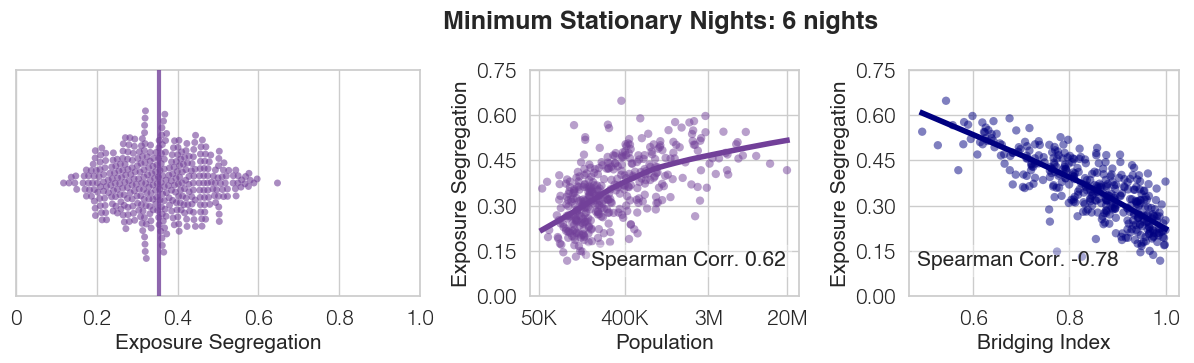}
   \caption{}
\end{subfigure}

\begin{subfigure}[t]{0.9\textwidth}
\includegraphics[width=\textwidth,keepaspectratio]{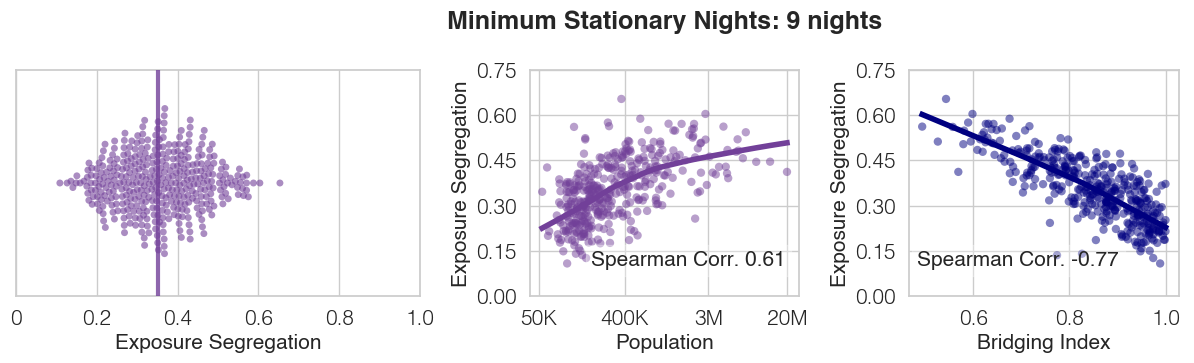}
   \caption{}
\end{subfigure}

\begin{subfigure}[t]{0.9\textwidth}
\includegraphics[width=\textwidth,keepaspectratio]{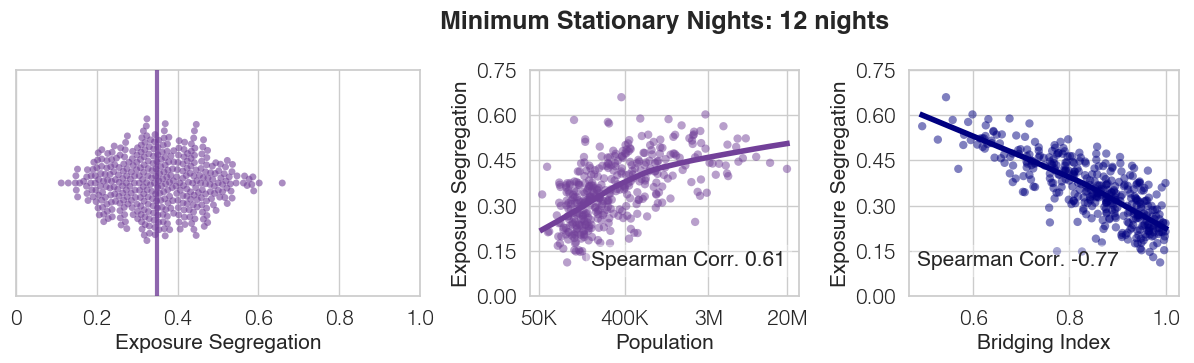}
   \caption{}
\end{subfigure}

\caption{\new{\textbf{Varying minimum stationary nights.} We find that our primary study findings that, (1) large, dense cities are more segregated and (2) exposure hub locations accessible to diverse individuals may mitigate segregation, are robust to varying thresholds of minimum stationary nights required, when identifying individual home locations. We increase the threshold from 3 (primary measure) to 6, 9, and 12 nights and find that are primary results remain unchanged.}
}
\label{fig:robust_11}
\end{figure}

\begin{table}[ht]
\centering
\begin{footnotesize}
\setlength\tabcolsep{2pt} %

\begin{tabular}{@{\extracolsep{5pt}}lcccccc}
\\[-1.8ex]\hline
\hline \\[-1.8ex]
& \multicolumn{6}{c}{\textit{Dependent variable:}} \ Exposure Segregation
\cr \cline{6-7}
\\[-1.8ex] & (1) & (2) & (3) & (4) & (5) & (6) \\
\hline \\[-1.8ex]
 Intercept & 0.355$^{***}$ & 0.355$^{***}$ & 0.355$^{***}$ & 0.356$^{***}$ & 0.355$^{***}$ & 0.355$^{***}$ \\
  & (0.004) & (0.004) & (0.004) & (0.003) & (0.003) & (0.003) \\
 log(POI Density) & 0.055$^{***}$ & & 0.035$^{***}$ & 0.039$^{***}$ & 0.020$^{***}$ & 0.022$^{***}$ \\
  & (0.004) & & (0.004) & (0.005) & (0.004) & (0.004) \\
 Gini Index (Estimated Rent) & & 0.064$^{***}$ & 0.051$^{***}$ & 0.052$^{***}$ & 0.046$^{***}$ & 0.047$^{***}$ \\
  & & (0.004) & (0.004) & (0.004) & (0.003) & (0.003) \\
 Political Alignment (\% Democrat in 2016 Election) & & & & 0.004$^{}$ & & 0.003$^{}$ \\
  & & & & (0.005) & & (0.004) \\
 Racial Demographics (\% non-Hispanic White) & & & & -0.003$^{}$ & & 0.002$^{}$ \\
  & & & & (0.004) & & (0.003) \\
 Mean SES (Estimated Rent) & & & & -0.015$^{***}$ & & -0.006$^{}$ \\
  & & & & (0.004) & & (0.004) \\
 Walkability (Walkscore) & & & & & 0.003$^{}$ & 0.003$^{}$ \\
  & & & & & (0.003) & (0.004) \\
 Commutability (\% Commute to Work) & & & & & -0.010$^{***}$ & -0.010$^{**}$ \\
  & & & & & (0.004) & (0.004) \\
 Conventional Segregation (NSI) & & & & & 0.045$^{***}$ & 0.043$^{***}$ \\
  & & & & & (0.003) & (0.003) \\
\hline \\[-1.8ex]
 Observations & 382 & 382 & 382 & 376 & 382 & 376 \\
 $R^2$ & 0.307 & 0.419 & 0.526 & 0.540 & 0.690 & 0.688 \\
 Adjusted $R^2$ & 0.305 & 0.417 & 0.524 & 0.534 & 0.685 & 0.681 \\
\hline
\hline \\[-1.8ex]
\textit{} & \multicolumn{6}{r}{$^{*}$p$<$0.1; $^{**}$p$<$0.05; $^{***}$p$<$0.01} \\
\end{tabular}

\end{footnotesize}
\caption{\textbf{POI density is significantly associated with \metric, after controlling for MSA income inequality (Gini Index), political alignment (\% Democrat in 2016 election), racial demographics (\% non-Hispanic White), mean ES, walkability (Walkscore\cite{walkscore}), commutability (\% of residents commuting to work), and residential segregation (NSI)}. This table is from an analogous regression to the regression shown in Extended Data Table \ref{tab:segregation_population}, using density of POIs (average number of POIs within 10km of a resident) instead of population size (we look at each separately due to co-linearity between population size and POI density). Here we show the coefficients (after normalizing via z-scoring to have mean 0 and variance 1) from the primary specifications estimating the effect of POI density on \metric across all MSAs. Columns (1-5) are models specified with different subsets of covariates; Column 6 shows model specification with all covariates. Differences between sample size in models is due to missing data for several covariates in a small number of MSAs (Walkscores were not available for all MSAs). (*p $<$ 0.1; **p $<$ 0.05; *** p $<$ 0.01).}
\label{tab:segregation_density_2}
\end{table}

\begin{figure}[htbp]
\vspace*{-20mm}
   \centering
\begin{subfigure}[t]{0.75\textwidth}
\includegraphics[width=\textwidth,keepaspectratio]{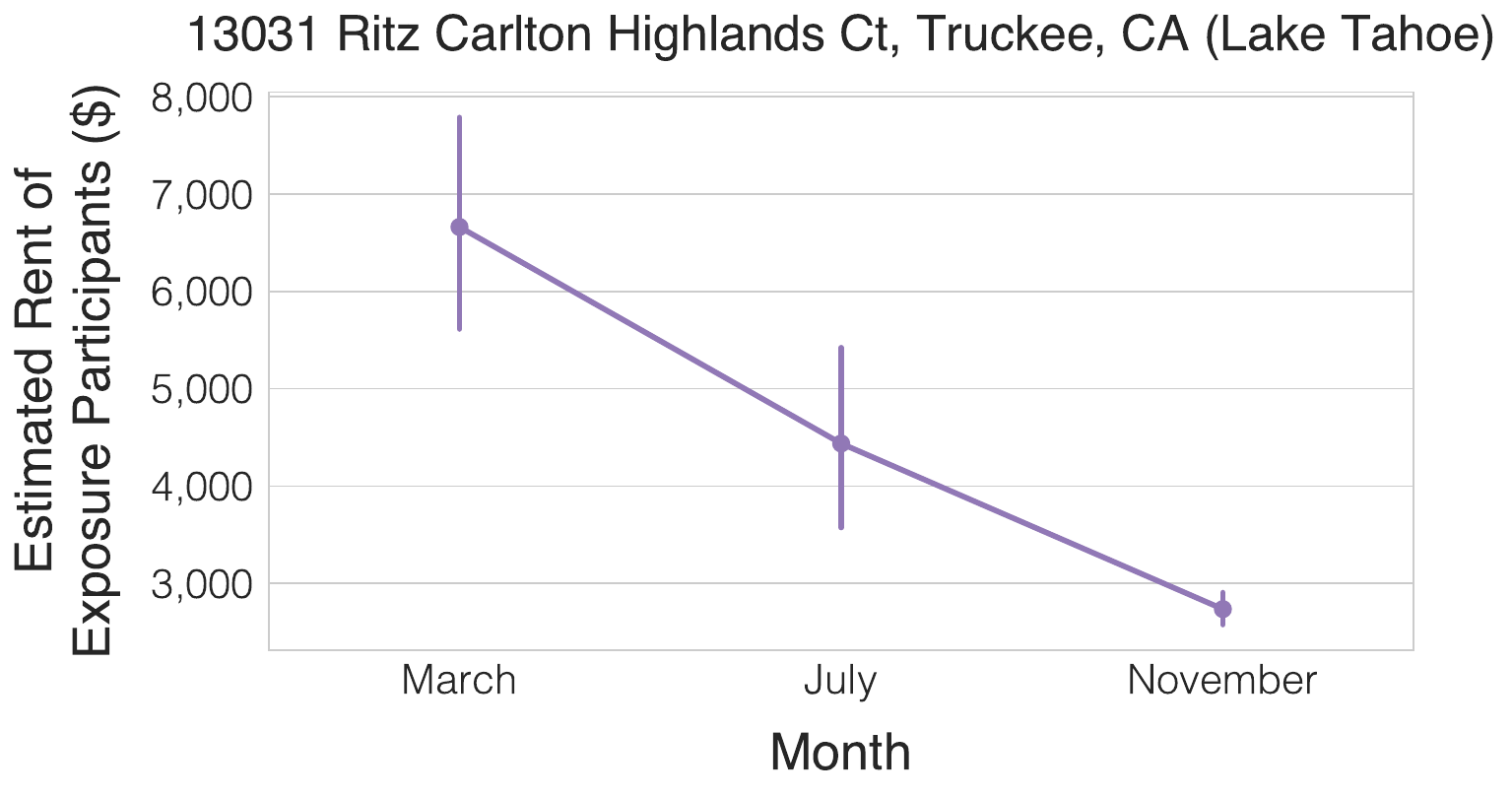}
\end{subfigure}
\begin{subfigure}[t]{0.75\textwidth}
\includegraphics[width=\textwidth,keepaspectratio]{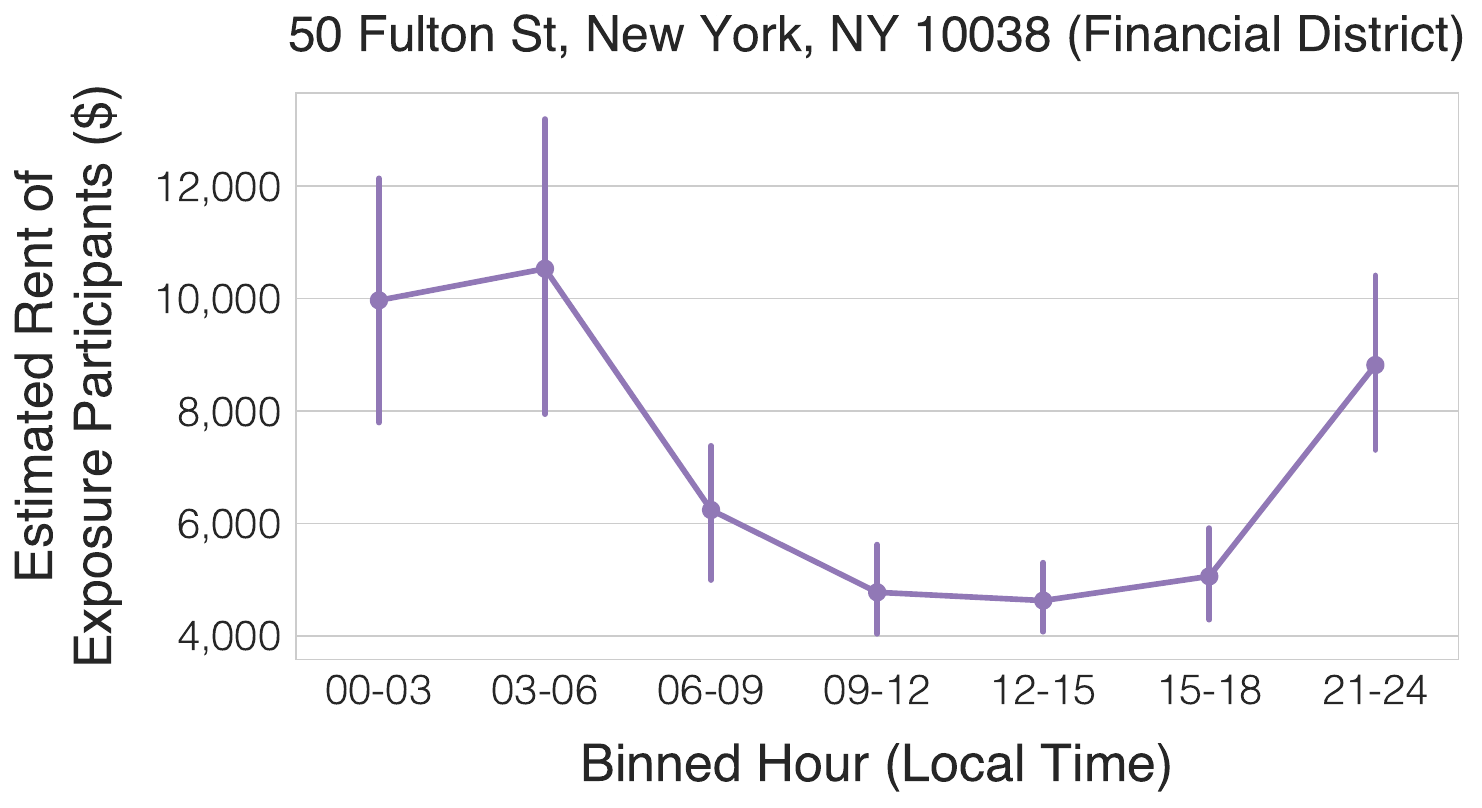}
\end{subfigure}
   \caption{\new{\textbf{Temporal heterogeneity in exposure participants at the same location: illustrative examples}. SES of exposure participants is plotted as a function of calendar month (top) and hour of day (bottom). Top: a high-end hotel in the Lake Tahoe region of California. SES is highest during March due to high hotel demand (March is prime ski season in North Lake Tahoe). SES is lowest in November (weather is cold, but ski resorts are either closed or have little snow). Bottom: A multi-story building in the Manhattan Financial District. SES is lowest during daytime due to out-of-neighborhood visitors (there is a fast food restaurant, Papaya Dog, on the ground floor). SES is highest during nighttime hours due to exposures between residents (the cost of living in the Financial District is extremely high).}}
   \label{fig:timestory1}
\end{figure}

\begin{figure}[htbp]
\vspace*{-20mm}
   \centering
\begin{subfigure}[t]{0.75\textwidth}
\includegraphics[width=\textwidth,keepaspectratio]{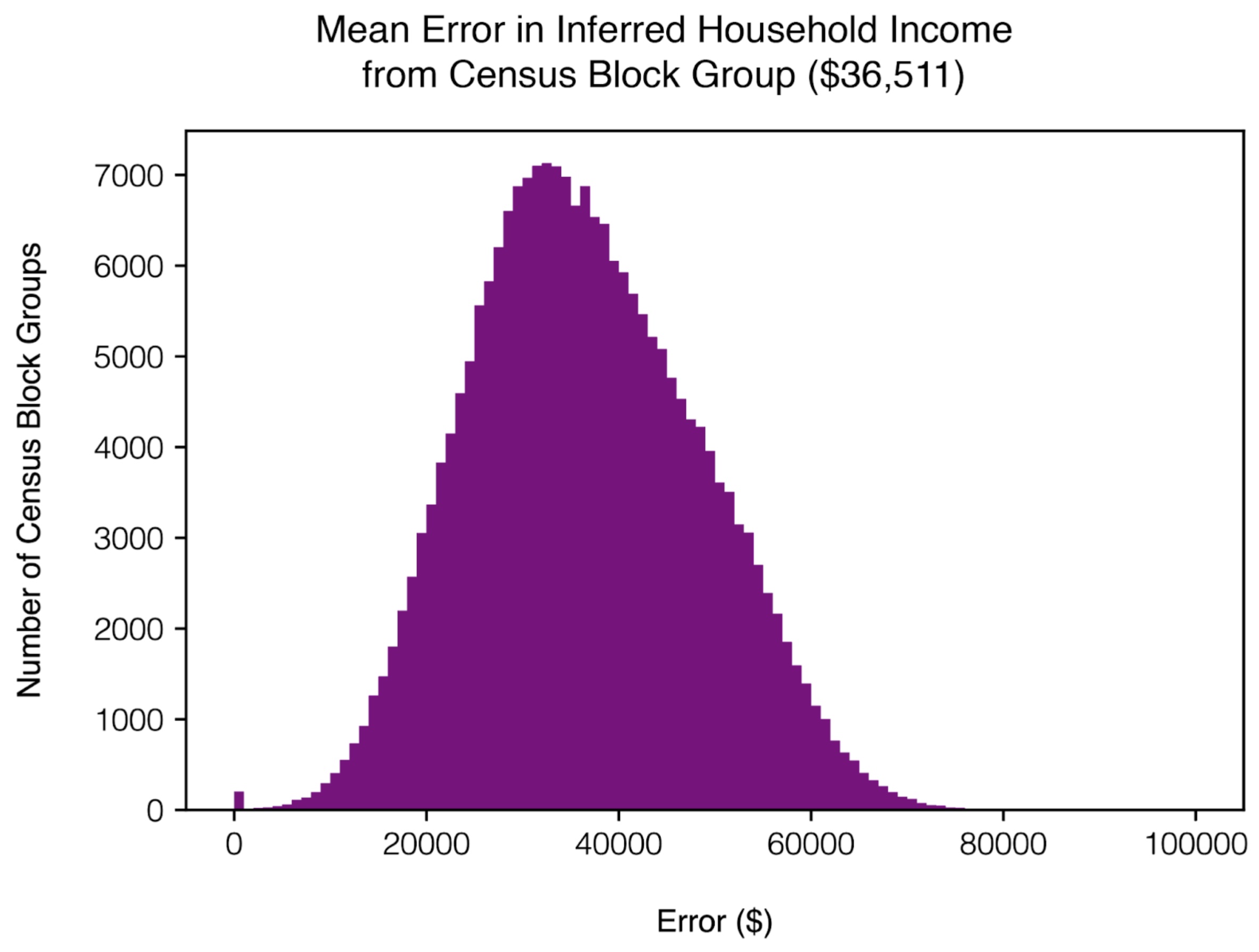}
\end{subfigure}
   \caption{\new{\textbf{Error from using census block group (CBG) income as a proxy for SES}. A histogram of the estimated error from a inferring each individual's SES (operationalized as household income) from the mean value of their census block group (CBG). To compute the error, we use the counts of household income by block group provided by the US census. Consistent with prior work\cite{soobader2001using,geronimus1998use}, we find that there is significant heterogeneity within each CBG, resulting in a mean error of \$36,511.}}
   \label{fig:timestory2}
\end{figure}

\begin{table}[!htbp] \centering
\begin{tabular}{@{\extracolsep{5pt}}lccc}
\\[-1.8ex]\hline
\hline \\[-1.8ex]
& \multicolumn{3}{c}{\textit{Dependent variable: Log(FB Social Connectedness)}} \
\cr \cline{3-4}
\\[-1.8ex] & (1) & (2) & (3) \\
\hline \\[-1.8ex]
 const & 11.123$^{***}$ & 11.123$^{***}$ & 11.123$^{***}$ \\
  & (0.003) & (0.002) & (0.002) \\
 Log(Distance km) & -1.163$^{***}$ & & -0.253$^{***}$ \\
  & (0.003) & & (0.003) \\
 Log(Exposure Network Social Connectness) & & 1.383$^{***}$ & 1.190$^{***}$ \\
  & & (0.002) & (0.003) \\
\hline \\[-1.8ex]
 Observations & 118,559 & 118,559 & 118,559 \\
 $R^2$ & 0.539 & 0.763 & 0.773 \\
 Adjusted $R^2$ & 0.539 & 0.763 & 0.773 \\
\hline
\hline \\[-1.8ex]
\textit{} & \multicolumn{3}{r}{$^{*}$p$<$0.1; $^{**}$p$<$0.05; $^{***}$p$<$0.01} \\
\end{tabular}
   \caption{\new{\textbf{Our exposure network strongly predicts friendship formation (between counties)}. Here we show the coefficients (after normalizing via z-scoring to have mean 0 and variance 1) and $R^2$ from predicting FB network friendship strengths between counties. Column (1) uses only county distance, Column (2) uses only exposure network social connectedness, and Column (3) uses the combination of distance and the exposure network. We find that our exposure network alone explains 76.3\% of the variance in friendship formation between counties and is a stronger predictor of friendship formation than distance ($p<10^{-4}$, Steiger’s Z-test).}}

      \label{tab:friendship_distance1}

\end{table}

\begin{table}[!htbp] \centering
\begin{tabular}{@{\extracolsep{5pt}}lccc}
\\[-1.8ex]\hline
\hline \\[-1.8ex]
& \multicolumn{3}{c}{\textit{Dependent variable: Log(FB Social Connectedness)}} \
\cr \cline{3-4}
\\[-1.8ex] & (1) & (2) & (3) \\
\hline \\[-1.8ex]
 const & 12.372$^{***}$ & 12.372$^{***}$ & 12.372$^{***}$ \\
  & (0.001) & (0.001) & (0.001) \\
 Log(Distance km) & -0.965$^{***}$ & & -0.175$^{***}$ \\
  & (0.001) & & (0.002) \\
 Log(Exposure Network Social Connectness) & & 1.183$^{***}$ & 1.051$^{***}$ \\
  & & (0.001) & (0.002) \\
\hline \\[-1.8ex]
 Observations & 1,038,424 & 1,038,424 & 1,038,424 \\
 $R^2$ & 0.335 & 0.503 & 0.508 \\
 Adjusted $R^2$ & 0.335 & 0.503 & 0.508 \\
\hline
\hline \\[-1.8ex]
\textit{} & \multicolumn{3}{r}{$^{*}$p$<$0.1; $^{**}$p$<$0.05; $^{***}$p$<$0.01} \\
\end{tabular}
   \caption{\new{\textbf{Our exposure network strongly predicts friendship formation (between zip codes)}. Here we show the coefficients (after normalizing via z-scoring to have mean 0 and variance 1) and $R^2$ from predicting FB network friendship strengths between counties. Column (1) uses only zip code distance, Column (2) uses only exposure network social connectedness, and Column (3) uses the combination of distance and the exposure network. We find that our exposure network alone explains 50.3\% of the variance in friendship formation between zip codes and is a stronger predictor of friendship formation than distance ($p<10^{-4}$, Steiger’s Z-test).}}

      \label{tab:friendship_distance2}

\end{table}
\FloatBarrier
\clearpage

\end{document}